\documentclass[a4paper,11pt]{book}

\usepackage[british]{babel}
\usepackage[T1]{fontenc}
\usepackage[utf8]{inputenc}

\usepackage[usenames,dvipsnames,pdftex]{color}
\usepackage[a4paper,colorlinks,pagebackref,linkcolor=blue,citecolor=BrickRed,urlcolor=OliveGreen]{hyperref}
\usepackage{a4}
\usepackage{a4wide}
\usepackage{import}
\usepackage{amsmath,amssymb,amsfonts,stmaryrd,dsfont,upgreek,mathrsfs}  
\usepackage{makeidx}
\usepackage{lmodern}\normalfont 
\DeclareFontShape{T1}{lmr}{bx}{sc} { <-> ssub * cmr/bx/sc }{}

\usepackage{minitoc}
\usepackage[pdftex]{graphicx}
\usepackage[all]{xypic}

\usepackage{chngcntr}
\counterwithout{figure}{chapter}

\usepackage{Common/Mybook}
\usepackage{Common/theorems}
\usepackage{Common/prooftree}
\usepackage{Common/appendix}
\capturecounter{theorem}
\capturecounter{definition}
\capturecounter{figure}

\long\def\ignore#1{\relax}

\newcommand\struto[1][15pt]{{\raise #1 \hbox{\strut}}}%
\newcommand\strutb[1][15pt]{{\raise-#1 \hbox{\strut}}}%

\newcommand\upline{\hline\struto[5pt]}
\newcommand\midline[1][5pt]{\\[#1]\hline\struto[5pt]}

\newcommand\downline[1][5pt]{\\[#1]\hline}

\makeatletter
\def\@boxfigurewith[#1]{\figure[#1]\vbox\bgroup\hrule height.1em}
\def\@boxfigurewithout{\figure\vbox\bgroup\hrule height.1em}
\newenvironment{bfigure}{\@ifnextchar[\@boxfigurewith\@boxfigurewithout}{\vskip.2em\hrule height.1em\egroup\endfigure}
\makeatother

\newtheorem{warning}{Notice}

 \newcommand\toaux[1]{\immediate\write\@auxout{#1}}

\newcommand\flush{\vrule width0pt height0pt\hfill}

\newcommand\olditem{}
\newcommand\olditemize{}
\newcommand\oldenditemize{}
\newcommand\oldenumerate{}
\newcommand\oldendenumerate{}
\let\olditem\item
\let\olditemize\itemize
\let\oldenditemize\enditemize
\let\oldenumerate\enumerate
\let\oldendenumerate\endenumerate

\newbox\itemlabbox
\newdimen\itemlabwd
\newdimen\itemizespacing
\newcommand\myitem{}
\makeatletter
\def\myitem{%
\@ifnextchar[\@myitemwith\@myitemwithout}
\long\def\@myitemwith[#1]{\hrule height0pt%
  \olditem[]%
  \removelastskip\vskip\itemizespacing%
  \leavevmode%
  \hskip-\leftmargin
  \setbox\itemlabbox=\hbox\bgroup\hfil#1\hfil\egroup%
  \ifdim\wd\itemlabbox>\itemlabwd%
  \box\itemlabbox%
  \else\hbox to\itemlabwd{#1\hfill}\fi%
  \rule[0pt]{0pt}{0pt}}
\long\def\@myitemwithout{\removelastskip\hrule height0pt\olditem\unskip\vskip\itemizespacing\rule[0pt]{0pt}{0pt}}
\makeatother

\renewenvironment{itemize}[1][]{%
  \removelastskip%
  \setbox\itemlabbox=\hbox\bgroup\hfil#1\hfil\egroup%
  \setlength\itemlabwd{\wd\itemlabbox}%
  \def\item{\myitem}%
  \widowpenalty=3000%
  \nobreak%
  \advance\leftmargini-10pt%
  \advance\leftmarginii-5pt%
  \olditemize%
  \unskip%
  \bgroup%
  \setlength\parskip{\itemizespacing}%
  \setlength\topsep{0pt}%
  \setlength\partopsep{0pt}%
  \setlength\parsep{0pt}%
  \setlength\itemsep{0pt}%
  \setlength\itemindent\parindent%
  \setlength\listparindent\parindent%
}
{\removelastskip\hrule height0pt\nobreak\egroup\oldenditemize\unskip\vskip\itemizespacing%
  \def\item{\olditem}%
}

\renewenvironment{enumerate}[1][0]{%
  \removelastskip%
  \setbox\itemlabbox=\hbox\bgroup\hfil#1\hfil\egroup%
  \setlength\itemlabwd{\wd\itemlabbox}%
  \def\item{\myitem}%
  \widowpenalty=3000%
  \nobreak%
  \advance\leftmargini-10pt%
  \advance\leftmarginii-5pt%
  \oldenumerate%
  \unskip%
  \bgroup%
  \setlength\parskip{\itemizespacing}%
  \setlength\topsep{0pt}%
  \setlength\partopsep{0pt}%
  \setlength\parsep{0pt}%
  \setlength\itemsep{0pt}%
  \setlength\itemindent\parindent%
  \setlength\listparindent\parindent%
}
{\removelastskip\hrule height0pt\nobreak\egroup\oldendenumerate\unskip\vskip\itemizespacing%
  \def\item{\olditem}%
}

\newcommand\oldparforMain{}
\let\oldparforMain\par
\newcommand\topequationskip{.2\baselineskip}
\newcommand\botequationskip{.2\baselineskip}

\makeatletter
\renewcommand\[[1][\topequationskip]{\begingroup\let\mysavedpar\par\let\par\oldparforMain\vskip#1\nopagebreak\hbox to\hsize\bgroup\hfil\(}
\renewcommand\][1][\botequationskip]{\)\hfil\egroup\hrule height 0pt\endgroup\@afterindentfalse\@afterheading\vskip#1}
\makeatother

\newbox\columnsbox
\newbox\tmpbox
\newdimen\columnsheight
\newdimen\columnwidth
\newdimen\remainingwidth
\newdimen\textwidthsave
\def\mycolumnsheight{}

\newcommand\columns[1]{%
  \def\mycolumnsheight{}%
  \setlength\remainingwidth\textwidth%
  \setbox\columnsbox=\vbox\bgroup\vskip0pt\vfil\hbox to\textwidth\bgroup#1\egroup\vfil\egroup%
  \columnsheight=\ht\columnsbox%
  \def\mycolumnsheight{to\columnsheight}%
  \hrule height 0pt\vtop{\hbox to\wd\columnsbox\bgroup#1\egroup}%
}

\makeatletter

\def\commonpart{%
  \setlength\columnwidth{\wd\tmpbox}%
  \vtop{\vskip0pt\hbox to\columnwidth{{\box\tmpbox}}}%
  \advance\remainingwidth-\columnwidth%
  \setlength\textwidth\textwidthsave%
  \hsize\textwidthsave%
}
\def\column{\unskip\setlength\textwidthsave\textwidth\@ifnextchar[\@columnwith\@columnwithout}
\long\def\@columnwith[#1]#2{%
  \def\newhsize{#1\dimexpr\textwidth\relax}%
  \hsize\newhsize%
  \ifdim\hsize<0.1pt\hsize\remainingwidth\fi%
  \setlength\textwidth\hsize%
  \setbox\tmpbox=\hbox to\hsize\bgroup\hfil\vtop\mycolumnsheight{\vskip0pt#2\vskip0pt}\hfil\egroup%
  \commonpart%
}
\long\def\@columnwithout#1{%
  \hsize\remainingwidth%
  \setlength\textwidth\hsize%
  \setbox\tmpbox=\hbox\bgroup\vtop\mycolumnsheight{\vskip0pt#1\vskip0pt}\egroup%
  \commonpart%
}
\makeatother

\newcommand\xtextwidth[1]{#1\dimexpr\hsize\relax}

\newenvironment{centre}{\begin{center}\unskip}{\end{center}\unskip}
\newcommand\ctr[1]{\begin{centre}#1\end{centre}}

\newcommand\abs[1]{{\left|#1\right|}}        
\newcommand\M[1]{{\mathcal{#1}}}                     

\newcommand\unit{\textsf{unit}}

\newcommand\N{\mathbb{N}}

\newcommand\ith{${}^{\text{th}}$}
\renewcommand{\iff}{if and only if}

\newcommand{\ie}{i.e.~}
\newcommand{\eg}{e.g.~}
\newcommand{\cf}{cf.~}

\newcommand{\wrt}{w.r.t.~}
\newcommand{\resp}{resp.~}
\newcommand{\aka}{a.k.a.~}

\newcommand\dom[1]{\textsf{Dom}{(#1)}}

\newcommand{\eqdef}{:=\ }
\newcommand{\recdef}{::=\ }
\newcommand{\sqin}{\textsf{\footnotesize{E}}}

\newcommand\powerset[1]{\mathds P(#1)}

\renewcommand\l{\lambda}

\newcommand{\Gam}{\Gamma}

\newcommand{\Del}{\Delta}

\newcommand\gGamma{\mathfrak{G}}

\newcommand\gA{\mathfrak{A}}
\newcommand\gB{\mathfrak{B}}

\newcommand\NJ{\textsf{NJ}}

\newcommand\LK{\textsf{LK}}

\newcommand\Giii{\textsf{G3ii}}

\newcommand\LJQ{\textsf{LJQ}}
\newcommand\LJF{\textsf{LJF}}
\newcommand\LKT{\textsf{LKT}}
\newcommand\LKQ{\textsf{LKQ}}
\newcommand\LKF{\textsf{LKF}}
\newcommand\LAF[1][]{$\textsf{LAF}_{#1}$}
\newcommand\LAFcf[1][]{$\textsf{LAF}^{\textsf{cf}}_{#1}$}

\newcommand\Lmm{$\overline{\lambda}\mu\widetilde{\mu}$}

\newcommand\mathFomega{F_\omega}
\newcommand\Fomega{\ifmmode\mathFomega\else$\mathFomega$\fi}
\newcommand\mathFomegaC{F_\omega^{\mathcal C}}
\newcommand\FomegaC{\ifmmode\mathFomegaC\else$\mathFomegaC$\fi}
\newcommand\mathDNE{\mathrm{DNE}}
\newcommand\DNE{\ifmmode\mathDNE\else$\mathDNE$\fi}

\newcommand\Coq{{\textsf{Coq}}}

\newcommand\Psyche{\textsc{Psyche}}

\newcommand\SysL{{\sf L}}
\newcommand\SysF{{\sf F}}
\newcommand\simpC{{\textsf{stC}}}
\newcommand\simplambda{{\textsf{st$\lambda$}}}
\newcommand\miniFH{{\textsf{FH$\imp$}}}

\newcommand\LKzN{\mbox{$\textsf{LK}^{\sf N}$}}
\newcommand\LKzV{\mbox{$\textsf{LK}^{\sf V}$}}
\newcommand\LKzF{\mbox{$\textsf{LK}^{\sf F}$}}
\newcommand\LC{{\sf LC}}
\newcommand\NJi{\textsf{NJ}_\imp}

\newcommand\LAFTh[2][]{\LAF[#1]($#2$)}

\newcommand\LKThp[1][\mathcal T]{\textsf{LK}$^p$($#1$)}

\newcommand\size[1]{\sharp(#1)}

\newcommand\pfunspace{\rightharpoonup}

\newcommand\co[2]{{#1}\cdot{#2}}
\newcommand\Id[1][]{\textsf{Id}_{#1}}

\newcommand\homclass[2]{{\sf hom}(#1,#2)}

\newcommand\morph[3]{\hbox{$#1\colon #2\longrightarrow #3$}}

\newcommand\termobj{1}
\newcommand\termmorph[1][]{1_{#1}}
\newcommand\initmorph[1][]{0_{#1}}
\newcommand\pair[2]{\langle #1,#2\rangle}
\newcommand\projmorph[2]{\pi_{#1/#2}}

\newcommand\evalmorph{{\sf eval}}

\newcommand\lammorph[1]{\Lambda#1}

\newcommand\Or{\raisebox{5pt}{\rotatebox{180}{\&}}}

\newcommand\SN[1]{\textsf{SN}^{#1}}

\newcommand{\multiset}[1]{\{\!\!\{#1\}\!\!\}}

\newcommand\emptylist{{[\,]}}
\newcommand\el{{[\,]}}
\newcommand\cons[2]{#1\unskip\colon\hskip-.4em\colon\unskip#2}

\newcommand\BNF{{\textsf{BNF}}}

\newcommand{\sep}{\mbox{\;\rule[-.35\ht\strutbox]{.7pt}{1.3\ht\strutbox}\;}}    

\newcommand\FV[2][{}]{\textsf{FV}_{#1}(#2)}

\newcommand\rulenamed[1]{(#1)}
\newcommand{\subst}[3]{ \left\{{}^{#3}\hspace{-.2em}\diagup\hspace{-.2em}_{#2} \right\} #1 }

\newcommand{\Rew}[2][]{\stackrel{#1}{\longrightarrow}_{#2}\;}
\newcommand{\Rewn}[2][*]{{\longrightarrow}^{#1}_{#2}\;}
\newcommand{\Rewplus}[2][+]{{\longrightarrow}^{#1}_{#2}\;}
\newcommand{\Rewsn}[2][*]{{\longleftrightarrow}^{#1}_{#2}\;}

\newcommand\lami[3]{\lambda {#1}^{#2}.{#3}}
\newcommand\st[2]{\cons{#1}{#2}}

\newcommand\name[2]{[#1]#2}
\newcommand\nott[1]{#1^\star}
\newcommand\TV[1]{ftv(#1)}

\newcommand\isnull[1]{\textsf{isnull}(#1)}

\newcommand\shft[1]{#1}

\newcommand\Seqmm[3][]{#2\seq[{#1}] #3}

\newcommand\stCTostL[1]{\overline{#1}}

\newcommand\encv[1]{\overline{#1}^{\textsc v}}

\newcommand\semcol{\mbox{$\ ;\ $}}

\newcommand\DerOSPos[3][]{\Seq[#1]{}{#2\Downarrow #3}}
\newcommand\DerOSNeg[3][]{\Seq[#1]{}{#2\Uparrow #3}}

\newcommand\DerOSPost[4][]{\Seq[#1]{}{[#4\col #3]\ #2}}
\newcommand\DerOSNegt[3][]{#3\col(\Seq[#1]{}{ #2})}

 \newcommand\Seqf[4][]{#3\seqf[{#1}]{#2} #4}
 \newcommand\Derif[5][]{\Seqf[#1]{#2}{#3}{{#4}\col{#5}}}

\newcommand{\seq}[1][]{\seqf[#1]{}}
\newcommand\Seq[3][]{#2\seq[{#1}] #3}
\newcommand\Deri[4][]{\Seq[#1]{#2}{{#3}\col{#4}}}

 \newcommand\Derili[4][]{\Derif[]{#1}{#2}{#3}{#4}}

\renewcommand\gGamma{\Gamma}

\renewcommand\gA{{A}}
\renewcommand\gB{{B}}

\newcommand\abort[1]{\textsf{abort}(#1)}
\newcommand\abortrule{\flat}
\newcommand\callCC{\textsf{cc}}
\newcommand\mC{\M C}

\newcommand\muet[4]{\mu(\inj[1]{#1}.#2,\inj[2]{#3}.#4)}
\newcommand\muou[3]{\mu({#1},{#2}).#3}
\newcommand\mufa[2]{\mu(\Lambda{#1}).#2}

\newcommand\church[1]{\underline{#1}}

\newcommand\su{\textsf {\bf s}}
\newcommand\rec{\textsf {\bf rec}}

\newcommand\stp{\textsf{\bf stop}}
\newcommand\goon{\textsf{\bf go}}
\newcommand\ifte{\textsf{\bf ifz}}

\newcommand\express[1]{\overline{#1}}
\newcommand\zerE{\textsf{0}}
\newcommand\sE[1]{\textsf{s}(#1)}

\newcommand{\m}[3][{}]{\mu {#2}^{#1}\!.{#3}}

\newcommand\nnot[1]{{#1}^{\bullet}}

\newcommand\CPS{\textsf{CPS}}
\newcommand\CBV{\textsf{CBV}}
\newcommand\CBN{\textsf{CBN}}
\newcommand\CBF{\textsf{F}}
\newcommand\CBNr{\textsf{CBNr}}

\newcommand\HStrans[1]{\underline{\underline{#1}}}

\newcommand\ssembasis[3]{\left\llbracket{#3}\right\rrbracket_{#1}^{#2}}
\newcommand\sembasis[3]{\left[{#3}\right]_{#1}^{#2}}

\newcommand{\sem}[2][]{{\ssembasis {#1} {} {#2}}}

\newcommand{\SemTy}[2][]{{\ssembasis {#1} {} {#2}}}
\newcommand{\SemTyP}[2][]{{\ssembasis {#1} + {#2}}}
\newcommand{\SemTyN}[2][]{{\ssembasis {#1} - {#2}}}
\newcommand{\SemTyr}[2]{{\sembasis {#2} {} {#1}}}
\newcommand{\SemTyrP}[2][]{{\sembasis {#1} + {#2}}}
\newcommand{\SemTyrN}[2][]{{\sembasis {#1} - {#2}}}

\newcommand\SemTe[2]{\sem[#2] {#1}}

\newcommand\semV[2][]{\ssembasis{\textsf{V}}{#1}{#2}}
\newcommand\semN[2][]{\ssembasis{\textsf{N}}{#1}{#2}}

\newcommand\ValDom{\mathcal D}
\newcommand\ValDomL{\mathcal D^*}
\newcommand\ValDomE{\mathcal E}

\newcommand{\mU}{\mathcal U}
\newcommand{\mV}{\mathcal V}
\newcommand{\fp}[2]{\Phi_{#1}{#2}}
\newcommand{\satur}[2]{\textsf{satur}({#1},{#2})}

\newcommand\ap{@}

\newcommand\orth[2]{{#1\mathrel{{\perp}}#2}}
\newcommand\uniorth[1]{#1^\bot}
\newcommand\biorth[1]{#1^{\bot\bot}}
\newcommand\triorth[1]{#1^{\bot\bot\bot}}

\newcommand\arr{{\rightarrow}}

\newcommand\col{\unskip{\colon}\hskip-.2em}
\newcommand\XcolY[2]{#1\col#2}
\newcommand\varRead[2][]{#2\left[#1\right]}

\newcommand\unitt{\texttt{unit}}
\newcommand\uniti{()}

\newcommand{\seqg}[3]{\mbox{$\ {#1}_{#2}^{#3}\ $}}

\newcommand{\seqf} [2][]{\seqg{\vdash}     {#1}{#2}}

\newcommand{\satf} [2][]{\seqg{\models}    {#1}{#2}}

\newcommand{\decf} [2][]{\seqg{\Vdash}    {#1}{#2}}

\newcommand{\decs} [2][]{\seqg{\Vvdash}    {#1}{#2}}

\newcommand\DerF[4][]{{#2}\seqf[{#1}]{}{[#3]}{#4}}
\newcommand\Der[3][]{{#2}\seqf[{#1}]{}{#3}}

\newcommand\Ders[3][]{{#2}\decs[{#1}]{}{#3}}

\newcommand\DerDec[4][]{{#3}\decf[{#1}]{#2}{#4}}

\newcommand\DerTF[4][]{{#2}\satf[{#1}]{}{[#3]}{#4}}
\newcommand\DerT[3][]{{#2}\satf[{#1}]{}{#3}}

\newcommand\DerFLKF[3][]{\seqf[{#1}]{} #2\Downarrow #3}
\newcommand\DerLKF[3][]{\seqf[{#1}]{}#2\Uparrow #3}
\newcommand\DerFlLJF[4][]{[#2]\stackrel{#3}{\longrightarrow}[#4]}
\newcommand\DerFrLJF[3][]{[#2]{-}_{#3}{\rightarrow}}

\newcommand\DerLJF[4][]{[#2]#3\longrightarrow [#4]}

\newcommand\cut{\textsf{cut}}

\newcommand\daggerL{\raise3pt\hbox{\rotatebox{-40}{$\dagger$}}}
\newcommand\daggerR{\raise0pt\hbox{\rotatebox{40}{$\dagger$}}}

\newcommand\Chole{[\;]}

\newcommand\et{{\wedge}}
\newcommand\andP{{\wedge^+}}
\newcommand\andN{{\wedge^-}}
\newcommand\ou{{\vee}}
\newcommand\orP{{\vee^+}}
\newcommand\orN{{\vee^-}}
\newcommand\trueP{{\top^+}}
\newcommand\trueN{{\top^-}}
\newcommand\falseP{{\bot^+}}
\newcommand\falseN{{\bot^-}}
\newcommand\negP{{\neg^+}}
\newcommand\negN{{\neg^-}}
\newcommand\imp{{\Rightarrow}}

\newcommand\EX[3][]{\exists #2^{#1} #3}
\newcommand\FA[3][]{\forall #2^{#1} #3}

\newcommand{\non}[1]{{#1}^{\perp}}

\newcommand\lv[1]{{\bf #1}}
\newcommand\Sorts{\mathbb S}
\newcommand\Terms[1][]{\mathbb T_{#1}}
\newcommand\SoContexts{\mathbb C}

\newcommand\RCextend[1][]{;}

\newcommand\DecompType[1][]{\mathbb D_{#1}}
\newcommand\Dstruct{\DecompType[\textsf{st}]}

\newcommand\SubsType[1][]{\bullet\bullet\bullet}

\newcommand\var[1][]{\textsf{Lab}_{#1}}
\newcommand\vare[1][]{\var[e]}

\newcommand\domP[1]{\textsf{dom}^+(#1)}
\newcommand\domN[1]{\textsf{dom}^-(#1)}
\newcommand\domE[1]{\textsf{dom}^e(#1)}
\newcommand\domAll[1]{\textsf{dom}(#1)}

\newcommand\Contexts[1][]{\mathcal G_{#1}}
\newcommand\Cextend[1][]{;}
\newcommand\Cfun[2][]{#2^{#1}}
\newcommand\Cfune[1]{\Cfun[e]{#1}}

\newcommand\TContexts[1][]{\mathsf{Co}_{#1}}
\newcommand\Textend[2][{{\bf r}}]{;(#2,{#1})}

\newcommand\Atms[1][]{\mathbb A_{#1}}
\newcommand\atmEq[2]{#1\equiv#2}
\newcommand\Moles[1][]{\mathbb M_{#1}}
\newcommand\TDecs[1][]{\mathbb D_{#1}}

\newcommand\IAtms{\Atms[\downarrow]}
\newcommand\IMoles{\Moles[\downarrow]}
\newcommand\ITDecs{\TDecs[\downarrow]}

\newcommand\Crename[2]{\pi_{#2}^{#1}}
\newcommand\Cst[2]{\textsf{st}_{#2}^{#1}}
\newcommand\CMap[2]{(#1)\circ#2}
\newcommand\Iproj[3]{\downarrow^{#2}_{#1}#3}
\newcommand\Cstaux[3]{\textsf{st}(#2,#1,#3)}

\newcommand{\Lit}{\mathbb L}
\newcommand{\Forms}{\mathbb F}
\newcommand{\Just}{\mathbb J}
\newcommand{\Data}{\mathsf{Pat}}
\newcommand{\Datast}[1]{\abs {#1}}
\newcommand{\PTerms}{\mathsf{Terms}}
\newcommand{\Decomp}{\mathsf{Terms}^{\mathsf d}}

\newcommand\Drefute[1]{{\sim}#1}
\newcommand\Dand{,}
\newcommand\Dunit{\bullet}
\newcommand\Dex[3][]{{#1}.{#3}}

\newcommand\Tunit{\bullet}
\newcommand\Tand{,}
\newcommand\Tex[3][]{{#2}^{#1}.{#3}}
\newcommand\THO[2][]{{#2}}

\newcommand\cutc[2]{\left\langle#1\mid#2\right\rangle}
\newcommand\just[1]{\{#1\}}

\newcommand{\Acc}[2][]{\textsf{Im}^{#1}(#2)}

\newcommand\Ppos{\_^+}
\newcommand\Pneg{\_^-}
\newcommand\Ptrue{\bullet}
\newcommand\inj[2]{\textsf{inj}_{#1}{(#2)}}
\newcommand\paire[2]{({#1},{#2})}
\newcommand\pexists[1]{\exists{#1}}

\newcommand\project[2]{\pi_{#1}{(#2)}}
\newcommand\caseanal[2]{[{#1},{#2}]}
\newcommand\switchl[1]{{\curvearrowleft}{(#1)}}
\newcommand\switchr[1]{{\curvearrowright}{(#1)}}

\newcommand\flatten[1]{\overline{#1}}

\newcommand\lefths{\textsf{l}}
\newcommand\righths{\textsf{r}}

\newcommand\STerms[1][]{\mathscr T_{#1}}
\newcommand\SSoContexts{\mathscr C}
\newcommand\SPrim{\mathscr L}
\newcommand\SPos{\mathscr P}
\newcommand\SNeg{\mathscr N}
\newcommand\SNegV{\mathscr N^{\textsf V}}

\newcommand\spat[1]{\tilde{#1}}
\newcommand\SContexts{\tilde{\TContexts}}
\newcommand\SCextend[1][]{;}

\newcommand\SDecs{\tilde{\TDecs}}

\newcommand\MolesIneq{\leqslant}

\newcommand\halfsubst[2][]{\langle\hskip-.5em\langle\,#2\,\rangle\hskip-.5em\rangle_{#1}}
\newcommand\closure[2][]{\langle\hskip-.5em\langle\,#2\,\mid\,#1\,\rangle\hskip-.5em\rangle}
\newcommand\machine[2][]{\langle\hskip-.5em\langle\ #2\ \mid\ #1\ \rangle\hskip-.5em\rangle}
\newcommand\headred[1][]{{\sf head}_{#1}}
\newcommand\Closures{\mathds C}
\newcommand\Values[1][]{\mathds V_{#1}}
\newcommand\ValuesE{\mathds T}
\newcommand\ValuesSC{\mathds S}
\newcommand\FL[1]{{\sf FL}(#1)}
\newcommand\rename[2]{#1\cdot#2}
\newcommand\rembeds[1][]{\sqsubseteq_{#1}}
\newcommand\RCA{\mathds R}
\newcommand\SCA{\mathds S}
\newcommand\substitute[2]{#1\cdot#2}
\newcommand\sembeds[1][]{\sqsubseteq_{#1}}
\newcommand\CwR[2]{#1\circ#2}
\newcommand\NormTo{\Downarrow}

\newcommand\emptyenv{()}

\newcommand\backstrict[1]{\llbracket #1\rrbracket}
\newcommand\forget[1]{|#1|}

\newcommand\unsat{\textsf{UNSAT}}

\newcommand\DPLL{\textsf{DPLL}}

\newcommand\DPLLTh{\textsf{DPLL}($\mathcal T$)}

\newcommand\mSat[1]{\textsf{Sat} (#1)}

\newcommand\atm[1]{\textsf{lit} (#1)}
\newcommand\atmCtxt[1]{{#1}_{\textsf{lit}}}
\newcommand\atmCtxtP[2]{\textsf{lit}_{#1} {(#2)} }
\newcommand\polar[2][\mathcal P]{#1;#2}
\newcommand\UP[1][\mathcal P]{\textsf{U}_{#1}}

\newcommand\nSat[2]{\textsf{Sat}_{#1}{(#2)}}

\newcommand\Init[1][]{\textsf{Init}_{#1}}
\newcommand\Release{\textsf{Release}}
\newcommand\Select{\textsf{Select}}
\newcommand\Store{\textsf{Store}}
\newcommand\Pol{\textsf{Pol}}

\renewcommand\sqin{\raise1pt\hbox{\large $\,\epsilon\,$}}

\newcommand\DerPos[4]{{#1}   \seqf{#4}   {[#2]}}
\newcommand\DerNeg[4]{{#1}   \seqf{#4}   {#2}}

\newcommand\DerPosTh[5][]{{#2}   \seqf[#1]{#5}   {[#3]}}
\newcommand\DerNegTh[5][]{{#2}   \seqf[#1]{#5}   {#3}}

\newcommand\DerNegLKpp[4][]{{#2}   \seqf[#1]{#4}   {#3}}

\newcommand\Decide{\textsf{Decide}}
\newcommand\Propagate[1][]{\textsf{Propagate}_{#1}}
\newcommand\Fail[1][]{\textsf{Fail}_{#1}}
\newcommand\Backtrack[1][]{\textsf{Backtrack}_{#1}}
\newcommand\Backjump[1][\mathcal T]{\mbox{$#1$-\textsf{Backjump}}}
\newcommand\Learn[1][\mathcal T]{\mbox{$#1$-\textsf{Learn}}}
\newcommand\Forget[1][\mathcal T]{\mbox{$#1$-\textsf{Forget}}}

\makeindex

\begin{document}

\dominitoc
\dominilof
\dominilot

\frontmatter

\pagestyle{empty}
\addtolength\topmargin{-1cm}
\addtolength\textheight{5cm}

\hbox to \textwidth{%
  \vbox{%
    \hbox{\includegraphics[width=\xtextwidth{.22}]{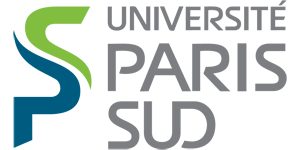}}%
    \vfill%
    \hbox{\rule[0pt]{0pt}{0pt}}
  }%
  \hfill%
  \vbox{%
    \hbox{\includegraphics[width=\xtextwidth{.1}]{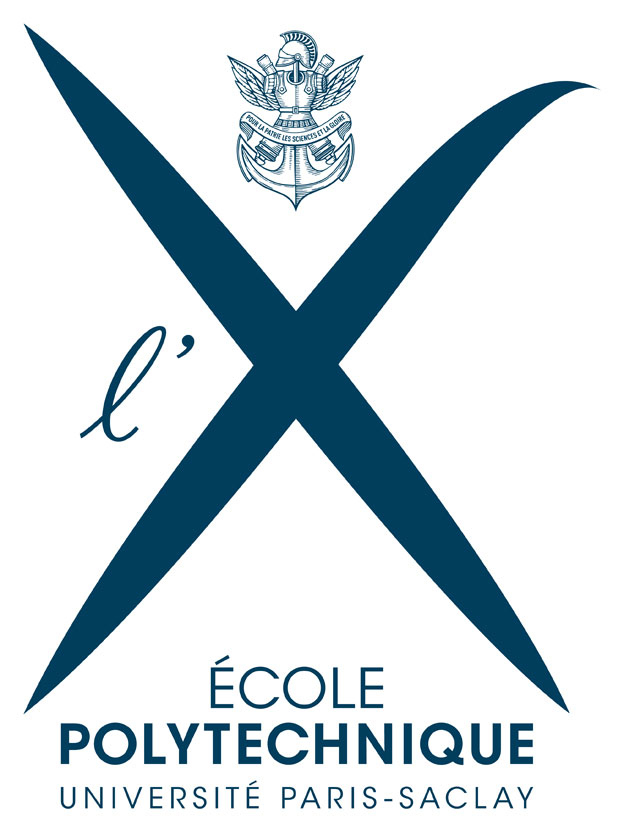}%
      \vbox{%
        \hbox{\includegraphics[width=\xtextwidth{.1}]{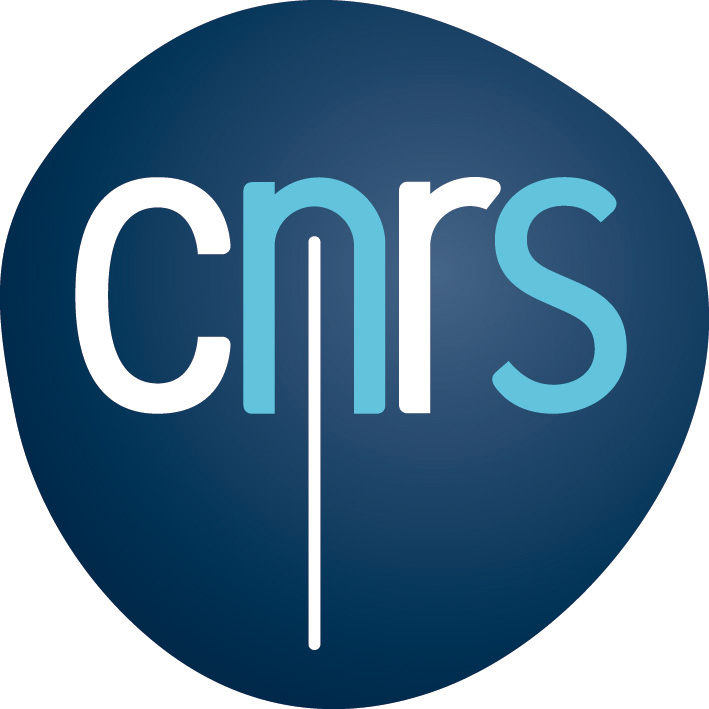}}%
        \hbox{\includegraphics[width=\xtextwidth{.13}]{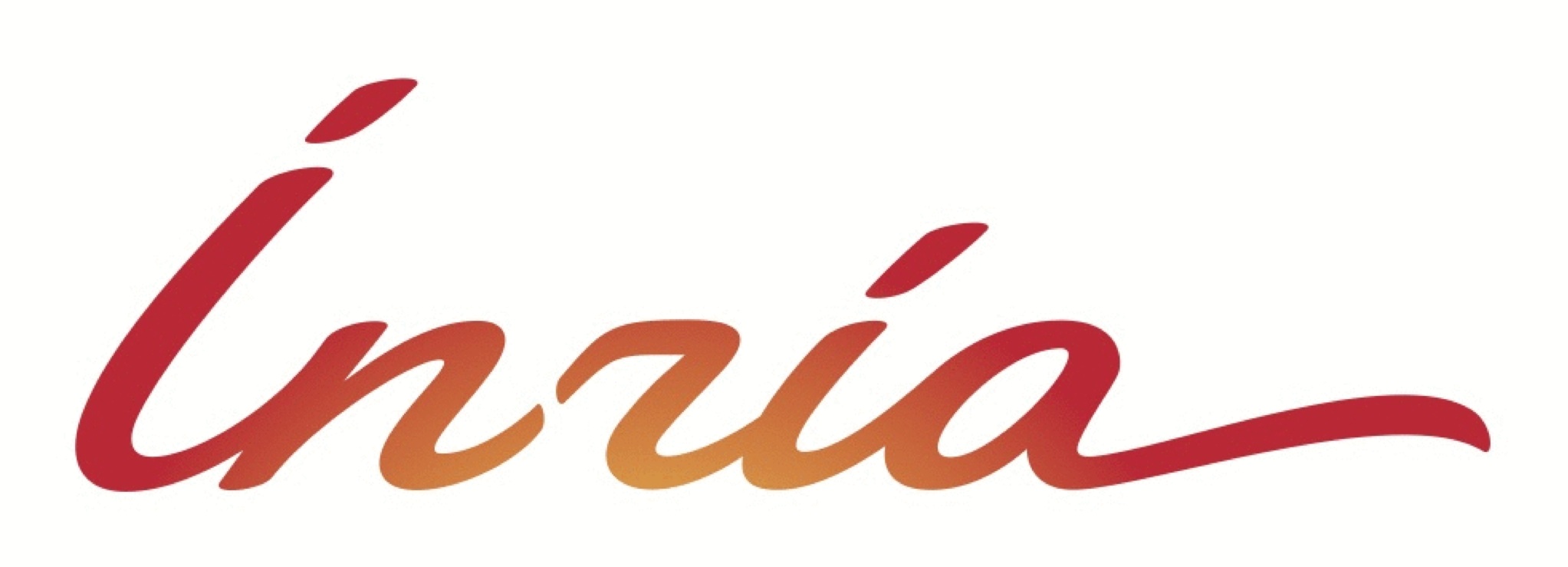}}%
      }%
    }%
  }%
}

\begin{center}
  \vspace{35pt}

  {\Huge \textbf{\textsc{Polarities \& Focussing:}}}\\[5pt] 
  {\huge \textbf{\textsc{a journey from\\[15pt] Realisability to Automated Reasoning}}}\\[50pt] 
  {\huge \textbf{\textsc{St\'ephane Graham-Lengrand}}}\\[15pt]
  {\large \textsf{}}\\[50pt]
 
  {\LARGE \textsc{Dissertation submitted towards the degree of}}\\[10pt]
  {\LARGE \textsc{Habilitation à Diriger des Recherches}}\\[10pt] 
  {\LARGE \textsc{Universit\'e Paris-Sud}}\\[50pt]

  {\large
     \textbf{\textsf{Thesis prepared at \'Ecole Polytechnique,
        with support from CNRS and INRIA,}\\[10pt]
      \textsf{and publicly defended on}
      \textsf{17${}^{\mbox{\small{th}}}$ December 2014}
      \textsf{before}}\\[15pt]
  \begin{tabular}{ll}
    {\sc Laurent Regnier}&\textsf{Chairman}\\
    {\sc Wolfgang Ahrendt} &\textsf{Referee}\\
    {\sc Hugo Herbelin} &\textsf{Referee}\\
    {\sc Frank Pfenning} &\textsf{Referee}\\
    {\sc Sylvain Conchon}&\textsf{Examiner}\\
    {\sc David Delahaye}&\textsf{Examiner}\\
    {\sc Didier Galmiche}&\textsf{Examiner}\\
    {\sc Christine Paulin-Mohring}&\textsf{Examiner}
  \end{tabular}}
\end{center}

\newpage

\addtolength\topmargin{1cm}
\addtolength\textheight{-5cm}
\strut

\newpage

\section*{Abstract}

This dissertation explores the roles of polarities and
focussing in various aspects of Computational Logic.

These concepts play a key role in the the interpretation of proofs as
programs, \aka the Curry-Howard correspondence, in the context of
classical logic. Arising from linear logic, they allow the
construction of meaningful semantics for cut-elimination in classical
logic, some of which relate to the Call-by-Name and Call-by-Value
disciplines of functional programming. The first part of this
dissertation provides an introduction to these interpretations,
highlighting the roles of polarities and focussing. For instance:
proofs of positive formulae provide structured data, while proofs of
negative formulae consume such data; focussing allows the description
of the interaction between the two kinds of proofs as pure
pattern-matching. This idea is pushed further in the second part of
this dissertation, and connected to realisability semantics, where the
structured data is interpreted algebraically, and the consumption of
such data is modelled with the use of an orthogonality relation. Most
of this part has been proved in the \Coq\ proof assistant.

Polarities and focussing were also introduced with applications to
logic programming in mind, where computation is proof-search. In the
third part of this dissertation, we push this idea further by
exploring the roles that these concepts can play in other applications
of proof-search, such as theorem proving and more particularly
automated reasoning. We use these concepts to describe the main
algorithm of SAT-solvers and SMT-solvers: \DPLL. We then describe the
implementation of a proof-search engine called \Psyche. Its
architecture, based on the concept of focussing, offers a platform
where smart techniques from automated reasoning (or a user interface)
can safely and trustworthily be implemented via the use of an API.

\newpage
\section*{\strut Acknowledgements}

It is difficult to determine when to write an habilitation thesis,
what it will include and which format it will have. In the case of a
Ph.D. thesis, such issues are often entirely resolved by the end of
the Ph.D.\ funding and the input of the Ph.D.\ adviser. Therefore,
encouragements to write an habilitation thesis are all the more
important, and for this reason I am grateful to Olivier Bournez, head
of our research laboratory (LIX), who first planted the idea in my
head, as well as to Benjamin Werner, head of Polytechnique's
C.S.\ department, for his friendly support.

In fact I surprisingly found, in the habilitation process, no other
obstacles than those I encountered on my own, as everybody that I
interacted with helped me overcome them:

In particular, I am indebted to my team leader Dale Miller and to my
Ph.D.\ adviser Roy Dyckhoff, who recommended me when I enrolled. I am
grateful to Stéphanie Druetta and Marie-Christine
Mignier\footnote{who successfully convinced me I once was a
  student at Paris-Sud in the 90s, which I had no recollection of} at
the Paris-Sud administration, as well as to Dominique Gouyou-Beauchamps,
whose work at the C.S.\ School of Doctoral Studies guarantees the high
standards of the degree towards which this dissertation is
submitted. They all worked impressively efficiently, especially given
the tight schedules that characterised my application process.

Christine Paulin-Mohring was kind enough to sponsor my application and
be one of the first people to look into my dissertation. I thank her,
as well as Sylvain Conchon, David Delahaye, Didier Galmiche and Laurent
Regnier, for accepting to sit on the defence panel. For the same
reason, but also for having reported on my dissertation, I
wish to thank Wolfgang Ahrendt, Hugo Herbelin and Frank
Pfenning: they honoured me with their time and interest.


The work described in this dissertation benefitted from many years of
interactions within my departement and elsewhere: I am very grateful
to all past and present assistants at LIX\footnote{This includes
  Martine Thirion for her help in making the defence happen.} and of
course I am also very grateful to all of my colleagues (whether or not
they are inclined towards Logic) for making Polytechnique's C.S.\ department
and my research community such a great working environment. I am
particularly thankful to Assia Mahboubi for the many years of
scientific interaction and friendship that have passed since we
studied at ENS Lyon.

I understand that an \emph{Habilitation à Diriger des Recherches} also
has to do with student supervision. Students in general have been a
very important part of my work ever since I came back to France. The
first part of this dissertation results from my teaching at MPRI. The
\Psyche\ engine was inspired by a few lines of code from a
undergraduate students' project at ESIEA, while its latest development
results from Damien Rouhling's internship at LIX. I also learnt a lot
from interacting, in very different styles, with my two
Ph.D.\ students Mahfuza Farooque and Alexis Bernadet; congratulations
again for your work and for having successfully defended your theses
before mine. In brief, I am very grateful to all of the students I
have worked with.

Besides Academia, I thank my parents and my friends who supported me in the
habilitation process, particularly the Rémi(e)s who repeatedly dared
to ask me for news updates, no matter how uninteresting to them the
topics of \emph{Polarities and Focussing} must have been. Finally,
no-one supported me more than my wife Claire, whose patience I
challenged by opening my laptop most nights and on every holiday: for
your unwavering love and cups of coffee I am forever grateful.










\newpage
\strut

\vspace{5.5cm}
\strut\hfill\emph{To all of those who are not computer scientists or logicians,}\\
\strut\hfill\emph{here is to inverted witch-hats and squash courts.}

\thispagestyle{empty}
\pagestyle{fancy}

\renewcommand{\contentsname}{Table of Contents} 
\tableofcontents

\mainmatter

\newpage

\chapterno{Introduction}

This dissertation concerns two fundamental ways in which mathematical
proofs relate to computation: \Index{proof-normalisation} and
\Index{proof-search}.
\begin{itemize}
\item The key idea, in the view of ``computation as
  proof-normalisation'', is that mathematical proofs can be composed
  in a modular way and that composed proofs can (sometimes) be
  ``simplified'' into \Index{normal forms} by a normalisation
  procedure. The most well-known tool to compose proofs (though not
  the only one) is a specific reasoning step known as \Index{cut} in
  the proof formalism of \Index{Sequent Calculus} and known as
  \Index{cut} or \Index{detour} in the proof formalism of
  \Index{Natural Deduction}~\cite{Gentzen35}. The normalisation
  process that turns proofs with cuts into cut-free proofs, known as
  \Index{cut-elimination}, strongly relates to the computational
  paradigm of Functional Programming, as shown by the Curry-Howard
  correspondence~\cite{CurryFeys:1958,How:fortnc}.
\item The view of ``computation as proof-search'', on the other hand,
  considers a mathematical formula as the input of computation, and a
  proof of that formula as its output. This strongly relates to the
  computational paradigm of Logic programming, as described for
  instance by the seminal paper on \Index{uniform
    proofs}~\cite{Miller91apal}.
\end{itemize}

Interestingly enough, investigating normal forms for proofs is useful
for both views: for the former, to understand to which proofs all
other proofs should reduce; for the latter, to only search for proofs
in normal form and thus restrict the search space in efficient
ways. For example, both kinds of computation are often taken to
produce cut-free proofs (though not always).

In fact, two key concepts in the study of normal forms have proved
useful for both views: \Index[polarity]{polarities} and \Index{focussing}. In
proof-search, they were used to design variants and generalisations of
logic programming
languages~\cite{AndreoliP89,andreoli92focusing,liang09tcs}. In
proof-normalisation, they were used to understand the semantics of,
and design meaningful variants of, cut-elimination
procedures~\cite{phdlaurent,LaurentQF05,MunchCSL09} (building on
previous work~\cite{DJS95,DJS97}).

Roughly speaking, polarities and focussing generalise the idea
that a formula of the form
\[
\forall x_1\exists y_1\forall x_2\exists y_2\ldots\forall
x_i\exists x_i\ldots
\] 
suggests a two-player game: the opponent gets to choose $x_1$, and
depending on $x_1$ the proponent gets to choose $y_1$, after which the
opponent gets to choose $x_2$, etc until some criterion (determined by
the final `\ldots') decides who has won, given all the choices that
have been made.  Of course, what the exact rules of the game are, what
a winning strategy is, etc depends on the logic considered and its
proofs (for instance in classical logic, one can
backtrack on a previous choice, using the input of the adversary).
Polarities and focussing generalise this idea to all connectives
(not only quantifiers), with some \Index[positive]{positive
  connectives} ``corresponding to'' proponent's moves and
\Index[negative]{negative connectives} ``corresponding to'' opponent's
moves.

\bigskip

The range of fields that build on, or benefit from, the two
computational aspects of mathematical proofs, is broad. This
dissertation engages in two of them which may seem distant from each
other: program semantics and theorem proving.  More specifically, it
proposes the use of polarities and focussing as the core concepts to
approach a field of topics ranging from \emph{realisability semantics}
to \emph{automated reasoning}. We can briefly illustrate how
polarities impact those two areas.

\bigskip

Realisability semantics (see
\eg\cite{KleeneSC:intint,vanOosten02Realiz}) is a way to interpret a
mathematical formula (in a broad sense, including a program type) as a
\Index{specification} that an object of a certain kind (such as a
computer program or a mathematical proof) may or may not satisfy. This
interpretation as specifications (\ie what it means for an object to
satisfy them) is defined by induction on the syntax of formulae, and
refers to the object either by its internal structure or by the way it
behaves when placed in a well-chosen environment. Realisability
semantics have been studied for various logics and systems, and a
particular approach emerged from classical logic and Girard's
\Index{linear logic}~\cite{girard-ll}, namely
\Index[orthogonality]{orthogonality}. This approach is sometimes
described as \Index{classical
  realisability}~\cite{DanosKrivine00,Krivine01} (even though it may
be used for other logics than classical logic).

This dissertation aims at the very essence of orthogonality-based
realisability, by building an abstract semantics only based on
polarities and focussing:
\begin{itemize}
\item if a formula starts with a positive connective, then the
  criterion determining whether an object satisfies the formula's
  specification refers primarily to the object's internal structure;
\item if a formula starts with a negative connective, then the
  criterion refers to the object's behaviour when placed in a
  well-chosen environment.
\end{itemize}

\bigskip

Automated reasoning (see \eg\cite{RobinsonV01}) concerns the numerous
algorithmic techniques by which the validity or the satisfiability of
mathematical formulae can be determined. Since a formula is valid
\iff\ it has a proof, an obvious approach to automated reasoning is
proof-search. The basic core of logic programming, for instance, can
be understood as proof-search on \emph{Horn clauses}, and in that
respect it can be seen as a very specific area of automated
reasoning. Now the reason why proof-search on Horn clauses also
provides a meaningful computational paradigm is because this class of
formulae makes a simple goal-directed proof-search strategy logically
complete, with well-identified backtrack points and a reasonable
covering of the proof-search space. This still holds when the class is
extended to \emph{hereditary Harrop formulae}~\cite{Miller91apal}, and
can hold on a wider class of formulae if logical connectives (and
atoms) are tagged with polarities:
\begin{itemize}
\item Negative connectives can be decomposed with \emph{invertible}
  inference rules: a goal-directed proof-search strategy performs the
  bottom-up application of those rules as basic proof-search steps,
  without loss of generality;\footnote{If the goal was provable, it
    remains provable after applying the step.} in other words, no
  backtracking is necessary on the application of such steps, even
  though other steps were possible.
\item Positive connectives are the (De Morgan's) duals of negative
  connectives, and their decomposition rules are not necessarily
  invertible, so a goal-directed proof-search procedure creates
  backtrack points when applying them bottom-up.
\end{itemize}
To what extent these ideas can be useful for a wider area of automated
reasoning (than logic programming) remains a recent field, with
numerous open questions but already with a couple of implementations
available, such as Imogen~\cite{McLaughlinP08} (using the
\emph{inverse method}) and Tac~\cite{BaeldeMS10} (using bottom-up
proof-search). This dissertation explores a particular take on this,
with its own implementation: \Psyche~\cite{Psyche}.

\bigskip

This dissertation is therefore a journey through the above topics,
trying to connect them with \eg common formalisms. It is organised in
three parts.

\medskip

Part~\ref{partI} of this dissertation is a short introduction to the
adaptation, to classical logic, of the Curry-Howard correspondence,
already mentioned above in the view of ``computation as
proof-normalisation''. Also known as the ``proofs-as-programs
paradigm'', the correspondence emerged with a strong flavour of
constructive mathematics, so its adaptation to classical logic only
emerged in the past 25 years~\cite{Griffin:popl90}. This part explores
(some of) the contributions that have been made in that period, where
we shall see the important roles of polarities and focussing. While it
starts from Parigot's $\lambda\mu$-calculus~\cite{Parigot92} and ends
with a Zeilberger-style system~\cite{ZeilbergerPOPL08,Zeilberger08},
this part mostly uses Curien and Herbelin's
System~\SysL~\cite{CurienHerbelinDuality99} as a common framework to
express and connect the concepts pertaining to the computational
interpretations of classical proofs.

Chapter~\ref{ch:CH} describes the basic set-up, viewing classical
proofs as programs. In particular, we give an overview of how classical
reasoning corresponds to the use of control
operators~\cite{Reynolds72,Strachey:2000,Felleisen:phd1987} that let
programs capture the contexts within which they are being evaluated.
We show standard ways of building meaningful operational and
denotational semantics for cut-elimination, which correspond to the
Call-by-Name and Call-by-Value evaluation strategies in
programming~\cite{Plotkin75}, and to control and co-control categories
in category theory~\cite{SelingerControlCat99}.

Chapter~\ref{ch:ortho} explores the concepts and techniques based on
orthogonality: orthogonality models form the classical version of
realisability semantics~\cite{DanosKrivine00,Krivine01}, as well as
providing methodology to prove strong normalisation
results~\cite{Parigot97,LM:APAL07}, \ie the termination of well-typed
programs. We also illustrate another use of orthogonality models for
extracting, out of a classical proof of an existential formula of
arithmetic (more precisely, a $\Sigma_1^0$-formula), a term witnessing
the existence; this technique is due to
Miquel~\cite{MiquelTLCA09,MiquelLMCS11}. Out of orthogonality
techniques we shall see the notion of polarity naturally emerge.

Chapter~\ref{ch:polarfocus} formalises this concept, inspired by a
discussion on $\eta$-conversion and observational equivalence. A new
semantics for the evaluation of classical programs is inferred from
the use of polarities (as in~\cite{MunchCSL09}), and three different
notions of normal forms are identified, out of which the concept of
focussing naturally emerges. The strongest version of focussing,
namely system \LKF~\cite{liang09tcs}, organises each proof into an
alternation of phases (similar to the alternation between proponent's
moves and opponent's moves in the intuitive view of the formula $\forall x_1\exists
y_1\forall x_2\exists y_2\ldots\forall x_i\exists x_i\ldots$). The
chapter then describes how each phase can be collapsed into one
inference step, giving rise to a presentation of \LKF\ in the style of
``big-step focussing''. It then describes the computational
interpretation of this in terms of pattern-matching, along the
lines of~\cite{ZeilbergerPOPL08,Zeilberger08}.

\medskip

Part~\ref{partII} of this dissertation takes this last idea further
and presents new material. Stripping focussed systems off the concept
of connective and off the inductive structure of mathematical
formulae, we only keep the core mechanisms of focussing to define a
highly abstract system for big-step focussing, called \LAF, whose
computational interpretation is pure pattern-matching. One of the main
goals is to formalise the strong ties between Zeilberger-style systems
and orthogonality models.

Chapter~\ref{ch:LAFwoQ} presents the syntax and the typing system of a
quantifier-free version of the \LAF\ abstract system, which is modular in
its syntax for atoms and formulae, in its logical connectives, in the
logic used, and in the implementation of variables. The chapter shows
how the abstract system can be instantiated to capture existing
focussed systems such as \LKF\ and its intuitionistic variant \LJF.

Chapter~\ref{ch:LAFwQ} presents the extension with quantifiers of that
abstract focussed system \LAF. Different approaches may lead to either
a treatment of quantifiers along the lines of the
$\omega$-rule~\cite{hilbert31,schuette50} (where a formula ``$\forall
n\in\N,A(n)$'' may be proved by providing a proof for each natural
number), or a treatment that forces to prove a universal formula in a
uniform way: for this we need to extend \LAF\ with a mechanism that
generalises eigenvariables.

Chapter~\ref{ch:real} presents the realisability models of the
abstract system \LAF, finally formalising the connection between
big-step focussing and orthogonality models: indeed we lift the
orthogonality models of Chapter~\ref{ch:ortho} to our abstract
framework, and we prove the Adequacy Lemma, that relates typing to
realisability, \ie syntax to semantics. We present instances of
orthogonality models such that the Adequacy Lemma immediately provides
the logical consistency of the \LAF\ system.

Chapter~\ref{ch:trans} explores proof transformations in the abstract
system \LAF; we start with an abstract machine to perform head-reduction, thus
revealing the actual pattern-matching mechanism of the proof-term
calculus. Using the realisability models of Chapter~\ref{ch:real}, we
show that head-normalisation terminates on typed proof-terms. We then
describe, via a notion similar to that of free variables, how to
identify the parts of a sequent that have actually been used in its
proof (which is useful if the proof is to be re-used for a sequent
that is similar). We also use this to extend the abstract machine into
a big-step operational semantics that evaluates a proof-term as a
normal form that is cut-free. Adapting the orthogonality model to this
big-step operational semantics, we prove that every typed proof-term
does evaluate as a cut-free form, and conclude the cut-elimination
result for the \LAF\ system.

\medskip

Part~\ref{partIII} of this dissertation concerns the roles that
polarities and focussing can have in automated reasoning and more
generally theorem proving (\ie proof construction may also be
interactive). Originally aiming at classical logic (and therefore
departing from the Imogen~\cite{McLaughlinP08} and
Tac~\cite{BaeldeMS10} provers), we investigated one of the most
popular automated reasoning techniques for classical propositional
logic (\aka SAT-solving): \DPLL~\cite{DavisP60,DavisLL62}, as well as
its extension known as \DPLLTh~\cite{Nieuwenhuis06} for solving
SAT-modulo-theories problems (SMT).

Chapter~\ref{ch:dpll} aims at describing and simulating \DPLLTh\ runs as
bottom-up proof-search in a focussed system for classical logic. We
therefore present a extension of system \LKF\ that allows atoms to be
assigned polarities \emph{on-the-fly} during proof-search, and that
integrates the possibility to call a procedure that decides whether a
conjunction of (ground) atoms is consistent with a given input theory
$\mathcal T$. The resulting system, \LKThp, is used to establish a
bisimulation result between proof-search and \DPLLTh\ runs. Based on
the fact that \LKF\ can be seen as an instance of \LAF, the chapter
then discusses how \LKThp\ could be seen as an instance of a
generalisation of \LAF\ that could work \emph{modulo} the theory
$\mathcal T$.

Chapter~\ref{ch:Psyche} describes a small prototype called
\Psyche~\cite{Psyche} implementing bottom-up proof-search in an
extension of \LKThp\ with quantifiers and meta-variables.  Highly
modular with respect to the decision procedure and the proof-search
strategy it can run with, \Psyche\ comes with a strategy \emph{plugin}
that implements the simulation of \DPLLTh\ described in
Chapter~\ref{ch:dpll}, and can also perform pure first-order
reasoning.

Chapter~\ref{ch:quant} concludes this dissertation, in particular by
giving an informal description of \Psyche's mechanisms for
quantifiers, and hinting at what could be achieved with them, in
particular in the combination of first-order reasoning with decision
procedures. It finally presents the \LAF\ system as the theoretical
foundations for the next version of \Psyche.

\bigskip

Note that the whole of Part~\ref{partII} is admittedly technical,
which is due to two reasons:
\begin{itemize}
\item The first reason is the systematic search for the greatest
  generality (and therefore strength) in the definitions and
  theorems. It was a goal in itself to determine exactly which
  ingredients are necessary and which are disposable for the system to
  make sense, for the models to be built, and for the theorems to be
  proved. Hypotheses are systematically weakened and structures are
  systematically parameterised to achieve this. The result of course
  is a highly parameterised framework with complex yet precise
  specifications. In order to digest this technicality with
  confidence, most of the proofs have been
  formalised~\cite{LengrandHDRCoq} in the proof assistant
  \Coq~\cite{Coq}, which was particularly useful to refine the
  definitions and theorems according to the above methodology.
\item The second reason is that the development of this abstract
  framework was not only done for the sake of it, but also to provide
  the foundations of (the next version of) our
  \Psyche\ implementation. Abstraction in the theoretical framework
  translates to genericity in the code, making the implementation more
  versatile, decomposing its architecture into smaller modules that
  could more easily be shown to be correct. Therefore, when
  introducing as a mathematical structure a tuple such as
  \[(\Sorts,\vare,\Terms,\decs{},\Atms,\Moles,\atmEq{}{},\var[+],\var[-],\RCA,\TContexts,\Data,\decf{})\]
  satisfying a long list of axioms, we really have in mind an OCaml
  module providing the corresponding types and functions and
  satisfying the corresponding specifications.  Hence the verbosity of
  our axiomatic structures in Part~\ref{partII}.
\end{itemize}

\section*{Personal note}

This section aims at relating this dissertation to the papers I have
published in the recent years.

Firstly, this dissertation lies within the very broad field of
\emph{Computational Logic}, on which Didier Galmiche and I edited a
special issue of the Journal of Logic and Computation, in honour of
Roy Dyckhoff~\cite{DyckhoffJLC}.

My interest for the topics developed in this dissertation can be
traced back to the first paper I wrote as the sole
author~\cite{lengrand03call-by-value}, relating Curien and Herbelin's
work on the computational interpretations of classical
logic~\cite{CurienHerbelinDuality99} to
Urban's~\cite{UrbThes}.\footnote{Despite its critical typos, it
  surprisingly appears to be my most cited paper.} However, such
topics stayed in the slow-cooker at the back of my mind, as my next
contributions mostly concerned intuitionistic systems, which could
more simply be related to the $\l$-calculus, the Curry-Howard
correspondence, and Type Theory: 

In~\cite{kikleng08} we explored a Call-by-Name cut-elimination
procedure for the intuitionistic sequent calculus, in~\cite{DL:JLC07}
we explored the focussed sequent calculus \LJQ, and
in~\cite{lengrand07tr} I proved some conjecture about the termination
of a Call-by-Value $\lambda$-calculus. Although these contributions
are not directly included in this dissertation, the work that I did
around that time greatly contributed to my understanding of focussing
and cut-elimination strategies.

Still in intuitionistic logic, two more recent contributions broached
the topics that this dissertation approaches under the focussing
angle, namely realisability and automated reasoning:
In~\cite{bernadetlengrand12b} we develop a simple presentation of
Hyland's effective topos~\cite{HylandJ:efft} which is based on
realisability concepts; in~\cite{lengrand11lmcs} we developed a
focussed sequent calculus that can describe proof-search in the type theory
behind the proof-assistant Coq~\cite{Coq}.

More directly included in this dissertation are the
publications~\cite{LM:APAL07}
and~\cite{bernadetleng11b,bernadetleng11blong,bernadetlengrand13}, all
of which formalise proofs of strong normalisation with orthogonality
techniques, aiming at genericity. In~\cite{LM:APAL07} we compare
Barbanera and Berardi's technique based on symmetric reducibility
candidates~\cite{BBsymm} with the basic orthogonality
technique. In~\cite{bernadetleng11b,bernadetleng11blong,bernadetlengrand13},
we formalise an abstract notion of orthogonality model and describe,
as instances of this notion, several variants of proofs for the strong
normalisation of System~$F$~\cite{Girard72}.

Chapter~\ref{ch:ortho} of this dissertation covers these contributions
with a systematic orthogonality model construction. It also uses the
same orthogonality framework to describe an interesting application of
classical realisability to witness extraction (due to
Miquel~\cite{MiquelTLCA09,MiquelLMCS11}).

Over the recent years, a greater proportion of my research was devoted
to the use of focussing for proof-search, in the context of our
ANR-funded project on \emph{Proof-Search Control in Interaction with
  domain-specific methods}~\cite{PSI}. Since the concept of focussing
emerged with motivations for logic programming, it was natural to
explore whether the concept could also impact automated reasoning. We
first explored propositional problem solving, which can be seen either
as proof-search or as satisfiability (SAT) solving: More precisely, we
described in~\cite{farooqueTR12,farooqueTR12b,farooque13} how one of
the main procedures, namely \DPLL~\cite{DavisP60,DavisLL62}, can be
seen as the gradual construction of proof-trees in a focussed sequent
calculus. We actually did this in combination with decision
procedures, so as to describe SMT-solving in terms of proof-search;
this required the extension of sequent calculus with such
procedures~\cite{farooqueTR11,farooqueTR13}. All this was put together
in Farooque's Ph.D. thesis~\cite{FarooquePhD} which I supervised, and where
another class of automated reasoning techniques, namely
\emph{tableaux} methods, are also simulated in the same focussed
sequent calculus.

In the present dissertation, the above contributions are not developed
in as many details as in~\cite{FarooquePhD}. However, they form the
theoretical basis of the \Psyche\ prototype~\cite{Psyche}, of which I
am the main developer; and the software is the topic of
Chapter~\ref{ch:Psyche}, which covers the system
description~\cite{GLPsyche13}.

The material presented in this dissertation does not only come from
publications. Some of it relates to an active teaching activity: in
particular, Part~\ref{partI} approximately covers the material that I teach at
M.Sc. level in Paris, with the most advanced parts inspired by
Munch-Maccagnoni's work relating focussing and classical
realisability~\cite{MunchCSL09}, and Zeilberger's work on big-step
focussing and pattern-matching
interpretations~\cite{ZeilbergerPOPL08,Zeilberger08}.

Part~\ref{partII} presents entirely new material, rather than
published work (or survey thereof), that builds on those two
inspirational topics: Zeilberger's framework seemed particularly
appropriate to relate focussing and classical realisability at a
particulary abstract level. The proposal is to make this the
theoretical foundation of \Psyche's next version, and in that it
connects to Part~\ref{partIII} this dissertation.

On the other hand, several publications do not (yet) relate (or only
very remotely) to this dissertation, since they are too disconnected
from its topic: In~\cite{GabbayLengrand08,gablen08}, we developed a
$\l$-calculus inspired by Nominal Logic~\cite{PittsAM:nomlfo-jv}, with
a special construct in order to represent binding in data-structures;
the goal is to allow incomplete terms within the scope of binders,
without blocking $\alpha$-conversion or
computation. In~\cite{bernadetleng11} we studied intersection type
systems~\cite{CD78:newtal} for the $\lambda$-calculus, in a
non-idempotent version similar to de
Carvalho's~\cite{Carvalho05,Carvalho09corr}, but such that the length
of the longest $\beta$-reduction sequence starting from a strongly
normalising term, can be directly read from its typing tree.

Finally, the careful reader will note that this dissertation not only
has little material in common with my
Ph.D.\ thesis~\cite{LengrandPhD}, but it is not even in its direct
continuation: I do not present here refinements or developments of its
contributions, but rather a thesis that complements my doctoral work
in the topics of my interest.

\chapterno{Notations and prerequisites}

In this dissertation, we assume the reader to be already familiar with
some areas and concepts of logic and computer science. Unless
specifically given, the notations and definitions used in this
dissertation are rather standard, and formally
follow~\cite{LengrandPhD}. The areas and concepts are:

\begin{itemize}
\item Set theory; see \eg \cite{KrivineSetTheory71}.\\ In particular
  we will use the concepts of, and notations for, subsets, power sets,
  union, intersection and difference of sets, relations, functions,
  injectivity, surjectivity, etc. Our notation for the power set of
  $A$ is $\powerset A$. Our notation for the set of total functions
  from $A$ to $B$ is denoted $A\rightarrow B$; the set of partial
  functions from $A$ to $B$ is denoted $A\pfunspace B$. We also assume
  the reader to be familiar with natural numbers, lists and trees.
\item The standard difference between object-level and
  meta-level.\\ In particular, variables of the meta-level are called
  meta-variables and (unless otherwise stated) ``rules'' and
  ``systems'' are meta-level devices (\ie a rule has no existence at
  the object level, but its instances do -and the collection of them,
  for example).
\item Trees and derivations.\\ We use inference rules and systems to
  define sets of (valid) derivations and derivability of judgements,
  as well as partial derivations; when we state that a rule is
  derivable/admissible/invertible (in a system) we actually mean that
  its instances are derivable/admissible/invertible (in the collection
  of derivations defined by the system).
\item Rewriting (first-order and higher-order); see \eg
  \cite{Terese03}.\\ In particular, the notations $\Rewn[n]{}$,
  $\Rewplus{}$, $\Rewn{}$, $\Rewsn{}$, denote the composition $n$
  times of a (binary) relation $\Rew{}$, the transitive closure, the
  transitive and reflexive closure, and the transitive, reflexive and
  symmetric closure, of the relation $\Rew{}$, respectively.

  We assume that the reader is familiar with the properties of
  confluence and Church-Rosser, weak normalisation, strong
  normalisation, and the usual techniques to prove them, in particular
  the simulation techniques.

  Following~\cite{LengrandPhD}, the notation
  \[
  \rulenamed\gamma\quad M\Rew{}N
  \] 
  introduces a rewrite rule whose contextual closure (or more
  precisely, the contextual closure of its instances) is denoted
  $M\Rew{\gamma}N$. We also use this notation when $\gamma$ is a
  system of rules.

  Our languages will often be made of terms whose syntax is defined by
  a \BNF-grammar. Some of its syntactic categories may contain
  \Index[variable]{variables}. We assume the reader is familiar with
  variable binding, $\alpha$-conversion and \Index{equivariance};
  specifying binders and their scopes automatically defines what the
  \Index[free variable]{free variables} of a term, denoted $\FV t$,
  are; \Index{capture-avoiding substitution} of $u$ for $x$ in $t$ is
  denoted
  \[
  \subst t x u
  \]
  where $x$ is a variable of some syntactic category with variables
  and $u$ is a term of that syntactic category.
\item Basic proof theory; see \eg \cite{TS}. In particular, standard
  proof formalisms such as Natural Deduction and Sequent Calculus, for
  intuitionistic and classical logic (propositional and first-order).
\end{itemize}

\part{The Curry-Howard view of classical logic - a short introduction}
\label{partI}
\chapter{Classical proofs as programs}
\label{ch:CH}

\minitoc 

The Curry-Howard correspondence~\cite{CurryFeys:1958,How:fortnc} has
been one of the most fruitful connections between proofs and
computation: As one of the embodiments of constructivism, where
mathematical proofs bear computational content, the correspondence
naturally emerged in the context of minimal and intuitionistic logic,
and gave rise to the field of Type Theory~\cite{ML82,ML84}.

Despite the non-constructive character of proofs in classical logic,
arising from the Law of Excluded Middle, or the Double Negation
Elimination etc, it is natural to investigate what part of the
Curry-Howard correspondence can still be built for that logic.

In this chapter we review the foundations of the correspondence in the
framework of classical logic, along the main lines of investigation
that were explored over the past 25 years since Griffin's seminal
work~\cite{Griffin:popl90}.

The first step in this programme is to turn a proof format for
classical logic into a typing system for a language. For such a
language to be of computational nature, an operational semantics
and/or a denotational semantics has to be designed.

Section~\ref{sec:CHreview} reviews the basic concepts of the
Curry-Howard correspondence, both in their original framework and at a
more abstract level. Section~\ref{sec:continuationscontrol} present
some concepts in programming, namely continuations and control, which
will prove useful to understand classical proofs as
programs. Section~\ref{sec:90s} presents early formalisations of the
above concepts as proof-term calculi for classical logic while
Section~\ref{sec:SysL} presents in more details one of the most
convenient ones, which relates to Gentzen's classical sequent
calculus. Section~\ref{sec:cbncbv} uses continuations to describe the
evaluation strategies known as Call-by-Name and Call-by-Value while
Section~\ref{sec:classicalCBNCBV} explains how this can be used to
build semantics for classical proofs.

\section{Curry-Howard correspondence: concepts and instances}
\label{sec:CHreview}

The correspondence relates logic to programming languages, and is
sometimes taken to involve a third aspect, namely category theory (as
it forms a popular framework to build the semantics of programming
languages). Table~\ref{tab:CH} gives a high-level view of the
correspondence,\footnote{We use the expression (Curry-Howard)
  ``correspondence'' rather than the popular (Curry-Howard)
  ``isomorphism'', as it is difficult to specify what the isomorphism
  exactly is before specifying exactly what formal systems we intend
  to relate.} which operates at several levels: mathematical formulae,
or propositions, correspond to the types of a given programming
language; proofs of such propositions correspond to programs that can
be given the corresponding type; the way proofs can be composed
corresponds to the way programs can be composed/applied; finally (and
this is where we adopt the view of computation as
proof-normalisation), cut-elimination corresponds to program
execution.

\begin{table}[!h]
  \begin{centre}
    \begin{tabular}{|l|l|l|}
      \hline
      \textbf{Logic} & \textbf{Programming language} &\textbf{Categories}\\\hline
      Propositions& Types& Objects\\\hline
      Proofs&Typed programs &Morphisms\\\hline
      Cut/Composition&Program composition &Morphism composition\\\hline
      Cut-elimination&Program execution &
      \begin{tabular}l Equality of morphisms\\
        (commuting diagrams)\end{tabular}
      \\\hline
    \end{tabular}
  \end{centre}
  \caption{High-level view of the Curry-Howard correspondence}
  \label{tab:CH}
\end{table}

The rest of this section gives a brief overview of the correspondence
in the framework of minimal and intuitionistic logic. An in-depth
presentation of the correspondence can be found in the book~\cite{sor:CH}.

\subsection{Simply-typed combinators}\label{sec:combinators}

The original instance of the correspondence was given in the study of
combinators~\cite{CurryFeys:1958}, which yields a simple language made
of three basic programs $\textbf I$, $\textbf K$, $\textbf S$ and with
program application as its only construct:
\begin{definition}[The ($\textbf I$, $\textbf K$, $\textbf S$)-combinatoric system]

  The syntax is given by the following grammar:
  \[
  M, N,\ldots\recdef \textbf I\sep \textbf K \sep \textbf S\sep M\;N
  \]
  The last construct, \Index{program application}, is associative to the left, \ie $(M\;N)\;P$ can
  be abbreviated as $M\;N\;P$.

  and its operational semantics is given by the following first-order rewrite
  system
  \[
  \begin{array}{lll}
    \textbf I\;M&\Rew{}M\\
    \textbf K\;M\;N&\Rew{}M\\
    \textbf S\;M\;N\;P&\Rew{}M\;P\;(N\;P)\\
  \end{array}
  \]
\end{definition}
Clearly, the operational semantics defines $\textbf I$ as the identity,
while $\textbf K$ provides erasure and $\textbf S$ provides duplication.
The reduction relation is confluent and defines a model of computation
that turns out to be Turing-complete. At the cost of losing that
property, the language can be given an intuitive typing system, using
\Index[simple type]{simple types}:
\begin{definition}[Simple types]
  \Index[simple type]{Simple types} are defined by the following grammar:
  \[
  A,B,\ldots \recdef a \sep A\arr B
  \]
  where $a$ ranges over a fixed set of elements called \Index[atomic
    type]{atomic types}.
  The symbol $\arr$ is associative to the right, \ie $A\arr (B\arr C)$ can
  be abbreviated as $A\arr B\arr C$.
\end{definition}

The typing system is defined as follows:
\begin{definition}[Simple types for the ($\textbf I$, $\textbf K$, $\textbf S$)-combinatoric system]

  Typing is a binary relation between terms and simple types, denoted
  with expressions such as $\Deri{}{M} A$.
  That relation is defined for the combinators as follows:
  \[
  \begin{array}{lll}
    \Deri{}{\textbf{I}}{A\arr A}\\
    \Deri{}{\textbf{K}}{A\arr B\arr A}\\
    \Deri{}{\textbf{S}}{(A\arr (B\arr C))\arr(A\arr B)\arr (A\arr C)}
  \end{array}
  \]
  and program application is typed by the following rule:
  \[
  \infers{
    \Deri{}{M\;N} B
  }{
    \Deri{} M {A\arr B}\quad \Deri{} N A
  }
  \]

  Derivability of the typing statement $\Deri{}{M} A$, from the above
  axioms and using the program application rule, is denoted
  $\Deri[\simpC]{}{M} A$.
\end{definition}

The reduction defined by the rewrite system preserves types, a property
called Subject Reduction:
\begin{theorem}[Subject reduction for simply-typed combinatoric system]

  If $\Deri{} M A$ and $M\Rew{}N$ then $\Deri{} N A$.
\end{theorem}
\begin{proof}
  By induction on the derivation of $M\Rew{}N$, with the base cases
  corresponding to the 3 rewrite rules themselves.
\end{proof}

The essence of the Curry-Howard correspondence, is the simple remark
that, viewing the functional type construct $\arr$ as the logical
symbol for implication, simples types are isomorphic\footnote{The
  isomorphism with simple types assumes that atomic types are
  isomorphic to atomic formulae.} to the syntax of formulae for
propositional minimal logic~\cite{johansson36minimal} and that typing
derivations are isomorphic to proofs in a particular Frege-Hilbert
system~\cite{Frege:1879,hilbert28} for minimal logic.

\begin{definition}[Propositional minimal logic]

  \Index[formula]{Formulae} of minimal logic are defined by the following grammar:
  \[
  A,B,\ldots \recdef a \sep A\imp B
  \]
  where $a$ ranges over a fixed set of elements called \Index[atomic formula]{atomic formulae}.

  \Index[proof]{Proofs} are the derivations built with the \Index{Modus Ponens} rule
  \[
  \infers{
    \Der{} B
  }{
    \Der{} {A\imp B}\quad \Der{} A
  }
  \]
  from the axioms:
  \[
  \begin{array}{l}
    A\imp A\\
    A\imp B\imp A\\
    (A\imp (B\imp C))\imp(A\imp B)\imp (A\imp C)
  \end{array}
  \]
  The symbol $\imp$ is associative to the right.

  Derivability of $\Der{} A$, from the above axioms and using the
  program application rule, is denoted $\Der[\miniFH]{} A$.
\end{definition}

Via the correspondence, the Subject Reduction property allows the view
of the reduction relation as a proof-transforming procedure.

\subsection{Simply-typed $\lambda$-calculus}
\label{sec:CHlambda}

Curry's view about minimal logic was extended by Howard to
intuitionistic first-order arithmetic~\cite{How:fortnc}.  A different
format was also proposed, both for proofs and for programs, and became
perhaps the most popular setting for the Curry-Howard correspondence:
Natural Deduction~\cite{Gentzen35} was the formalism used for proofs,
and the $\lambda$-calculus~\cite{Church} was the formalism used for
programs. This instance of the Curry-Howard correspondence that we
present is a version of natural deduction using
\Index[sequent]{sequents} and a version of the simply-typed
$\lambda$-calculus using typing contexts.

\begin{definition}[$\lambda$-calculus]
  The syntax of the $\lambda$-calculus is given by the following grammar:
  \[ M,N, \ldots \recdef x \sep \lambda x.M\sep M\;N\]
  where $x$ ranges over a denumerable set of
  \Index[variable]{variables}, and the construct $\lambda x.M$ binds
  $x$ in $M$.\footnote{As mentioned in the section about notations,
    specifying binders and their scopes automatically defines free
    variables, $\alpha$-conversion, capture-avoiding substitution, etc.}

  Standard conventions are used for parentheses~\cite{Bar84}: the
  scopes of binders extend as much as parentheses allow (\ie $\l
  x. M\;N$ abbreviates $\l x .(M\;N)$); program application is
  associative to the left (\ie $M\;N\;P$ abbreviates $(M\;N)\;P$);
  moreover, binder can be grouped, so that $\l xy.M$ abbreviates $\l
  x. \l y.M$.

  The following rewrite rules
  \[\begin{array}{llll}
  \rulenamed\beta&(\l x.M)\;N&\Rew{}\subst M x N\\
  \rulenamed\eta&\l x.M\ x&\Rew{}M&\mbox{if
    $x\notin\FV M$}
  \end{array}\]
  define the reduction relations $\Rew\beta$, $\Rew\eta$ and $\Rew{\beta\eta}$.
\end{definition}

As for the combinatoric system from Section~\ref{sec:combinators}, the
reduction relations are confluent:
\begin{theorem}[Confluence]
$\Rew\beta$, $\Rew\eta$ and $\Rew{\beta\eta}$ are confluent.
\end{theorem}
\begin{proof}
  See for instance~\cite{Bar84}.
\end{proof}

\begin{definition}[Simply-typed $\lambda$-calculus]\label{def:stlambdacalc}
  \Index[typing context]{Typing contexts} are finite maps from
  variables to simple types, with $\emptyenv$ denoting the empty
  context (sometimes the notation $\emptyenv$ is completely omitted),
  $\Gamma,\Gamma'$ denoting the union of contexts $\Gamma$ and
  $\Gamma'$ (assuming it is defined), and $\XcolY x A$ denoting the
  singleton context mapping variable $x$ to the simple type $A$.

  The typing rules of the simply-typed $\lambda$-calculus are given in
  Fig.~\ref{fig:stlambda}. 

  Derivability of the typing statement $\Deri{\Gamma}{M} A$ in that
  system is denoted $\Deri[\simplambda]{\Gamma}{M} A$.
\end{definition}
\begin{bfigure}[!h]
  \[
  \begin{array}{c}
    \infers{\Deri{\Gamma,\XcolY{x}A}{x}{A}}{}\\\\
    \infers{
      \Deri\Gamma{\l x.M} {A\arr B}
    }{
      \Deri{\Gamma,\XcolY {x} A} M B
    }
    \qquad
    \infers{
      \Deri\Gamma{M\;N} B
    }{
      \Deri\Gamma M {A\arr B}\quad \Deri\Gamma N A
    }
  \end{array}
  \]
  \caption{Simply-typed $\lambda$-calculus}
  \label{fig:stlambda}
\end{bfigure}

As for the combinatoric system from Section~\ref{sec:combinators}, the
reduction relations satisfy Subject Reduction:

\begin{theorem}[Subject reduction for simply-typed $\l$-calculus]
  \begin{enumerate}
  \item If $\Deri{\Gamma,\XcolY x A} M B$ and $\Deri{\Gamma} N A$ then $\Deri\Gamma {\subst M x N} B$.
  \item If $\Deri\Gamma M A$ and $M\Rew{\beta\eta}N$ then $\Deri\Gamma N A$.
  \end{enumerate}
\end{theorem}
\begin{proof}
  See for instance~\cite{Bar84}.
\end{proof}

Our second instance of the Curry-Howard correspondence relates the
simply-typed $\lambda$-calculus with the Natural Deduction system
$\NJi$ for minimal logic.

\begin{definition}[Natural Deduction for minimal logic - $\NJi$]

  System $\NJi$ is the inference system given in Fig.~\ref{fig:NJi},
  where
  \begin{itemize}
  \item $A$, $B$ range over formulae of minimal logic;
  \item $\Gamma$ stands for a ``collection'' of formulae.  By
    collection we mean either set or multiset,\footnote{That choice
      will change the number of proofs of a given formula.} with
    $\Gamma,\Gamma'$ denoting the union of $\Gamma$ and $\Gamma'$, $A$
    denoting either the formula $A$ itself or the singleton $\{A\}$
    (or $\multiset A$), while the empty set (or multiset) is sometimes
    ommitted;
  \item $\Der\Gamma A$ is a structure called \Index{sequent}.
  \end{itemize}

  Derivations in that system are called \Index[proof]{proofs} in $\NJi$.

  Derivability in $\NJi$ of a sequent $\Der\Gamma A$ is denoted $\Der[\NJi]\Gamma A$.
\end{definition}

\begin{bfigure}[!h]
  \[
  \begin{array}c
  \infer{\Der{\Gamma,A}{A}}{}\\\\
  \infer{\Der\Gamma{A\imp B}}{\Der{\Gamma,A}B}
  \qquad
  \infer{
    \Der{\Gamma} B
  }{
    \Der{\Gamma} {A\imp B}\quad \Der{\Gamma} A
  }
  \end{array}
  \]
  \caption{Natural Deduction for minimal logic - $\NJi$}
  \label{fig:NJi}
\end{bfigure}

Comparing Fig.~\ref{fig:stlambda} and Fig.~\ref{fig:NJi} reveals our
second instance of the Curry-Howard correspondence, although the exact
meaning of the word `correspondence' is in this case more subtle than
for our first instance: 

Clearly, the bijective aspect that pertains to the word
``isomorphism'' is jeopardised as $\infer{\Deri{\XcolY x A,\XcolY y
    A}x A}{}$ and $\infer{\Deri{\XcolY x A,\XcolY y A}y A}{}$ are
clearly two distinct typing derivations which would both `correspond
to' the proof $\infer{\Der{A,A} A}{}$ (whether we use sets or
multisets). Moreover, binding introduces an ambiguity in the way we
count typing derivations: is there one or infinitely many derivations
of $\Deri{}{\l x.x}{A\arr A}$?\footnote{Formally, the typing system
  allows infinitely many premisses for that typing judgement,
  depending on the variable that we pick to place in the typing
  context with type $A$.}

For this reason, this dissertation takes the view that the interesting
aspects of the Curry-Howard correspondence do not include the
bijective aspect of an encoding from one system into another, but
rather its compositionality (for trees), and the soundness and
completeness properties:

In the present case, the forgetful encoding that maps every typing
derivation to a proof is compositional with respect to the
tree-structure of derivations; its surjectivity provides completeness
of type inhabitation -whether there exists a $\lambda$-term of a given
type- with respect to the provability of the corresponding formula;
soundness is simply the fact that the tree obtained by forgetting
variables and terms from a typing derivation is a correct proof.

These are the properties that we will aim at when investigating the
variants of the Curry-Howard correspondence.

\medskip

As for the combinatoric system from Section~\ref{sec:combinators}, the
Subject Reduction property allows the view of the reduction relations
$\Rew\beta$, $\Rew\eta$ and $\Rew{\beta\eta}$ as proof-transforming
procedures.

\medskip

In summary, the most well-known settings for the Curry-Howard correspondence are:\\
\begin{centre}
  \begin{tabular}{l@{\quad$\leftrightarrow$\quad}ll}
    Frege-Hilbert system& Combinators (S,K,I)&\cite{CurryFeys:1958} \\
    Natural Deduction&Typed $\l$-terms&\cite{How:fortnc}
  \end{tabular}
\end{centre}

\subsection{The categorical aspect}

We now briefly mention what is sometimes considered a third aspect of
the Curry-Howard corespondence, in category theory.

Categories can be used to shed a semantical light on the Curry-Howard
correspondence. In our case, a particular kind of category provides
models of the simply-typed $\lambda$-calculus: cartesian closed
categories (CCC). In brief, CCC feature a terminal object, products,
and exponential objects. We start with a few notational conventions:

\begin{notation}[Category]

  The class of morphisms from object $A$ to object $B$ is denoted
  $\homclass A B$, and the expression $\morph f A B$ denotes that $f$
  is a morphism from $A$ to $B$. Identity morphisms are denoted
  $\Id[A]$, and the composition of $\morph f A B$ and $\morph g B C$
  is denoted $\morph{\co f g}A C$.

  In a cartesian closed category (CCC),
  \begin{itemize}
  \item the \Index{terminal object} is denoted $\termobj$, with morphisms
    $\morph{\termmorph[A]}A\termobj$
  \item the \Index{product} of $A$ and $B$ is denoted $A\times B$, with
    projections denoted $\morph{\pi_1}{A\times B} A$ and
    $\morph{\pi_2}{A\times B} B$ (and more generally, the $i$\ith\ 
    projection from $n$ objects is denoted $\projmorph i n$) and
    morphism pairing denoted $\morph{\pair{f_1}{f_2}}C{A\times B}$ for
    every $\morph{f_1}CA$ and $\morph{f_2}CB$.
  \item the \Index{exponential} of $A$ and $B$ is denoted $B^A$,
    with morphisms $\morph{\evalmorph}{B^A\times A} B$ and a currified
    morphism $\morph {\lammorph g}X{B^A}$ for every $\morph{g}
    {X\times A} {B}$.
  \end{itemize}
\end{notation}

\begin{toappendix}[]
  \chapter{Basic definitions for categories}
\end{toappendix}

\begin{toappendix}[For a formal definition of the above concepts, see Appendix~\thisappendix.]
\begin{definition}[Category]
    A \Index{category} is the combination of
    \begin{itemize}
    \item a class of elements called \Index[object]{objects}, denoted
      $A$, $B$,\ldots
    \item for every pair of objects $A$ and $B$, a class $\homclass A B$
      of elements called \Index[morphism]{morphisms from $A$ to $B$};
      the expression $\morph f A B$ denotes that $f$ is a morphism from
      $A$ to $B$;
    \item for every object $A$, a morphism $\Id[A]$ called
      \Index{identity};
    \item for every objects $A$, $B$, $C$, a binary operation called
      \Index{composition} mapping every $\morph f A B$ and $\morph g B
      C$ to a morphism $\morph{\co f g}A C$
    \end{itemize}
    such that the following properties holds
    \begin{itemize}
    \item composition is associative ($\co{(\co f g)}h=\co f{(\co g
      h)}$)
    \item identities are units for composition ($\co{\Id[A]}f=\co
      f{\Id[A]}=f$).
    \end{itemize}

    Given a category, we often use \Index[diagram]{diagrams} to
    represent a collection of objects -represented as vertices- and
    morphisms -represented as labelled arrows between vertices. Using
    morphism composition, each path between any two given vertices
    unambiguously represents a morphism. A diagram
    \Index[commute]{commutes} when for each pair of vertices $A$ and
    $B$, all paths from $A$ to $B$ represent equal morphisms.

    Two objects $A$ and $B$ are \Index{isomorphic} when there are two
    morphisms $\morph f A B$ and $\morph g B A$ such that $\co f g
    =\Id[A]$ and $\co g f = \Id[B]$.
  \end{definition}

  Standard examples of categories are: the categories of sets and
  functions (objects are sets and morphisms from $A$ to $B$ are
  functions from $A$ to $B$), the category of sets and relations
  (objects are sets and morphisms from $A$ to $B$ are relations from $A$
  to $B$), etc. Groups form a particular kind of categories (where there
  is only one object, and elements of the group are the morphisms from
  that object to itself). Partially ordered sets form another particular
  kind of categories (where there is at most one morphism between any
  two objects), etc.

  \begin{definition}[Cartesian Closed Category]\label{def:CCC}\\
    A \Index{Cartesian Closed Category} (CCC), is a category such that
    \begin{itemize}
    \item there is an object, denoted $\termobj$ and called
      \Index{terminal object}, such that, for each object $A$, there is
      a unique morphism $\morph{\termmorph[A]}A 1$;
    \item for every two objects $A$ and $B$, there is an object, denoted
      $A\times B$ and called the \Index{product} of $A$ and $B$,
      together with two morphisms $\morph{\pi_1}{A\times B} A$ and
      $\morph{\pi_2}{A\times B} B$, called the first and second
      \Index[projection]{projections}, satisfying the following
      property:\\ for every object $C$ and morphisms $\morph{f_1}CA$ and
      $\morph{f_2}CB$, there is a unique $\morph{f}C{A\times B}$,
      denoted $\pair{f_1}{f_2}$, such that the following diagram
      commutes
      \[
      \xymatrix{
        &C\ar[d]^f\ar[dl]_{f_1}\ar[dr]^{f_2}&\\
        A&A\times B\ar[l]_{\pi_1}\ar[r]^{\pi_2}&B
      }
      \]
    \item for every two objects $A$ and $B$, there is an object, denoted
      $B^A$ and called the \Index{exponential} of $A$ and $B$, together
      with a morphism $\morph{\evalmorph}{B^A\times A} B$, satisfying
      the following property:\\ for every object $X$ and morphism
      $\morph g {X\times A} {B}$ there is a unique $\morph fX{B^A}$,
      denoted $\lammorph g$, such that the following diagram commutes
      \[
      \xymatrix{
        X\times A\ar[d]_{\pair{\lammorph g}{\Id[A]}}\ar[dr]^{g}&\\
        B^A\times A\ar[r]_{\quad\evalmorph}&B
      }
      \]
    \end{itemize}

    We choose the convention that products are associative to the
    left, \ie $(A\times B)\times C$ can be abbreviated as $A\times
    B\times C$. A family of morphisms $\morph{\projmorph i
      n}{A_1\times\cdots\times A_n}{A_i}$, for $1\leq i \leq n$, can
    be defined by composing the two projections in the obvious way:
    \[\begin{array}{lll}
    \projmorph 1 1 &\eqdef \Id[A_1]\\
    \projmorph n n &\eqdef \pi_2&\mbox{ when }1<n\\
    \projmorph p n &\eqdef \co{\pi_1}{\projmorph p {n-1}}&\mbox{ when }p<n
    \end{array}
    \]
  \end{definition}

  \begin{remark}
    One can quickly check that the terminal object, products and
    exponentials are unique up to isomorphism (\ie two objects satisfying
    the property of the terminal object, or the product / exponential
    object for a given $A$ and $B$, are isomorphic); hence the notations
    $\termobj$, $A\times B$, $B^A$.
  \end{remark}

\end{toappendix}

  \begin{definition}[Semantics of the simply-typed $\lambda$-calculus in a CCC]

  Consider a cartesian closed category.

  Consider a mapping that interprets every atomic type $a$ as an
  object $\sem a$ of the CCC, and extend it to all simple types by
  defining $\sem {A\arr B}$ as ${\sem B}^{\sem A}$.

  Consider a total order on the $\lambda$-calculus variables; when
  writing a typing context as\linebreak
  $\XcolY{x_1}{A_1},\ldots,\XcolY{x_n}{A_n}$ we now follow the
  convention that $x_1,\ldots,x_n$ is an increasing sequence; we then
  define the semantics of any typing context by
  \[\sem{\XcolY{x_1}{A_1},\ldots,\XcolY{x_n}{A_n}}\eqdef\termobj\times\sem{A_1}\times\cdots\times
  \sem{A_n}\]

  The semantics of a typing derivation $\pi$ for the typing judgement
  $\Deri\Gamma M A$ is defined according to Fig.~\ref{fig:stlcCCC}, by
  induction on $\pi$, as a morphism $\morph{\sem\pi}{\sem
    \Gamma}{\sem A}$.
\end{definition}
\begin{bfigure}[!h]
\[
  \begin{array}{ll}
    \sem{\infers{\Deri{\XcolY{x_1}{A_1},\ldots,\XcolY{x_n}{A_n}}{x_i}{A_i}}{}} 
    &\eqdef \morph{\projmorph {i+1} {n+1}}{\termobj\times\sem{A_1}\times\cdots\times \sem{A_n}}{\sem{A_i}}\\\\
    \sem{\infers{
      \Deri\Gamma{\l x.M} {A\arr B}
    }{
      \concludes{
        \Deri{\Gamma,\XcolY {x} {A}} M B
      }[\pi]
      {}
    }}&\eqdef \morph{\lammorph g}{\sem{\Gamma}}{{\sem B}^{\sem A}}\\
    \hfill\mbox{where}&
    \morph{
      g=\sem{\concludes{
        \Deri{\Gamma,\XcolY {x} {A}} M B
      }[\pi]
      {}}
    }{\sem\Gamma\times \sem A}{\sem B}
    \\\\
    \sem{
      \infers{
        \Deri\Gamma{M\;N} B
      }{
        \concludes{\Deri\Gamma M {A\arr B}}[\pi_1]{}
        \quad 
        \concludes{\Deri\Gamma N A}[\pi_2]{}
      }
    }
    &\eqdef\morph{\co{\pair{g_1}{g_2}}\evalmorph}{\sem\Gamma}{\sem B}\\
    \hfill\mbox{where}&
    \morph{
      g_1=\sem{\concludes{
        \Deri{\Gamma} M {A\arr B}
      }[\pi_1]
      {}}
    }{\sem\Gamma}{{\sem B}^{\sem A}}\\\\
    \hfill\mbox{and}&
    \morph{
      g_2=\sem{\concludes{
        \Deri{\Gamma} N A
      }[\pi_2]
      {}}
    }{\sem\Gamma}{\sem A}
  \end{array}
  \]
  \caption{Semantics of the simply-typed $\lambda$-calculus in a CCC}
  \label{fig:stlcCCC}
\end{bfigure}

Note that in the case of a $\lambda$-abstraction, we assume that $x$
is (strictly) greater than any variable in $\Gamma$. Any derivation in
which this is not the case can easily be turned into one satisfying
this condition: variables can always be renamed\footnote{Using
  equivariance of typing derivations.} so that they are introduced in
the typing context in increasing order.\footnote{Alternative
  presentations of the simply-typed $\lambda$-calculus may be more
  convenient to define its semantics in a CCC: the use of De Bruijn
  indices (see \eg\cite{Bar84}) instead of named variables provides a
  natural way of ordering the objects in the interpretation of a
  typing environment (without resorting to ordering the set of
  variables); if variables carry their own type (or when each type
  comes with its own set of variables), the interpretation can be
  defined on the terms themselves rather than their typing
  derivations.}

Now as mentioned earlier, we can relate the reductions in the simply-typed $\lambda$-calculus
to the equality of morphisms in CCC:

\begin{theorem}[Soundness and completeness]\label{th:SoundCompleteCCC}

  Assume $\concludes{\Deri[\simplambda] \Gamma M A}[\pi]{}$ and 
  $\concludes{\Deri[\simplambda] \Gamma N A}[\pi']{}$.\\\\
  $M\Rewsn{\beta\eta}N$ \iff\ in every CCC we have $\sem \pi=\sem{\pi'}$.
\end{theorem}

The equality theorems that can be derived from the axioms of CCC (and
that therefore hold in every CCC) are reflected syntactically in the
simply-typed $\lambda$-calculus, and the simply-typed $\l$-calculus is
therefore said to form an \Index{internal language} for CCC.

Note that it is easy to define the semantics of the ($\textbf I$,
$\textbf K$, $\textbf S$)-combinatoric system (with simple types) in a
CCC, for instance by encoding combinators as simply-typed
$\lambda$-terms:

\begin{theorem}[Semantics of the ($\textbf I$, $\textbf K$, $\textbf S$)-combinatoric system]

  The encoding of Fig.~\ref{fig:stCTostL} satisfies satisfies the
  following properties:
  \begin{itemize}
  \item If $\Deri[\simpC]{}M A$ then $\Deri[\simplambda]{}{\stCTostL
    M}A$, with a function $\pi\mapsto\stCTostL \pi$ transforming a
    derivation of the former into a derivation of the latter.
  \item If $M\Rew{}M'$ then $\stCTostL M\Rew{\beta}M'$.
  \end{itemize}
  The above properties allow the definition of the semantics $\sem\pi$
  of a typing derivation $\pi$ of $\Deri[\simpC]{}M A$ as the morphism
  $\morph{\sem{\stCTostL\pi}}{\termobj}{\sem A}$, such that the
  following holds:\\ If $\pi\Rewsn{}\pi'$ then $\sem\pi=\sem{\pi'}$.
\end{theorem}
\begin{bfigure}[!h]
\[
\begin{array}{ll}
  \stCTostL{{\sf I}}&\eqdef \lambda x. x\\
  \stCTostL{{\sf K}}&\eqdef \l xy.x\\
  \stCTostL{{\sf S}}&\eqdef \l xyz.x\;z\;(y\;z)\\
  \stCTostL{M\;N}&\eqdef \stCTostL M\;\stCTostL N
\end{array}
\]
\caption{($\textbf I$, $\textbf K$, $\textbf S$)-combinators as
  $\l$-terms}
\label{fig:stCTostL}
\end{bfigure}

\subsection{Applying the methodology to other systems}

The approach of the Curry-Howard correspondence can be, and has been,
generalised with the following methodology:
\begin{itemize}
\item The first step is to decorate proofs with proof-terms:
  $\Gamma\vdash A$ becomes $\Deri {\Gamma'} M A$, with $\Gamma'$ being
  a typing context whose co-domain (\ie the types which have been
  assigned to variables) is $\Gamma$;
\item the second is to express proof transformations in terms of
  proof-term reduction, denoted $M\Rew{\mathcal S}N$, often given by a
  rewrite system $\mathcal S$.
\end{itemize}
The desired properties of reduction are
\begin{itemize}
\item \emph{Progress}, \ie any term containing ``undesirable structures'' can be reduced.
\item \emph{Subject reduction} property, \ie preservation of typing:\\
  If $\Deri\Gamma M A$ and $M\Rew{\mathcal S}N$ then $\Deri\Gamma N A$
\item possibly \emph{Confluence}, programs are deterministic.
\item possibly \emph{Normalisation}, \ie the fact that the execution of programs terminates.
\end{itemize}

The notion of ``undesirable structures'' is of course one of the
concepts to identify in an interesting way; for instance in the
simply-typed $\lambda$-calculus, a structure of the form $(\lambda
x.M)\;N$ corresponds to the introduction of implication followed by
the elimination of the introduced implication, a situation which we
may consider undesirable from a proof-theoretic point of view.

To illustrate this methodology, we show how the correspondence from
Section~\ref{sec:CHlambda} can be extended to intuitionistic logic
with both the implication connective and the logical constant $\bot$.

First note that by identifying $\bot$ simply as one of the atomic
formulae, intuitionistic negation can be defined as
follows: $\neg A\eqdef A\imp\bot$. With this definition, the following
rules are instances of those of the simply-typed $\l$-calculus:
\[
\infer{\Deri\Gamma{\l x.M}{\neg A}}{\Deri{\Gamma,\XcolY x A} M \bot}
\qquad
\infer{\Deri\Gamma{M\;N}\bot}{\Deri\Gamma{M}{\neg A}\quad\Deri\Gamma N A}
\]
and these reflect the usual Natural Deduction rules for negation, but
what is missing, to have intuitionistic logic, is the rule named
\emph{Ex falso quodlibet} (EFQ):
\[
\infer{\Der\Gamma A}{\Der\Gamma \bot}
\]
We now see what that rule may become in the Curry-Howard
correspondence.

\begin{example}[Extension of the Curry-Howard correspondence for $\NJ_{\imp,\bot}$]

  We extend the syntax of the $\lambda$-calculus with the following construct:
  \[ M,N,\ldots\recdef \ldots\sep \abort M \]
  and we add to the simply-typed $\lambda$-calculus a typing rule corresponding to EFQ:
  \[\infer{\Deri\Gamma{\abort M}A}{\Deri\Gamma{M}\bot}\]

  We may then add to the $\lambda$-calculus a rule such as
  \[\rulenamed\abortrule\quad\abort M \ N\Rew{}\abort M\]
  to computationally interpret the new construct as a greedy consumer
  of arguments, and $\Rew{\abortrule}$, $\Rew{\beta\abortrule}$,
  $\Rew{\eta\abortrule}$, and $\Rew{\beta\eta\abortrule}$ are all
  confluent.

  With these definitions, we still have Subject Reduction:\\
  If $\Deri\Gamma M A$ and $M\Rew{\beta\eta\abortrule}N$ then $\Deri\Gamma N A$.

  We can also interpret EFQ in category theory by requiring from a CCC the extra axiom that
  there is an \Index{initial object} $\bot$, \ie an object such that,
  from every object $A$ there is a unique morphism $\morph
  {\initmorph[A]} \bot A$ (which is the dual of the terminal object $\termobj$ of
  the CCC).
\end{example}

From there, it is natural to try to extend the above correspondence to
classical logic, which can be obtained from intuitionistic logic by
adding any one of the three axiom schemes:
\[\begin{array}{lc}
\mbox{\Index{Elimination of double negation} (EDN):}& (\neg\neg A) \imp A\\
\mbox{\Index{Peirce's law} (PL):}&((A\imp B)\imp A)\imp A\\
\mbox{\Index{Law of excluded middle} (LEM):}&A\vee \neg A
\end{array}
\]

In presence of EFQ (in short, the axiom scheme \(\bot\imp A\)), the above
schemes are all equivalent in terms of formula
provability. Interestingly enough and as noted
in~\cite{HerbelinAriolaH03}, without EFQ, we only have the following
implications between the schemes:
\begin{centre}EDN$\imp$PL$\imp$LEM\\
EDN$\imp$EFQ\end{centre}

Alternatives to adding axiom schemes is to add inference rules such as
\ctr{\(\infer{\Seq \Gamma A}{\Seq{\Gamma,\neg A}\bot}\) for EDN} or
even to change the structure of the proof formalism, for instance by
using the classical sequent calculus~\cite{Gentzen35} with
\emph{right-contraction}.

When thinking about classical logic, we have a tendency to identify a
formula $A$ with $\neg\neg A$, as suggested not only by the
elimination of double negation but also by models of classical
provability in boolean algebras.

Now, attempts to apply the Curry-Howard methodology to, say, the above
axiom schemes or inference rule, are limited by the following fact:

A CCC with initial object $\bot$ and such that every object $A$ is
naturally isomorphic to $\bot^{\bot^A}$, collapses to a boolean
algebra: there is at most 1 morphism between any 2 objects (see the
proof in~\cite{LambekJ:inthoc} or~\cite{StrassburgerHDR}).

That means that such a category would not distinguish two proofs of
the same theorem, which is rather useless for a theory of proofs, or
for the proofs-as-programs paradigm.

At that point, the natural question to ask is whether classical logic
has computational content? To that question, and based on the above
remarks, the book \emph{Proofs and Types}~\cite{GTL:prot} answers in
1989:

\emph{``[The Curry-Howard] interpretation is not possible with classical
logic: there is no sensible way of considering proofs as
algorithms. In fact, classical logic has no denotational semantics,
except the trivial one which identifies all the proofs of the same
type.''}

In the rest of this chapter we explore the alternative answers that
have been given to the question since then.

\section{Continuations and control}\label{sec:continuationscontrol}

For this we start with some concepts which at first sight may seem
unrelated: continuations and control.

A freeze-frame shot taken at one point of a program's execution flow
could be represented, in a high-level view, as follows:
\[
\begin{array}{ll}
\downarrow&\mbox{code $P$ that has been executed, producing data $v$}\\
v&\mbox{its output}\\
\downarrow&\mbox{code $E$ that remains to be executed, consuming data $v$}
\end{array}
\]

The code $E$ that remains to be executed, and more generally the
programming environment or programming context within which some code
is executed, is called \Index{continuation}. 

The concept is also useful for compiling recursive calls: consider the
following pseudo code\\
\noindent
\texttt{
  myfunction(a1,...,an)\{\\
  \strut\quad some code;\\
  \strut\quad x = myfunction(a1',...,an');\\
  \strut\quad some code possibly using x;\\
  \}
}\\
When executing the recursive call, the code that remains to be
executed (\ie \texttt{some code possibly using x}), together with the
values of the local variables, needs to be stored in order to resume
computation after the recursive call has returned with a value for
\texttt x. But this is not needed in the case of \emph{tail
  recursion}, in which \texttt{some code possibly using x} just
returns (the value for) \texttt{x}.\\
The above code can be transformed into a tail-recursive code by
modelling the remaining code \texttt{some code possibly using x} as a
``continuation'' function \texttt{c'} taking the value of \texttt x
as input:\\
\noindent
\texttt{
  myfunction(a1,...,an,c)\{\\
  \strut\quad some code;\\
  \strut\quad return myfunction(a1',...,an',c');\\
  \}
}

Now we see what these concepts become in the case where the
programming language is the $\l$-calculus. An instance of the above
program execution flow picture
\[
\begin{array}{ll}
  \downarrow&\mbox{code $P$ that has been executed, producing data $v$}\\
  v&\mbox{its output}\\
  \downarrow&\mbox{code $E$ that remains to be executed, consuming data $v$}
\end{array}
\]
can be seen by considering
\begin{itemize}
\item $P$ to be a $\l$-term that is reduced,
\item $v$ to be the value to which $P$ reduces,
\item $E$ to be the context, in the syntactic sense: a term with a
  hole $E[\ ]$ (with the original $\lambda$-term being $E[P]$, \ie the
  context $E[\ ]$ whose hole has been filled with the $\l$-term $P$).
\end{itemize}

This is only a general idea: whether that view accurately describes
program execution depends on the evaluation strategy for
$\lambda$-terms; in particular, whether $\l$-terms are reduced
inside-out, what notion of ``value'' is considered (is it a normal
form?), what grammar for contexts $E[\ ]$ ranges over, etc.

But in pure $\lambda$-calculus, it is clear that $P$ has no knowledge
of $E[\ ]$ while being evaluated.

\Index[control]{Control} is about letting a program know and manipulate its
evaluation context. Originally, the concept was used to model
\texttt{goto} instructions, and other features that are not pure
functional programming.

In the case of $\l$-calculus, the evaluation context $E[\ ]$ within
which a term is evaluated gives rise to a continuation function $\l
x.E[x]$ (for a fresh variable $x$) that could be passed as an
argument.

Reynolds~\cite{Reynolds72}, Strachey-Wadsworth~\cite{Strachey:2000}
(re-edition of 74) explored continuations and control along those
lines, letting a program capture its evaluation context with a feature
known as \Index{call-with-current-continuation} (call-cc):
$\callCC$. This was added to the programming language \texttt{Scheme}.

Felleisen's PhD work~\cite{Felleisen:phd1987} on the Syntactic Theory of
Control introduced another control operator: $\mC$.

The general idea of these control operators is given by the following
reduction rules:
\[\begin{array}{ll}
E[\callCC\ M]&\Rew{}E[M\ (\l x.{E[x]})]\\
E[\mC\ M]&\Rew{} M\ (\l x.{E[x]})\\
\end{array}\]

In presence of $\abort{\ }$ and its (slightly modified) rule
\[E[\abort{M}]\Rew{}M\]
$\callCC$ and $\mC$ are interdefinable:
\[\begin{array}{lll}
\mC\ M&\eqdef\callCC\ (\l k.\abort {M\ k})&k\not\in\FV M\\
\callCC\ M&\eqdef\mC\ (\l k.k\ (M\ k))&k\not\in\FV M
\end{array}\]

Indeed,
\[
\begin{array}{lcl}
  E[\callCC\ M]&=&E[\mC\ (\l k.k\ (M\ k))]\\
  &\Rew{}&(\l k.k\ (M\ k))\ (\l x.{E[x]})\\
  &\Rew{}&(\l x.{E[x]})\ (M\ \l x.{E[x]})\\
  &\Rew{}&E[M\ (\l x.{E[x]})]\\
  E[\mC\ M]&=&E[\callCC\ (\l k.\abort {M\ k})] \\
  &\Rew{}&E[(\l k.\abort {M\ k})\ (\l x.{E[x]})]\\
  &\Rew{}&E[\abort {M\ \l x.{E[x]}}]\\
  &\Rew{}&M\ (\l x.{E[x]})
\end{array}
\]

Of course, the above rules are not ``standard'' rewrite rules (clearly
not first-order rewrite rules) and remain informal because we have not
specified what $E[\ ]$ exactly stands for or ranges over.

More fundamentally, a central question about control is: what kind of
continuation can be captured by a control operator and how? Is the
capture delimited? undelimited? etc.

But what is interesting is that the above intuitions are sufficient to
start seeing a connection with classical logic, as initiated by
Griffin in~\cite{Griffin:popl90}: 
\ctr{\begin{tabular}{ll}
    $\callCC$ can be typed by PL:& $((A\arr B)\arr A)\arr A$\\
    $\mC$ can be typed by EDN: & $(\neg\neg A)\arr A$
  \end{tabular}
}

With these types, the rewrite rules satisfy the following Subject
Reduction properties:\\
The following (again, informal) typing tree for $E[\callCC\ M]$
\[
\infer{\Deri\Gamma{E[\callCC\ M]}{B}}{
  \infer{\Deri\Gamma{\l x.E[x]}{A\arr B}}{}
  \quad
  \infer{\Deri\Gamma{\callCC\ M}{A}}{
    \infer{\Deri{\Gamma}\callCC{((A\arr B)\arr A)\arr A}}{}
    \quad
    \Deri\Gamma M{(A\arr B)\arr A}
  }
}
\]
can be transformed into a typing tree for $E[M\ (\l x.{E[x]})]$
\[
\infer{\Deri\Gamma{E[M\ (\l x.{E[x]})]}{B}}{
  \infer{\Deri\Gamma{\l x.E[x]}{A\arr B}}{}
  \quad
  \infer{\Deri\Gamma{M\ (\l x.{E[x]})}{A}}{
    \Deri\Gamma M{(A\arr B)\arr A}
    \quad
    \infer{\Deri\Gamma{\l x.E[x]}{A\arr B}}{}
  }
}
\]
while the following (again, informal) typing tree for $E[\mC\ M]$
\[
\infer{\Deri\Gamma{E[\mC\ M]}{\bot}}{
  \infer{\Deri\Gamma{\l x.E[x]}{A\arr \bot}}{}
  \quad
  \infer{\Deri\Gamma{\mC\ M}{A}}{
    \infer{\Deri{\Gamma}\mC{((A\arr \bot)\arr \bot)\arr A}}{}
    \quad
    \Deri\Gamma M{(A\arr \bot)\arr \bot}
  }
}
\]
can be transformed into a typing tree for $E[M\ (\l x.{E[x]})]$
\[
\infer{\Deri\Gamma{M\ (\l x.{E[x]})}{\bot}}{
  \infer{\Deri\Gamma M{(A\arr \bot)\arr \bot}}{}
  \quad
  \infer{\Deri\Gamma{\l x.E[x]}{A\arr \bot}}{}
}
\]

We can already see that, for the Subject Reduction property to hold in
the case of $\mC$, the context $E[\ ]$ cannot be any context: it has
to produce something of type $\bot$. Similarly, for the generalised
abort rule to satisfy Subject Reduction, the context $E[\ ]$ also needs
to produce something of type $\bot$.

In fact, $\mC$ generalises $\abort{\_}$, since we can define
\[
\abort M \eqdef \mC\ (\lambda x.M)
\]
where $x$ is a fresh (and therefore dummy) variable.

This reflects what we have already seen in pure logic:
EDN $\Leftrightarrow$ (PL $\wedge$ EFQ)

\section{Contributions in the 90s}\label{sec:90s}

One possible formalisation of the above informal concepts was proposed
by Parigot~\cite{Parigot92} in the form of the $\lambda\mu$-calculus.

\begin{definition}[$\lambda\mu$-calculus]
  The syntax of terms extends that of $\lambda$-calculus as follows:
  \[
  \begin{array}{lrl}
    \mbox{Terms}&M,N,P\ldots&\recdef x\sep \l x.M\sep M\ N\sep \m\alpha c\\
    \mbox{Commands}&c&\recdef \name\alpha M
  \end{array}
  \]
  where $\alpha$ ranges over a new set of variables called
  \Index[continuation variable]{continuation variables}, and $\m\alpha
  c$ binds $\alpha$ in $c$. The scope of this binder, as well as that
  of the unique \Index{command} construct $\name\alpha M$, extend as
  much as parentheses allow, so that $\m\alpha M\;N$ stands for
  $\m\alpha (M\;N)$ and $\name\alpha M\;N$ stands for
  $\name\alpha (M\;N)$.

  The typing rules extend those of $\lambda$-calculus as follows:
  \[
  \begin{array}c
    \infer{\Seq{\Gamma,x\col A}{x\col A;\Delta}}{\strut}\\\\
    \infer{\Seq{\Gamma}{\l x.M\col A\arr B;\Delta}}{\Seq{\Gamma,x\col A}{M\col B;\Delta}}
    \quad      
    \infer{\Seq{\Gamma}{M\ N\col B;\Delta}}{\Seq{\Gamma}{M\col A\arr B;\Delta}\quad\Seq{\Gamma}{N\col A;\Delta}}\\\\
    \infer{\Seq{\Gamma}{\m \alpha c\col A;\Delta}}{c\col(\Seq{\Gamma}{;\alpha\col A,\Delta})}
    \quad      
    \infer{\name \alpha M\col(\Seq{\Gamma}{;\alpha\col A,\Delta})}{\Seq{\Gamma}{M\col A;\alpha\col A,\Delta}}
  \end{array}
  \]
  where $\Gamma$ is a typing context for term-variables and $\Delta$
  is a typing context for continuation-variables.  Derivability of
  sequents in this system is respectively denoted
  $\Seq[\l\mu]{\Gamma}{{M\col A}{\semcol \Delta}}$ and $\XcolY
  c{(\Seq[\l\mu]\Gamma{;\Delta})}$.

  The reduction rules extend the $\beta$-reduction of
  $\lambda$-calculus as follows:
  \[
  \begin{array}{lll}
    (\l x.M)\;N&\Rew{}\subst M x N\\
    (\m\alpha c)\ N&\Rew{}\m\beta{\subst c{\name \alpha M}{\name \beta {M\ N}}}\\
    \name \beta{\m\alpha c}&\Rew{}\subst c \alpha\beta&
  \end{array}
  \]
  where $\subst c{\name \alpha M}{\name \beta {M\ N}}$ is an
  unconventional substitution operation, consisting in replacing, in
  $c$, every subcommand (\ie subterm that is a command) of the form
  $\name \alpha M$ by $\name \beta {M\ N}$, with the usual
  capture-avoiding conditions pertaining to substitution.

  The rules define a reduction relation $\Rew{\l\mu}$ on both terms
  and commands.
\end{definition}

A basic intuition of the syntax is that each continuation variable
$\alpha$ represents a ``place'' where various sub-terms of a given
type (that of $\alpha$) can be ``stored'' with a construct such as
$\name \alpha M$. The construct $\m\alpha c$ retrieves what is stored
under the continuation variable $\alpha$ and presents it as if it was
a simple term.  The second rewrite rule distributes for instance an
argument to every sub-term stored under the variable $\alpha$.

This calculus provides a computational interpretation of classical
logic. Indeed, the typing system, when forgetting variables and terms,
turns into the proof system of Fig.~\ref{fig:NK}, where $\Gamma$ and
$\Delta$ now stand for sets or multisets of formulae. We see that the
system generalises $\NJi$, in particular with a more general form of
sequent: $\Der{A_1,\ldots,A_n}{A;B_1,\ldots,B_m}$, and a new form of
sequent $\Der{A_1,\ldots,A_n}{;B_1,\ldots,B_m}$.
\begin{bfigure}[!h]
  \[
  \begin{array}c
    \infer{\Seq{\Gamma,A}{A;\Delta}}{\strut}\\\\
    \infer{\Seq{\Gamma}{A\imp B;\Delta}}{\Seq{\Gamma, A}{B;\Delta}}
    \quad
    \infer{\Seq{\Gamma}{B;\Delta}}{\Seq{\Gamma}{ A\imp B;\Delta}\quad\Seq{\Gamma}{ A;\Delta}}\\\\
    \infer{\Seq{\Gamma}{A;\Delta}}{\Seq{\Gamma}{; A,\Delta}}
    \quad
    \infer{\Seq{\Gamma}{;A,\Delta}}{\Seq{\Gamma}{A;A,\Delta}}
  \end{array}
  \]
  \caption{The proof system corresponding to the simply-typed $\l\mu$-calculus}
  \label{fig:NK}
\end{bfigure}

Just like a sequent $\Der{A_1,\ldots,A_n}A$ of $\NJi$ can be
interpreted as the formula $(A_1\wedge\cdots\wedge A_n)\imp A$, the
two sequent forms above can respectively be interpreted as the
formulae $(A_1\wedge\cdots\wedge A_n)\imp (A\vee B_1\vee\cdots\vee
B_m)$ and $(A_1\wedge\cdots\wedge A_n)\imp (B_1\vee\cdots\vee B_m)$.
With this interpretation, the system of Fig.~\ref{fig:NK} can be
easily checked to be sound with respect to classical logic, and for
completeness we can see that
\begin{itemize}
\item the rules of $\NJi$ are particular instances of the first three rules;
\item the system features a right-contraction rule, which allows
  Peirce's Law to be proved, as we see below.
\end{itemize}

As with the simply-typed $\l$-calculus, the rewrite rules satisfy Subject Reduction, which allows 
the $\l\mu$-calculus to describe a proof-transforming procedure for the system of Fig.~\ref{fig:NK}:
\begin{theorem}[Subject reduction for the simply-typed $\l\mu$-calculus]
  \begin{enumerate}
  \item If $\Deri[\l\mu]\Gamma M {A;\Delta}$ and $M\Rew{\l\mu}M'$ then $\Deri[\l\mu]\Gamma {M'} {A;\Delta}$.
  \item If $\XcolY c {(\Der[\l\mu]\Gamma {;\Delta})}$ and $c\Rew{\l\mu}c'$ then $\XcolY {c'} {(\Der[\l\mu]\Gamma {;\Delta})}$.
  \end{enumerate}
\end{theorem}

\begin{remark}
  This calculus integrates Peirce's law: By defining
  \[
  \callCC \eqdef \l x.\m \alpha {\name \alpha{(x\ \l y.\m\beta{\name\alpha y})}}
  \]
  we can build the following typing tree:
  \[
  \infer{\Deri{}{\callCC}{((A\arr B)\arr A)\arr A;}}{
    \infer{\Deri{\XcolY x{(A\arr B)\arr A}}{
        \m \alpha {\name \alpha{(x\ \l y.\m\beta{\name\alpha y})}}}{A;}
    }{
      \infer{\XcolY{\name \alpha{(x\ \l y.\m\beta{\name\alpha y})}}
        {(\Der{\XcolY x{(A\arr B)\arr A}}{;\XcolY\alpha A})}
      }{
        \infer{\Deri{\XcolY x{(A\arr B)\arr A}}{x\ \l y.\m\beta{\name\alpha y}}{A;\XcolY\alpha A}}{
          \infer{\Deri{\XcolY x{(A\arr B)\arr A}}{x}{(A\arr B)\arr A;\XcolY\alpha A}}{}
          \quad
          \infer{\Deri{\XcolY x{(A\arr B)\arr A}}{\l y.\m\beta{\name\alpha y}}{A\arr B;\XcolY\alpha A}}{
            \infer{\Deri{\XcolY x{(A\arr B)\arr A},\XcolY y A}{\m\beta{\name\alpha y}}{B;\XcolY\alpha A}}{
              \infer{\XcolY{\name\alpha y}{(\Der{\XcolY x{(A\arr B)\arr A},\XcolY y A}{;\XcolY\alpha A,\XcolY\beta B})}}{
                \infer{\Deri{\XcolY x{(A\arr B)\arr A},\XcolY y A}{y}{A;\XcolY\alpha A,\XcolY\beta B}}{ 
                }
              }                         
            }             
          }       
        } 
      }
    }
  }
  \]

  Now, consider that contexts are of the form $E[\ ]={\name {\gamma}{([\ ]\ N_1\ldots N_n})}$.
  We can perform the following reduction:
  \[\begin{array}{lcl}
  E[\callCC\ M]&=&{\name {\gamma}{(\l x.\m \alpha {\name \alpha{(x\ \l y.\m\beta{\name\alpha y})}})\ M\ N_1\ldots N_n}}\\
  &\Rew{}&{\name {\gamma}{(\m \alpha {\name \alpha{(M\ \l y.\m\beta{\name\alpha y})}})\ N_1\ldots N_n}}\\
  &\Rew{}&{\name {\gamma}{(\m \alpha {\name \alpha{(M\ \l y.\m\beta{\name\alpha y\ N_1})\ N_1}})\ N_2\ldots N_n}}\\
  &\Rew{}&\ldots\\
  &\Rew{}&{\name {\gamma}{\m \alpha {\name \alpha{(M\ \l y.\m\beta{\name\alpha y\ N_1\ldots N_n})\ N_1\ldots N_n}}}}\\
  &\Rew{}&{\name {\gamma}{(M\ \l y.\m\beta{\name{\gamma} y\ N_1\ldots N_n})\ N_1\ldots N_n}}\\
  &=&E[M\ (\l y.{\m\beta{E[y]}})]
  \end{array}
  \]
\end{remark}

Notice that what is passed to $M$ as an argument is not exactly $\l
y.{{E[y]}}$, since $E[\ ]$ forms a command and $\l y.{{E[y]}}$ is not
correct syntax, but $\m\beta{E[y]}$ turns the command $E[y]$ into a
term (of any type).

\begin{remark}If given a top-level continuation variable $\textsf{top}\col\bot$ (Ariola-Herbelin~\cite{HerbelinAriolaH03,AriolaHerbelin07}), then the $\l\mu$-calculus
integrates \emph{Ex falso quodlibet} and the elimination of double negation: 

We can build the following typing tree:
  \[
  \infer{\Deri{}{\l x.\m \alpha {\name {\textsf{top}}{x}}}{\bot\arr A;}}{
    \infer{\Deri{\XcolY x{\bot}}{
        \m \alpha {\name {\textsf{top}}{x}}}{A;}
    }{
      \infer{
        \XcolY{\name {\textsf{top}}{x}}{(\Der{\XcolY x{\bot}}{;\XcolY\alpha A})}
      }{
        \infer{\Deri{\XcolY x{\bot}}x{\bot;\XcolY\alpha A}}{}
      }
    }
  }
  \]

  and perform the following reduction:
  \[\begin{array}{lcl}
  E[(\l x.\m \alpha {\name {\textsf{top}}{x}})\ M]&=&{\name {\gamma}{(\l x.\m \alpha {\name {\textsf{top}}{x}})\ M\ N_1\ldots N_n}}\\
  &\Rew{}&{\name {\gamma}{(\m \alpha {\name {\textsf{top}} M})\ N_1\ldots N_n}}\\
  &\Rew{}&{\name {\gamma}{(\m \alpha {\name {\textsf{top}} M})\ N_2\ldots N_n}}\\
  &\Rew{}&\ldots\\
  &\Rew{}&{\name {\gamma}{(\m \alpha {\name {\textsf{top}} M})}}\\
  &\Rew{}&{\name {\textsf{top}} M}
  \end{array}
  \]

  And by defining
  \[
  \mC \eqdef \l x.\m \alpha {\name {\textsf{top}}{(x\ \l y.\m\beta{\name\alpha y})}}
  \]
  we can build the following typing tree:
  \[
  \infer{\Deri{}{\l x.\m \alpha {\name {\textsf{top}}{(x\ \l y.\m\beta{\name\alpha y})}}}{(\neg\neg A)\arr A;}}{
    \infer{\Deri{\XcolY x{\neg\neg A}}{
        \m \alpha {\name {\textsf{top}}{(x\ \l y.\m\beta{\name\alpha y})}}}{A;}
    }{
      \infer{\XcolY{\name {\textsf{top}}{(x\ \l y.\m\beta{\name\alpha y})}}
        {(\Der{\XcolY x{\neg\neg A}}{;\XcolY\alpha A})}
      }{
        \infer{\Deri{\XcolY x{\neg\neg A}}{x\ \l y.\m\beta{\name\alpha y}}{\bot;\XcolY\alpha A}}{
          \infer{\Deri{\XcolY x{\neg\neg A}}{x}{\neg\neg A;\XcolY\alpha A}}{}
          \quad
          \infer{\Deri{\XcolY x{\neg\neg A}}{\l y.\m\beta{\name\alpha y}}{\neg A;\XcolY\alpha A}}{
            \infer{\Deri{\XcolY x{\neg\neg A},\XcolY y A}{\m\beta{\name\alpha y}}{\bot;\XcolY\alpha A}}{
              \infer{\XcolY{\name\alpha y}{(\Der{\XcolY x{\neg\neg A},\XcolY y A}{;\XcolY\alpha A,\XcolY\beta \bot})}}{
                \infer{\Deri{\XcolY x{\neg\neg A},\XcolY y A}{y}{A;\XcolY\alpha A,\XcolY\beta \bot}}{ 
                }
              }                         
            }             
          }       
        } 
      }
    }
  }
  \]
  and we can perform the following reduction:
  \[\begin{array}{lcl}
  E[\mC\ M]&=&{\name {\gamma}{(\l x.\m \alpha {\name {\textsf{top}}{(x\ \l y.\m\beta{\name\alpha y})}})\ M\ N_1\ldots N_n}}\\
  &\Rew{}&{\name {\gamma}{(\m \alpha {\name {\textsf{top}}{(M\ \l y.\m\beta{\name\alpha y})}})\ N_1\ldots N_n}}\\
  &\Rew{}&{\name {\gamma}{(\m \alpha {\name {\textsf{top}}{(M\ \l y.\m\beta{\name\alpha y\ N_1})}})\ N_2\ldots N_n}}\\
  &\Rew{}&\ldots\\
  &\Rew{}&{\name {\gamma}{\m \alpha {\name {\textsf{top}}{M\ \l y.\m\beta{\name\alpha y\ N_1\ldots N_n}}}}}\\
  &\Rew{}&{\name {\textsf{top}}{M\ \l y.\m\beta{\name{\gamma} y\ N_1\ldots N_n}}}\\
  &=&\name{\textsf{top}}{M\ (\l y.\m\beta{E[y]})}
  \end{array}
  \]
\end{remark}
  
Notice that what is eventually produced by the rewrites is not $M$ and
$M\;(\l y.\m\beta{E[y]})$, respectively, but $\name {\textsf{top}}M$
and $\name {\textsf{top}}{M\;(\l y.\m\beta{E[y]})}$, since reducing a
command has to produce a command. But since $M$ is of type $\bot$
(respectively produces a term of type $\bot$), it can be stored in the
top-level continuation ${\textsf{top}}$.

\medskip

Now, when thinking about classical logic, we often have in mind
concepts of symmetry or duality:

Inversing the order in a boolean algebra provides another boolean
algebra where \eg the top and bottom elements have been swapped, the
meet and join operations have been swapped.

Very related to this are De Morgan's rules, which show a duality, via
negation, between $\et$ and $\ou$:
\[\begin{array}{ll}
\neg(A\et B)&={\neg A}\ou{\neg B}\\ 
\neg(A\ou B)&={\neg A}\et{\neg B}
\end{array}\]

In terms of proof formalisms, the classical sequent calculus
\LK~\cite{Gentzen35} shows a symmetry between the left-hand side and
the right-hand of sequents, of the form
$\Der{A_1,\ldots,A_n}{B_1,\ldots,B_m}$: whatever can be done on the
left-hand side can be done on the right-hand side, and vice versa. For
instance, the left-introduction rule for $\et$ is symmetric to the
right-introduction rule for $\vee$ and vice versa; left-contraction
symmetric to right-contraction (and this is very different from
intuitionistic logic).

But so far, such symmetries and dualities are not explicitly reflected
in our proof-term approach to classical proofs.

However, before even Griffin made the connection between control
operators and classical logic, Filinski~\cite{Filinski:1989}
formalised a duality between
\begin{itemize}
\item functions as values
\item functions as continuations
\end{itemize}
in the form of a ``symmetric $\l$-calculus'', with explicit
conversions from one view of functions to the other.  Yet there was no
explicit connection with classical logic.

In~\cite{BBsymm}, Barbanera and Berardi formalised their own symmetric
$\l$-calculus, with a typing system providing a Curry-Howard
interpretation of classical proofs. The classical proof system
depicted by their calculus is a one-sided version of the classical
sequent calculus~\cite{Gentzen35}, with a proof of normalisation for
typed terms (we will see such a proof in
Chapter~\ref{ch:ortho}).

Since then, two calculi emerged to provide Curry-Howard
interpretations of the two-sided sequent calculus \LK\ (or variants
thereof), with the reduction rules describing the famous
proof-transformation procedure known as \Index{cut-elimination}:
\begin{itemize}
\item
  Urban's calculus~\cite{UrbThes},
\item
  Curien and Herbelin's \Lmm~\cite{CurienHerbelinDuality99} for
  $\imp$,\\ later extended by Wadler~\cite{Wadlerdual} for $\et{}{}$
  and $\ou{}{}$ (explicitly connecting the symmetries of the calculus
  to De Morgan's duality).
\end{itemize}

\ignore{  { 
    \dis2{Types: atoms, $\wedge$, $\vee$, involutive negation, {\color{white}{\disloc[\gray]2{+ $\bot$}}}
      \[\begin{array}c
        \infer{\Gamma\vdash \mu x.M: A^\bot}{\Gamma,x:A\vdash M {\color{white}{\disloc[\gray]2{: \bot}}}}
        \qquad\qquad
        \infer{\Gamma\vdash \cutc M N {\color{white}{\disloc[\gray]2{: \bot}}}}{\Gamma\vdash M : A^\bot\quad\Gamma\vdash N : A}\\\\
        \infer{\Gamma\vdash \paire M N\col \et A B}{\Gamma\vdash M \col A\quad \Gamma\vdash N \col B}
        \qquad
        \infer{\Gamma\vdash \textsf{inj}_1(M) \col \ou A B}{\Gamma\vdash M \col A}
        \qquad
        \infer{\Gamma\vdash \textsf{inj}_2(M) \col \ou A B}{\Gamma\vdash M \col B}
      \end{array}
      \]}
    \dis4{\[\begin{array}{lll}
        \cutc{(\mu x.M)} N&\Rew{}\subst M x N\\
        \cutc{\paire{M_1}{M_2}}{\textsf{inj}_i(N)}&\Rew{}\cutc{M_i} N
      \end{array}
      \]
      +symmetric rules}}
}

These two independent (sets of) contributions had different aims:
Curien and Herbelin's was to expose, as the syntactic symmetry of the
classical sequent calculus, a \emph{duality} in computation based on
Filinski's ideas about continuations and on the call-by-value and
call-by-name evaluation strategies; they gave semantics to their
calculus, but with no proof of normalisation. Urban's aim was to have
a typing system as close as possible to \LK\ and have a reduction
system as close as possible to basic cut-elimination procedures; his
Ph.D.\ adapted Barbanera and Berardi's proof of strong normalisation
to his calculus, but gave no (denotational) semantics.

Several papers formalise the links between the various proof calculi
for classical logic: in particular,~\cite{lengrand03call-by-value}
relates \Lmm\ and Urban's calculus,~\cite{Rocheteau05} relates
\Lmm\ and $\lambda\mu$, and~\cite{HerbHDR} presents an extensive
exploration of the relations between the various calculi.

In the rest of this chapter, we focus on Curien and Herbelin's
calculus to explore some more semantical concepts, but many of them
can be transposed to other calculi for classical logic (in particular,
Parigot's $\l\mu$-calculus).

\section{System \SysL}\label{sec:SysL}

System \SysL\ is the new name of Curien and Herbelin's \Lmm, extended
with other connectives. We start with its syntax:

\begin{definition}[Syntax]
The syntax of \SysL\ is made of three syntactic categories:
    \[
    \begin{array}{lll}
      \mbox{terms}&t&\recdef x \sep \mu\beta.c \sep  \lambda x.t \sep  \paire {t_1}{t_2} \sep  \inj i t\\
      \mbox{continuations}&e&\recdef \alpha \sep  \m x c \sep  \cons t e \sep  \paire {e_1}{e_2} \sep  \inj i e\\
      \mbox{commands}&c&\recdef\cutc t e
    \end{array}
    \]
    where $i$ ranges over $\{1,2\}$, $x$ and $\alpha$ respectively
    range over term variables and continuation variables,\footnote{As in Parigot's $\l\mu$-calculus~\cite{Parigot92}.} $\mu\beta.c$
    binds $\beta$ in $c$, $\mu x.c$ binds $x$ in $c$, and $\l x.t$
    binds $x$ in $t$. The scope of binders extends as much as
    parentheses allow.
\end{definition}

A (somewhat shallow) intuition of the syntax can be given as follows:\\
\begin{tabular}{ll}
  Term variables $x$, $y$, \ldots &denote inputs\\
  Continuation variables $\alpha$, $\beta$, \ldots &denote outputs\\
  A \emph{term} has &one main output\\
  &some inputs\hfill (free term variables)\\
  &some ``alternative'' outputs\hfill (free continuation variables)\\
  A \emph{continuation} has &one main input\\
  &some ``additional'' inputs\hfill (free term variables)\\
  &some possible outputs\hfill (free continuation variables)\\
  A \emph{command} is & a term facing a continuation (the interaction is computation)
\end{tabular}

\begin{definition}[Typing]

  We consider the following grammar for \Index[type]{types} (extending
  that of simple types):
  \[
  A,B,\ldots \recdef a \sep A\arr B \sep A\wedge B \sep A\vee B
  \]
  where $a$ ranges over a fixed set of elements called \Index[atomic
    type]{atomic types}.
  The symbol $\arr$ is associative to the right, \ie $A\arr (B\arr C)$ can
  be abbreviated as $A\arr B\arr C$.

  The typing system for System \SysL\ is given for three kinds of sequents corresponding to the three syntactic categories of the syntax: 
\[
\Seq{\Gamma}{{t\col A}{\semcol \Delta}}\qquad\qquad
\Seq{\Gamma\semcol {e\col A}}\Delta\qquad\qquad
\XcolY c{(\Seq{\Gamma}{\Delta})}
\]
where $\Gamma$ is a typing context for term-variables and $\Delta$ is a typing context for continuation-variables.

  The system is presented in Fig.~\ref{fig:Ltyping}. Derivability of sequents in this system is respectively denoted $\Seq[\SysL]{\Gamma}{{t\col A}{\semcol \Delta}}$,\qquad\qquad $\Seq[\SysL]{\Gamma\semcol {e\col A}}\Delta$,\qquad\qquad and $\XcolY c{(\Seq[\SysL]{\Gamma}{\Delta})}$.
\end{definition}

\begin{bfigure}[!h]
  \[
  \begin{array}{c}
      \infer{\Seq{\Gamma,x\col A}{{x\col A}{\semcol \Delta}}}{}
    \qquad\qquad
      \infer{\Seq{\Gamma\semcol {\alpha\col A}}{\alpha\col A,\Delta}}{}
    \\\\
      \infer{\Seq{\Gamma }{{\l x. t\col A\arr B}{\semcol\Delta}}}
      {\Seq{\Gamma,x\col A}{{t\col B}{\semcol\Delta}}}
    \qquad\qquad
      \infer{\Seq{\Gamma\semcol{\cons t e\col A\arr B}}{\Delta}}
      {
        \Seq\Gamma{{t\col A}\semcol\Delta}
        \quad
        \Seq{\Gamma\semcol {e\col B}}{\Delta}}
    \\\\
      \infer{\Seq{\Gamma }{{\paire {t_1}{t_2}\col {A_1}\wedge {A_2}}{\semcol\Delta}}}
      {\Seq\Gamma{{{t_1}\col {A_1}}{\semcol\Delta}}
        \quad
        \Seq{\Gamma}{{{t_2}\col {A_2}}{\semcol\Delta}}
      }
    \qquad\qquad
      \infer{\Seq{\Gamma\semcol{\inj i e\col {A_1\wedge A_2}}}{\Delta}}
      {
        \Seq{\Gamma\semcol {e\col A_i}}{\Delta}
      }
    \\\\
      \infer{\Seq{\Gamma}{{\inj i t\col {A_1\vee A_2}}{\semcol\Delta}}}
      {
        \Seq{\Gamma}{{t\col A_i}{\semcol\Delta}}
      }
    \qquad\qquad
      \infer{\Seq{\Gamma\semcol{\paire {e_1}{e_2}\col {A_1}\vee {A_2}} }{\Delta}}
      {\Seq{\Gamma\semcol{{e_1}\col {A_1}}}{\Delta}
        \quad
        \Seq{\Gamma\semcol{{e_2}\col {A_2}}}{\Delta}
      }
    \\\\
      \infer{\Seq{\Gamma }{{\m \alpha c\col A}\semcol\Delta}}
      {c\col(\Seq{\Gamma}{\alpha\col A,\Delta})}
    \qquad\qquad
      \infer{\Seq{\Gamma\semcol {\m x c\col A}}{\Delta}}
      {c\col(\Seq{\Gamma,x\col A}{\Delta})}
    \\\\
      \infer{\cutc{t}{e}\col(\Seq{\Gamma}{\Delta})}
      {\Seq{\Gamma}{{{t}\col {A}}\semcol\Delta}
        \quad
        \Seq{\Gamma\semcol{{e}\col {A}}}{\Delta}
      }
  \end{array}
  \]
  \caption{Typing system for \SysL}
  \label{fig:Ltyping}
\end{bfigure}

As we can see, forgetting about variables and proof-terms does not
give the sequent calculus \LK\ exactly as we know it
from~\cite{Gentzen35} or as the popular variants described
in~\cite{TS} (for this one can look at Urban's Ph.D.~\cite{UrbThes}),
if only because there are three types of sequents. However, it is a
variant with a bit more structure, which defines the same notion of
provability as \LK, and which will prove useful for the computational
interpretation of classical logic.

An intuition about this interpretation can be given as follows:
similarly to the Curry-Howard correspondence in intuitionistic logic,
each connective in the syntax of formulae corresponds to a type
construct in programming; term constructs offer basic ways in which
such types can be inhabited, while continuation constructs offer basic
ways in which inhabitants of such types are consumed:
\begin{itemize}
\item A conjunction $A_1\wedge A_2$ corresponds to a product type, so
  basic inhabitants are pairs $\paire {t_1}{t_2}$ of terms (with the
  first component inhabiting $A_1$ and the second inhabiting $A_2$);
  basic continuations that consume such a pair start by extracting
  either the first or the second component (in other words, they start
  with one of the two projections), which corresponds to the
  continuation constructs $\inj1e$ and $\inj2e$.
\item A disjunction $A_1\vee A_2$ corresponds to a sum type, so
  basic inhabitants are the injections $\inj1t$ and $\inj2t$ (with $t$
  inhabiting $A_1$ or inhabiting $A_2$, respectively); basic
  continuations that consume such an injection must handle both cases,
  so the case analysis leads to providing a pair $\pair{e_1}{e_2}$ of
  two continuations: the former can consume inhabitants of $A_1$ and the
  latter can consume inhabitants of $A_2$.
\item An implication $A_1\imp A_2$ corresponds to a function type,
  with the basic inhabitants being constructed with
  $\lambda$-abstractions just like in the $\lambda$-calculus; we do
  not have the construct that directly applies a function to an
  argument, but a basic way in which a continuation consumes a
  function is to offer an argument $t$ as the input of the function,
  together with a continuation $e$ that can consume the output of the
  function; hence the continuation construct $\cons t e$ (which is
  simply the usual stacking construct that can be found in abstract
  machines to implement computation in the $\lambda$-calculus).
\end{itemize}

This intuition will be strengthened by the reduction rules for System~\SysL,
but we first start with an example.

The following story is borrowed from Phil Wadler~\cite{Wadlerdual}
(who might have borrowed it from Peter Selinger), and illustrates the
computational contents of classical proofs:

\begin{example}[The devil, the fool, and the \$1,000,000]

The Devil meets a man and says:\\
``- I have an offer for you! I promise you that
  \begin{centre}
    {\em either I offer you \$1,000,000 or, if you give me \$1,000,000, then I will grant you any wish.}
  \end{centre}
  Actually, I choose to offer you the latter.''

The man then goes back home and, motivated by the Devil's
promise, strives to gather \$1,000,000. Ten years later, he finally
succeeds; he goes back to the Devil and, handing him the money, says:

  ``- Here's \$1,000,000! I want immortality.''\\
The Devil takes the money and says:\\
``- Well done and thank you!\\
  Actually, I've changed my mind. I've now decided to fulfil my promise by offering you \$1,000,000. Here is your money back!''
\end{example}

The reason why that short story illustrates the computational contents
of classical logic is that the Devil behaves as a proof of the Law of
Excluded Middle: Imagine that
\begin{itemize}
\item the money (\$1,000,000) can be seen as an atomic proposition $a$
\item the part of the promise ``If you give me \$1,000,000, I'll grant you any wish'' can be seen as 
  the formula $a\Rightarrow\bot$, \ie $\neg a$;
\end{itemize}
the Devil is then the proof of $a\lor\neg a$ shown in Fig.~\ref{fig:devil}.
\begin{bfigure}[!h]
  \[
  \infer{\Seq  {} {\XcolY{\m \alpha{\cutc{\inj 2{\l y.\m \beta{\cutc{\inj 1{y}}{\alpha}}}}{\alpha}}} {a\ou\neg a}\semcol}}
        {
          \infer{\XcolY{\cutc{\inj 2{\l y.\m \beta{\cutc{\inj 1{y}}{\alpha}}}}{\alpha}}{(\Seq {} {\alpha\col a\ou \neg a})} }
                {
                  \infer{\Seq {} {\XcolY{\inj 2{\l y.\m \beta{\cutc{\inj 1{y}}{\alpha}}}} {a\ou\neg a}\semcol \alpha\col a\ou \neg a}}
                        {
                          \infer{\Seq {} {\XcolY{\l y.\m \beta{\cutc{\inj 1{y}}{\alpha}}} {\neg a}\semcol \alpha\col a\ou \neg a}}
                                {
                                  \infer{\Seq {y\col a} {\XcolY{\m \beta{\cutc{\inj 1{y}}{\alpha}}}{\bot}\semcol \alpha\col a\ou \neg a}}
                                        {
                                          \infer{\XcolY{\cutc{\inj 1{y}}{\alpha}}{(\Seq {y\col a} {\alpha\col a\ou \neg a,\beta\col \bot})}}
                                                {
                                                  \infer{\Seq {y\col a} {\XcolY{\inj 1{y}} {a\ou\neg a}\semcol\alpha\col a\ou \neg a,\beta\col \bot}}
                                                        {
                                                          \infer{\Seq {y\col a} {y\col a\semcol\alpha\col a\ou \neg a,\beta\col \bot}}
                                                                {}
                                                        }
                                                        \infer{\Seq  {y\col a\semcol\alpha\col a \ou \neg a}{\alpha\col a\ou \neg a,\beta\col \bot}}{}
                                                }
                                        }
                                }
                        }
                  \infer{\Seq  {\semcol\alpha\col a \ou \neg a}{\alpha\col a\ou \neg a}}{}
                }
        }
        \]
        \caption{A proof of LEM}
        \label{fig:devil}
\end{bfigure}
Indeed, following the bottom-up construction of the left-hand branch,
we see that
\begin{itemize}
\item the proof (the Devil) starts by choosing to prove $\neg a$, as
  reflected by the $\inj2{\_}$ construct;
\item that requires an input of type $a$ (the $\$1,000,000$ earned by
  the fool), namely $y$, as reflected by the $\lambda y.\_$ construct;
\item given the impossibility to prove $\bot$ directly (the
  immortality wish, or for that matter, any wish), the proof
  re-attacks the original formula to prove, namely $a\ou \neg a$ (the
  Devil returns to his original promise), but this time with the input
  $\XcolY y A$ (the $\$1,000,000$ that the fool gave him);
\item this time, the proof chooses to prove $a$, which is trivially
  done by returning $y$ (the Devil chooses to give $\$1,000,000$, by
  returning the money that the man earned).
\end{itemize}

We see here that the proof works because of the possibility to
construct an inhabitant of $a\ou \neg a$, twice along the same branch
(we inhabit it the first time with the second injection, then with the
first one), which is technically allowed by the
\emph{right-contraction} implicitly featured in the bottom two steps
of the proof. While in intuitionistic logic it is possible to contract
on the left but not contract on the right, classical logic allows both
symmetrically.

This allows to also build a proof-term of type PL and, allowing again
(as in Parigot's $\lambda\mu$) a top-level continuation variable
{\texttt{top}} of type $\bot$, we can build proof-terms for \emph{Ex
  falso quodlibet} and the elimination of double negation.

\medskip

In summary, we have seen that it is easy enough to introduce
proof-terms to represent classical proofs, such that the symmetry of
classical logic reflects the symmetry between programs and
continuations.

The use of classical reasoning corresponds to the use of control
features allowing programs to capture their continuation, as we now
see by looking at reductions:

\begin{definition}[Reductions]
  The reductions are given by the following rewrite system:
  \[
  \begin{array}{lclc}
    \rulenamed\rightarrow &\cutc{\lambda x.t_1}{\cons{t_2} e}&\Rew{} &\cutc{t_2}{\m x{\cutc{t_1}{e}}}\\
    \rulenamed\wedge&\cutc{\paire {t_1}{t_2}}{\textsf{inj}_i(e)}&\Rew{} &\cutc{t_i}{e}\\
    \rulenamed\vee&\cutc{\textsf{inj}_i(t)}{\paire {e_1}{e_2}}&\Rew{} &\cutc{t}{e_i}\\
    \rulenamed{\stackrel\leftarrow\mu}&\cutc{\mu \beta.c}{e}&\Rew{} &\subst c \beta e\\
    \rulenamed{\stackrel\rightarrow\mu}&\cutc{t}{\m x c}&\Rew{} &\subst c x t
  \end{array}
  \]
\end{definition}

Now, while it was very clear that Parigot's $\l\mu$ forms an extension
of $\lambda$-calculus, we should emphasise the fact that the
$\lambda$-calculus can be encoded in System~\SysL:

\begin{definition}[Encoding of $\l$-calculus]\label{def:lambdatoL}

We encode $\lambda$-terms as terms of System~\SysL\ by first encoding values, then all terms:
  \[
  \begin{array}{lll}
    \encv x&\eqdef& x\\
    \encv {\l x.M}&\eqdef& \l x.\overline M
  \end{array}
  \qquad
  \qquad
  \begin{array}{lll}
  \overline {V\ M_1\ \ldots\ M_n}\eqdef \m\alpha{\cutc{\encv V}{\st{\overline{M_1}}{\ldots\st{\overline{M_n}}{\alpha}}}}\\
  \flush\mbox{where $V$ is not an application and $n\geq 0$}
  \end{array}
  \]
\end{definition}

\begin{lemma}[Simulation of $\l$-calculus]\label{lem:lambdatoL}
  \begin{enumerate}
  \item $\m\alpha{\cutc{\overline M}{\st{\overline {M_1}}{\ldots\st{\overline {M_n}}\alpha}}}\Rew{}\overline{M\ M_1\ \ldots\ M_n}$.
  \item $\subst {\overline M} x {\overline N}\Rewn{}\overline{\subst M x N}$.
  \item If $M\Rew{\beta}N$ then $\overline M\Rewn{}\overline N$.
  \end{enumerate}
\end{lemma}
\begin{proof}
  The first point is a simple $\stackrel\leftarrow\mu$-reduction, the
  second point is by induction on $M$, the third point is by induction
  on the rewrite derivation.
\end{proof}

\begin{lemma}[Preservation of simple types]
  \begin{enumerate}
  \item If $\Deri[\simplambda] \Gamma V A$ then $\Seq[\SysL] \Gamma {\XcolY {\encv M} A\semcol}$
  \item If $\Deri[\simplambda] \Gamma M A$ then $\Seq[\SysL] \Gamma {\XcolY {\overline M} A\semcol}$
  \end{enumerate}
\end{lemma}

Since System~\SysL\ contains cuts, a proof of LEM, and it can encode
simply-typed $\lambda$-terms, it is clearly complete for classical
logic (in the same sense as for the $\l\mu$-calculus: EDN and EFQ
require the presence of a top-level continuation variable
$\textsf{top}\col\bot$). Soundness can be trivially checked by
checking that all the inference rules are sound when forgetting about
variables and proof-terms.

Substitution behaves well with respect to typing:
\begin{theorem}[Substitution Lemma]\label{lem:substL}
  \begin{enumerate}
  \item If $\XcolY c {(\Seq[\SysL]{\Gamma,\XcolY x A} \Delta)}$ and $\Seq[\SysL]{\Gamma} {\XcolY t A\semcol\Delta}$ then $\XcolY {\subst c x t} {(\Seq[\SysL]{\Gamma} \Delta)}$.
  \item If $\XcolY c {(\Seq[\SysL]{\Gamma} {\XcolY \alpha A,\Delta})}$ and $\Seq[\SysL]{\Gamma\semcol \XcolY e A} {\Delta}$ then $\XcolY {\subst c \alpha e} {(\Seq[\SysL]{\Gamma} \Delta)}$.
  \end{enumerate}
\end{theorem}
\begin{proof}
  By induction on $c$, simultaneously proving the two analogous properties for both terms and continuations.
\end{proof}

And the reduction relation satisfies Subject Reduction:
\begin{theorem}[Subject reduction for System \SysL]
  \begin{enumerate}
  \item If $\XcolY c {(\Seq[\SysL]{\Gamma} \Delta)}$ and $c\Rew{}c'$ then $\XcolY {c'} {(\Seq[\SysL]{\Gamma} \Delta)}$.
  \item If $\Seq[\SysL]{\Gamma} {\XcolY t A\semcol\Delta}$ and $t\Rew{}t'$ then $\Seq[\SysL]{\Gamma} {\XcolY {t'} A\semcol\Delta}$.
  \item If $\Seq[\SysL]{\Gamma\semcol\XcolY e A} {\Delta}$ and $e\Rew{}e'$ then $\Seq[\SysL]{\Gamma\semcol\XcolY {e'} A} {\Delta}$.
  \end{enumerate}
\end{theorem}
\begin{proof}
  Straightforward induction on the rewrite derivations.
\end{proof}

Again, Subject Reduction allows the rewrite system to describe a proof
transformation procedure in the classical sequent calculus, and in
this case it is \emph{cut-elimination}~\cite{Gentzen35}.

Let us see the other properties we mentioned when introducing the
Curry-Howard methodology:

Progress depends of course on what we consider an ``undesirable
structure''. In the case of sequent calculus, the natural concept of
undesirable structure is the cut, which in the typing system of
\SysL\ is (at least at first sight) represented as the bottom-most
rule of Fig.~\ref{fig:Ltyping}. And at this point we notice that some
cuts cannot be reduced, as no rewrite rule applies to their
proof-terms, namely those of the form $\cutc x e$ and $\cutc t \alpha$
when $e$ is not of the form $\m x c$ and $t$ is not of the form
$\m\alpha c$. We may think progress fails (in terms of
cut-elimination), but we should also notice that cuts of that form are
very peculiar: they do nothing but respectively implement a
left-contraction or a right-contraction, two rules that the extra
structure of the system requires for completeness (compared to \eg
\Giii~\cite{TS}):
\[
\infer{\cutc{x}{e}\col(\Seq{\Gamma,x\col A}{\Delta})}{
  \infer{\Seq{\Gamma,x\col A}{{{x}\col {A}}\semcol\Delta}}{}
  \quad
  \Seq{\Gamma,x\col A\semcol{{e}\col {A}}}{\Delta}
}
\qquad
\infer{\cutc{t}{\alpha}\col(\Seq{\Gamma}{\alpha\col A,\Delta})}{
  \Seq{\Gamma}{{{t}\col {A}}\semcol\alpha\col A,\Delta}
  \quad
  \infer{\Seq{\Gamma\semcol{{\alpha}\col {A}}}{\alpha\col A,\Delta}}{}
}
\]

We actually used two of these special ``cuts'' in the proof of LEM
showed in Fig.~\ref{fig:devil}, and we would not expect to eliminate
them (unless we had specific constructs for contractions and for the
axiom represented as $\cutc x \alpha$).

Concerning normalisation, it can be proved that typed commands (\resp
terms, continuations) are strongly normalising. This was inferred from
Urban's calculus in~\cite{lengrand03call-by-value}, but can be more
simply obtained as the direct application of Barbanera and Berardi's
technique, as shown in~\cite{Polo04} for a variant of
System \SysL\ with explicit substitutions. This will be the topic of
Chapter~\ref{ch:ortho}.

Finally, we look at the confluence property.

\section{Non-confluence of cut-elimination in classical logic}

As the reduction relation of System \SysL\ specifies a cut-elimination
procedure, we should note that cut-elimination in classical logic, at
a purely logical level, can easily be defined as a non-confluent
transformation procedure. A typical example of this is Lafont's
example, which we first express in the original sequent calculus \LK,
with explicit rules for weakenings and contractions and
``context-splitting'' rules (see \eg\cite{TS}):

\begin{example}[Lafont's example for non-confluence]

Consider the following cut that we would like to eliminate
  \[
  \infer{\Seqmm {\Gamma,\Gamma'}{\Delta,\Delta'}}{
    \infer{\Seqmm{\Gamma}{\Delta,A}}{\concludes{\Seqmm \Gamma\Delta}[\pi]{}}
    \quad 
    \infer{\Seqmm{\Gamma', A}{\Delta'}}{\concludes{\Seqmm {\Gamma'}{\Delta'}}[\pi']{}}
  }
  \]

  There are two ways to eliminate the cut:
  \[
  \infer{\Seqmm {\Gamma,\Gamma'}{\Delta,\Delta'}}{
    \concludes{\Seqmm{\Gamma}\Delta}[\pi]{}
  }
  \mbox{\quad or \quad}
  \infer{
    \Seqmm {\Gamma,\Gamma'}{\Delta,\Delta'}
  }{
    \concludes{\Seqmm{\Gamma'}{\Delta'}}[\pi']{}
  }\]
\end{example}

This obviously leads to non-confluence as soon as $\pi$ and $\pi'$ are
two distinct proofs (say, cut-free).  Note that we could, somewhat
artificially, avoid the choice between $\pi$ and $\pi'$ by considering
the following \Index{mix} rule~\cite{FleuryR94}:
\[\infer{\Seqmm {\Gamma,\Gamma'}{\Delta,\Delta'}}
        {\Seqmm\Gamma\Delta\quad\Seqmm{\Gamma'}{\Delta'}}
\]
which would allow the symmetric combination of $\pi$ and $\pi'$ into a
single proof (and what would be the semantics of this
combination?). The question is whether we accept this derivation
as a normal proof. Let us look at the same example in a sequent
calculus (such as \Giii~\cite{TS}) where rules are
``context-sharing'':

\begin{example}[Lafont's example in a context-sharing sequent calculus]\label{ex:lafont2}

The following cut:
\[\infer{\Seqmm {\Gamma}{\Delta}}{
  \infer{\Seqmm{\Gamma}{\Delta,A}}{\concludes{\Seqmm \Gamma\Delta}[\pi]{}}
  \quad  
  \infer{\Seqmm{\Gamma, A}{\Delta}}{\concludes{\Seqmm {\Gamma}{\Delta}}[\pi']{}}
}
\]
can be reduced to:
\[\concludes{\Seqmm {\Gamma}\Delta}[\pi]{}
\mbox{\quad or \quad}
\concludes{\Seqmm {\Gamma}\Delta}[\pi']{}
\]
\end{example}
What is even more striking in this example is that $\pi$ and $\pi'$
are two proofs of the same sequent, which we probably do not want to
consider denotationally equal and whose combination via the following
rule 
\[\infer{\Seqmm{\Gamma}\Delta}
{\Seqmm{\Gamma}\Delta\quad\Seqmm{\Gamma}\Delta}
\]
looks even more artificial than with the context-splitting mix. Again, do
we want this derivation as a normal proof?

Unsatisfying though the mix may seem, it does technically solve
Lafont's non-confluence problem based on two
weakenings. Unfortunately, it cannot solve the even more problematic
example obtained with contractions instead of weakenings:

\begin{example}[Example with contractions]
  The following cut
\[
\infer{\Seqmm {\Gamma}{\Delta}}{
  \infer{\Seqmm{\Gamma}{\Delta,A}}{
    \concludes{\Seqmm \Gamma{\Delta,A,A}}[\pi]{}
  }
  \quad  
  \infer{\Seqmm{\Gamma, A}{\Delta}}{
    \concludes{\Seqmm {\Gamma, A, A}{\Delta}}[\pi']{}
  }
}
\]
can be reduced to
\[
\concludes{\Seqmm\Gamma\Delta}[\pi{(\simeq)}]{
  \concludes{\bullet}[\pi']{}
  \quad
  \concludes{\bullet}[\pi']{}
}
\mbox{\qquad or\qquad}          
\concludes{\Seqmm\Gamma\Delta}[\pi'{(\simeq)}]{
  \concludes{\bullet}[\pi]{}
  \quad
  \concludes{\bullet}[\pi]{}}
\]
where $\pi{(\simeq)}$ (\resp $\pi'{(\simeq)}$) denotes the proof $\pi$
(\resp $\pi'$) modified by the propagation of $\pi'$ (\resp $\pi$) into
its structure.

We can give a concrete instance of the above:
\[
\infer{\Seqmm{(A\rightarrow B)\rightarrow A,A\rightarrow C,A\rightarrow D}{C\wedge D}}{
  \Seqmm{(A\rightarrow B)\rightarrow A}{A}
  \qquad\qquad
  \Seqmm{A,A\rightarrow C,A\rightarrow D} {C\wedge D}
}
\]
Peirce's Law requires a right-contraction on the cut-formula $A$ while the right-hand side
proof requires a left-contraction on the cut-formula $A$.
\end{example}      

Coming back to the proof-term side, both examples would appear in
System \SysL\ as instances of the following scheme:
\[
\infer{{\cutc {\m \alpha c}{\m x {c'}}}\col(\Seq\Gamma\Delta)}{
  \infer{\Seq\Gamma{{\m \alpha c\col A\semcol\Delta}}}{
    c\colon(\Seq{\Gamma}{\alpha\col A,\Delta})}
  \qquad
  \infer{\Seq{\Gamma\semcol\m x {c'}\col A}{\Delta}}{
    c'\col(\Seq{\Gamma,x\col A}{\Delta})}
}
\]

$\alpha$ (\resp $x$) could be used 0 ({weakening}), 1, or several
({contraction}) times in $c$ (\resp $c'$).

That cut could be reduced to
\[
\winfer{{\subst c \alpha{\m x {c'}}}\col(\Seq\Gamma\Delta)}{
  {c\colon(\Seq{\Gamma}{\alpha\col A,\Delta})}
  \qquad
  \infer{\Seq{\Gamma\semcol\m x {c'}\col A}{\Delta}}{
    c'\col(\Seq{\Gamma,x\col A}{\Delta})}
}
\qquad\mbox{ or }\qquad
\winfer{{\subst {c'} x{\m \alpha {c}}}\col(\Seq\Gamma\Delta)}{
  \infer{\Seq\Gamma{{\m \alpha c\col A\semcol\Delta}}}{
    c\colon(\Seq{\Gamma}{\alpha\col A,\Delta})
  }
  \qquad
      {c'\col(\Seq{\Gamma,x\col A}{\Delta})}
}
\]
where dotted lines do not represent primitive inference rules, but
inference rules that have been shown admissible in the typing system
(Lemma~\ref{lem:substL}).

In the case of weakening, and reflecting Example~\ref{ex:lafont2},
$\alpha\notin\FV c$ and $x\notin\FV{c'}$ and we can reduce ${\cutc {\m
    \alpha c}{\m x {c'}}}$ to two arbitrary commands $c$ and $c'$ with the same
type.

This makes it hard to give a denotational semantics of classical
proofs or of typed proof-terms: if we require $\cutc {\m \alpha c}{\m
  x {c'}}$, $\subst c \alpha{\m x {c'}}$, and $\subst {c'} x{\m \alpha
  {c}}$, to have the same denotation, Example~\ref{ex:lafont2} leads
to giving the same denotation to every proof of the same sequent.

This of course relates to the fact that we have already mentioned: a
CCC with initial object where every object $A$ naturally isomorphic to
$\neg\neg A$ collapses to a boolean algebra. As identified
in~\cite{StrassburgerHDR}, there are three natural ways to resolve
this:
\begin{enumerate}
\item Break the symmetry between $\et{}{}$ and $\ou{}{}$
\item Break the cartesian product (as studied for instance in~\cite{PymFu04,LamStr05LICS,lamarche07,StrassburgerHDR})
\item Break curryfication (as studied for instance in~\cite{DosenPetric04book,CockettS09})
\end{enumerate}

In this dissertation, we break the symmetry between $\et{}{}$ and
$\ou{}{}$, since out of the three solutions it is the one for which
the Curry-Howard correspondence with programs is best understood.

One way of breaking the non-confluence problem
\[
\infer{\Seq \Gamma \Delta}{
  \concludes{\Seq \Gamma {\Delta,A}}[\pi]{}
  \qquad
  \concludes{\Seq {\Gamma,A} \Delta}[\pi']{}
}
\]
is simply to give systematic priority to 
\begin{itemize}
\item the right (push $\pi$ into $\pi'$)
\item or to the left (push $\pi'$ into $\pi$)
\end{itemize}

Almost by definition, both solutions make the calculus confluent. 

They also break the $\wedge\vee$ symmetry: Giving systematic priority
to the right, say, makes a term $t$ of type $A\et B$ have the same
behaviour as $\paire{\inj1 t}{\inj2 t}$, whereas a continuation e of
type $A\ou B$ will not necessarily have the same behaviour as $\paire{\inj1
  e}{\inj2 e}$. 

More details on this will be given in Chapter~\ref{ch:polarfocus}, but
we shall also see by semantical means that the $\wedge\vee$ symmetry
is broken. The two reduction strategies suggest to construct two
denotational semantics $\semN c$ and $\semV c$ with the hope
that:\\ $\semN {c_0}=\semN {c_1}$\qquad iff\qquad ``$c_0\Rewsn{}c_1$
with systematic priority to the right''\\ $\semV {c_0}=\semV
{c_1}$\qquad iff\qquad ``$c_0\Rewsn{}c_1$ with systematic priority to
the left''

The use of the letters ${\sf N}$ and ${\sf V}$ reflects the fact that
the strategies relate to the notions of {Call-by-name} and
{Call-by-value}, as investigated for instance by
Plotkin~\cite{Plotkin75}, Moggi~\cite{Moggi:lics1989}, and others.

\medskip

In conclusion of this section, we have seen that it is easy enough to
give a rewrite system on proof-terms to represent cut-elimination (and
the system follows the intuitions of continuations and control), but
it gives a non-confluent calculus because cut-elimination is
non-confluent in classical logic (via the Curry-Howard correspondence,
because programs and continuations fight for the control of
computation).

The rest of this chapter is devoted to the construction of the above
\CBN\ and \CBV\ semantics.

\section{Continuations, Call-by-Name and Call-by-Value}\label{sec:cbncbv}

Call-by-name and call-by-value are two strategies for evaluating
programs. Imagine the definition of a function (in pseudo-code):
\begin{verbatim}
MyFavoriteFunction(x){
    ... x ...
}
\end{verbatim}
and later a call to that function with argument \verb=A=
\begin{verbatim}
MyFavoriteFunction(A)
\end{verbatim}
The main question is whether \verb=A= should be evaluated before
entering the (code of the) function (\CBV) or when it is actually used
(\CBN)? This is a question of interpretation or compilation of
programs and, especially in presence of side-effects, knowing which of
the two the compiler implements, is vital for the determinism of evaluation.

In general, we call \Index[value]{values} what evaluation should
produce (\eg booleans \textsf{true}, \textsf{false}). In functional
programming, functions are particular values and can be passed as
arguments. In general, functions are therefore not reduced.

The $\lambda$-calculus is both a core functional language and a theory
of functions. 

As a core functional language, it is equipped with an operational
semantics, close to implementation, which can be expressed by an
evaluation strategy that selects a unique $\beta$-redex to reduce:
\begin{itemize}
\item Never reduce a $\l$-abstraction, as it is a ``value''
  \flush (this is called \Index{weak reduction})
\item Always reduce $M$ first in an application $M\;N$. Then:
  \begin{itemize}
  \item If $M$ is an abstraction:\hfill
    reduce the $\beta$-redex first (\CBN)\\
    ${} $\hfill reduce $N$ first (\CBV)
    \item Otherwise, reduce $N$
      \flush{(never happens with closed terms)}
  \end{itemize}
\end{itemize}
We denote those strategies $\Rew{\CBN}$ and $\Rew{\CBV}$.\footnote{For
  instance, Haskell implements \CBN\ while OCaml implements \CBV.}

As a theory of functions, the $\lambda$-calculus is equipped with a
denotational semantics close to the mathematical notion of functions:
in particular, equalities are congruences (\eg if $M=N$ then $\l
x.M=\l x.N$) and reductions are congruences (this is called
\Index{strong reduction}). In~\cite{Plotkin75}, Plotkin investigated
the concepts of call-by-name and call-by-value by identifying
particular $\lambda$-terms as values:
\begin{definition}[Value, $\beta_v$, Call-by-Name and Call-by-Value]

  $\lambda$-terms of the form $\l x.M$ and $x$ are called
  \Index[value]{values}, and denoted $V$, $V'$, etc, while
  $\lambda$-terms of the form $M N$ are not values.\footnote{The
    intuition is that, by evaluating $M N$, you may get a
    $\lambda$-term of a completely different shape.}
  \begin{itemize}
  \item ``Call-by-name'' evaluation is given by general
    $\beta$-reduction
    \[\rulenamed \beta\qquad(\l x.M)\;N\Rew{}\subst M x N\]
  \item
    ``Call-by-value'' evaluation is given by the restriction of
    $\beta$-reduction where the argument is a value
    \[\rulenamed {\beta_v}\qquad (\l x.M)\;V\Rew{}\subst M x V\]
  \end{itemize}
\end{definition}

Now, natural questions to raise are
\begin{itemize}
\item[\CBN:] whether there is a relation between $\Rew{\CBN}$ and
  $\Rew{\beta}$;
\item[\CBV:] whether there is a relation between $\Rew{\CBV}$ and
  $\Rew{\beta_v}$?
\end{itemize}
Clearly,\hfil $\Rew{\CBN}\ \subseteq\ \Rew{\beta}$ and
$\Rew{\CBV}\ \subseteq\ \Rew{\beta_v}$, but what about the
converse? A bridge between weak and strong reductions was given by
Plotkin~\cite{Plotkin75}:
\begin{theorem}[\CBN\ and \CBV: weak and strong reductions]
  \begin{itemize}
  \item[\CBN:]
    $\Rewn{\beta}$ is the closure of $\Rewn{\CBN}$ under \hfill
    $\infer{M_1\Rew{}C[M_3]}{M_1\Rewn{\CBN}C[M_2]\qquad M_2\Rew{}M_3}$\\
  \item[\CBV:]
    $\Rewn{\beta_v}$ is the closure of $\Rewn{\CBV}$ under \hfill 
    $\infer{M_1\Rew{}C[M_3]}{M_1\Rewn{\CBV}C[M_2]\qquad M_2\Rew{}M_3}$
  \end{itemize}
  where $C[\ ]$ ranges over any kind of context (\ie $\l$-term with a
  hole).
\end{theorem}

The point of this result is that we shall now call \CBN\ and \CBV, not
some operational semantics of some functional programming language,
but some rewriting theories in $\lambda$-calculus.

As we have already mentioned compilation in reference to \CBN/\CBV, it
is interesting to see that $\l$-calculus can be compiled into (a
fragment of) itself: this is based on the idea of the program
transformation presented in Section~\ref{sec:continuationscontrol}
using continuations. As continuations are passed as an extra argument
to every call, such transformations are known as \Index{Continuation
  Passing Style} (\CPS)-translations.

\begin{definition}[\CPS-translations]\label{def:CPS1}

Two important \CPS-translations were defined for \CBN\ and \CBV:
\[
\begin{array}{c}
  \mbox{\CBN-translation (Plotkin~\cite{Plotkin75})}\\
  \begin{array}{clll}
    \underline{x}&\eqdef\l k.x\;k\\
    \underline{\lambda x.M}&\eqdef\lambda k.k\;(\lambda x.\underline{M})\\
    \underline{M\;N}&\eqdef\lambda k. \underline M\;(\l y.y\;\underline N\;k)
  \end{array}
\end{array}
\qquad
\begin{array}{c}
  \mbox{\CBV-translation (Reynolds~\cite{Reynolds72})}\\
  \begin{array}{clll}
    \overline{x}&\eqdef\lambda k.k\;x\\
    \overline{\lambda x.M}&\eqdef\lambda k.k\;(\lambda x.\lambda k'.\overline{M}\ k')\\
    \overline{M\;N}&\eqdef\lambda k. \overline M\;(\l y.\overline N\;(\l z.y\;
    z\; k))
  \end{array}
\end{array}
\]
where the variables $k$ and $k'$ are always chosen to be fresh.
\end{definition}

One main feature of these translations is their target fragment of the
$\l$-calculus: in this fragment, arguments are always values!  This
fragment is stable under $\Rew{\beta}$ and $\Rew{\beta_v}$, which
actually coincide. The evaluation of a \CPS-translated term is
\Index{strategy-indifferent}.  How this evaluation relates to the
evaluation of the original term is given by the following simulation
properties:

\begin{theorem}[\CPS-translations preserve reductions]\label{th:CPSsimul}
  \begin{enumerate}
  \item[Soundness:]
    \begin{itemize}
    \item[\CBN] If $M\Rew{\beta}N$ then $\underline M\Rewn{\beta}\underline N$
    \item[\CBV] If $M\Rew{\beta_v}N$ then $\overline M\Rewn{\beta}\overline N$
    \end{itemize}
  \item[Completeness:]
    \begin{itemize}
    \item[\CBN] If $\underline M\Rewsn{\beta}\underline N$ then $M\Rewsn{\beta}N$
    \item[\CBV] It is \textbf{not} the case for \CBV, that if $\overline M\Rewsn{\beta}\overline N$ then $M\Rewsn{\beta_v}N$.
    \end{itemize}
  \end{enumerate}
\end{theorem}
\begin{proof}
  It is interesting to look at soundness to see how or why the
  \CPS-translations make sense; for complete proofs,
  see~\cite{Plotkin75}.
  \[
  \begin{array}{lcl}
    \underline{(\l x.M)\;N}&=&\lambda k. (\lambda k'.(k'\;(\lambda x.\underline{M})))\;(\l y.y\;\underline N\;k)\\
    &\Rew{}&\lambda k. (\l y.y\;\underline N\;k)\;(\lambda x.\underline{M})\\
    &\Rew{}&\lambda k. (\lambda x.\underline{M})\;\underline N\;k\\
    &\Rew{}&\lambda k. (\subst{\underline{M}}x {\underline N})\;k\\
    &=&\lambda k. (\underline{\subst{M}x N})\;k\\
    &\Rew{}&\underline{\subst{M}x N}
  \end{array}
  \]
  The second equality of course relies on the property that the
  \CPS-translation behaves well with substitution:
  ($\subst{\underline{M}}x {\underline N}=\underline{\subst{M}x
    N}$). The last rewrite is an instance of $\beta$-reduction because
  $\underline{\subst{M}x N}$ necessarily starts with a
  $\lambda$-abstraction (\ie we are not using $\eta$-reduction).

  \[
  \begin{array}{lcl}
    \overline{(\l x.M)\;V}&=&\lambda k. (\lambda k'.(k'\;(\lambda xk''.\overline{M}\;k'')))\;(\l y.\overline V\;(\l z.y\;z\;k))\\
    &\Rew{}&\lambda k. (\l y.\overline V\;(\l z.y\;z\;k))\;(\lambda xk''.\overline{M}\;k'')\\
    &\Rew{}&\lambda k. \overline V\;(\l z.(\lambda xk''.\overline{M}\;k'')\;z\;k)\\
    &\Rew{}&\lambda k. \overline V\;(\l z.(\lambda k''.\subst{\overline{M}}x z\;k'')\;k)\\
    &=&\lambda k. \overline V\;(\l x.(\lambda k''.\overline{M}\;k'')\;k)\\
    &\Rew{}&\lambda k. \overline V\;(\l x.\overline{M}\;k)\\
    &\Rew{}&\lambda k. (\l x.\overline{M}\;k)\; H\hfill\mbox{where }\overline V=\lambda k.k\;H\\
    &\Rew{}&\lambda k. (\subst{\overline{M}}x H)\;k\\
    &=&\lambda k. (\overline{\subst{M}x V})\;k\\
    &\Rew{}&\overline{\subst{M}x V}
  \end{array}
  \]
  The second equality is simply the renaming of $z$ into $x$; the third
  one relies again on the property that the \CPS-translation behaves
  well with substitution by values ($\subst{\underline{M}}x
  {H}=\underline{\subst{M}x V}$ if $\underline V=\lambda k.k H$). The
  last rewrite is again an instance of $\beta$-reduction, not
  $\eta$-reduction.

  The point here is to realise that if $V$ had not been a value, then
  $\overline V$ would not be of the form $\lambda k.k H$, and the
  simulation of this specific $\beta$-reduction would be stuck.
\end{proof}

The above simulations give some intuition about the encodings: the
translation of any term $M$ starts with a $\l$-abstract on a fresh
variable $k$ that is used exactly once. The variable $k$ stands for
the current continuation (hence the expression
\emph{continuation-passing style}). In the encoding of an abstraction
$\l x.M$ (which is a value), the current continuation is applied to
the encoding of the body $M$ under a $\lambda$-abstraction on $x$. In
case of an application $M\;N$, the current continuation is not
directly applied, but wrapped in a bigger continuation that is passed
as an argument to the encoding of $M$; what this wrapping exactly is
depends on whether we do \CBN\ or \CBV\ and will determine whether we
reflect the evaluation of $N$ as a value $V$ before we reflect the
reduction of $M\;N$.

Now, the fact that $\overline M\Rewsn{\beta}\overline N$ does not
imply $M\Rewsn{\beta_v}N$ is slightly disappointing: one way to look
at it is to consider that $M\Rewsn{\beta_v}N$ is too weak, or
incomplete, for Call-by-Value. Indeed, the inspiration from monads,
and Moggi's monadic $\lambda$-calculus~\cite{Moggi:lics1989}, has
allowed the extension of the Call-by-Value $\lambda$-calculus into a
sound and complete calculus with respect to the
\CBV\ \CPS-translation (see for instance~\cite{LengrandPhD}).

We now turn to the behaviour of the \CPS-translations with respect to
typing: Assume we have $\Deri \Gamma M A$. Do we have: $\Deri
{\Gamma'} {\underline M} {A'}$ (for some $\Gamma'$, $A'$) and $\Deri
{\Gamma''} {\overline M} {A''}$ (for some $\Gamma''$, $A''$)?

The \CPS-translations reveal two classes of terms in the target:
\emph{values} \& \emph{continuations} (like $k$). The types of values
and continuations in the translated terms depend on \CBN\ or \CBV:

\begin{definition}[\CPS-translations of simple types]\label{def:CPStypes}

  We choose or we add a particular atomic type $R$, called the
  \Index{response type}, then we define
  \[\begin{array}{c}
  \CBN\\
  \begin{array}{ll}
    \underline a&\eqdef a\\
    \underline{A\rightarrow B}&\eqdef((\underline A\arr R)\arr R)\arr (\underline B\arr R)\arr R
  \end{array}
  \end{array}
  \qquad
  \begin{array}{c}
    \CBV\\
    \begin{array}{ll}
      \overline a&\eqdef a\\
      \overline{A\rightarrow B}&\eqdef\overline A\arr (\overline B\arr R)\arr R
    \end{array}
  \end{array}
  \]
\end{definition}

Intuitively, a type $A$ in the original calculus will give rise to a
type $\underline A$ (\resp $\overline A$) of ``$A$-values'';
continuations are functions consuming those and returning something in
the response type $R$ (which is abstract in the sense that we will
never need to know what it is), so continuations will therefore be of
type $\underline A\arr R$ (\resp $\overline A\arr R$). 

The encoding $\underline M$ (\resp $\overline M$) of a term $M$ of
type $A$ will take the current continuation, of type $\underline A\arr
R$ (\resp $\overline A\arr R$), and using that continuation, it will
eventually output a response in the response type (as we have seen,
the encoding starts with $\lambda k.\ldots$). It will therefore be of
type $(\underline A\arr R)\arr R$ (\resp $(\overline A\arr R)\arr R$).

This is formalised as the following theorem:

\begin{theorem}[\CPS-translations preserve types]\label{th:CPStyping}

  If $\Deri \Gamma M A$ then
  $\Deri {(\underline\Gamma\arr R)\arr R} {\underline M} {(\underline A\arr R)\arr R}$ and
  $\Deri {\overline\Gamma} {\overline M} {(\overline A\arr R)\arr R}$,

  where $\overline{\XcolY{x_1}{A_1},\ldots,\XcolY{x_n}{A_n}}$ stands
  for $\XcolY{x_1}{\overline{A_1}},\ldots,\XcolY{x_n}{\overline{A_n}}$\\
  and $((\underline{\XcolY{x_1}{A_1},\ldots,\XcolY{x_n}{A_n}})\arr
    R)\arr R$ stands for $\XcolY{x_1}{(\underline{A_1}\arr R)\arr
      R},\ldots,\XcolY{x_n}{(\underline{A_n}\arr R)\arr R}$.
\end{theorem}
\begin{proof}
  Straightforward induction on $M$.
\end{proof}

Variants of \CPS-translations exist, of which we mention two that are
related to \CBN\ and \CBV:
\begin{definition}[Variants]\label{def:CPS2}
  \begin{itemize}
  \item Fischer's translation for \CBV~\cite{Fischer:pacp1972}
    \[
    \begin{array}{clll}
      \overline{x}&\eqdef\lambda k.k\;x\\
      \overline{\lambda x.M}&\eqdef\lambda k.(k\;({ \lambda k'}.\lambda x.\overline{M}{\ {k'}}))\\
      \overline{M\;N}&\eqdef\lambda k. \overline M\;(\l y.\overline N\;(\l z.y\;
      {k\; z}))
    \end{array}    
    \hfill\hfill
    \begin{array}{ll}
      \overline a&\eqdef a\\
      \overline{A\rightarrow B}&\eqdef{(\overline B\arr R)\rightarrow \overline A\arr R}
    \end{array}
    \]
  \item Hofmann \& Streicher's translation for \CBN~\cite{HofmanStreicher97}, using product types
    \[
    \begin{array}{clll}
      \underline{x}&\eqdef\l k.x\;k\\
      \underline{\lambda x.M}&\eqdef\lambda \paire x k.\underline{M}\ k\\
      \underline{M\;N}&\eqdef\lambda k. \underline M\;\paire{\underline N}k
    \end{array}
    \hfill\hfill
    \begin{array}{ll}
      \HStrans{ a}&\eqdef a\arr R\\
      \HStrans{A\rightarrow B}&\eqdef(\HStrans A\arr R)\times \HStrans B
    \end{array}
    \]
\end{itemize}
\end{definition}

Fischer's \CBV-translation is very similar to Reynolds's: they only
differ in the order in which arguments are passed in the encoding of
$\l$-abstractions and applications (\eg for the abstraction,
Reynolds's translation binds $x$ first, then binds the continuation
variable $k'$, whereas Fischer's binds $k'$ first, then $x$). This is
reflected in the encoding of the function type $A\arr B$: the two
arguments are swapped.

Hofmann \& Streicher's \CBN-translation differs more importantly from
Plotkin's, as fewer ``continuation wrappings'' are introduced,
reflected in the number of $\cdots\arr R$ in the encoding of types:
that encoding works ``negatively'', as $\HStrans A$ is directly the
type of $A$-continuations which are not necessarily functions
consuming an $A$-value and returning in the response type.

Again, these translations allow the same simulations as Plotkin's and
Reynolds's, and of course preserve typing, with a slightly different
formulation in the case of \CBN:

\begin{theorem}[Hofmann \& Streicher's \CBN-translation preserves types]\label{th:HSCPStyping}

  If $\Deri \Gamma M A$  then  $\Deri {\HStrans\Gamma\arr R} {\underline M} {\HStrans A\arr R}$

  where $(\HStrans{\XcolY{x_1}{A_1},\ldots,\XcolY{x_n}{A_n}})\arr
    R$ stands for $\XcolY{x_1}{\HStrans{A_1}\arr R},\ldots,\XcolY{x_n}{\HStrans{A_n}\arr R}$.
\end{theorem}

Let us now look at \CPS-translations with respect to denotational
semantics: Remember that simply-typed $\lambda$-terms have a semantics
in a Cartesian Closed Category.  \CPS-translations compile the
simply-typed $\l$-calculus into itself (preserving types in the sense
of Theorem~\ref{th:CPStyping}), so we can now assign to a
simply-typed $\l$-term $M$, the semantics (in a CCC) of $\underline M$
or $\overline M$ (so that semantics now depends on \CBN/\CBV). By the
simulation theorem (Theorem~\ref{th:CPSsimul}), reductions are sound
\wrt their corresponding semantics.

More interestingly, notice that we do not need the whole structure of
a CCC to build those two semantics, as $\underline M$ or $\overline M$
live in a fragment of the simply-typed $\lambda$-calculus (the
\CPS-fragment), where in particular the types of $\underline M$ or
$\overline M$ are functional types. More than this, every functional
type that we ever need for that fragment is of the form $A\arr
R$.\footnote{To be precise, we did use types such as $A_1\arr \cdots
  \arr A_n\arr R$, but if we have products we can consider this to be
  the type $(A_1\times \cdots \times A_n)\arr R$.} Therefore, in order
to build the categorical semantics of the \CPS-fragment, we do not
need as strong axioms as those of a CCC: on top of asking for
cartesian products we only require the existence of exponentials
objects of the form $R^A$. This is called a \Index{response category}.

Now given a response category, the sub-category made of the objects
of the form $R^A$ is called a \Index{continuation category}, \aka
\Index{control category} (Selinger~\cite{SelingerControlCat99}).  Such
a category turns out to have a rich structure that proves very useful
for classical logic: not only it is a CCC (with exponential objects
${(R^A)}^{(R^B)}$ defined as $R^{A\times{(R^B)}}$) but objects of the form
$R^{A\times B}$, denoted $R^A\Or R^B$, will play an important role.

\section{Classical logic and \CBN/\CBV}\label{sec:classicalCBNCBV}

We now relate classical logic to the above notions. We first review
known translations from classical logic into intuitionistic logic: The
intuition is that we can always turn $P$ into $P'$ by adding (enough)
double negations, to get the property that \ctr{If $\vdash_c P$ then
  $\vdash_i P'$.}  where $\vdash_c$ denotes classical provability and
$\vdash_i$ denotes intuitionistic provability.
Obviously, $\vdash_c P\leftrightarrow P'$, since the two formulae only
differ by some double negations.

A potential question is then: If it suffices to add double negations
in a classically provable formula to make it intuitionistically
provable, are the two logics \emph{really} different? Well, they
differ at least in the sense that double negations break some of the
nice properties of intuitionistic logic:

\ctr{If $\Seq[i]{}{A_1\vee A_2}$ then either $\Seq[i]{}{A_1}$ or $\Seq[i]{}{A_2}$.\\
If $\Seq[i]{}{\exists x A}$ then there is $t$ such that $\Seq[i]{}{\subst A x t}$}
Getting $t$ from the proof of $\Seq[i]{}{\exists x A}$ is called \Index{witness extraction}.
This can also be done in some theories, like (Heyting) arithmetics:
\ctr{If $\Seq[i]{HA}{\exists x A}$ then there is $t$ such that $\Seq[i]{HA}{\subst A x t}$.}

But in the most general case we cannot have the same properties when
$\Seq[i]{}{\neg\neg(A_1\vee A_2)}$ or $\Seq[i]{}{\neg\neg{\exists x
    A}}$. So what to do with a classical proof of $\Seq{}{{\exists x
    A}}$ is unclear. However, it is known that if $A$ satisfies some
specific property, a witness may be obtained from a classical proof of
$\Seq{}{{\exists x A}}$; this is called classical witness extraction,
and we will see this in Chapter~\ref{ch:ortho}.

The principle of inserting double negations gives rise to double negation translations (or
$\neg\neg$-translations), of which we present two, remembering that $\neg A$ is $A\imp\bot$:

\begin{definition}[Double negation translations]
\[
\begin{array}{ll}
  \nnot a&\eqdef a\\
  \nnot{(A\imp B)}&\eqdef((\nnot A\imp\bot)\imp\bot)\imp 
  (\nnot B\imp\bot)\imp\bot
\end{array}
\qquad
\begin{array}{ll}
\nott a&\eqdef a\\
\nott{(A\imp B)}&\eqdef\nott A\imp (\nott B\imp\bot)\imp \bot
\end{array}
\]
\end{definition}

We realise here that these translations, via the Curry-Howard
correspondence, are exactly the translations of types from
Definition~\ref{def:CPStypes} that make Plotkin's and Reynolds's
translations ``preserve types'': The response type previously denoted
$R$ corresponds to the formula $\bot$, and a continuation is a proof
of negation.

\subsection{Identifying \CBN\ and \CBV\ in System~\SysL}

The fact that double negation translations allow the construction of
an intuitionistic proof of $\nnot A$ (\resp $\nott A$) from a
classical proof of $A$, suggests that we can adapt the
\CPS-translations of Definitions~\ref{def:CPS1} and~\ref{def:CPS2} to
encode classical proof-terms, say of System \SysL, into the
simply-typed $\lambda$-calculus. If this encoding not only preserves
types but also reductions (as in \eg Theorem~\ref{th:CPSsimul}), then
we could assign to a classical proof-term the categorical semantics of
its \CPS-encoding (which, as a simply-typed $\lambda$-term, is
well-understood).

It remains to identify which reductions of System \SysL\ will be
reflected in the \CPS-encoding.

Inspired by Theorem~\ref{th:CPSsimul}, we remark that the
$\beta$-reductions that can be reflected by the \CBV-encoding are of
the form $\beta_v$, \ie those reductions where every substitution that
is computed substitute a variable by a value. In System \SysL, we can
impose similar restrictions: a \CBV-reduction should only allow a
substitution $\subst c x t$ to be computed if $t$ is a ``value''; and
by symmetry, we could expect \CBN-reduction to only allow a
substitution $\subst c \alpha e$ to be computed if $e$ is a
``continuation value''. But we still need to identify what the notions
of values and continuation values are for System~\SysL. Considering
the non-confluence situation $\cutc{\m\alpha c}{\m x {c'}}$ described
in Section~\ref{sec:SysL} (which causes so much difficulty for
building semantics for System~\SysL), ruling out $\m\alpha c$ as value
and ruling out $\m x {c'}$ as continuation value solves the problem:
\CBN-reduction would allow the reduction $\cutc{\m\alpha c}{\m x
  {c'}}\Rew{}\subst{c'}x{\m\alpha c}$ and disallow $\cutc{\m\alpha
  c}{\m x {c'}}\Rew{}\subst{c}\alpha{\m x {c'}}$, while \CBV-reduction
would allow the reduction $\cutc{\m\alpha c}{\m x
  {c'}}\Rew{}\subst{c}\alpha{\m x {c'}}$ and disallow $\cutc{\m\alpha
  c}{\m x {c'}}\Rew{}\subst{c'}x{\m\alpha c}$.
In other words, \CBV-reduction gives priority to the right while 
\CBV-reduction gives priority to the left.

We can formalise this as the following definition:
\begin{definition}[\CBN\ and \CBV\ for System \SysL\ -first attempt]\label{def:CBNCBVshallow}\nopagesplit

  We identify the following notions of term values and continuation values:
  \[
  \begin{array}{lll}
    \mbox{Term values}&V
    &\recdef x\sep \lambda x.t \sep \paire{t_1}{t_2} \sep \textsf{inj}_i(t)\\ 
    \mbox{Continuation values}&E
    &\recdef \alpha \sep \cons t e \sep \paire{e_1}{e_2} \sep \textsf{inj}_i(e)
  \end{array}
  \]
  The reduction relations $\Rew{\CBN}$ and $\Rew{\CBV}$ are defined as
  the contextual closures of the (groups of) rules in Fig.~\ref{fig:LCBNCBV}.
\end{definition}
\begin{bfigure}[!h]
  \[
  \begin{array}{c}
    \begin{array}{lclc}
      \rulenamed\rightarrow &\cutc{\lambda x.t_1}{\cons{t_2} e}&\Rew{} &\cutc{t_2}{\m x{\cutc{t_1}{e}}}\\
      \rulenamed\wedge&\cutc{\paire {t_1}{t_2}}{\textsf{inj}_i(e)}&\Rew{} &\cutc{t_i}{e}\\
      \rulenamed\vee&\cutc{\textsf{inj}_i(t)}{\paire {e_1}{e_2}}&\Rew{} &\cutc{t}{e_i}\\
    \end{array}
    \\\\
  \begin{array}{c}
    \begin{array}{cc@{}l@{}l}
      \rulenamed{\stackrel\leftarrow\mu_{\sf N}}&\cutc{\mu \beta.c}{ E}&\Rew{} &\subst c \beta { E}\\
      \rulenamed{\stackrel\rightarrow\mu}&\cutc{t}{\m x c}&\Rew{} &\subst c x t\\\\
    \end{array}\\
    \CBN
  \end{array}
  \begin{array}{c}
    \begin{array}{cc@{}l@{}l}
      \rulenamed{\stackrel\leftarrow\mu}&\cutc{\mu \beta.c}{e}&\Rew{} &\subst c \beta e\\
      \rulenamed{\stackrel\rightarrow\mu_{\sf V}}&\cutc{{V}}{\m x c}&\Rew{} &\subst c x {V}\\\\
    \end{array}\\
    \CBV
  \end{array}
  \end{array}
  \]
  \caption{\CBN\ and \CBV\ reduction in System~\SysL\ (first attempt)}
  \label{fig:LCBNCBVshallow}
\end{bfigure}

The fact that \CBV-reduction keeps
$\rulenamed{\stackrel\leftarrow\mu}$ and restricts
$\rulenamed{\stackrel\rightarrow\mu}$ into
$\rulenamed{\stackrel\rightarrow\mu_{\sf V}}$ (and vice versa for
\CBN), is the formalisation of what we described before. 

Doing this ``works'' in the sense that both \CBN\ and \CBV\ reductions
are confluent systems (as higher-order orthogonal rewrite systems).

Unfortunately, these restrictions are not sufficient to build the
denotational semantics of those systems according to methodology of
\CPS-translations, at least if we are to re-use the \CPS-translations
of types that we have seen in Section~\ref{sec:cbncbv}: Indeed, the
simulation property that would be, for System~\SysL, the equivalent of
Theorem~\ref{th:CPSsimul}, fails.

This was noticed in an erratum of~\cite{CurienHerbelinDuality99},
which also notices that the simulation does work on two specific
fragments of System~\SysL: Concentrating on the implicational
fragment,
\begin{itemize}
\item \CBN-reduction can be simulated in the $\lambda$-calculus
  (according to Hofmann and Streicher's translation of
  types~\cite{HofmanStreicher97}) when every continuation of the form
  $\cons t e$ is such that $t$ is a term value;
\item \CBV-reduction can be simulated in the $\lambda$-calculus
  (according to Reynold's or Fischer's translation of
  types~\cite{Reynolds72,Fischer:pacp1972}) when every continuation of
  the form $\cons t e$ is such that $e$ is a continuation value.
\end{itemize}

Note that this makes sense because the former and the latter fragments
are stable under \CBN\ and \CBV\ reduction, respectively.

This also suggests how to refine the notions of term values and
continuation values as follows: instead of these notions concerning
only the top-level construct of a term or a continuation, our new and
more appropriate notions of values will be recursively defined.

This \emph{strong} notion of value is taken primarily
from~\cite{Wadlerdual}:\footnote{Inspired by~\cite{MunchCSL09}, we
  make a change about implication in the case of \CBV, for which we
  \emph{also} restrict continuation values, since this will make
  \CBV\ normal forms correspond to proofs in \LKQ~\cite{DJS95,DJS97},
  as we shall see in Chapter~\ref{ch:polarfocus}.}

\begin{definition}[\CBN\ and \CBV\ for System \SysL]\label{def:CBNCBV}

In \CBN, we identify the following notion of \Index[continuation value]{continuation
  values}:
    \[
    \begin{array}{lll}
      \mbox{\CBN\ continuation values}&E
      &\recdef \alpha \sep \cons t E \sep \paire{E_1}{E_2} \sep \textsf{inj}_i(E)\\
    \end{array}
    \]
    In \CBV, we identify the following notion of \Index[term
      value]{term values} and \Index[continuation value]{continuation
      values}:
    \[
    \begin{array}{lll}
      \mbox{\CBV\ term values}&V
      &\recdef x\sep \lambda x.t \sep \paire{V_1}{V_2} \sep \textsf{inj}_i(V)\\ 
      \mbox{\CBV\ continuation values}&F
      &\recdef \alpha \sep \cons V F \sep \paire{F_1}{F_2} \sep \textsf{inj}_i(F)\sep \m x c\\
    \end{array}
    \]
    The reduction relations $\Rew{\CBN}$ and $\Rew{\CBV}$ are the contextual closures of
    the rules in Fig.~\ref{fig:LCBNCBV}.
\end{definition}

\begin{bfigure}[!h]
  \[
  \begin{array}{c}
  \begin{array}{cc@{}l@{}l}
    \rulenamed{\stackrel\leftarrow\mu_{\sf N}}&\cutc{\mu \beta.c}{ E}&\Rew{} &\subst c \beta { E}\\
    \rulenamed{\stackrel\rightarrow\mu}&\cutc{t}{\m x c}&\Rew{} &\subst c x t\\\\
    \rulenamed{\zeta_{\sf N}}&R[e]&\Rew{}&\cutc{\m \alpha{R[\alpha]}}{e}\\\\\\
    \rulenamed{\rightarrow_{\sf N}} &\cutc{\lambda x.t_1}{\cons{t_2} E}&\Rew{} &\cutc{\subst{t_1}x{t_2}}{E}\\
    \rulenamed{\wedge_{\sf N}}&\cutc{\paire {t_1}{t_2}}{\textsf{inj}_i(E)}&\Rew{} &\cutc{t_i}{E}\\
    \rulenamed{\vee_{\sf N}}&\cutc{\textsf{inj}_i(t)}{\paire {E_1}{E_2}}&\Rew{} &\cutc{t}{E_i}
\strut
  \end{array}\\\\
    \CBN
  \end{array}
  \begin{array}{c}
  \begin{array}{cc@{}l@{}l}
    \rulenamed{\stackrel\leftarrow\mu}&\cutc{\mu \beta.c}{e}&\Rew{} &\subst c \beta e\\
    \rulenamed{\stackrel\rightarrow\mu_{\sf V}}&\cutc{{V}}{\m x c}&\Rew{} &\subst c x {V}\\\\
    \rulenamed{\zeta_{\sf V}}& S[t]&\Rew{}&\cutc{t}{\m  x{S[x]}}\\
    \rulenamed{\zeta_{\sf V}}&T[e]&\Rew{}&T[\m x{\cutc x e}]\\\\
    \rulenamed{\rightarrow_{\sf V}} &\cutc{\lambda x.t_1}{\cons{V} e}&\Rew{} &\cutc{\subst{t_1}x{V}}{e}\\
    \rulenamed{\wedge_{\sf V}}&\cutc{\paire {V_1}{V_2}}{\textsf{inj}_i(e)}&\Rew{} &\cutc{V_i}{e}\\
    \rulenamed{\vee_{\sf V}}&\cutc{\textsf{inj}_i(V)}{\paire {e_1}{e_2}}&\Rew{} &\cutc{V}{e_i}
  \end{array}\\\\
    \CBV
  \end{array}
  \]
  where $R$, $S$, and $T$ range over contexts of the following grammar:
  \[
  \begin{array}{lll}
    \mbox{\CBN\ continuation contexts}&R
    &\recdef \cutc t{\cons {t'} \Chole} \hfil\sep\hfil \cutc t{\paire{\Chole}{e}} \hfil\sep\hfil \cutc t{\paire{E}{\Chole}} \hfil\sep\hfil \cutc t{\textsf{inj}_i(\Chole)}\\
    \mbox{\CBV\ term contexts}&S
    &\recdef \cutc {V}{\cons {\Chole} e}\hfil\sep\hfil \cutc{\paire{\Chole}{t}}e \hfil\sep\hfil \cutc{\paire{V}{\Chole}}e \hfil\sep\hfil \cutc{\textsf{inj}_i(\Chole)}e \\ 
    \mbox{\CBV\ continuation contexts}&T
    &\recdef \cutc {V}{\cons {V'} \Chole} \hfil\sep\hfil \cutc V{\paire{\Chole}{e}} \hfil\sep\hfil \cutc V{\paire{F}{\Chole}} \hfil\sep\hfil \cutc V{\textsf{inj}_i(\Chole)}
  \end{array}
  \]
  the $\rulenamed{\zeta_{\sf N}}$ only applies under the condition that $e$ is not
  a (\CBN-) continuation value,\\
  the $\rulenamed{\zeta_{\sf V}}$-rules only apply under the condition that $t$ is not a (\CBV-) term value and $e$
  is not a (\CBV-) continuation value.
  \caption{\CBN\ and \CBV\ reduction in System~\SysL}
  \label{fig:LCBNCBV}
\end{bfigure}

In this version, we kept the \CBV-rules
$\rulenamed{\stackrel\leftarrow\mu}$ and
$\rulenamed{\stackrel\rightarrow\mu_{\sf V}}$, and the \CBN-rules
$\rulenamed{\stackrel\leftarrow\mu_{\sf N}}$ and
$\rulenamed{\stackrel\rightarrow\mu}$.

The $\rulenamed{\zeta_{\sf V}}$-rules (\resp the
$\rulenamed{\zeta_{\sf N}}$-rule) are new: they were introduced in a
slightly more general version in~\cite{Wadlerdual}, while the version
we take here more closely follows~\cite{MunchCSL09}. These rules are
due to our strong restriction on term values (\resp continuation
values): the fact that a term is a value is not just the fact that it
is not of the form $\m \alpha c$, as term values are recursively
defined. Therefore if a term $t$ is not of the form $\m \alpha c$ but
one of its (say direct) subterms is, then $t$ is not a value and there
is no \CBV-rule to reduce $\cutc{t}{\m x c}$. Progress then fails if
we do not add the $\rulenamed{\zeta_{\sf V}}$-rules to pull the first
subterm of $t$ that is not a value to the top-level.

These $\zeta$ rules also impact rules $\rulenamed\rightarrow$,
$\rulenamed\wedge$, and $\rulenamed\vee$, which now have to be
restricted in order to preserve confluence: for instance in \CBV, the
fact that $\paire{\m\alpha c}{t}$ is not a term value means that, when
facing a continuation $e$, $\m\alpha c$ will be extracted from the
pair and will have the control of computation
\[
\cutc{\paire{\m\alpha c}{t}}{e}
\Rew{\zeta_{\sf N}}\cutc{\m\alpha c}{\m x{\cutc{\paire x t}e}}
\Rew{\stackrel\leftarrow\mu}
\subst c \alpha{\m x{\cutc{\paire x t}e}}
\]
and therefore it is clear that, should $e$ be of the form $\inj 2
{e'}$, the original application of rule $\rulenamed\wedge$ would have a
totally different semantics:
\[
\cutc{\paire{\m\alpha c}{t}}{e}
\Rew{\wedge}
\cutc{t}{e'}
\]

Hence the restriction of $\rulenamed\rightarrow$, $\rulenamed\wedge$,
and $\rulenamed\vee$ to $\rulenamed{\rightarrow_{\sf N}}$,
$\rulenamed{\wedge_{\sf N}}$, and $\rulenamed{\vee_{\sf N}}$ in the \CBN\ case, and to 
$\rulenamed{\rightarrow_{\sf V}}$,
$\rulenamed{\wedge_{\sf V}}$, and $\rulenamed{\vee_{\sf V}}$ in the \CBV\ case.

Note that in \CBN, we decided to make $\rulenamed{\rightarrow_{\sf
 N}}$ collapse the two reduction steps
\[\cutc{\lambda x.t_1}{\cons{t_2} e}
\Rew{\rightarrow}\cutc{t_2}{\m
  x{\cutc{t_1}{e}}}
\Rew{\stackrel\rightarrow\mu}
\cutc{\subst{t_1}x{t_2}}{e}
\] 
into one step, because $\rulenamed{\stackrel\rightarrow\mu}$ has priority
anyway.\footnote{We shall see that it makes the $\m x \_$ construct
  superfluous (in the sense that the fragment without this construct
  is logically complete, and stable under reduction).}  The rule
$\rulenamed{\rightarrow_{\sf V}}$ is designed by symmetry, collapsing the two steps
\[\cutc{\lambda x.t_1}{\cons{V} e}
\Rew{\rightarrow}\cutc{V}{\m
  x{\cutc{t_1}{e}}}
\Rew{\stackrel\rightarrow\mu}
\cutc{\subst{t_1}x{V}}{e}
\] 
and noticing that if $t_2$ is not a term value, then the original rule
$\rulenamed\rightarrow$ is recovered as follows:
\[
\cutc{\lambda x.t_1}{\cons{t_2} e}
\Rew{\zeta_{\sf V}}
\cutc{t_2}{\m y{\cutc{\lambda x.t_1}{\cons{y} e}}}
\Rew{\rightarrow_{\sf V}}
\cutc{t_2}{\m x{\cutc{t_1}{e}}}
\]

Of course the extra rules satisfy Subject Reduction, so that we have:
\begin{theorem}[Subject reduction for System \SysL: \CBN\ \& \CBV]

  If $\XcolY c {(\Seq{\Gamma} \Delta)}$ and either $c\Rew{\CBN}c'$ or
  $c\Rew{\CBV}c'$ then $\XcolY {c'} {(\Seq{\Gamma} \Delta)}$.

  And similarly for terms and continuations.
\end{theorem}
\begin{proof}
  Straightforward induction on the rewrite derivation.
\end{proof}

\begin{theorem}[Confluence]
  $\Rew{\CBN}$ and $\Rew{\CBV}$ are confluent.
\end{theorem}
\begin{proof}
  They are orthogonal higher-order rewrite systems.\footnote{The
    rewrite system presented in Fig.~\ref{fig:LCBNCBV} is a standard
    (higher-order) rewrite system: we did use a non-standard
    formulation for the $\zeta$ rules based on a grammar for
    continuation contexts and term contexts as well as on
    side-conditions (``$t$ is not a term value and $e$ is not a
    continuation value''), but we could equally have formulated all
    the cases as standard (but numerous!) rewrite rules.}
\end{proof}

Finally, one could be puzzled by what seems like an asymmetry between
\CBN\ and \CBV, the latter having more rules and requiring a notion of
continuation value while the former does not need a notion of term
value.  This asymmetry is not due to \CBN\ vs.\ \CBV, but is due to the
implication: its main continuation construct $\cons t e$ has
a term as a direct sub-term, while no term contruct has a continuation
as a direct sub-term. It would be the case if we considered the De
Morgan dual of implication, namely \Index{subtraction} (see for
instance~\cite{Crolard04}), which would make \CBN\ completely
symmetric to \CBV.

\subsection{Two stable fragments}

Now it is easy to connect the $\Rew{\CBN}$ and $\Rew{\CBV}$ reduction
relations of Definition~\ref{def:CBNCBV} to those of our first attempt
in Definition~\ref{def:CBNCBVshallow}, if we concentrate on the two
fragments:\footnote{Namely, those fragments where the reductions of
  Definition~\ref{def:CBNCBVshallow} are actually simulated by (the
  adaptation to System~\SysL\ of) the \CPS translations.}

\begin{definition}[\LKzN\ and \LKzV]\label{def:LKzNLKzV}
  Let \LKzN\ and \LKzV\ be the fragments of System~\SysL\ consisting of
  $\Rew{\zeta_{\textsf N}}$-normal forms and $\Rew{\zeta_{\textsf
      V}}$-normal forms, respectively.
\end{definition}

\begin{remark}
  \begin{enumerate}
  \item Concerning implication only, \LKzN\ is the fragment where every continuation of
    the form $\cons t e$ is such that $e$ is a continuation value.
  \item Concerning implication only, \LKzV\ is the fragment where every continuation of the form
    $\cons t e$ is such that $t$ is a term value;
  \item
    Also, $\Rew{\zeta_{\textsf N}}$ and $\Rew{\zeta_{\textsf V}}$ are
    terminating reduction relations, so it is easy to normalise a
    command into one of these fragments, using cuts.
  \item Moreover, \LKzN\ and \LKzV\ are respectively stable under
    $\Rew{\CBN}$ and $\Rew{\CBV}$, so the cuts can be reduced while
    staying in the fragments.
  \item Furthermore, in \LKzN\ and \LKzV, $\Rew{\CBN}$ and
    $\Rew{\CBV}$ of Definition~\ref{def:CBNCBV} respectively coincide
    with those of Definition~\ref{def:CBNCBVshallow}.\footnote{Almost,
      since in Definition~\ref{def:CBNCBV} the rule
      $\rulenamed{\rightarrow_{\sf N}}$ (\resp
      $\rulenamed{\rightarrow_{\sf V}}$) collapses the two steps
      $\co{\Rew{\rightarrow}}{\Rew{\stackrel\rightarrow\mu}}$ (\resp
      $\co{\Rew{\rightarrow}}{\Rew{\stackrel\rightarrow\mu_{\sf V}}}$)
      of $\Rew{\CBN}$ (\resp $\Rew{\CBV}$) from
      Definition~\ref{def:CBNCBVshallow}: but in \LKzN\ (\resp \LKzV),
      there is no other choice than $\Rew{\stackrel\rightarrow\mu}$
      (\resp $\Rew{\stackrel\rightarrow\mu_{\sf V}}$) for a top-level
      reduction that can follow $\Rew{\rightarrow}$ in \CBN-reduction
      (\resp\CBV-reduction).}

  \item Notice that the encoding of $\l$-calculus in System~\SysL\ in
    Definition~\ref{def:lambdatoL}, actually only reaches the fragment
    \LKzN, and the simulation lemma (Lemma~\ref{lem:lambdatoL}) only
    involves \CBN-reduction.
  \end{enumerate}
\end{remark}

\subsection{Denotational semantics of \CBN\ and \CBV}

As anticipated, it is now possible to define \CPS-translations of terms,
continuations, and commands, respectively denoted $\underline t$,
$\underline e$, $\underline c$ for \CBN, and $\overline t$, $\overline
e$, $\overline c$ for \CBV, in a way that preserves reductions:

\begin{theorem}[Preservation of reduction]\label{th:CPSreducL}
  \begin{itemize}
  \item[\CBN:] If $c_1\Rew{\CBN}c_2$ then $\underline {c_1}\Rewn{\beta}\underline {c_2}$
  \item[\CBV:] If $c_1\Rew{\CBV}c_2$ then $\overline {c_1}\Rewn{\beta}\overline {c_2}$
  \end{itemize}
\end{theorem}

We do not give the details here, which are just technical, but they
can be found in \eg \cite{Wadlerdual}.

And these encodings also preserve typing, if Hofmann and Streicher's
encoding of types for \CBN, and Fischer's encoding of types for \CBV,
are considered not just for $\arr$ (\ie $\imp$) but also $\times$ (\ie
$\et$) and $+$ (\ie $\ou$):

\begin{theorem}[Preservation of typing]

  Assume
  \[
  \begin{array}{c}
    \Seq{\Gamma}{{t\col A}\semcol \Delta}\\
    \Seq{\Gamma\semcol{e\col A}}{\Delta}\\
    c\col(\Seq{\Gamma}{\Delta})
  \end{array}
  \]

  Then
  \[
  \begin{array}{c}
    \begin{array}{l}
      \Deri{\HStrans\Gamma\arr R,\HStrans\Delta} {\underline t}{\HStrans A\arr R}\\
      \Deri{\HStrans\Gamma\arr R,\HStrans\Delta} {\underline e}{(\HStrans A\arr R)\arr R}\\
      \Deri{\HStrans\Gamma\arr R,\HStrans\Delta} {\underline c}{R}
    \end{array}\\\\
    \mbox{Using Hofmann \& Streicher's}\\
    \mbox{translation of types~\cite{HofmanStreicher97}}
  \end{array}
  \qquad
  \begin{array}{c}
    \begin{array}{l}
      \Deri{\overline \Gamma,\overline \Delta\arr R} {\overline t}{(\overline A\arr R)\arr R}\\
      \Deri{\overline \Gamma,\overline \Delta\arr R} {\overline e}{\overline A\arr R}\\
      \Deri{\overline \Gamma,\overline \Delta\arr R} {\overline c}{R}
    \end{array}\\\\
    \mbox{Using Fischer's}\\
    \mbox{translation of types~\cite{Fischer:pacp1972}}
  \end{array}
  \]
\end{theorem}

As already mentioned, we can now use these \CPS-translations to define
categorical semantics for classical proofs:

\begin{definition}[Semantics of System \SysL\ in a response category]

  Assume \(c\col(\Seq{x_1\col A_1,\ldots,x_n\col A_n}{\alpha_1\col B_1,\ldots,\alpha_m\col B_m})\).

  Define the semantics $\semN[r] c\eqdef\sem{\underline c}$ and $\semV[r]
  c\eqdef\sem{\overline c}$, where $\sem t$ is the semantics, in a
  response category, of a $\l$-term $t$ in the \CPS-fragment, as
  defined by the rules of Fig.~\ref{fig:stlcCCC}.

  Writing $K_A$ for the object corresponding to $\HStrans A$, and $C_A$ for $R^{K_A}$, we have
  \[
  \morph{\semN[r] c}
        {(C_{A_1}\times\ldots\times C_{A_n}\times K_{B_1}\times\ldots\times K_{B_m})}
        {R}
  \]

  Writing $V_A$ for the object corresponding to $\overline A$, $K_A$ and for $R^{V_A}$, we have
  \[
  \morph{\semV[r] c}
        {(V_{A_1}\times\ldots\times V_{A_n}\times K_{B_1}\times\ldots\times K_{B_m})}
        {R}
  \]
\end{definition}

Now, remember that a \Index{control category} is the sub-category of a
response category $\mathcal C$ whose objects are in $\{R^A | A\in
\mathcal C\}$, and that $R^A\Or R^B$ denotes $R^{A\times B}$.

\begin{definition}[Semantics of System \SysL\ in control and co-control categories]

  Assume \(c\col(\Seq{x_1\col A_1,\ldots,x_n\col A_n}{\alpha_1\col B_1,\ldots,\alpha_m\col B_m})\).

  \begin{itemize}
  \item[\CBN] The semantics $\semN[r] c$ in a response category gives rise, by curryfication, to a morphism
    \[
    \morph{\semN c}{C_{A_1}\times\ldots\times C_{A_n}}{C_{B_1}\Or\ldots\Or C_{B_m}}
    \]
    in a control category.
  \end{itemize}

  \begin{itemize}
  \item[\CBV] The semantics $\semV[r] c$ in a response category gives rise, by curryfication, to a morphism
  \[
  \morph{\semV c}{K_{B_1}\times\ldots\times K_{B_m}}{K_{A_1}\Or\ldots\Or K_{A_n}}
  \]
  in a control category.

  As the arrow looks ``reversed'', from the original typing of $c$, it
  is more natural to interpret $c$ as the corresponding morphism
  \[
  \morph{\semV c}{K_{A_1}\otimes\ldots\otimes K_{A_n}}{K_{B_1}+\ldots+ K_{B_m}}
  \]
  in a \Index{co-control category}, the dual of a control category where
  $\otimes$ denotes the dual of $\Or$ (and the co-product $+$ denotes
  the dual of the product $\times$).
  \end{itemize}
\end{definition}

This formalises the idea that \CBN-reduction corresponds to a
denotational semantics in control categories, while \CBV-reduction
corresponds to a denotational semantics in co-control categories.

Indeed, from Theorem~\ref{th:CPSreducL} we get that the semantics
validate the reductions:
\begin{theorem}[Soundness of \CBN\ and \CBV\ in control and co-control categories]
\label{th:SoundCompleteCBNCBV}%
\noindent If $c\Rew{\CBN}c'$ then $\semN c=\semN
  {c'}$

If $c\Rew{\CBV}c'$ then $\semV c=\semV {c'}$
\end{theorem}
And we did this by breaking the symmetry between $\wedge$ and $\vee$:
Indeed in a control category, $\Or$ is not the dual of $\times$
(equivalently in a co-control category, $+$ is not the dual of
$\otimes$).

\sectionno{Conclusion} 
 
The work on control and co-control categories is due to
Selinger~\cite{SelingerControlCat99} for Parigot's $\l\mu$-calculus,
showing a duality between \CBN\ and \CBV\ in the categorical sense of
duality, and even before a syntactic duality between \CBN\ and
\CBV\ was displayed by System \SysL\ and its variants.
It follows preliminary works by Hofmann, Streicher,
Reus~\cite{HofmanStreicher97,StreicherReus98}, etc, on the semantics
of continuations (where the question of duality between \CBV\ and
\CBN\ -in $\lambda\mu$- is conjectured).

In conclusion, many variants of classical proof calculi have been
studied; in particular,
\begin{itemize}

\item the De Morgan dual of implication in classical logic, namely
  \Index{subtraction}, can also be given a computational
  interpretation, as shown for instance by~\cite{Crolard04};

\item variants of Parigot's $\lambda\mu$ have different properties
  with respect to observational equivalence, separation and
  $\eta$-conversion,
  etc\ldots~\cite{Saurin05,Saurin08,HerbelinZimmermann09,Saurin10,Saurin10b,Saurin12};

\item \Index[control delimiter]{control delimiters} can be used to
  limit the scope of the context that can be captured by a term via
  control operators, allowing for instance the capture of the
  \emph{shift} and \emph{reset} operators
  of~\cite{danvy1989functional,danvy1990abstracting}; these give rise
  to \Index[delimited continuations]{delimited continuations} and can
  be given a proof-theoretic
  interpretation~\cite{AriolaHerbelinSabry07,HerbelinGhilezan08,Saurin10c};

\item other reduction strategies than \CBV\ and \CBN\ have been
  investigated under the light of the duality of computation, such as
  Call-by-Need~\cite{AriolaF97,AriolaHS11,AriolaDownenHerbelinNakataSaurin12}.

\end{itemize}

\chapter[Orthogonality, normalisation and witness extraction]{Orthogonality: models for normalisation and witness extraction}
\label{ch:ortho}

\minitoc 

In this chapter we present the concept of \Index{orthogonality} and
two applications of it that are useful for classical proof-term
calculi: strong normalisation and witness extraction.

Orthogonality was used by Girard in the context of \Index{linear
  logic}~\cite{girard-ll} to prove normalisation of cut-elimination
and it lies at the heart of its proof semantics based on
\Index[coherent space]{coherent spaces}.

The concept of orthogonality has also proved a key concept in the
proof theory of classical logic, as it features, just like linear
logic does, a duality that is most immediately seen in the form of De
Morgan's laws.  Indeed, orthogonality is the basis of \Index{classical
  realisability}~\cite{DanosKrivine00,Krivine01}, which can be seen as
a semantical approach to the Curry-Howard correspondence for classical
logic. It also provided a new tool for models of classical proofs, and
for proving properties of
programs~\cite{Parigot97,mellies05recursive,LM:APAL07}, most
notoriously the strong normalisation property. Furthermore,
orthogonality shed a new light on the theory of \Index{polarisation}
and \Index{focussing} for classical logic, as revealed for instance
in~\cite{MunchCSL09} and explored in Chapter~\ref{ch:polarfocus}.

In this chapter we start by illustrating how proofs of strong
normalisation relate to
orthogonality. Summarising~\cite{bernadetleng11b},
Section~\ref{sec:SysF} rephrases and modularises, with the notion of
\Index[orthogonality model]{orthogonality models}, the well-known
techniques by Tait~\cite{tait67,tait75}, Reynolds~\cite{Reynolds72}
and Girard~\cite{Girard72} for proving the strong normalisation of the
simply-typed $\lambda$-calculus and
System~$F$. Section~\ref{sec:OrthoClassic} shows how such models allow
the adaptation, to classical proof-term calculi, of strong
normalisation proofs, both in the case of a confluent calculus and
non-confluent calculus. Section~\ref{sec:WitnessClassic} then shows
how orthogonality models can be used for classical witness extraction,
using a technique due to Miquel~\cite{MiquelTLCA09,MiquelLMCS11}.

\section{Revisiting Proofs of Strong Normalisation for System~$F$}
\label{sec:SysF}

This section presents concepts and a methodology developed
in~\cite{bernadetleng11b}: in particular, we approach the strong
normalisation of System~$F$ with the notion of \Index{orthogonality
  model}, adapting the Tait-Girard
methodology~\cite{tait75,Girard72}. Although System~$F$ is an
intuitionistic system, this approach will form the starting point from
which the adaptation to classical logic will be explored.

\begin{definition}[System~$F$]

  The types of System~$F$ are given by the following grammar:
  \[
  A,B,\ldots \recdef \alpha \sep A\arr B \sep \forall\alpha A
  \]
  where $\alpha$ ranges over a denumerable set of elements called
  \Index[type variable]{type variables}, and the construct
  $\forall\alpha A$ binds $\alpha$ in $A$.

  \Index[typing context]{Typing contexts} are defined as in
  Definition~\ref{def:stlambdacalc}, using System $F$ types instead of
  simple types; they will be denoted $\Gamma$, $\Delta$, etc.

  The \Index[free type variable]{free type variables} of a type $A$
  (\resp a typing context $\Gamma$) will be denoted $\TV A$ (\resp $\TV\gGamma$).

  The typing system of System~$F$ is given in Fig.~\ref{fig:SystemF}.
  Derivability of a sequent in System~$F$ is denoted $\Deri[\SysF] \Gamma M A$.
\end{definition}

\begin{bfigure}[!h]
 \[
    \begin{array}{c}
      \infer{\Derili{\gGamma,x\col \gA} x \gA}{\strut}
      \qquad
      \infer{\Derili{\gGamma} {\lambda x.M} {\gA\arr \gB}}{\Derili{\gGamma,x\col \gA} M \gB}
      \\\\
      \infer{\Derili{\gGamma} {M\ N} \gB}{\Derili\gGamma M {\gA\arr \gB}\quad\Derili\gGamma N \gA}
      \\\\
        \infer[\alpha\notin\TV\gGamma]{\Derili \gGamma M {\forall \alpha \gA}}{\Derili \gGamma M {\gA}}
        \qquad
        \infer{\Derili \gGamma M {\subst \gA \alpha \gB}}{\Derili \gGamma M {\forall \alpha \gA}}
    \end{array}
    \]
\caption{System~$F$}
\label{fig:SystemF}
\end{bfigure}

The method to prove strong normalisation is to build a model with
\begin{itemize}
\item an interpretation for terms as elements of a set $\ValDomE$
\item an interpretation for types as (interesting) subsets of $\ValDomE$
\item such that
  \begin{itemize}
  \item the interpretation of a term of type $A$ is in the interpretation of $A$
  \item if the interpretation of a term is in there, the term is strongly normalising.
  \end{itemize}
\end{itemize}


\subsection{Orthogonality models and the Adequacy Lemma}

\begin{definition}[Orthogonality models]\label{def:orthomodel}\nopagesplit

  An \Index{orthogonality model} is a 4-tuple
  $(\ValDom,\orth{}{},\ValDomE,\SemTe \_ \_)$ where
  \begin{itemize}
  \item $\ValDom$ is a set of elements called \Index[value]{values};
  \item $\orth{}{}$ is a relation between values and lists of values
    called the \Index{orthogonality relation};
  \item $\ValDomE$ is a superset of $\ValDom$;
  \item $\SemTe {\_ }{\_}$ is a function mapping every $\lambda$-term
    $M$ (typed or untyped) to an element $\SemTe {M}{\rho}$ of
    $\ValDomE$, where $\rho$ is a parameter called \Index{semantic
    context} mapping term variables to values.
  \end{itemize}

  \begin{itemize}
  \item the following axioms are satisfied:
    \begin{enumerate}
    \item[(A1)]
      For all $\rho$, $\vec v$, $x$,
      if $\orth{\rho(x)} {\vec v}$ then $\orth {\SemTe x\rho} {\vec v}$.
    \item[(A2)]
      For all $\rho$, $\vec v$, $M_1$, $M_2$,
      if $\orth {\SemTe {M_1}\rho}{(\cons {\SemTe {M_2}\rho}{\vec v})}$ then $\orth {\SemTe {M_1\ M_2}\rho}{\vec v}$. 
    \item[(A3)]
      For all $\rho$, $\vec v$, $x$, $M$ and for all values $u$,
      if $\orth {\SemTe {M}{\rho,x\mapsto u}}{\vec v}$ then $\orth {\SemTe {\lambda x.M}\rho}{(\cons u{\vec v})}$. 
    \end{enumerate}
  \end{itemize}
\end{definition}

In fact, $\ValDom$ and $\orth{}{}$ are already sufficient to interpret
any System~$F$ type $A$ as a set $\SemTyP{A}{}$ of values (see
Definition~\ref{def:TypeInterpretation} below): if types are seen as
logical formulae, we can see this construction as a way of building
some of their realisability / set-theoretical models.

There is no notion of computation pertaining to values, but the
interplay between the interpretation of terms and the orthogonality
relation is imposed by the axioms so that the Adequacy Lemma (which
relates typing to semantics) holds:

\begin{centre}If $\Deri[\SysF] {} M A$ then $\SemTe M{}\in\SemTyP A{}$\end{centre}

We now assume that we are given an orthogonality model
$(\ValDom,\orth{}{},\ValDomE,\SemTe \_ \_)$.

\begin{notation}
  By $\ValDomL$ we denote the set of lists of values.

  If $X\subseteq\ValDom$ and $Y\subseteq\ValDomL$ let
    \[\begin{array}{ll}
    \cons X Y &\eqdef \{\cons u {\vec v}\mid u\in X,\vec v \in Y \}\\
    \uniorth X &\eqdef \{ \vec{v}\in \ValDomL \mid \forall u \in X, \orth u {\vec{v}} \}\\
    \uniorth Y &\eqdef \{ u\in \ValDom \mid \forall \vec{v} \in Y, \orth u {\vec{v}} \}
    \end{array}
    \]
\end{notation}

\begin{remark}The usual properties of orthogonality hold:\\
  $X \subseteq \biorth X$, $\triorth X= \uniorth X$, and if $X\subseteq X'$ then $\uniorth{X'}\subseteq\uniorth X$
\end{remark}

We now define the interpretation of types. The intuition is the same
as that of Krivine's classical
realisability~\cite{DanosKrivine00,Krivine01}:
\begin{itemize}
  \item we first interpret a formula $A$ as a set of
    ``counter-proofs'', with the basic constructs that we expect to
    use in order to refute the formula: for instance the basic way to
    refute $A_1\arr A_2$ is to provide a ``proof'' of $A_1$ and a
    ``counter-proof'' of $A_2$; similarly, the basic way to refute
    $\forall \alpha A_1$ is to find a suitable interpretation of
    $\alpha$ and produce a ``counter-proof'' of $A_1$ under this
    interpretation; the set of counter-proofs for atomic formulae is then
    naturally given by a \emph{valuation};
  \item we then define the ``proofs'' (``realisers'' would be a better
    term) of a formula $A$ as any value that is able to ``face all
    counter-proofs'', this latter concept being what the orthogonality
    relation is precisely there to specify.
\end{itemize}

\begin{definition}[Interpretation of types]\label{def:TypeInterpretation}\nopagesplit

  A \Index{valuation} is a function, denoted $\sigma$, $\sigma'$,\ldots, from type variables to subsets of $\ValDomL$.

  Two interpretation of types are defined by simultaneous induction of types, a \Index{positive interpretation} and a \Index{negative interpretation}:
  \[\SemTyP[\sigma] \gA  \eqdef\uniorth {\SemTyrN[\sigma] \gA}
    \qquad\qquad \begin{array}{lll}
      \SemTyrN[\sigma] \alpha &\eqdef \sigma(\alpha)\\
      \SemTyrN[\sigma] {\gA \arr \gB} &\eqdef \cons{\SemTyP[\sigma] \gA}{\SemTyrN[\sigma] \gB}\\
      {\SemTyrN[\sigma] {\forall \alpha \gA}} &{\eqdef \bigcup_{Y\subseteq\ValDomL} \SemTyrN[{\sigma, \alpha \mapsto Y}] \gA }\\
    \end{array}
    \]

    We then define the interpretation of typing contexts:
  \[\SemTy[\sigma] \gGamma\eqdef \{ \rho\mid \forall (x\col \gA)\in\gGamma,\rho(x)\in\SemTyP[\sigma] \gA\}\]
\end{definition}

This approach begs the question: why is it the case that
counter-proofs are defined more primitively than proofs? As described
for instance in~\cite{MunchCSL09} (and in the rest of this thesis),
this is simply a coincidence about the two type constructs we use in
System~$F$: they both have a ``negative polarity'', if we see them in
the more general context of \emph{polarised logic}, as we shall
discuss in Chapter~\ref{ch:polarfocus}. Second-order quantification is
often used in the field of realisability (and elsewhere) to encode
other logical connectives, which made the negative approach prevalent
in that field, with counter-proofs being defined first and proofs
being defined by orthogonality. With primitive connectives that have a
``positive polarity'', such as intuitionistic disjunction, proofs
would be defined first and counter-proofs would be defined by
orthogonality. We shall come back to that discussion in the next
chapters.

\begin{remark}We have the usual properties of substitutions:
\[\SemTyrN[\sigma]{\subst\gA\alpha\gB}=\SemTyrN[{\sigma,\alpha\mapsto\SemTyrN[\sigma]{\gB}}]{\gA}\quad
\mbox{and}\quad
\SemTyP[\sigma]{\subst\gA\alpha\gB}=\SemTyP[{\sigma,\alpha\mapsto\SemTyrN[\sigma]{\gB}}]{\gA}\]
Also notice that the \emph{for all} quantifier is interpreted as an intersection:
\[\SemTyP[\sigma]{\forall \alpha \gA} = \bigcap_{Y\subseteq\ValDomL} \SemTyP[{\sigma, \alpha \mapsto Y}] \gA \]
\end{remark}

With these definitions we can prove the Adequacy Lemma:

\begin{lemma}[Adequacy Lemma]\label{lem:AdequacyF}\nopagesplit

  If $\Deri[\SysF] \gGamma M \gA$, then for all valuations $\sigma$ and for
  all mappings $\rho\in\SemTy[\sigma]\gGamma$ we have $\SemTe
  M\rho\in\SemTyP[\sigma] \gA$.
\end{lemma}
\begin{proof}
  By induction on the derivation of $\Deri \gGamma M \gA$, using
  axioms (A1), (A2) and (A3). See~\cite{bernadetleng11b}.
\end{proof}

\subsection{Applicative orthogonality models and Strong Normalisation}

\begin{definition}[Applicative orthogonality model]\strut\\
  An {\em applicative orthogonality model} is a 4-tuple
  $(\ValDom,\ValDomE,\ap,\SemTe {\_ }{\_} )$ where:
  \begin{itemize}
  \item $\ValDom$ is a set, $\ValDomE$ is a superset of $\ValDom$,
    $\ap$ is a (total) function from $\ValDomE\times\ValDomE$ to
    $\ValDomE$, and $\SemTe {\_ }{\_}$ is a function (parameterised by
    a semantic context) from $\lambda$-terms to $\ValDomE$.
  \item $(\ValDomE,\ValDom,\orth{}{},\SemTe {\_ }{\_} )$ is an
    orthogonality model,\\ where the relation $\orth u {\vec v}$ is
    defined as $(u\ap \vec{v})\in \ValDom$\\ (writing $u\ap\vec v$ for
    $(\ldots(u\ap v_1)\ap\ldots \ap v_n)$ if $\vec
    v=\cons{v_1}{\ldots\cons{v_n}\el}$).
  \end{itemize}
\end{definition}

\begin{remark} Axioms (A1) and (A2) are ensured provided that $\SemTe {M\ N}\rho=\SemTe {M}\rho\ap \SemTe {N}\rho$ and $\SemTe {x}\rho=\rho(x)$.
  These conditions can hold by definition (as in the following
  example), or can be proved.
\end{remark}

We now give an applicative orthogonality model to conclude strong
normalisation of System~$F$; this will capture, in essence, the
Tait-Girard proof methodology~\cite{tait75,Girard72}. The model is
here a \Index{term model}, in that $\ValDomE$ is the set of all
$\lambda$-terms and a $\lambda$-term is interpreted as itself.

\begin{example}[A term-model for Strong Normalisation]\label{ex:SNF}\nopagesplit

  Let $\ValDom$ be the set of strongly-normalising $\lambda$-terms,
  and let $\ValDomE$ be set of all $\lambda$-terms.  We define $\orth
  u {\vec v}$ as $(u\ap \vec{v})\in \SN{}$, and the interpretation of
  terms as follows:
  \[\begin{array}{lll}
  \SemTe x \rho&\eqdef \rho(x)\\
  \SemTe {M_1\ M_2} \rho&\eqdef \SemTe {M_1} \rho\ \SemTe {M_2} \rho\\
  \SemTe {\lambda x.M} \rho&\eqdef \lambda x.\SemTe {M} {\rho,x\mapsto x}
  \end{array}
  \]
  Requirement 3 is a consequence of anti-reduction:\\
  If $\subst M x P\ \vec N \in \SN{}$ and $P\in \SN{}$ then $(\lambda x.M)\ P\ \vec N \in \SN{}$.\\
  Note that for all $\vec N\in{\SN{}}^*$ and all term variables $x$, $\orth x {\vec N}$.\\
  Hence, for all valuations $\sigma$ and all types $\gA$, $x\in\SemTyP[\sigma] \gA$.\\
  We apply the Adequacy Lemma (Lemma~\ref{lem:AdequacyF}):\\
  If $\Deri \gGamma M \gA$, then for all valuation $\sigma$ and all mapping $\rho\in\SemTy[\sigma]\gGamma$ we have $\SemTe M\rho\in \SN{}$.\\
  Hence, $M\in \SN{}$.
\end{example}

In summary, we have defined a family of models for the
(polymorphically) typed $\lambda$-calculus, and presented one instance
with which strong normalisation could be
inferred. In~\cite{bernadetleng11b} we presented other instances of
orthogonality models, based for instance on intersection types.
Unlike usual models (\eg CCC), orthogonality models do not necessarily
equate terms up to $\beta$-reduction (if $M\Rew{\beta}N$, we do not
necessarily have $\SemTe M {}=\SemTe N{}$).  This allows us to build a
model where $\SemTe M {}=M$, from which we can infer strong
normalisation of typed terms (an instance of CCC would be useless for
this).

\section{Adapting the approach to classical calculi}
\label{sec:OrthoClassic}

Orthogonality was used by Parigot to prove strong normalisation of
\CBN\ $\lambda\mu$-calculus~\cite{Parigot97}.  For their non-confluent
calculus, Barbanera \& Berardi~\cite{BBsymm} adapted the Tait-Girard
reducibility technique with ``symmetric reducibility candidates''.
The key idea in both cases is still that a type $A$ is interpreted as
a pair of two orthogonal sets:\footnote{Two sets $\mathcal U$ and
  $\mathcal V$ are \Index{orthogonal} if $\forall t\in\mathcal
  U,\forall u \in\mathcal V, \orth t u$.}
\begin{itemize}
\item a set $\SemTyP \gA {}$ of proof(-terms)
\item a set $\SemTyN \gA {}$ of counter-proof(-terms)
\end{itemize}
\ldots satisfying some saturation property (like reducibility candidates do).

In the proof of strong normalisation of System~$F$ that we presented
in the previous section, the notion of saturation that holds for
$\SemTyP \gA {}$ is that it is closed under bi-orthogonal
($\biorth{\SemTyP \gA {}}=\SemTyP \gA {}$).  In particular, in an
applicative term model, the fact that $\SemTyP \gA {}$ is closed under
bi-orthogonal allows to derive, from axiom (A3) of the orthogonality
relation, the property that if $(\subst M x N)\;N_1\ldots
N_n\in\SemTyP \gA {}$ and $N, N_1,\ldots,N_n\in\mathcal D$, then $(\l
x.M) N\;N_1\ldots N_n\in\SemTyP \gA {}$.  Although not explicitly used
in our proof of strong normalisation (Example~\ref{ex:SNF}), this
property lies in the background and is often explicitly used in more
traditional presentations of the reducibility
technique~\cite{tait75,Girard72}. In brief, the technique works
because the interpretation of a type is closed under
``head-anti-reduction''.

This is also the approach for classical proof-term calculi, in
particular for a confluent calculus such as the Parigot's
$\lambda\mu$~\cite{Parigot97}.

\subsection{The case of a confluent calculus} 
\label{sec:OrthoClassicConfluent}

In this section we take the example of the \LKzN\ fragment of System
\SysL\ (Definition~\ref{def:LKzNLKzV}), with $\imp$ as the only
connective, and considering the reduction relation \CBN. We can also
prove strong normalisation by building a term model based on
orthogonality:

Three rewrite rules apply:
\[
\begin{array}{lcll}
  \rulenamed\rightarrow &\cutc{\lambda x.t_1}{\cons{t_2} E}&\Rew{} &\cutc{\subst{t_1}x{t_2}}{E}\\
  \rulenamed{\stackrel\leftarrow\mu_{\sf N}}&\cutc{\mu \beta.c}{E}&\Rew{} &\subst c \beta E\\
  \rulenamed{\stackrel\rightarrow\mu}&\cutc{t}{\m x c}&\Rew{} &\subst c x {t}\\
\end{array}
\]

Therefore we can adapt the axiom (A3) of Definition~\ref{def:orthomodel} as follows:
\begin{definition}[Orthogonality model for System \LKzN]

  An \Index{orthogonality model} for System \LKzN\ is given by
  $(\mathcal D_t,\mathcal D_e,\orth{}{})$ where $\mathcal D_t$ is a set
  of terms, $\mathcal D_e$ is a set of continuations, and $\orth{}{}$ is
  a relation between $\mathcal D_t$ and $\mathcal D_e$, which can be
  seen as a set of commands, and satisfying the following \Index[saturation]{saturation requirements}:\\
  \begin{tabular}{ll}
    If $\cutc{\subst{t_1}x{t_2}}{e}\rho\in\orth{}{}$ &then $\cutc{\lambda x.t_1}{\cons{t_2} e}\rho\in\orth{}{}$\\
    If $(\subst c \beta E)\rho\in\orth{}{}$ and $E\in\mathcal D_e$ &then $\cutc{\mu \beta.c}{E}\rho\in\orth{}{}$\\
    If $(\subst c x t)\rho\in\orth{}{}$ and $t\in\mathcal D_t$ &then $\cutc{t}{\m x c}\rho\in\orth{}{}$
  \end{tabular}\\
  where $c\rho$ denotes the capture-avoiding application, to $c$, of a
  substitution $\rho$ (mapping term variables to terms and
  continuation variables to continuations).
\end{definition}

\begin{definition}[Interpretation of types for System \LKzN]\label{def:typeinterpretSysL}\nopagesplit

  A \Index{valuation} is a function, denoted $\sigma$, $\sigma'$,\ldots, from type variables to subsets of $\mathcal D_e$ that only contain value continuations.

  Two interpretation of types are defined by simultaneous induction of types, a \Index{positive interpretation} and a \Index{negative interpretation}:
  \[\begin{array}{cc}
  &\begin{array}{lll}
     \SemTyrN[\sigma] \alpha &\eqdef \sigma(\alpha)\\
     \SemTyrN[\sigma] {\gA \arr \gB} &\eqdef \cons{\SemTyP[\sigma] \gA}{\SemTyN[\sigma] \gB}\\
   \end{array}\\
  \SemTyP[\sigma] \gA  \eqdef\uniorth {\SemTyrN[\sigma] \gA}
  &
  \SemTyN[\sigma] \gA  \eqdef\biorth {\SemTyrN[\sigma] \gA}
  \end{array}
  \]
  where $\cons X Y$ denotes $\{\cons u E\mid u\in X,E \in Y \}$ for
  any $X\subseteq\mathcal D_t$ and $Y\subseteq\mathcal D_e$.
  
  We then define the interpretation of a typing context (\ie a pair of
  a typing context for term variables and a typing context for
  continuation variables):
  \[
  \SemTy[\sigma] {\gGamma,\Delta}\eqdef \{ \rho\mid \forall (x\col
  \gA)\in\gGamma,\rho(x)\in\SemTyP[\sigma] \gA,\mbox{ and }\forall
  (\alpha\col \gA)\in\Delta,\rho(\alpha)\in\SemTyN[\sigma] \gA\}
  \]
\end{definition}

\begin{lemma}[Adequacy Lemma for System  \LKzN]\label{lem:adequacyLCBV}\nopagesplit
  \begin{enumerate}
  \item If $\Seq[\SysL] \gGamma {\XcolY t\gA\semcol\Delta}$, then for all valuations $\sigma$ and for
    all $\rho\in\SemTy[\sigma]{\gGamma,\Delta}$ we have $t\rho\in\SemTyP[\sigma] \gA$.
  \item If $\Seq[\SysL] {\gGamma\semcol \XcolY e\gA}\Delta$, then for all valuations $\sigma$ and for
    all $\rho\in\SemTy[\sigma]{\gGamma,\Delta}$ we have $e\rho\in\SemTyN[\sigma] \gA$.
  \item If $\XcolY c{(\Seq[\SysL] {\gGamma}\Delta)}$, then for all valuations $\sigma$ and for
    all $\rho\in\SemTy[\sigma]{\gGamma,\Delta}$ we have $c\rho\in\orth{}{}$.
  \end{enumerate}
  where $t\rho$, $e\rho$, and $c\rho$ denotes the capture-avoiding
  application of $\rho$, seen as a substitution, to $t$, $e$, and $c$,
  respectively.
\end{lemma}
\begin{proof}
  By simultaneous induction on the typing derivations, using the axioms about $\orth{}{}$.
\end{proof}

\begin{example}[Strong Normalisation of System \LKzN]

  We define $\mathcal D_t$ to be the set of strongly normalising terms
  and $\mathcal D_e$ to be the set of strongly normalising
  continuations.  We define the orthogonality relation $\orth{}{}$
  between $\mathcal D_t$ and $\mathcal D_e$ as those commands that are
  strongly normalising.\footnote{The notion of strong normalisation in
    the definition of $\mathcal D_t$, $\mathcal D_e$ and $\orth{}{}$
    is of course considered for $\Rew\CBN$.}

  We can check that the saturation requirements are met by purely
  syntactical/rewriting reasoning, but it only works because there is
  at most one way to reduce the top-level command.

  Take $\sigma$ to map every type variable to $\mathcal D_e$. Notice
  that term variables are in every $\SemTyP[\sigma] \gA$ and
  continuation variables are in every $\SemTyN[\sigma] \gA$, and that
  the identity substitution $\rho$ is in every $\SemTy[\sigma]
  {\gGamma,\Delta}$.

  We then apply the Adequacy Lemma with $\sigma$ and $\rho$, and get
  that every typed term, continuation, and command is strongly
  normalising for $\Rew{\CBN}$.
\end{example}

In this section, we have proved the strong normalisation of the
confluent calculus \LKzN\ for classical logic, a \CBN-fragment of
System~\SysL. We could have done it along the same lines for the full
syntax of System~\SysL\ (but still with the confluent reduction
$\Rew{\CBN}$), but dealing with the extra constructs and extra
reductions $\rulenamed{\zeta_{\sf N}}$ would have meant a heavier
machinery (along the lines of~\cite{MunchCSL09,Munch11,MunchPhD}).  We
aimed instead at simplicity, which emphasises the connection with the
orthogonality models for System $F$, and those that we present in the
next section.

In summary, in a confluent calculus such as the \LKzN, building the
positive and negative interpretations of a type $A$ can be described
as follows:
\[
\begin{array}{lc@{\quad}c@{\quad}c@{\quad}c}
  &\mbox{Sets of terms}&&\mbox{Sets of continuations}\\
  \mbox{Stage 1}&&& Y_0\eqdef\SemTyrN \gA {}\\
  \mbox{Stage 2}&X_1\eqdef\uniorth{Y_0}&&Y_1\eqdef\biorth{Y_0}\\
  \mbox{Finished}&\SemTyP \gA {}\eqdef X_1&&\SemTyN \gA {}\eqdef Y_1
\end{array}
\]
The construction is finished in 2 steps, because the sets $X_1$ and
$Y_1$, which are orthogonal, already have all of the saturation
properties required to contain all the terms and continuations of type
$A$, which is checked when proving the Adequacy Lemma.

In other words, closure under bi-orthogonality provides adequate
saturation properties.

\subsection{The case of a non-confluent calculus} 

Let us now consider the situation of a non-confluent calculus such as
System \SysL\ with its original reduction system
\[
\begin{array}{lcll}
  \rulenamed\rightarrow &\cutc{\lambda x.t_1}{\cons{t_2} e}&\Rew{} &\cutc{t_2}{\m x{\cutc{t_1}{e}}}\\
  \rulenamed{\stackrel\leftarrow\mu}&\cutc{\mu \beta.c}{e}&\Rew{} &\subst c \beta e\\
  \rulenamed{\stackrel\rightarrow\mu}&\cutc{t}{\m x c}&\Rew{} &\subst c x t
\end{array}
\]

The Adequacy Lemma might still work if we had the saturation requirements:\\
\begin{tabular}{ll}
  If $\cutc{t_2}{\m x{\cutc{t_1}{e}}}\rho\in\orth{}{}$ &then $\cutc{\lambda x.t_1}{\cons{t_2} e}\rho\in\orth{}{}$\\
  If $(\subst c \beta e)\rho\in\orth{}{}$ and $e\in\mathcal D_e$ &then $\cutc{\mu \beta.c}{e}\rho\in\orth{}{}$\\
  If $(\subst c x t)\rho\in\orth{}{}$ and $V\in\mathcal D_t$ &then $\cutc{t}{\m x c}\rho\in\orth{}{}$
\end{tabular}\\
But in any case, because of non-confluence, these requirements are not met if we
define $\mathcal D_t$ to be the set of strongly normalising terms and
$\mathcal D_e$ to be the set of strongly normalising continuations,
and $\orth{}{}$ the set of strongly normalising commands.\footnote{The
  notion of strong normalisation in the definition of $\mathcal D_t$,
  $\mathcal D_e$ and $\orth{}{}$ is now considered for the full
  reduction relation $\Rew{}$.}

This means that because of non-confluence, we need to change our
notion of saturation, so that $\SemTyP \gA {}$ and $\SemTyN \gA {}$
respectively contain enough terms and continuations for the Adequacy
Lemma to hold, and because of that change, the pair $(\SemTyP \gA
{},\SemTyN \gA {})$ will not be constructed in 2 steps as in the
confluent case, but in infinitely many steps:
\[
\begin{array}{lc@{\quad}c@{\quad}c@{\quad}c}
  &\mbox{Sets of terms}&&\mbox{Sets of continuations}\\
\mbox{Stage 1}&  X_0&\orth{}{}& Y_0&\mbox{not {saturated}}\\
\mbox{Stage 2}&  X_1&\orth{}{}&Y_1&\mbox{not {saturated}}\\
\mbox{Stage 3}&  X_2&\orth{}{}&Y_2&\mbox{not {saturated}}\\
 & \ldots&\orth{}{}&\ldots&\ldots\\
\mbox{Stage $\infty$}&  X_\infty&\orth{}{}&Y_\infty&\mbox{{saturated}}\\
\mbox{Finished}&  \SemTyP \gA {}&\orth{}{}&\SemTyN \gA {}&\mbox{{saturated}}
\end{array}
\]
We get a saturated pair of sets in infinitely many steps (via a
fixpoint construct).
In~\cite{LM:APAL07}, we showed that the fixpoint construct could not
be captured by a bi-orthogonal completion step.

We now see the details of the technique. In the rest of this section,
we fix $\orth{}{}$ to be the set of strongly normalising commands.

\begin{definition}[Orthogonality and saturation]

  Let $\var[t]$ denote the set of term variables and $\var[e]$ denote the set of continuation variables. 

  Given a set $\mathcal U$ of terms and a set $\mathcal V$ of continuations, the pair $(\mathcal U,\mathcal V)$ is
  \begin{itemize}
  \item \Index{orthogonal} if $\forall t\in\mathcal U,\forall u \in\mathcal V, \orth t u$
  \item \Index{saturated} if the following two conditions hold 
    \begin{enumerate}
    \item $\var[t]\subseteq\mathcal U$ and $\var[e]\subseteq\mathcal V$
    \item $\{\m \alpha c\sep \forall e\in\mathcal V, \subst
      c x v\in\orth{}{}\}\subseteq \mathcal U$ and\\
      $\{\m x c\sep \forall t\in\mathcal U, \subst
      t x v\in\orth{}{}\}\subseteq \mathcal V$.
    \end{enumerate}
  \end{itemize}

  A set of terms (\resp continuations) is said to be \Index[simple set]{simple} if
  it is non-empty and it contains no term of the form $\m \alpha c$
  (\resp $\m x c$).

  For every set $X$ of terms (set $ Y$ of continuations), we define a function
  \[
  \fp {X}{(\mathcal W)}\eqdef X \cup \var[t] \cup \{\m \alpha c\sep \forall e\in\mathcal W, \subst c \alpha e\in\orth{}{}\}
  \]
  \resp
  \[
  \fp { Y}{(\mathcal W)}\eqdef Y \cup \var[e] \cup \{\m x c\sep \forall t\in\mathcal W, \subst c x t\in\orth{}{}\}
  \]
\end{definition}

\begin{lemma}
  Given a set of terms $X_0$ and a set of continuations $Y_0$,
  \begin{enumerate}
  \item
  $\fp{X_0}{}$ and $\fp{Y_0}{}$ are anti-monotonic.\footnote{In other words for $\fp{X_0}{}$, if $\mathcal W\subseteq\mathcal W'$ then $\fp{X_0}{(\mathcal W)}\supseteq\fp{X_0}{(\mathcal W')}$. And similarly for $\fp{Y_0}{}$.}
  \item Hence, $\fp  {X_0} {\circ \fp  {Y_0} {}}$ is monotonic and admits a fixpoint $X_\infty$, with $\fp {X_0} {(\fp  {Y_0} {(X_\infty)})}= X_\infty$.
  \item Writing $Y_\infty$ for $\fp  {Y_0} {( X_\infty)}$, we clearly have 
    \[\begin{array}l
       X_\infty= X \cup \var[t] \cup \{\m \alpha c\sep \forall e\in Y_\infty, \subst c \alpha e\in\orth{}{}\}\\
       Y_\infty= Y \cup \var[e] \cup \{\m x c\sep \forall t\in X_\infty, \subst c x t\in\orth{}{}\}
    \end{array}
    \]
  \item So $(X_\infty, Y_\infty)$ is saturated, and a quick case analysis shows that it is orthogonal if $X_0$ and $Y_0$ are simple and orthogonal to each other.
  \item Finally, $X_0\subseteq X_\infty$ and $Y_0\subseteq Y_\infty$.
\end{enumerate}
  We finally define $\satur{X_0}{Y_0}$ as $(X_\infty, Y_\infty)$.
\end{lemma}

\begin{definition}[Interpretation of types for System \SysL]\nopagesplit

  A \Index{valuation} is a function, denoted $\sigma$, $\sigma'$,\ldots, from type variables to orthogonal pairs of simple sets.

  Two interpretation of types are defined by simultaneous induction of types, a \Index{positive interpretation} and a \Index{negative interpretation}:
  \[
  \begin{array}{cll}
     (\SemTyrP[\sigma] a,\SemTyrN[\sigma] a) &\eqdef \sigma(\alpha)\\
     (\SemTyrP[\sigma] {\gA \arr \gB},\SemTyrN[\sigma] {\gA \arr \gB})
     &\eqdef 
     (\{ \l x.t \mid \l x.t\in \uniorth{(\cons{\SemTyP[\sigma] \gA}{\SemTyN[\sigma] \gB})}\},\cons{\SemTyP[\sigma] \gA}{\SemTyN[\sigma] \gB})\\\\
  (\SemTyP[\sigma] {\gA},\SemTyN[\sigma] {\gA}) &\eqdef\satur{\SemTyrP[\sigma] {\gA}}{\SemTyrN[\sigma] {\gA}}
  \end{array}
  \]
  
  Again, we define
  \[
  \SemTy[\sigma] {\gGamma,\Delta}\eqdef \{ \rho\mid \forall (x\col \gA)\in\gGamma,\rho(x)\in\SemTyP[\sigma] \gA,\mbox{ and }\forall (\alpha\col \gA)\in\Delta,\rho(x)\in\SemTyN[\sigma] \gA\}
  \]
\end{definition}

Now notice the difference with Definition~\ref{def:typeinterpretSysL}:
the definition of $\SemTyrN[\sigma] {\gA \arr \gB}$ is the same but if
we just took $\SemTyP[\sigma] {\gA \arr \gB}$ to be its orthogonal,
the pair $(\SemTyP[\sigma] {\gA \arr \gB},\SemTyrN[\sigma] {\gA \arr
  \gB})$ would not be saturated, as we have already seen; so instead
we take all of the abstractions in the orthogonal of $\SemTyrN[\sigma]
{\gA \arr \gB}$ and form an orthogonal (but not saturated) pair of
simple sets $(\SemTyrP[\sigma] {\gA \arr \gB},\SemTyrN[\sigma] {\gA
  \arr \gB})$. Then we saturate that pair into $(\SemTyP[\sigma] {\gA
  \arr \gB},\SemTyN[\sigma] {\gA \arr \gB})$, which is orthogonal and
saturated:

\begin{lemma}[Interpretations of types are orthogonal and saturated]\label{lem:saturNortho}

For all valuations $\sigma$ and all types $A$, $(\SemTyP[\sigma]
{\gA},\SemTyN[\sigma] {\gA})$ is orthogonal and saturated.
\end{lemma}

The rest is now just as in the \CBN\ case:

\begin{lemma}[Adequacy Lemma for System \SysL]\nopagesplit
  \begin{enumerate}
  \item If $\Seq[\SysL] \gGamma {\XcolY t\gA\semcol\Delta}$, then for all valuations $\sigma$ and for
    all $\rho\in\SemTy[\sigma]{\gGamma,\Delta}$ we have $t\rho\in\SemTyP[\sigma] \gA$.
  \item If $\Seq[\SysL] {\gGamma\semcol \XcolY e\gA}\Delta$, then for all valuations $\sigma$ and for
    all $\rho\in\SemTy[\sigma]{\gGamma,\Delta}$ we have $e\rho\in\SemTyN[\sigma] \gA$.
  \item If $\XcolY c{(\Seq[\SysL] {\gGamma}\Delta)}$, then for all valuations $\sigma$ and for
    all $\rho\in\SemTy[\sigma]{\gGamma,\Delta}$ we have $c\rho\in\orth{}{}$.
  \end{enumerate}
  where $t\rho$, $e\rho$, and $c\rho$ denotes the capture-avoiding
  application of $\rho$, seen as a substitution, to $t$, $e$, and $c$,
  respectively.
\end{lemma}
\begin{proof}
  By simultaneous induction on the typing derivations, using the Lemma~\ref{lem:saturNortho}.
\end{proof}

\begin{theorem}[Strong Normalisation of System \SysL]

  Take $\sigma$ to map every type variable to the orthogonal pair
  $(\var[t],\var[e])$ of simple sets. Notice again that the identity
  substitution $\rho$ is in every $\SemTy[\sigma] {\gGamma,\Delta}$.

  We then apply the Adequacy Lemma with $\sigma$ and $\rho$, and get
  that every typed term, continuation, and command is strongly
  normalising for $\Rew{}$.
\end{theorem}

The points to remember are
\begin{itemize}
\item As for System~$F$, we have proved strong normalisation by building a term model
  \begin{itemize}
  \item which does not equate terms up to reduction\flush (non-confluence would make that very problematic)
  \item where axiom (A3) is replaced by a saturation property.
\end{itemize}
\item Because of non-confluence,
  \begin{itemize}
  \item saturation has to be a property of pairs $(\SemTyP[\sigma]{A},\SemTyN[\sigma]{A})$, not a property of each component separately;
  \item saturating is difficult (adding terms in one component of the pair affects the other component), and obtained by a fixpoint construction.
  \end{itemize}
  As shown in~\cite{LM:APAL07}, the saturation process is not just a bi-orthogonality completion:\\
  if $(\mU,\mV)$ is orthogonal, then $(\biorth{\mU},\biorth{\mV})$ is orthogonal but not necessarily saturated.
\end{itemize}

\ssection[Orthogonality models for witness extraction]{Orthogonality models for extracting witnesses from classical proofs}
\label{sec:WitnessClassic}

We now show how to extract a witness from a classical proof of a
$\Sigma_1^0$-formula, \ie a closed formula of the form $\exists a
A(a)$ where $A(a)$ is a quantifier-free formula of arithmetics.

The technique is due to Miquel~\cite{MiquelTLCA09,MiquelLMCS11}, we
simply adapted it to our proof-term calculus for classical logic, and
somewhat simplified it using the concepts and notations of the
previous sections.

We work in a particular setting where such a formula is expressed in
the shape of $\neg\FA a {\neg\isnull{e(a)}}$, the grammar of formulae
being defined as follows:

\begin{definition}[Expressions and Formulae]
  \[\begin{array}{l@{\quad}ll}
  \mbox{\bf Expressions}&u,u',\ldots&\recdef a\sep \zerE\sep \sE u\sep u+u'\sep u\times u'\sep u\leq u'\\
  \mbox{\bf Formulae}&A,B,\ldots&\recdef \isnull u\sep A \arr B\sep \FA a A\\
  \end{array}
  \]

  We represent integers as expressions: let $\express 0\eqdef \zerE$
  and, for all integers $n$, let $\express {n+1}\eqdef \sE{\express n}$.

  We define $\neg A\eqdef A\arr\isnull{\express 1}$.
\end{definition}

This shape for a $\Sigma_1^0$-formula brings no loss of generality.
Moreover, such an expression as $u(a)$, with one free variable $a$,
expresses a primitive recursive function from $\N$ to $\N$.

We will now build an orthogonality model that we will use for witness
extraction.  As in the previous sections, each formula $A$ will be
interpreted as a set $\SemTyP[\sigma]{A}$ of terms and a set
$\SemTyN[\sigma]{A}$ of continuations, terms and continuations being
those of \LKzN.

The extraction mechanism itself will be given by the reductions of
\LKzN, and more precisely by {\bf root} \CBN-reduction, which
we denote $\Rew{\CBNr}$.\footnote{The fact that we use \CBN-reduction
  is important to make sure that reduction {\bf can} produce a
  witness; the fact that we only use root reduction is not, but in
  order to implement the extraction mechanism deterministically, it is
  convenient to never have to choose the next redex to reduce.}

In other words, from a proof of $\neg\FA a {\neg\isnull{u(a)}}$ in
(the extension to arithmetic of) System~\SysL, we will perform
$\Rew{\CBNr}$-reduction until we reach (in a provably finite number of
steps) a command where we can directly read a witness.

For this we need to express numerals as proof-terms. We simply use
Church's numerals in $\l$-calculus (see \eg\cite{Bar84}) and encode them in
\LKzN\ with Definition~\ref{def:lambdatoL}:

\begin{definition}[Church's numerals as terms]
\[
\begin{array}{ll}
c_0&\eqdef \cutc {x} {\alpha}\\
c_{n+1}&\eqdef \cutc{f}{\st{(\m {\alpha}{c_n})}{\alpha}}\\
\church n&\eqdef\lami {x}{}{\lami {f}{}{\m {\alpha}{c_n}}}
\end{array}
\]
\end{definition}

\begin{remark}
  Doing the same thing with the $\l$-terms for the successor function
  and the recursion function, we get two \LKzN\ terms $\su$ and
  $\rec$ such that, for all $t$, $u_0$, $u_1$, for all value
  continuations $E$, and all integer $n$,
  \[
  \begin{array}{ll}
    \cutc\su{\st {\church n}{\st{t}{E}}}&\Rewn{\CBNr} \cutc{t}{\st{\church{n+1}}{E}}\\
    \cutc\rec{\st {u_0}{\st{u_1}{\st{\church 0}{E}}}}&\Rewn{\CBNr} \cutc{u_0}{E}\\
    \cutc\rec{\st {u_0}{\st{u_1}{\st{\church {n+1}}{E}}}}&\Rewn{\CBNr} \cutc{u_1}{\st{\church n}{\st{(\m {\alpha}{\cutc{\rec}{\st{u_0}{\st{u_1}{\st{\church n}{\alpha}}}}})}{E}}}\\
  \end{array}
  \]
  using the simulation of $\beta$-reduction by $\Rew{\CBN}$.\footnote{And the fact that we can do this with root-reduction only is rather clear.}

  Let $\ifte\eqdef\lami{n x_0 x_1}{}{\m {\alpha}{\cutc{\rec}{
        \st{x_0}{
          \st{(\lami{y_0y_1}{}{x_1})}{
            \st{n}{\alpha}
          }
        }
      }
    }
  }$, so that
  \[\begin{array}{cl}
  \cutc{\ifte}{\st{\church 0}{\st{u_0}{\st{u_1}{E}}}}&\Rewn{\CBNr}\cutc{u_0}{E}\\
  \cutc{\ifte}{\st{\church {n+1}}{\st{u_0}{\st{u_1}{E}}}}&\Rewn{\CBNr}\cutc{u_1}{E}
  \end{array}
  \]
\end{remark}

\begin{definition}[Orthogonality semantics]\label{def:InterpretTypesWitness}

  Let $\orth{}{}$ be an arbitrary set of commands, stable under anti-reduction
  (if $c\Rew{\CBNr}c'$ and $c'\in\orth{}{}$ then $c\in\orth{}{}$).

  A valuation $\sigma$ is a mapping from expression variables ($a$, etc) to integers.

  Given a valuation $\sigma$, Fig.~\ref{fig:WitExt} defines the
  interpretation of an expression $u$ as an integer $\SemTy[\sigma]u$
  and a formula $A$ as a set $\SemTyP[\sigma]{A}$ of terms
  and a set $\SemTyN[\sigma]{A}$ of continuations.
\end{definition}
\begin{bfigure}[!h]
  \[
  \begin{array}{lll}
    {    \SemTy[\sigma]{a}}             &\eqdef {\sigma(a)}&\\
    {    \SemTy[\sigma]{\zerE}}         &\eqdef 0&\\
    {    \SemTy[\sigma]{\sE u}}         &\eqdef \SemTy[\sigma]{u}+1&\\
    {    \SemTy[\sigma]{u_1+u_2}}       &\eqdef \SemTy[\sigma]{u_1}+\SemTy[\sigma]{u_2}&\\
    {    \SemTy[\sigma]{u_1\times u_2}} &\eqdef \SemTy[\sigma]{u_1}\times \SemTy[\sigma]{u_2}&\\
    {    \SemTy[\sigma]{u_1\leq u_2}}   &\eqdef 1 &\mbox{if $\SemTy[\sigma]{u_1}\leq \SemTy[\sigma]{u_2}$}\\
    {    \SemTy[\sigma]{u_1\leq u_2}}   &\eqdef 0 &\mbox{if $\SemTy[\sigma]{u_1}> \SemTy[\sigma]{u_2}$}
  \end{array}\]
  \medskip
  \[
  \begin{array}{l@{\qquad}lllll}
    &      {    \SemTyrN[\sigma]{\isnull u}}    &\eqdef \mathcal E\flush\mbox{if $\SemTy[\sigma] u\neq 0$}\\
    &      {    \SemTyrN[\sigma]{\isnull u}}    &\eqdef \emptyset\flush\mbox{if $\SemTy[\sigma] u =0$}\\
    &      {    \SemTyrN[\sigma]{A \arr{B}}}    &\eqdef \st{\SemTyP[\sigma] A}{\SemTyN [\sigma]B}\\
    &      {    \SemTyrN[\sigma]{\FA{a}{A}}}    &\eqdef \bigcup_{n\in\N}(\st{\{\church n\}}{\SemTyN[\sigma,a\mapsto n] A})\\\\
    \SemTyP[\sigma] A      \eqdef\uniorth{\SemTyrN[\sigma] {A}}&
    \SemTyN[\sigma] A &     \eqdef\biorth{\SemTyrN[\sigma] {A}}
  \end{array}
  \]
  where $\mathcal E$ is the set of all value continuations.
  \caption{Semantics of expressions and formulae}
  \label{fig:WitExt}
\end{bfigure}

\begin{remark}
  \begin{enumerate}
  \item
    Clearly, $\SemTy[\sigma]{\express n}=n$ for all $n$ and $\sigma$.
  \item By induction on $u$ we get $\SemTy[\sigma]{\subst {u} a {\express n}}=\SemTy[\sigma,a\mapsto n]{u}$, and by induction on $A$ we get
    $\SemTyrN[\sigma] {\subst A a {\express n}}=\SemTyrN[\sigma,a\mapsto n] {A}$ and $\SemTyP[\sigma] {\subst A a {\express n}}=\SemTyP[\sigma,a\mapsto n] {A}$.
  \end{enumerate}
\end{remark}

Now, for simplicity we do not specify the exact proof system for
arithmetic, nor do we give a typing system corresponding to it through
the Curry-Howard correspondence. We assume that it could be built as
an extension of Fig.~\ref{fig:Ltyping}, and that the Adequacy Lemma
can be proved (along the lines of Lemma~\ref{lem:adequacyLCBV} for
\LKzN): 
\ctr{A closed proof $t_0$ of a formula
$\neg\FA a {\neg\isnull{u(a)}}$ is such that,\\ for all possible
$\orth{}{}$ closed under ``anti-reduction'' (the inverse of $\Rew{\CBNr}$), $t_0\in\SemTy {\neg\FA a {\neg\isnull{u(a)}}}{}$.}

We thus start with such a term $t_0$.

\ignore{
  with the following rules for
  quantifiers
  \[
  \infers{\Seq{\Gamma }{{\l x. t\col \FA a A}{\semcol\Delta}}}[a\not\in\FV{\Gamma,\Delta}]{
    \Seq{\Gamma,\XcolY x{\textsf{nat} a}}{{t\col A}{\semcol\Delta}}
  }
  \qquad\qquad
  \infers{\Seq{\Gamma\semcol{\cons t e\col \FA a A}}{\Delta}}{
    \Seq{\Gamma}{\XcolY t {\textsf{nat} u}\semcol\Delta}
    \quad
    \Seq{\Gamma\semcol {e\col \subst A a u}}{\Delta}
  }
  \]
  where $\textsf{nat}(e)$ is a type formula such that the only closed proofs of 
}

We now define a term that can check whether an integer is a witness of
the property and, depending on this check, continue with one term or
another:

\begin{definition}[Witness checker]

  Let $f$ be the primitive recursive function defined by: for any
  integer $n$, $f(n)\eqdef\SemTy[a\mapsto n]{u(a)}$.

  Let $\church f$ be a term representing $f$ in the sense that, for
  any integer $n$, and term $t$ and any continuation $E$,
  \[\cutc{\church f}{\st{\church n}{\st{t}{E}}}
  \Rewn{\CBNr}\cutc{t}{\st{\church{f(n)}}{E}}
  \]
  Such a term can be constructed from $\su$ and $\rec$, as the
  projections, composition, etc are all available in System~\SysL.
  
  We define the \Index{witness checker} as follows:
  \[d_f\eqdef 
  \lami {nxy} {} 
        {\m {\alpha}{\cutc
            {\church f}
            {\st{n}
              {\st{(\lami{p}{}{\m {\alpha_1}{\cutc{\ifte}{\st{p}{\st
                          {x}
                          {\st {y}{\alpha_1}}}}}})}{\alpha}}
          }}
        }\]
\end{definition}

\begin{lemma}[Witness checker property]

  For any integer $n$, any $u_0$ and $u_1$ and $E$, we have

  $\cutc{d_f}{\st{\church n}{\st{u_0}{\st{u_1}{E}}}}\Rewn{\CBNr}\cutc{u_0}{E}$ if $f(n)=0$

  $\cutc{d_f}{\st{\church n}{\st{u_0}{\st{u_1}{E}}}}\Rewn{\CBNr}\cutc{u_1}{E}$ if $f(n)\neq0$
\end{lemma}
\begin{proof}
  \[\begin{array}{lll}
  \cutc{d_f}{\st{\church n}{\st{u_0}{\st{u_1}{E}}}}
  &\Rewn{\CBNr}&\cutc{{\m {\alpha}{\cutc
        {\church f}
        {\st{\church n}
          {\st{(\lami{p}{}{\m {\alpha_1}{\cutc{\ifte}{\st{p}{\st
                      {u_0}
                      {\st {u_1}{\alpha_1}}}}}})}{\alpha}}
      }}
  }}{E}\\
  &\Rewn{\CBNr}&\cutc
       {\church f}
       {\st{\church n}
         {\st{(\lami{p}{}{\m {\alpha_1}{\cutc{\ifte}{\st{p}{\st
                     {u_0}
                     {\st {u_1}{\alpha_1}}}}}})}{E}}
       }\\
       &\Rewn{\CBNr}&\cutc{\lami{p}{}{\m {\alpha_1}{\cutc{\ifte}{\st{p}{\st
                 {u_0}
                 {\st {u_1}{\alpha_1}}}}}}}
       {\st{\church {f(n)}}
         {{E}}
       }\\
       &\Rewn{\CBNr}&\cutc{{\m {\alpha_1}{\cutc{\ifte}{\st{\church {f(n)}}{\st
                 {u_0}
                 {\st {u_1}{\alpha_1}}}}}}}
       {
         {{E}}
       }\\
       &\Rewn{\CBNr}&\cutc{\ifte}{\st{\church {f(n)}}{\st
           {u_0}
           {\st {u_1}{E}}}}\\
  \end{array}
  \]
  If $f(n)=0$, this reduces to $\cutc{u_0}{E}$.
  Otherwise, this reduces to $\cutc{u_1}{E}$.
\end{proof}

\begin{definition}[Orthogonality and contradicter]

  Let $\stp$ be an arbitrary term and $\goon$ be an arbitrary continuation.

  We now take a particular orthogonality set defined by
  \[\orth{}{}\eqdef\{c \sep \mbox{ there exists $n$ such that $f(n)=0$ and $c \Rewn{\CBNr}\cutc\stp{\st{\church n}{\goon}}$}\}\]
  It is closed under anti-\CBNr-reduction.

  We now define a ``contradicter'':\footnote{In the sense that it will contradict what the proof $t_0$ claims.}
  Let $t_1\eqdef 
  \l nx.\m\alpha{
    \cutc{d_f}
         {\st{n}{\st{(\m {\alpha_0}{\cutc\stp{\st{n}{\goon}}})}{\st x{\alpha}}}}}$.
\end{definition}

\begin{lemma}[Behaviour of the contradicter]\label{lem:t1behaviour}

  For all integer $n$, and all continuation $E$ in $\SemTyN{\neg\isnull{u(\express n)}}{}$, we have $\orth{t_1}{\st {\church n}E }$.
\end{lemma}
\begin{proof}We have
  \[
  \cutc{t_1}{\st {\church n}E}
  \Rewn{\CBNr}
  \cutc{\l x.\m\alpha{\cutc{d_f}{\st{\church n}{\st{(\m {\alpha_0}{\cutc\stp{\st{\church n}{\goon}}})}{\st x{\alpha}}}}}}E
  \]
  To prove that this is an orthogonal command, we only have to show, as
  $E\in\biorth{\SemTyrN{\neg\isnull{u(\express n)}}{}}$, that the
  left-hand side term is orthogonal to every continuation in
  $\SemTyrN{\neg\isnull{u(\express n)}}{}$.  

  Consider such a continuation; it is of the form $\st t {E'}$ with
  $t\in\SemTyP{\isnull{u(\express n)}}$.

  If $f(n)\neq 0$ then $\SemTyr{{u(\express n)}}{}\neq 0$, so
  $\SemTyrN{\isnull{u(\express n)}}{}=\mathcal E$ and $t$ is orthogonal to every continuation, in particular $E'$. So we have
  \[\begin{array}{cl}
  &\cutc{\l x.\m\alpha{\cutc{d_f}{\st{\church n}{\st{(\m {\alpha_0}{\cutc\stp{\st{\church n}{\goon}}})}{\st x{\alpha}}}}}}{\st {t}{E'}}\\
  \Rewn{\CBNr}&\cutc{d_f}{\st{\church n}{\st{(\m {\alpha_0}{\cutc\stp{\st{\church n}{\goon}}})}{\st t{E'}}}}\\
  \Rewn{\CBNr}&\cutc{t}{E'}
  \end{array}
  \]
  which is in $\orth{}{}$.

  If $f(n)= 0$ then we have
  \[\begin{array}{cl}
  &\cutc{\l x.\m\alpha{\cutc{d_f}{\st{\church n}{\st{(\m {\alpha_0}{\cutc\stp{\st{\church n}{\goon}}})}{\st x{\alpha}}}}}}{\st {t}{E'}}\\
  \Rewn{\CBNr}&\cutc{d_f}{\st{n}{\st{(\m {\alpha_0}{\cutc\stp{\st{\church n}{\goon}}})}{\st t{E'}}}}\\
  \Rewn{\CBNr}&\cutc{\m {\alpha_0}{\cutc\stp{\st{\church n}{\goon}}}}{E'}\\
  \Rewn{\CBNr}&\cutc\stp{\st{\church n}{\goon}}
  \end{array}
  \]
\end{proof}

\begin{corollary}[Classical witness extraction]

  $\cutc{t_0}{\st {t_1}\goon}\Rewn{\CBNr}\cutc\stp{\st{\church n}{\goon}}$ for some integer $n$ such that $f(n)=0$.
\end{corollary}
\begin{proof}

  From Lemma~\ref{lem:t1behaviour} we get that
  $t_1\in\SemTyP{\FA a\neg\isnull{u(a)}}{}$ and therefore $\st
  {t_1}\goon\in\SemTyrN{\neg\FA a {\neg\isnull{u(a)}}}{}$.

  Since we have assumed $t_0\in\SemTyP{\neg\FA a {\neg\isnull{u(a)}}}{}$, we have
  $\orth{t_0}{\st{t_1}\goon}$, from which we conclude.
\end{proof}

In other words, once given a classical proof, we match it against the
continuation $\st {t_1}\goon$ and we are certain that $\CBNr$ will
produce $\cutc\stp{\st{\church n}{\goon}}$ in a finite number of
steps, with $n$ being a witness.

For a comparison with other techniques of classical witness
extraction, see~\cite{MiquelTLCA09,MiquelLMCS11}.

\sectionno{Conclusion}

In summary, we have seen in this chapter a fundamental concept for
model construction, namely orthogonality.  We built several
orthogonality models for various purposes: rephrase strong
normalisation proofs for System~$F$, prove the strong normalisation of
a confluent proof-term calculus for classical logic as well as a
non-confluent calculus (thereby proving cut-elimination), and finally
extract witnesses from classical proofs of $\Sigma_1^0$-formulae.

In each of those model constructions, we have interpreted formulae
first with basic inhabitants (terms or continuations), and then closed
their interpretation by a completion process that could simply be
taking the bi-orthogonal, in the case of confluent calculi, or a more
complex fixpoint, in the case of a non-confluent calculus.

Whether in those constructions we first define the negative
interpretation of a formula (as a set of ``counter-proofs'') or its
positive interpretation (as a set of ``proofs''), is a question that
depends on the formula's \Index{polarity}. This is the topic of the
next chapter.

\chapter{Polarisation and focussing}
\label{ch:polarfocus}

\minitoc 

In the previous chapters, we have seen that
\begin{itemize}
\item A proof-term syntax, together with a typing system, can be used
  to represent classical proofs (\eg System~\SysL), such that the
  symmetry of classical logic is reflected by the symmetry between
  programs and continuations. The use of classical reasoning
  corresponds to letting a program capture its continuation.
\item A rewrite system on proof-terms can be given to represent
  cut-elimination, following the intuitions of continuations and
  control. This gives a non-confluent calculus because (unrestricted)
  cut-elimination is non-confluent in classical logic, reflected by
  the fact that programs and continuations compete for the control of
  computation.
\item Still, the rewrite system is strongly normalising on typed
  proof-terms (\ie those representing real proofs), showing that cuts
  are admissible. The proof of strong normalisation was the occasion
  to introduce orthogonality techniques, although non-confluence
  requires more, namely specific saturation properties.
\item The semantics of classical proofs, or typed proof-terms, is
  problematic until confluence is recovered in some way.
\end{itemize}
Back to the main issue, a CCC with $\neg\neg A\simeq A$ collapses, and
out of the 3 natural ways to avoid the collapse, namely
\begin{enumerate}
  \item break the symmetry between $\et{}{}$ and $\ou{}{}$,
  \item break the cartesian product,
  \item break the curryfication,
\end{enumerate}
we investigate the breaking of symmetry between $\et{}{}$ and $\ou{}{}$.

In Chapter~\ref{ch:CH} we saw how to break the
$\wedge\vee$-symmetry by the \CBV/\CBN\ approach. In this chapter,
we break the $\wedge\vee$ symmetry by \Index{polarisation}.

\section{Recovering confluence by polarisation}

\subsection{Symmetry, asymmetry, and $\eta$-expansions}

We start this section by coming back to a fundamental question: What
is symmetrical about Classical Logic? There is definitely a symmetry
based on the duality of negation / De Morgan's duality.  It can be
seen in the truth semantics of formulae, in \eg truth tables or more
generally in a boolean algebra: meet / join and top / bottom are
swapped when flipping the order upside-down, and all the axioms of a
boolean algebra are preserved.

At the level of proofs, there is also a symmetry that can be seen for
instance in the two-sided sequent calculus: the left-introduction rule
of a connective is symmetric to the right-introduction rule of its
dual connective (in other words, the rules are preserved under duality
flipping).

Cut-elimination is symmetrical (\eg the rewrite system of
Fig.~\ref{fig:Ltyping}), but to make semantical sense of it, one
breaks the symmetry by making a choice between \CBN\ and \CBV\ that
is completely arbitrary.

More interestingly, the following example reveals something asymmetric
between the left-introduction of $\ou$ and the right-introduction
of $\ou$:
\[
\infer{\Seq{\Gamma,A\ou B}{\Del}}{\Seq{\Gamma, A}{\Del}\quad \Seq{\Gamma, B}{\Del}}
\quad
\infer{\Seq{\Gamma,A\ou B}{\Del}}{\Seq{\Gamma,A}{\Del}}
\quad
\infer{\Seq{\Gamma,A\ou B}{\Del}}{\Seq{\Gamma,B}{\Del}}
\]

Of course, we have never claimed that there is a symmetry between the
left-introduction of $\ou$ and the right-introduction of $\ou$, but an
interesting question is raised by the following situation: it is known
(see \eg\cite{TS}) that in the sequent calculus, the axiom rule
\[
\infer{\Seq{A}{A}}{}
\]
(say in a context-splitting setting) can be restricted, without losing
logical completeness, to the \emph{atomic} axiom rule
\[
\infer{\Seq{a}{a}}{}
\]
Every instance of the general instance can be replaced by a small
proof only using atomic axioms, which is proved by induction on $A$:
in particular, transforming the axiom $\infer{\Seq{A\ou B}{A\ou 
    B}}{}$ into a proof with atomic axioms, we produce
\[
\infer{\Seq{ A\ou B}{ A\ou B}}
      {\infer{\Seq{A}{ A\ou B}}
        {\infer{\Seq{A}{A}}{\ldots}}
        \quad
        \infer{\Seq{B}{ A\ou B}}
              {\infer{\Seq{B}{B}}{\ldots}}
      }
      \]
and then recursively transform $\Seq{A}{A}$ and $\Seq{B}{B}$ (until
all of the used axioms are atomic).

So the interesting question is whether there is a fundamental reason
why $\ou{}{}$ is decomposed on the left before being decomposed on the
right (looking at the bottom-up construction of the proof).
Starting the decomposition on the right would have failed.

A related situation occurs with $\eta$-expansion in $\l$-calculus:

In $\l$-calculus, the use of an axiom corresponds to a variable in the
proof-term. Typing the term
\[\lami z{A\rightarrow B} {z}\]
(where we indicate the types of variables as superscripts) uses an
axiom on $A\arr B$.
Typing its $\eta$-expansion 
\[
\lami z{A\rightarrow B} {{\lami y{A}{z\ y}}}
\]
uses, strictly speaking, an axiom on $A\arr B$ and an axiom on $A$,
but as $z$ is immediately applied and its type $A\arr B$ immediately
destructed, the $\eta$-expansion only uses, ``morally'' speaking,
axioms on the smaller formulae $A$ and $B$. Turning this moral
intuition into something formal can be done by taking a proof-term
calculus for sequent calculus (rather than natural deduction), as we
shall see below, but still: we first have the $\lambda$-abstraction,
and underneath it the application. Why again do they have to appear in
that order?

Indeed, in a classical calculus such a System~\SysL, the axiom on
$A\arr B$ is represented as
\[
\infer{\Seq{z\col A\arr B}{{z\col A\arr B}\semcol}}{}
\]
The $\eta$-expansion of $z$ is:
\[    \infer{\Seq {z\col A\arr B} {{\lambda y.\m \alpha {\cutc{z}{\st y \alpha}}\col A\arr B}\semcol }}
      {
        \infer{\Seq{{z\col A\arr B},y\col A}{{\m \alpha{\cutc{z}{\st y \alpha}}\col B}\semcol}}
              {
                \infer
                    {\cutc{z}{\st y \alpha}\col(\Seq{{z\col A\arr B},y\col A}{\alpha\col B})}
                    {
                      {{
                          \infer{\Seq{z\col A\arr B,y\col A}{{z\col A\arr B}\semcol{\alpha\col B}}}{}
                      }}
                      \quad
                      \infer{\Seq{{{z\col A\arr B,}}y\col A\semcol {(\st y \alpha)\col A\arr B}}{\alpha\col B}}
                            {
                              \infer{\Seq{{{z\col A\arr B,}}y\col A}{{y\col A}\semcol{{\alpha\col B}}}}{}
                              \quad
                              \infer{\Seq{{{z\col A\arr B,y\col A}}\semcol {\alpha\col B}}{\alpha\col B}}{}
                            }
                    }
              }
      }
      \]
and then we can recursively transform the axioms on $y\col A$ and
$\alpha\col B$ (until axioms are atomic).  Of course, this
$\eta$-expansion still features the use of $z\col A\arr B$, but only
to implement a contraction (or even more precisely to implement the
placing of the formula $A\arr B$ where it can be decomposed), not to
implement a proper axiom.

Now the $\eta$-expansion we used in the $\lambda$-calculus to
illustrate our point is only one particular instance of
$\eta$-expansion: the general form
\[M \Rew{\eta} \lami y{A}{M\ y}\qquad y\notin\FV M\]
can be recovered, by the capture avoiding substitution of $M$ for $z$,
from the axiomatic $\eta$-expansion on axiom $z$:
\[z \Rew{\eta} \lami y{A}{z\ y} \]

And in $\lambda$-calculus, no matter $M$ (where $y$ is not free), $M$
and $\lambda y.M\ y$ have the same computational behaviour (with
respect to $\beta$-reduction). In technical terms, $M$ and $\lambda
y.M\ y$ cannot be separated (even using untyped terms)~\cite{Bar84}.

In System~\SysL, the $\eta$-expansion on axiom $z$
\[z \Rew{\eta} \lambda y.\m \alpha {\cutc{z}{\st y \alpha}}\]
also provides, after instantiation of $z$ by $t$ (where $y$
and $\alpha$ are not free), a general form of $\eta$-expansion:
\[t \Rew{\eta} \lambda y.\m \alpha {\cutc{t}{\st y \alpha}} \]
which transforms
\[
{\Seq\Gamma{{t\col A\arr B}\semcol\Del}}
\]
into
\[
\infer{\Seq \Gam {{\lambda y.\m \alpha {\cutc{t}{\st y \alpha}}\col A\arr B}\semcol \Del}}
      {
        \infer{\Seq{\Gamma,y\col A}{{\m \alpha{\cutc{t}{\st y \alpha}}\col B}\semcol\Del}}
              {
                \infer{\cutc{t}{\st y \alpha}\col(\Seq{\Gamma,y\colon A}{\alpha\col B,\Del})}
                      {
                        {\Seq\Gamma{{t\col A\arr B}\semcol\Del}}
                              \quad
                              \infer{\Seq{{\Gam,}y\col A\semcol {(\st y \alpha)\col A\arr B}}{\alpha\col B{,\Del}}}
                                    {
                                      \infer{\Seq{{\Gam,}y\col A}{{y\col A}\semcol{\alpha\col B,\Del}}}{}
                                      \quad
                                      \infer{\Seq{{\Gam,y\col A}\semcol {\alpha\col B}}{\alpha\col B{,\Del}}}{}
                                    }
                      }
              }
      }
      \]

\subsection{Towards polarised System~\SysL}
\label{sec:toLpol}

Now the above general $\eta$-expansion can be instantiated with $t=\m
\beta c$:
\[\m \beta c \Rew{\eta} \lambda y.\m \alpha {\cutc{\m \beta c}{\st y \alpha}} \]


If we put those two terms in context, \eg facing a continuation $\m x {c'}$, we get that
\begin{itemize}
\item
  $\cutc{\m \beta c}{\m x {c'}}$ rewrites to
  \[
  \subst {c'} x {\m \beta {c}}
  \qquad\mbox{ or }\qquad
  \subst c \beta {\m x {c'}}
  \]
\item \textbf{but}
    $\cutc{\lambda y.\m \alpha {\cutc{\m \beta c}{\st y \alpha}}}{\m x {c'}}$ rewrites
    only to
    \[
    \subst {c'} x {\lambda y.\m \alpha {\cutc{\m \beta c}{\st y \alpha}}}
    \]
\end{itemize}

If $\eta$-convertible terms should have undistinguishable
computational behaviour, we must forbid $\cutc{\m[{A_1\rightarrow
      A_2}] \beta c}{\m[A_1\rightarrow A_2] x {c'}}\Rew{}\subst c
\beta {\m[A_1\rightarrow A_2] x {c'}}$

The grounds for breaking the symmetry in such a way is that
$\m[{A_1\rightarrow A_2}] \beta c$ can be $\eta$-expanded, but
$\m[A_1\rightarrow A_2] x {c'}$ cannot, which reflects the fact that
the right-introduction rule for $A_1\rightarrow A_2$ is invertible
while its left-introduction rule is not.

In short, when encountering
\[
\infer{\cutc {\m \beta c}{\m x {c'}}\col(\Seq\Gamma\Delta)}
      {\infer{\Seq\Gamma{{\m \beta c}\col A_1\arr A_2\semcol\Del}}
          {c\col(\Seq{\Gamma}{\beta\col A_1\arr A_2,\Del})}
        \qquad
        \infer{\Seq{\Gamma\semcol \m x {c'}\col A_1\arr A_2}{\Del}}{c'\col(\Seq{\Gamma,x\col A_1\arr A_2}{\Del})}}
      \]
we could consider that the term $\m \beta c$ is a ``cheater'' in the
sense that its type $A_1\arr A_2$ could be proved or inhabited in
another way (\eg with the $\eta$-expansion of $\m \beta c$), avoiding
the critical pair, and solving the non-confluence problem.

In particular, if $\beta$ is used 0 times in $c$, or more than once, we can understand the typing tree
\[\infer{\Seq\Gamma{{\m \beta c}\col A_1\arr A_2\semcol\Del}}{
  c\col(\Seq{\Gamma}{\beta\col A_1\arr A_2,\Del})
}
\] 
as finishing with a weakening or a contraction. What $\eta$-expansion
proves is that the proof can be transformed into a proof that finishes
with a proper introduction of the implication.

In our earlier example about the connective $\ou$, it is the
contrary: its left-introduction rule is invertible while its
right-introduction rules are not.

This leads to considering a notion that arose from linear
logic~\cite{girard-ll}: \Index[polarity]{polarities}.

The intuition for \Index[positive connective]{positive connectives} is
that we expect no particular property of their right-introduction
rules. These rules are called \Index{synchronous}. In particular for
goal-directed proof-search, applying such a rule bottom-up is \emph{a
  priori} a choice which we may have to backtrack to if we fail to
finish the proof. For logical completeness, (right-)weakenings or
(right-)contractions may be necessary on a formula with a positive
connective at its root.

The intuition for \Index[negative connective]{negative connectives} is
that their right-introduction rules \emph{are} invertible. 
These rules are called \Index{asynchronous}. In
goal-directed proof-search we may apply such rules without loss of
generality and therefore without creating backtrack points. Also,
(right-)weakenings and (right-)contractions (on formulae that have a
negative connective at their roots) are superfluous as far as logical
completeness is concerned. On the other hand, the right-introduction
rules ``must interact well with the left-introduction rules'' (or the
right-introduction rules of the dual connective), in cut-elimination
as well as in the expansion of axioms that we described in this section.

Just as in $\lambda$-calculus you can \emph{always} inhabit a
(non-empty) function type with a $\lambda$-abstraction, you can always
$\eta$-expand an inhabitant of a type whose main connective is negative.
Considering the $\eta$-expansion rules that we can apply in
System~\SysL, we can derive the polarities of the three connectives we
considered:
\[\begin{array}{lllll}
    {\mbox{negative }}&A\arr B&{t}&\Rew{}&\lambda y.\m \alpha {\cutc{{t}}{\st y \alpha}}\\
    {\mbox{negative }}&A\wedge B&{t}&\Rew{}&\paire{\m \alpha{\cutc{{t}}{\inj 1 \alpha}}}{\m \gamma{\cutc{{t}}{\inj 2 \gamma}}}\\
    {\mbox{positive }}&A\vee B&{e}&\Rew{}&\paire{\m x{\cutc{\inj 1 x}{{e}}}}{\m z{\cutc{\inj 2 z}{{e}}}}
\end{array}
\]

Now in order to solve the confluence problem, we also need to
determine how to reduce $\cutc {\m \beta c}{\m x {c'}}$ when the
cut-formula is atomic. This leads to splitting the set of the atomic
formulae into positives and negatives as well. Unlike non-atomic
formulae, the choice of polarity for each atom is arbitrary, and
sometimes called the \Index{bias}~\cite{liang09tcs}.

Now we can use these ideas to layer System~\SysL\ with polarities:

\begin{definition}[Polarised System~\SysL]
  The polarised syntax of formulae is defined as
  \[
  \begin{array}{lll}
    P,P',\ldots&\recdef a^+\sep A\ou B\\
    N,N',\ldots&\recdef a^-\sep A\et B\sep A\arr B\\
    A,B,\ldots&\recdef P\sep N
  \end{array}
  \]

  The syntax for proof-terms, together with their associated forms of typing judgements, is given below:
    \[\begin{array}{lll@{\qquad}l}
      -\mbox{terms}&t^-&\recdef x^-  \sep  \lambda x.t \sep  \paire {t_1}{t_2}{\color{white}{\sep\textsf{inj}_i(t)}}\sep \mu{\beta^-}.c &\Seq\Gam{{t^-\col N}\semcol\Del}\\
      +\mbox{terms}&t^+&\recdef x^+  {\color{white}{\sep  \lambda x.t \sep\paire {t_1}{t_2}}} \sep  \textsf{inj}_i(t)\sep \mu{\beta^+}.c &\Seq\Gam{{t^+\col P}\semcol\Del}\\
      \mbox{terms}&t&\recdef t^+\sep t^- &\Seq\Gam{{t\col A}\semcol\Del}\\\\[-5pt]
      -\mbox{continuations}&e^-&\recdef \alpha^-  \sep  \cons t e {\color{white}{ \sep  \paire {e_1}{e_2}}} \sep  \textsf{inj}_i(e) \sep  \m {x^-} c &\Seq{\Gam\semcol{e^-\col N}}\Del\\
      +\mbox{continuations}&e^+&\recdef \alpha^+\ {\color{white}{ \sep  \cons t e}} \sep  \paire {e_1}{e_2}  {\color{white}{ \sep  \textsf{inj}_i(e) } } \sep  \m {x^+} c&\Seq{\Gam\semcol{e^+\col P}}\Del \\
      \mbox{continuations}&t&\recdef e^+\sep e^-&\Seq{\Gam\semcol{e\col A}}\Del\\\\[-5pt]
      \mbox{commands}&c&\recdef\cutc {t^+} {e^+} \sep \cutc {t^-} {e^-}&c\col(\Seq\Gam\Del)
    \end{array}
    \]
    writing $x$ for either $x^+$ or $x^-$.
\end{definition}

Now that polarities explicitly appear in the syntax of proof-terms, it
is easy to reduce $\cutc{\m{\alpha} c}{\m {x} {c'}}$:
\[\cutc{\m{\alpha^+} c}{\m {x^+} {c'}}\Rew{}\subst c {\alpha^+}{\m {x^+} {c'}}\mbox{\quad and\quad} \cutc{\m{\alpha^-} c}{\m {x^-} {c'}}\Rew{}\subst {c'} {x^-}{\m {\alpha^-} {c}}\]

This turns into the following rewrite system:
\begin{definition}[Reductions in the polarised System~\SysL]\label{def:SysLPolRed}
  Again, we define values:
  \[
  \begin{array}{lll}
    \mbox{term values}&V
    &\recdef x\sep\textsf{inj}_i(V)\sep t^-\\ 
    \mbox{continuation values}&E
    &\recdef \alpha \sep \cons V E \sep \textsf{inj}_i(E)\sep e^+
  \end{array}
  \]

  The reduction relation $\Rew{\CBF}$ is defined as the contextual
  closure of the rules in Fig.~\ref{fig:SysLPolRed}.
\end{definition}

\begin{bfigure}[!h]
  \[
  \begin{array}{lclllllll}
    \rulenamed\rightarrow &\cutc{\lambda x.t}{\cons{V} E}&\Rew{} &\cutc{\subst t x V}{E}\\
    \rulenamed\wedge&\cutc{\paire {t_1}{t_2}}{\textsf{inj}_i(E)}&\Rew{} &\cutc{t_i}{E}\\
    \rulenamed\vee&\cutc{\textsf{inj}_i(V)}{\paire {e_1}{e_2}}&\Rew{} &\cutc{V}{e_i}\\\\
    \rulenamed{\stackrel\leftarrow\mu_{-}}&\cutc{\m {\beta^-}c}{E}&\Rew{} &\subst c {\beta^+} { E}\\
    \rulenamed{\stackrel\rightarrow\mu}&\cutc{t^-}{\m {x^-} c}&\Rew{} &\subst c {x^-} {t^-}\\
    \rulenamed{\stackrel\leftarrow\mu}&\cutc{\m {\beta^+}c}{e^+}&\Rew{} &\subst c {\beta^+} {e^+}\\
    \rulenamed{\stackrel\rightarrow\mu_{+}}&\cutc{{V}}{\m {x^+} c}&\Rew{} &\subst c {x^+} {V}\\\\
    \rulenamed{\zeta_{\textsf F}}&\cutc{t^-}{\cons {t^+} e}&\Rew{}&\cutc{t^+}{\m {x^+}{\cutc{t^-}{\cons {x^+} e}}}\\
    \rulenamed{\zeta_{\textsf F}}&\cutc{t^-}{\cons {V} {e^-}}&\Rew{}&\cutc{\m \alpha{\cutc{t^-}{\cons {V} \alpha}}}{e^-}\\
    \rulenamed{\zeta_{\textsf F}}&\cutc{t^-}{\inj i{e^-}}&\Rew{}&\cutc{\m \alpha{\cutc{t^-}{\inj i{\alpha}}}}{e^-}\\
    \rulenamed{\zeta_{\textsf F}}&\cutc{\inj i{t^+}}{e^+}&\Rew{}&\cutc{\m  {x^+}{\cutc{\inj i{x^+}}{e^+}}}{t^+}
  \end{array}
  \]
  where the $\rulenamed{\zeta_{\textsf F}}$-rules apply only under the condition
  that $t^+$ and $e^-$ are not values.
  \caption{Rewrite system for polarised System~\SysL}
  \label{fig:SysLPolRed}
\end{bfigure}

As in the \CBN\ and \CBV\ cases, we have:

\begin{theorem}[Confluence and Subject Reduction]

  $\Rew{\CBF}$ is confluent and satisfies Subject Reduction.
\end{theorem}

Notice that the notion of value is slightly different from that of
Definition~\ref{def:CBNCBV}: Indeed, if $\et$ is to be taken to be
negative, as the dual of (the obviously positive) $\ou$, we can take
every pair to be a value (in Definition~\ref{def:CBNCBV} we stuck to
Wadler's presentation~\cite{Wadlerdual}); this also removes the need for 
$\zeta$-rules for pairs. On the other hand, for a continuation $\cons t
e$ to be a value, we require it to be of the form $\cons V E$, as we no
longer recover confluence by opposing left vs.\ right (terms vs.\ 
continuations) but by opposing positives vs.\ negatives.

Precisely because we now no longer make any distinction based on the
left vs.\ right opposition (terms vs.\ continuations opposition), this
system could equally be given as a one-sided system, merging the
syntaxes of terms and continuations, but keeping of course the
distinction between positive terms and negative
terms.\footnote{Otherwise we would get back to Barbanera and Berardi's
  symmetric (and non-confluent) $\lambda$-calculus, with unclear
  denotational semantics.} At the level of formulae, we would get 4
connectives $\orP$ and $\andP$ of positive polarity, and $\orN$ and
$\andN$ of negative polarity:
\[\begin{array}{lll}
A\vee B&\mbox{becomes}& A\orP B\\ A\wedge B&\mbox{becomes}& A\andN B
\end{array}\qquad\qquad
\begin{array}{lll}
    A\arr B&\mbox{becomes}& \non A\orN B\\
    \non{(A\arr B)}&\mbox{becomes}& A\andP\non B
\end{array}
\]
where $\non{(A\arr B)}$ represents the dual of implication:
\Index{subtraction} (see \eg\cite{Crolard04}).

This is what we will do in Section~\ref{sec:LKF}.

\subsection{Focussing}

Now, the polarised System~\SysL\ presented above, which has been
studied at length by
Munch-Maccagnoni~\cite{MunchCSL09,Munch11,MunchPhD}, solves
non-confluence, not by giving systematic priority to the left (\CBV)
or to the right (\CBN), but by giving priority to the non-invertible
side (depending on the connective).

So the system takes advantage of the invertibility properties of the
asynchronous rules (right-introduction of negative connectives,
left-introduction of positive connectives). Invertibility entails
that, in terms of proof-search, you can \emph{chain} the decomposition
of every formula of the sequent that has an asynchronous introduction
rule, before doing anything else, without loss of generality (\ie
without losing logical completeness).

Now in~\cite{AndreoliP89,andreoli92focusing}, Andreoli proved a more
surprising result: \Index{focussing}, that says that, once you have
chosen to decompose by a synchronous rule a particular formula in the
sequent,\footnote{Positive formula on the right-hand side of a
  sequent, or a negative formula on its left-hand side} you can also
chain without loss of generality (\ie without losing logical
completeness) the recursive decomposition of its subformulae by
synchronous rules until you reveal a subformula of the opposite
polarity (whose decomposition can then be done by asynchronous rules
again).

This was in the context of linear logic, whence polarities have come,
but it is now understood in other polarised logics (classical or
intuitionistic). This result can be expressed as the completeness of a
sequent calculus with a \Index{focus} device, which syntactically
highlights a formula in the sequent and forces the next proof-search
step to decompose it with a synchronous rule, keeping the focus on its
newly revealed subformulae. In terms of proof-search, focussing
considerably reduces the search space, otherwise heavily redundant
when Gentzen-style inference rules are used.

Focussed proofs are proofs that implement such a chaining of
synchronous decompositions. The main idea is that focussed proofs are
those whose proof-terms systematically use term values and
continuation values, in other words, the normal forms for
$\zeta$-rules. Of course, such normal forms may feature cuts
($\zeta$-rules introduce cuts), but one should notice the following
properties:
\begin{remark}

  Just like $\Rew{\zeta_{\textsf N}}$ and $\Rew{\zeta_{\textsf V}}$
  (from Definition~\ref{def:CBNCBV}), the relation
  $\Rew{\zeta_{\textsf F}}$ is terminating.
\end{remark}

\begin{definition}[\LKzF]

Let \LKzF\ be the fragment of System~\SysL\ consisting of
$\Rew{\zeta_{\textsf F}}$-normal forms.
\end{definition}

\begin{remark}
  Just like \LKzN\ and \LKzV\ are stable under $\Rew{\CBN}$ and
  $\Rew{\CBV}$, the fragment \LKzF\ is stable under $\Rew{\CBF}$.
\end{remark}

\begin{remark} These normal form fragments relate to calculi of the literature:
  \begin{enumerate}
  \item 
    \LKzN\ is exactly the calculus called \LKT~\cite{DJS95,DJS97};
  \item 
    \LKzV\ is exactly the calculus called \LKQ~\cite{DJS95,DJS97};
  \item \LKzF\ relates to Liang and Miller's \LKF~\cite{liang09tcs},
    and this will be the object of Section~\ref{sec:LKF}.
  \end{enumerate}
\end{remark}

We therefore have 3 versions of the focussing result in classical
logic, an unfocussed proof $c$ can be turned into a focussed proof
$c'$ (in the sense of \LKzN, \LKzV, or \LKzF) by normalising it with
respectively $\Rew{\zeta_{\textsf N}}$, $\Rew{\zeta_{\textsf V}}$, or
$\Rew{\zeta_{\textsf F}}$, and normalising it by respectively
$\Rew{\CBN}$, $\Rew{\CBV}$, or $\Rew{\CBF}$ to eliminate cuts and
finally obtain a cut-free focussed proof.

\subsection{Weak $\eta$-conversion}

Now, the notion of $\eta$-conversion that we used as an introduction
to polarities in Section~\ref{sec:toLpol}, is a \emph{strong} notion
of $\eta$-conversion: 

Inspired by the way we can reduce an axiom on a non-atomic formula into
a proof using axioms on smaller formulae, we considered the
$\eta$-expansion of variables $x$ and $\alpha$:
\[\begin{array}{lllll}
    {x}&\Rew{}&\lambda y.\m \alpha {\cutc{x}{\st y \alpha}}\\
    {x}&\Rew{}&\paire{\m \alpha{\cutc{x}{\inj 1 \alpha}}}{\m \gamma{\cutc{x}{\inj 2 \gamma}}}\\
    {\alpha}&\Rew{}&\paire{\m x{\cutc{\inj 1 x}{\alpha}}}{\m z{\cutc{\inj 2 z}{\alpha}}}
\end{array}
\]
which we sought to generalise to
\[\begin{array}{lllll}
    {t}&\Rew{}&\lambda y.\m \alpha {\cutc{{t}}{\st y \alpha}}\\
    {t}&\Rew{}&\paire{\m \alpha{\cutc{{t}}{\inj 1 \alpha}}}{\m \gamma{\cutc{{t}}{\inj 2 \gamma}}}\\
    {e}&\Rew{}&\paire{\m x{\cutc{\inj 1 x}{{e}}}}{\m z{\cutc{\inj 2 z}{{e}}}}
\end{array}
\]
for \emph{any} term $t$ and any continuation $e$.

This led to a polarity-based reduction relation that contrasts with
the \CBN\ and \CBV\ reduction relations from Chapter~\ref{ch:CH}. But
that does not mean that \CBN\ and \CBV\ are incompatible with the
concept of $\eta$-conversion: they just require \emph{weaker} notions
of $\eta$-conversion than that discussed above.

The notion of $\eta$-conversion that is suitable for \CBN\ are
\[\begin{array}{lllll}
    {t}&\Rew{}&\lambda y.\m \alpha {\cutc{{t}}{\st y \alpha}}\\
    {t}&\Rew{}&\paire{\m \alpha{\cutc{{t}}{\inj 1 \alpha}}}{\m \gamma{\cutc{{t}}{\inj 2 \gamma}}}\\
    {E}&\Rew{}&\paire{\m x{\cutc{\inj 1 x}{E}}}{\m z{\cutc{\inj 2 z}{E}}}
\end{array}
\]
where $\alpha$ has not been substituted by any continuation $e$ but \emph{only} by a continuation value $E$.

The notion of $\eta$-conversion that is suitable for \CBV\ are
\[\begin{array}{lllll}
    {V}&\Rew{}&\lambda y.\m \alpha {\cutc{V}{\st y \alpha}}\\
    {V}&\Rew{}&\paire{\m \alpha{\cutc{V}{\inj 1 \alpha}}}{\m \gamma{\cutc{V}{\inj 2 \gamma}}}\\
    {e}&\Rew{}&\paire{\m x{\cutc{\inj 1 x}{{e}}}}{\m z{\cutc{\inj 2 z}{{e}}}}
\end{array}
\]
where $x$ has not been substituted by any term $t$ but \emph{only} by a term value $V$.

Including these notions of \CBN-$\eta$-conversion and
\CBV-$\eta$-conversion in the \CBN\ and \CBV\ notions of reduction, is
actually necessary if these are to capture the semantics of classical
proofs in control and co-control categories, respectively: just like
in Theorem~\ref{th:SoundCompleteCCC} we needed $\eta$-conversion to
make the simply-typed $\lambda$-calculus sound and complete with
respect to the semantics given by CCC, here we would need the above
notions of \CBN-$\eta$-conversion and \CBV-$\eta$-conversion in order
to turn the implications of Theorem~\ref{th:SoundCompleteCBNCBV}
(soundness) into equivalences (soundness and completeness). This is
actually what Selinger proved~\cite{SelingerControlCat99} in the
context of the $\lambda\mu$-calculus.

\subsection{Related works}

The role of polarities and focussing in classical proof theory has
been investigated by a substantial literature, inspired by Girard's linear
logic~\cite{girard-ll}. Following this work and Andreoli's on
focussing~\cite{AndreoliP89,andreoli92focusing}, Girard developed
in~\cite{Gir:newclc} a sequent calculus \LC\ for classical logic with
more structure than Gentzen's \LK~\cite{Gentzen35}, based on an
assignment of polarities to classical formulae. Danos, Joinet and
Schellinx~\cite{DJS95,DJS97} studied semantically meaningful ways to
make cut-elimination confluent in the classical sequent calculus,
introducing
\begin{itemize}
\item the calculi \LKT\ and \LKQ\ mentioned above
\item a version of the sequent
calculus called $\LK^{tq}$ where a \Index{colour} $t$ or $q$ on each formula
indicates whether a cut on that formula should be pushed to the right
or to the left,
\item more restricted versions thereof,
\end{itemize}
all inspired by the various translations of classical logic into
linear and intuitionistic logics. Out of that field, which includes
the duality betweenn \CBN\ and \CBV,\footnote{As revealed by Curien
  and Herbelin's System~\SysL~\cite{CurienHerbelinDuality99} and
  Selinger's control categories~\cite{SelingerControlCat99}.}
\Index{polarised classical logic} emerged, developed as such by
Laurent et al.~\cite{phdlaurent,LaurentQF05}. It develops and enriches
Girard’s work on \LC, in particular by explaining the proof theory of
classical formulae as given by \LC\ as a combination of
\begin{itemize}
\item an encoding from classical formulae to polarised classical
  formulae
\item a proof theory for polarised classical logic.
\end{itemize}

Closer to Andreoli's original line of research, which was motivated by
logic programming, Liang and Miller then formalised \LKF\ as a more
strongly focussed calculus than that called \LKzF\ above; we will
study it in the next section.

A useful introduction to that literature can be found in Chapter 2 of
Farooque's thesis~\cite{FarooquePhD}.

\medskip

More recently, Munch-Maccagnoni approached the concept of focussing
via orthogonality models~\cite{MunchCSL09}. He built for the polarised
version of System~\SysL\ the same kind of orthogonality model as the
one we presented in Section~\ref{sec:OrthoClassicConfluent} for the
\LKzN-case, with an interpretation $\SemTyP A$ of a formula $A$ built
as the orthogonal or bi-orthogonal of a more basic set of
(counter-)proof-terms. He essentially shows that the interpretation
$\SemTyP A$ of a formula is generated from its values, in the sense
that $\SemTyP A=\biorth{(\SemTyP A\cap\mathcal V)}$ where $\mathcal V$
denotes the set of term values (and symmetrically $\SemTyN
A=\biorth{(\SemTyN A\cap\mathcal E)}$ where $\mathcal E$ denotes the
set of continuation values).

This sheds an interesting light on our definition of 
\[
\SemTyrN[\sigma] {\gA \arr \gB} \eqdef \cons{\SemTyP[\sigma] \gA}{\SemTyN[\sigma] \gB}
\]
in our orthogonality models of
\LKzN\ (Definitions~\ref{def:typeinterpretSysL} and
\ref{def:InterpretTypesWitness}): While in \LKzN\ the above construct
only considers those inhabitants of $\SemTyN[\sigma] \gB$ that are
continuation values anyway, there is in the general case of
System~\SysL\ a question of whether we want continuations of the form
$\st t{\m x c}$,\footnote{(with $t\in\SemTyP[\sigma] \gA$ and $\m x
  c\in\SemTyN[\sigma] \gB$)} which are not focussed (in the sense of
\LKzN\ or \LKzF), in the interpretation $\SemTyrN[\sigma] {\gA \arr
  \gB}$. The result that $\SemTyN {\gA \arr \gB} = \biorth{(\SemTyN
  {\gA \arr \gB}\cap\mathcal E)}$ means that an ``unfocussed''
counter-proof such as $\st t{\m x c}$ would be accepted after the
bi-orthogonal completion (\ie in $\SemTyN[\sigma] {\gA \arr
  \gB}=\biorth{(\SemTyrN[\sigma] {\gA \arr \gB})}$).

\medskip

So far we have taken advantage of focussing, \ie the chaining of
synchronous rules, to identify complete fragments \LKzN, \LKzV, and
\LKzF\ of classical sequent calculus proofs.

Although we have discussed invertibility of asynchronous rules, in
order to introduce the notion of polarity, we have not forced our
proofs to apply asynchronous rules eagerly, before applying other
rules (in \LKzF, $\m {\alpha^-}{c}$ is still an accepted proof of a
negative formula such as $A\arr B$). 

This is the main difference with the focussed proof systems in the
style of \eg Liang and Miller~\cite{liang09tcs}, where \eg all proofs
of $A\arr B$ finish with the right-introduction of $\arr$. In terms of
proof-terms, it means that all proof-terms are in $\eta$-long normal
forms (we can transform every proof-term into a proof-term in that
form by a series of $\eta$-expansions, but it is always tricky to
control the termination of $\eta$-expansion without having the types
explicitly in the terms).

Coming back to the purely logical level, focussed proofs in the
tradition of Miller et al. can be described in terms of ``big-step
focussing'': going up a branch of the proof is an alternation of
sychronous and asynchronous phases, which we may consider to be
atomic. The next section shows a computational interpretation of that
strongly focussed formalism.

\newcommand\Slide[2][]{\subsection{#1}#2}
\newcommand\dis[2]{#2}
\newcommand\disloc[2]{#2}
\newcommand\bleu[1]{#1}
\newcommand\rouge[1]{#1}
\newcommand\orange[1]{#1}
\newcommand\bord[1]{#1}
\newcommand\gray[1]{#1}
\newcommand\remph[1]{#1}
\newcommand\bemph[1]{#1}
\newcommand\oemph[1]{#1}
\newcommand\Sectionslide[1]{\section{#1}}
\newcommand\subhead[1]{\textbf{#1}}
\newcommand\theo\relax
\newcommand\defi\relax
\newcommand\coro\relax
\newcommand\itemph[1]{\baam{#1}}

\section{Computational interpretation of a focussed calculus}
\label{sec:LKF}

The starting point of this section is Liang and Miller's
\LKF~\cite{liang09tcs}, a variant of the system \LKzF\ described in the
previous section that forces the asynchronous decomposition of
formulae. It is described in purely logical terms and we will see how,
by formalising the concept of \Index{big-step focussing}, a
Curry-Howard interpretation can be given to \LKF, following
Zeilberger's work~\cite{ZeilbergerPOPL08,Zeilberger08}.

We start with the formulae of \Index{polarised classical logic}.

\begin{definition}[Polarised formulae]

The syntax of formulae is given by the following grammar
    \[
    \begin{array}{lll}
      \mbox{\Index[positive formula (\LKF)]{Positive formulae}}&P&\recdef a\sep A_1\andP A_2 \sep A_1\orP A_2 \\
      \mbox{\Index[negative formula (\LKF)]{Negative formulae}}&N&\recdef \non a\sep A_1\andN A_2 \sep A_1\orN A_2\\
      \mbox{\Index[formula (\LKF)]{Formulae}}&A&\recdef P\sep N
    \end{array}
    \]
    where $a$ ranges over a fixed set of elements called
    \Index[positive atom (\LKF)]{positive atoms}, and $\non a$ ranges over a
    bijective copy of that set ($a\mapsto\non a$ is the bijection),
    whose elements are called \Index[negative atom (\LKF)]{negative atoms}.
 

    We extend the bijection between positive and negative atoms into
    an involutive bijection, called \Index[negation (\LKF)]{negation},
    between positive and negative formulae:
    \[
    \begin{array}c
    \begin{array}{|ll|ll|}
      \hline
      \non {(a)}&\eqdef \non a& \non {(\non a)}&\eqdef a  \\
      \non{(A_1\andP {A_2})}&\eqdef \non A_1 \orN\non A_2&\non{(A_1\andN {A_2})}&\eqdef \non A_1 \orP\non A_2\\
      \non{(A_1\orP {A_2})}&\eqdef \non A_1 \andN\non A_2&\non{(A_1\orN {A_2})}&\eqdef \non A_1 \andP\non A_2\\
      \hline
    \end{array}
    \end{array}
    \]
\end{definition}

Following the suggestion made in the previous section, we fold
\LKF\ into a 1-sided sequent calculus (hence our interest for the
involutive negation), as it is traditional in the field arising from
linear logic~\cite{girard-ll}.

\begin{example}For instance, Peirce's, law, which in the previous chapters and sections we wrote as
  $((a\arr b)\arr a)\arr a$, is now $\shft(\shft(\non a\orN\shft
  b)\andP \shft{\non a})\orN \shft a$.
\end{example}

\begin{definition}[Liang-Miller's \LKF]

  The rules of \LKF\ are given in Fig.~\ref{fig:LKF} for two kinds of sequents:\\
  \begin{tabular}{ll}
    $\DerOSPos{\Theta}{A}$&{ focussed sequent}\\
    $\DerOSNeg{\Theta}{\Gamma}$&{ unfocussed sequent}
  \end{tabular}\\
  where $A$ is an arbitrary formula, $\Theta$ is a multiset of either
  negative atoms or positive formulae and $\Gamma$ is a multiset of
  arbitrary formulae.

  Derivability in \LKF\ of the sequents $\DerOSPos{\Theta}{A}$ and
  $\DerOSNeg{\Theta}{\Gamma}$ is respectively denoted
  $\DerOSPos[\LKF]{\Theta}{A}$ and
  $\DerOSNeg[\LKF]{\Theta}{\Gamma}$.
\end{definition}
\begin{bfigure}[!h]
  \[
    \begin{array}{c}
      \mbox{{Synchronous phase}}
      \hfill
      \infer{\DerOSPos{\Theta}{A_1\andP A_2}}
      {\DerOSPos{\Theta}{A_1}\qquad \DerOSPos{\Theta}{A_2}}
      \qquad
      \infer{\DerOSPos{\Theta}{A_1\orP A_2}}
      {\DerOSPos{\Theta}{A_i}}
      \\\\
      \mbox{{End of synchronous phase}}
      \hfill
      \infer{\DerOSPos \Theta {\shft N}}{\DerOSNeg \Theta N}
      \qquad
      \infer[\non a\in\Theta]{\DerOSPos{\Theta}{a}}{\strut}
      \\\\
      \mbox{{Asynchronous phase}}
      \hfill
      \infer{\DerOSNeg{\Theta}{A_1\andN A_2,\Gamma}}
      {\DerOSNeg{\Theta}{A_1,\Gamma}\qquad \DerOSNeg{\Theta}{A_2,\Gamma}}
      \qquad
      \infer{\DerOSNeg{\Theta}{A_1\orN A_2,\Gamma}}
      {\DerOSNeg{\Theta}{A_1,A_2,\Gamma}}
      \\\\
      \mbox{{End of asynchronous phase}}
      \qquad\hfill
      \infer{\DerOSNeg \Theta {\shft P,\Gamma}}{\DerOSNeg {\Theta, {P}} { \Gamma}}
      \qquad
      \infer{\DerOSNeg \Theta {\non p,\Gamma}}{\DerOSNeg {\Theta,\non p} {\Gamma}}
      \qquad
      \infer{\DerOSNeg {\Theta, {P}} {}}{\DerOSPos {\Theta,{P}} P}
    \end{array}
    \]
    \caption{\LKF}
    \label{fig:LKF}
\end{bfigure}

Liang and Miller showed in~\cite{liang09tcs} that
\begin{itemize}
\item various cut-rules are admissible;
\item the polarities of atoms and connectives do not change the
  provability of an unfocussed sequent, but they change the shape of
  its proofs;
\item with the admissibility of cut-rules, the system is complete for
  classical logic, no matter which polarities are placed on connectives
  and literals.
\end{itemize}

To prove cut-admissibility, they do not explicitly formalise a
cut-elimination procedure, but it could probably be inferred from the
proof.

Note that soundness of the system with respect to classical logic is
trivially checked, rule by rule, forgetting about polarities and the
structure of sequents.

Polarities in classical logic raise interesting questions: $A\andP B$
and $A\andN B$ are equiprovable, and so are $A\orP B$ and $A\orN
B$. But, while the difference between the (direct) proofs of $A\orP B$
and (direct) proofs of $A\orN B$ is clear,\footnote{Direct proofs of
  $A\orP B$ choose one side and throw away the other, while direct
  proofs of $A\orN B$ keep the two sides.} one may wonder what the
real difference is between (direct) proofs of $A\andP B$ and (direct)
proofs of $A\andN B$, given that the two rules look very much alike.

The difference lies not in their structure, but in the way they will
behave in cut-elimination:
\begin{itemize}
\item from a proof of $A\andN B$ (facing a proof of $\non A\orP \non
  B$), only one sub-proof is used while the other is thrown away,
\item from a proof of $A\andP B$ (facing a proof of $\non A\orN \non
  B$), both sub-proofs are used.
\end{itemize}
Metaphorically, proving either conjunction is like picking 1 boy name
and 1 girl name, when your couple is pregnant: proving a negative
conjunction is picking the two names when you are expecting one baby
(not knowing whether it is a boy or a girl), while proving the
positive conjunction is picking the two names when expecting twins (a
boy and a girl). On the paper, you have the same job to do, but you
will probably approach the problem very differently.

\subsection{Informal relation to System~\SysL}

It may not be obvious, but system \LKF\ roughly expresses, without
proof-terms, some derivations of System~\SysL\ in a 1-sided format.\footnote{A 1-sided version of System~\SysL\ would merge terms and continuations into terms, so that $\cons V {V'}$ is a term value, and merge term variables and continuation variables into continuation variables.}

Indeed, think of 
\begin{itemize}
\item a focussed sequent $\DerOSPos{\Theta}{A}$ as a typing judgement for a term value $V$:\flush $\Seq{}{\XcolY V A\semcol \Theta}$.
\item an unfocussed sequent $\DerOSNeg{\Theta}{\Gamma}$ as a typing judgement for a command $c$:\flush $\XcolY c{(\Seq{}{\semcol \Theta,\Gamma})}$.
\end{itemize}
with $\Gamma$ being the part of the typing context for negative
continuation variables $\alpha^-$ with non-atomic types (which can be
asynchronously decomposed), and $\Theta$ the rest of it (typing negative continuation variables $\alpha^-$ with atomic types, and typing positive continuation variables $\alpha^+$).

To derive a focussed sequent $\DerOSPos{\Theta}{A}$:
\begin{itemize}
\item the two rules of the group
`Synchronous phase' correspond to the typing rules for $\cons V {V'}$ and for $\inj i {V'}$;
\item the first rule of the group `End of synchronous phase' does not
  correspond to a rule of System~\SysL\ but simply the realisation
  that the value $V$ is a negative term $t^-$;
\item the second rule of that group is when $V$ is a variable.
\end{itemize}
To derive an unfocussed sequent $\DerOSNeg{\Theta}{\Gamma}$:
\begin{itemize}
\item the two rules of the group `Asynchronous phase' correspond to
  the typing of $\paire{t_1}{t_2}$ and $\lambda {\alpha}.t$;
\item the first two rules of the group `End of asynchronous phase' are
  not reflected in System~\SysL\ (they just move formulae that cannot
  be asynchronously decomposed from $\Gamma$ to $\Theta$)
\item the third rule of that phase corresponds to the typing of
  $\cutc{V}{\alpha^+}$.
\end{itemize}

Roughly speaking, we should think of \LKF\ as typing those proof-terms
of (a 1-sided version of) System~\SysL\ that are $\eta$-long
$\Rew{\CBF}$-normal forms. These are described by the following grammar:
\[
\begin{array}{ll}
  V, V'&\recdef \alpha^+\sep \cons V {V'}\sep \inj i {V}\sep t^-\sep \m {\alpha^-}c\\
  t^-,\ldots&\recdef \paire{t_1}{t_2}\sep\lambda {\alpha} .t\\
  t,\ldots&\recdef \m {\alpha^+}c\sep t^-\sep \m {\alpha^-}c\\ 
  c,\ldots &\recdef \cutc V{\alpha^+}\sep\cutc {t^-}{\alpha^-}
\end{array}
\]
where the type of every $\m {\alpha^-}c$ is atomic.

It is only ``roughly speaking'', because in Liang-Miller's \LKF:
\begin{enumerate}
\item when a formula is asynchronously decomposed, no copy of the
  formula is kept in the sequent (which means in the above grammar that in every command
  $\cutc {t^-}{\alpha^-}$ we impose $\alpha^-\not\in\FV{t^-}$),
\item when the focus is placed on a formula, all the formulae that
  could be asynchronously decomposed have already been asynchronously
  decomposed (which means in the above grammar that in every command $\cutc V{\alpha^+}$,
  $V$ has no free variable of the form $\alpha^+$).
\item finally, the order in which formulae are decomposed in the
  asynchronous phase, is less deterministic than that imposed by the
  above grammar.
\end{enumerate}

This is because, in Liang and Miller's view of focussing, and more
generally in the tradition of linear logic, ``what happens in the
asynchronous phase stays in the asynchronous phase'', in the sense
that the details of the asynchronous phase (\eg the order in which
formulae are decomposed) are meaningless and should not impact the
semantics of the proof.

This is difficult to reflect at the level of System~\SysL's proof-terms.

Therefore, we will now develop a Curry-Howard interpretation for that
particular view of focussing, along the lines of Zeilberger's
work~\cite{ZeilbergerPOPL08,Zeilberger08,Zeilberger10}: in order to
forget about the inner details of the asynchronous phase, we formalise
the idea of compacting each phase (asynchronous and even synchronous)
into one atomic inference. This is called big-step focussing.


\subsection{Identifying phases as atomic steps}

We start by showing an example of how positive connectives are decomposed.

\begin{example}
Trying to prove ${\DerOSPos{\Theta}{\shft N_1\andP(a \orP\shft N_2)}}$ we can build:\\
either
  \[
  \infer{\DerOSPos{\Theta}{\shft N_1\andP(a \orP\shft N_2)}}
  {\infer{\DerOSPos{\Theta}{\rouge{\shft N_1}}}
      {\infer{\DerOSNeg{\Theta}{\rouge{N_1}}}
        {\mbox{End of synch phase}}}
    \quad
      \infer{\DerOSPos{\Theta}{a\orP \shft N_2}}
      {\infer{\DerOSPos{\Theta}{\rouge{a}}}
          {\infer{\rouge{\non a}\in\Theta}
            {\mbox{End of synch phase}}}}
  }
\]

or
\[\infer{\DerOSPos{\Theta}{\shft N_1\andP(a \orP\shft N_2)}}
        {\infer{\DerOSPos{\Theta}{\rouge{\shft N_1}}}
          {\infer{\DerOSNeg{\Theta}{\rouge{N_1}}}
            {\mbox{End of synch phase}}}
          \quad
          \infer{\DerOSPos{\Theta}{a\orP \shft N_2}}
                {\infer{\DerOSPos{\Theta}{\rouge{\shft N_2}}}
                  {\infer{\DerOSNeg{\Theta}{\rouge{N_2}}}
                    {\mbox{End of synch phase}}}
                }
        }      
  \]

  The whole synchronous phase can be expressed in just one step:
    \[
    \infer{\DerOSPos{\Theta}{\shft N_1\andP(a \orP\shft N_2)}}
    {\DerOSNeg{\Theta}{\rouge{N_1}}\quad \rouge{\non a}\in\Theta}
    \qquad\mbox{ or }\qquad
    \infer{\DerOSPos{\Theta}{\shft N_1\andP(a \orP\shft N_2)}}
    {\DerOSNeg{\Theta}{\rouge{N_1}}\quad \DerOSNeg{\Theta}{\rouge{N_2}}}
    \]

    In other words:
    \[
    \infer{\DerOSPos{\Theta}{\shft N_1\andP(a \orP\shft N_2)}}
    {\forall N\in{\Gamma},\quad \DerOSNeg{\Theta}{N}\qquad \forall a\in{\Gamma},\quad \non a\in\Theta}
    \]
    with $\Gamma=\rouge{N_1},\rouge{a}$ or $\Gamma=\rouge{N_1},\rouge{N_2}$

    In either case, we say that $\Gamma$ ``is a positive decomposition of''\quad $\shft N_1\andP(a \orP\shft N_2)$,\quad which we denote:
    $\DerDec+{\rouge{N_1},\rouge{a}}{\ \shft N_1\andP(a \orP\shft N_2)}$
    and
    $\DerDec+{\rouge{N_1},\rouge{N_2}}{\ \shft N_1\andP(a \orP\shft N_2)}$.
\end{example}

Now we generalise this example into a formal definition:

\begin{definition}[Decomposition of positive connectives]

  The \Index{positive decomposition relation} is the binary relation,
  defined by the rules of Fig.~\ref{fig:posdecomp}, where
  $\Gamma,\Gamma_1,\Gamma_2$ are sets of positive atoms or negative
  formulae.

  The \Index{one-step synchronous phase} is the rule:
  \[
  \infers{\DerOSPos{\Theta}{A}}[\textsf{synch}]
        {\DerDec+{\Gamma}{A}\qquad\forall N\in{\Gamma}, \DerOSNeg{\Theta}{N}\qquad \forall a\in{\Gamma},\quad \non a\in\Theta}
        \]
          {}
\end{definition}
\begin{bfigure}[!h]
    \[
    \infer{\DerDec+{N} {\shft N}}{}
    \qquad
    \infer{\DerDec+{a} {a}}{}
    \]
    \[
    \infer{\DerDec+{\Gamma_1,\Gamma_2}{A_1\andP A_2}}
    {\DerDec+{\Gamma_1}{A_1}
      \qquad
      \DerDec+{\Gamma_2}{A_2}
    }
    \qquad
    \infer{\DerDec+{\Gamma}{A_1\orP A_2}}
    {\DerDec+{\Gamma}{A_i}}
    \]
    \caption{Positive decomposition relation}
    \label{fig:posdecomp}
\end{bfigure}

Notice the syntax we use for the \textsf{synch} rule: the symbols
$\forall$ and $\in$ are \textbf{meta-level} symbols: the number of
premisses is the cardinal of $\Gamma$ (plus one if you count
$\DerDec+{\Gamma}{A}$).

We now show an example of how negative connectives are decomposed.

\begin{example}
  \[
  \infer{\DerOSNeg{\Theta}{\shft P_1\orN (\non a \andN\shft P_2)}}
  {\infer{\DerOSNeg{\Theta}{\shft P_1, (\non a \andN\shft P_2)}}
      {\infer{\DerOSNeg{\Theta}{\shft P_1, \non a}}
        {
          \infer{\DerOSNeg{\Theta,\rouge{P_1, \non a}}{}}
              {\mbox{End of asynch phase}}
          \qquad
          \infer{\DerOSNeg{\Theta}{\shft P_1, \shft P_2}}
            {  {\infer{\DerOSNeg{\Theta,\rouge{P_1, P_2}}{}}
                {\mbox{End of asynch phase}}}
            }
        }
      }
  }
  \]

  The whole asynchronous phase can be expressed in just one step:
  \[
  \infers{\DerOSNeg{\Theta}{\shft P_1\orN (\non a \andN\shft P_2)}}[\textsf{asynch}]
        {\DerOSNeg{\Theta,\rouge{P_1, \non a}}{}\qquad \DerOSNeg{\Theta,\rouge{P_1, P_2}}{}}
        \]

    In other words
    \[
    \infer{\DerOSNeg{\Theta}{\shft P_1\orN (\non a \andN\shft P_2)}}
    {\forall {\Delta},\quad \DerOSNeg{\Theta,{\Delta}}{}}
    \]
    where $\Delta$ ranges over $\{\ \rouge{\{P_1,\non a\}}\ ,\ \rouge{\{P_1, P_2\}}\ \}$
  
    In either case, we say that $\Delta$ ``is a negative decomposition of''\quad $\shft P_1\orN (\non a \andN\shft P_2)$,\quad which we denote
    $\DerDec-{\rouge{P_1},\rouge{\non a}}{\ \shft P_1\orN (\non a \andN\shft P_2)}$
    and
    $\DerDec-{\rouge{P_1},\rouge{P_2}}{\ \shft P_1\orN (\non a \andN\shft P_2)}$.
\end{example}

Now we generalise this example into a formal definition:

\begin{definition}[Decomposition of negative connectives]

  The \Index{negative decomposition relation} is the binary relation,
  defined by the rules of Fig.~\ref{fig:negdecomp}, where
  where $\Delta,\Delta_1,\Delta_2$ are sets of negative atoms or positive formulae.

    The \Index{one-step asynchronous phase} is the rule:
    \[
    \infer{\DerOSNeg \Theta {A}}
    {\forall \Delta, (\DerDec-{\Delta}{A})
      \Rightarrow
      (\DerOSNeg {\Theta,{\Delta}} {})}
    \]
\end{definition}
\begin{bfigure}[!h]
    \[
    \infer{\DerDec-{P} {\shft P}}{}
    \qquad
    \infer{\DerDec-{\non a} {\non a}}{}
    \]
    \[
    \infer{\DerDec-{\Delta}{A_1\andN A_2}}
    {\DerDec-{\Delta}{A_i}}
    \qquad
    \infer{\DerDec-{\Delta_1,\Delta_2}{A_1\orN A_2}}
    {\DerDec-{\Delta_1}{A_1}
      \qquad
      \DerDec-{\Delta_2}{A_2}
    }
    \]
    \caption{Negative decomposition relation}
    \label{fig:negdecomp}
\end{bfigure}

Again, notice that the syntax we use for the \textsf{asynch} rule uses
the \textbf{meta-level} symbols $\forall$ and $\imp$: the number of
premisses is the number of $\Delta$ satisfying $\DerDec-{\Delta}{A}$
for the given $A$.

We now put everything together in the style of
Zeilberger~\cite{ZeilbergerPOPL08,Zeilberger08,Zeilberger10}.

\begin{definition}[Big-step \LKF, v1]

  The big-step \LKF\ system is given in Fig.~\ref{fig:bstepLKF1},
  where $\Theta,\Delta$ are sets of negative atoms or positive
  formulae and $\Gamma$ is a set of positive atoms or negative
  formulae.
\end{definition}

\begin{bfigure}[!h]
  \[
  \begin{array}{c}
    \infers{\DerOSPos{\Theta}{A}}[\textsf{synch}]{
      \DerDec+{\Gamma}{A}\qquad\forall N\in{\Gamma}, \DerOSNeg{\Theta}{N}\qquad \forall a\in{\Gamma},\quad \non a\in\Theta
    }
    \\\\
    \infers{\DerOSNeg {\Theta, P} {}}[\textsf{focus}]{\DerOSPos {\Theta,P} P}
    \qquad\qquad
    \infers{\DerOSNeg \Theta {A}}[\textsf{asynch}]
          {\forall \Delta, (\DerDec-{\Delta}{A})
            \Rightarrow
            (\DerOSNeg {\Theta,{\Delta}} {})}
  \end{array}
  \]
  \caption{Big-step \LKF, v1}
  \label{fig:bstepLKF1}
\end{bfigure}

\begin{remark}

  Sequents of the form $\DerOSPos{\Theta}{N}$ and sequents of the form
  $\DerOSNeg{\Theta}{P}$ are never present in the premisses of the
  rules. Such sequents can only appear as the very conclusion of a
  whole proof-tree.

  Hence, we can equivalently present the big-step \LKF\ system as the
  system of Fig.~\ref{fig:bstepLKF2}, and declare $\DerOSNeg \Theta {P}$
  as syntactic sugar for $\DerOSNeg {\Theta,P} {}$, and $\DerOSPos
  \Theta {N}$ as syntactic sugar for $\DerOSNeg {\Theta} {N}$.
\end{remark}

\begin{bfigure}[!h]
\[
    \begin{array}{c}
      \infer{\DerOSPos{\Theta}{P}}
      {\DerDec+{\Gamma}{P}\qquad\forall N\in{\Gamma}, \DerOSNeg{\Theta}{N}\qquad \forall a\in{\Gamma},\quad \non a\in\Theta}
      \\\\
      {\infer{\DerOSNeg {\Theta, P} {}}{\DerOSPos {\Theta,P} P}}
      \qquad\qquad
      {\infer{\DerOSNeg \Theta {N}}
      {\forall \Delta, (\DerDec-{\Delta}{N})
        \Rightarrow
        (\DerOSNeg {\Theta,{\Delta}} {})}}
    \end{array}
    \]
  \caption{Big-step \LKF, v2}
  \label{fig:bstepLKF2}
\end{bfigure}

Now we should notice a complete symmetry, and therefore some redundancy, in the two decomposition relations that we have defined:

\begin{remark}
  $\DerDec+{\Gamma }P$ \iff\ $\DerDec-{\non{\Gamma}}{\non P}$.
\end{remark}

Hence, we can define in Fig.~\ref{fig:bstepLKF3} a simplified version of big-step \LKF, where this redundancy is eliminated.

\begin{bfigure}[!h]
    \[
    \begin{array}{c}
      \infer{\DerOSPos{\Theta}{P}}
      {\DerDec{}{\Gamma}{P}\qquad\forall N\in{\Gamma}, \DerOSNeg{\Theta}{N}\qquad \forall a\in{\Gamma},\quad \non a\in\Theta}
      \\\\
      {\infer{\DerOSNeg {\Theta, P} {}}{\DerOSPos {\Theta,P} P}}
      \qquad\qquad
      {\infer{\DerOSNeg \Theta {N}}
        {\forall \Gamma, (\DerDec{}{\Gamma}{\non N})
          \Rightarrow
          (\DerOSNeg {\Theta,\non{\Gamma}} {})}}
    \end{array}
    \]
    where $\DerDec{}{\Gamma}{P}$ is $\DerDec+{\Gamma}{P}$.
  \caption{Big-step \LKF, v3}
  \label{fig:bstepLKF3}
\end{bfigure}


Now notice that the rules for negative connectives are never used in
the system! Due to the duality in the syntax, given by the involutive
negation, we should be able to remove negative connectives altogether.
We just need to introduce a marker in the syntax of a formula, to
denote every change of polarity.

Let us write $\neg$ for this marker.
\begin{definition}[Syntax with positive connectives only]

Formulae are now defined by the following syntax:
\[\begin{array}{ll}
P&\recdef a\sep A_1\andP A_2 \sep A_1\orP A_2\\
A&\recdef P\sep \neg P
\end{array}
\]

with the following involutive
negation:
    \[\begin{array}{ll}
      \non {P}&\eqdef \neg P\\
      \non {(\neg P)}&\eqdef P
    \end{array}
    \]
\end{definition}

\begin{remark}
The previous grammar can be encoded into that one:
    \[\begin{array}{ll@{\qquad}ll}
      \overline a&\eqdef a\\
      \overline{A\andP B}&\eqdef\overline A\andP\overline B\\
      \overline{A\orP B}&\eqdef\overline A\orP\overline B\\
    \end{array}
    \qquad
    \overline N\eqdef\neg(\overline{\non N})
    \]
\end{remark}

In Fig.~\ref{fig:bstepLKF4} we reformulate big-step \LKF\ with this syntax for formulae.
\begin{bfigure}[!h]
\[
    \begin{array}{c}
      \infer{\DerOSPos{\Theta}{P}}
      {\DerDec{}{\Gamma}{P}\qquad\forall \neg P'\in{\Gamma}, \DerOSNeg{\Theta}{\neg P'}\qquad\forall a\in{\Gamma}, \neg a\in\Theta}
      \\\\
      \infer{\DerOSNeg {\Theta, P} {}}{\DerOSPos {\Theta, P} P}
      \qquad\qquad
      \infer{\DerOSNeg \Theta {\neg P}}
      {\forall \Gamma, (\DerDec{}{\Gamma}{P})
        \Rightarrow
        (\DerOSNeg {\Theta,\non{\Gamma}} {})}
    \end{array}
    \]
    where $\DerDec{}{\Gamma}{P}$ is $\DerDec+{\Gamma}{P}$.
  \caption{Big-step \LKF, v4}
  \label{fig:bstepLKF4}
\end{bfigure}

Finally, we notice that it is more natural to write $\Gamma$ on the
left-hand side of a sequent:

\begin{definition}[Big-step \LKF, v5]

  The big-step \LKF\ system v5 is given in Fig.~\ref{fig:bstepLKF5},
  where $\Gamma$ is a set of atoms $a$ or formulae of the form $\neg P$ and
  $\Theta$ is a set of negated atoms $\neg a$ or formulae of the form
  $P$.
\end{definition}
\begin{bfigure}[!h]
\[
\begin{array}{c}
  \infer{\Gamma_0\DerOSPos{}{P}}
        {\DerDec{}{\Gamma}{P}\qquad\forall \neg P'\in{\Gamma},\ \Gamma_0\DerOSNeg{}{\neg P'}\qquad\forall a\in{\Gamma},\ a\in\Gamma_0}
        \\\\
        \infer{\Gamma_0, \neg P\DerOSNeg {} {}}{\Gamma_0,\neg P\DerOSPos {} P}
        \qquad\qquad
        \infer{\Gamma_0\DerOSNeg {} {\neg P}}
              {\forall \Gamma, (\DerDec{}{\Gamma}{P})
                \Rightarrow
                (\Gamma_0,{\Gamma}\DerOSNeg {} {})}
\end{array}
\]
    where $\DerDec{}{\Gamma}{P}$ is $\DerDec+{\Gamma}{P}$.
  \caption{Big-step \LKF, v5}
  \label{fig:bstepLKF5}
\end{bfigure}

The lesson to be remembered from this formulation of big-step \LKF, is
that (the big-step version of) asynchronous rules happens to {\bf
  coincide} with a rule inferred from (the big-step version of)
synchronous rules.  This will make cut-elimination work, and it
formalises (at least in classical logic) the concept known in
philosophical logic as
\Index{harmony}~\cite{TennantNatLogic,Read00,Read10} (expressed
originally between the introduction rules and elimination rules of
Natural Deduction, or between left-introduction rules and
right-introduction rules of Sequent Calculus).

Now we can consider that both synchronous and asynchronous rules are
defined primitively, and notice the somewhat ``miraculous''
coincidence, or we can adopt the view that only synchronous rules are
defined primitively; asynchronous phases then work by duality from the
way synchronous phases work.

In other words,
\begin{itemize}
\item positive connectives are ``defined'' by their introduction rules;
\item negative connectives are ``defined'' by duality, from the
  introduction rules of their positive duals.
\end{itemize}
We may not even need to bother representing their rules.

In this view, and via the Curry-Howard correspondence, we should
define how to inhabit a type $A\arr B$ with (proof-)terms, from the
way we inhabit $A$ with terms and $B$ with continuations (with
$\arr$ being a negative connective). Writing $\l x.M$ with a variable
$\XcolY x A$ and a body $\XcolY M B$, would then only be a mere
representation for (or a mere even implementation of) an inhabitant of $A\arr
B$ that pre-exists the syntactical notation.

This is of course expressed \emph{semantically} in orthogonality
models (say in
Definitions~\ref{def:TypeInterpretation},~\ref{def:typeinterpretSysL}
and~\ref{def:InterpretTypesWitness}) by the fact that we first define
an interpretation for positive formulae (mentioning the syntax of
their basic inhabitants, such as the construct $\cons t e$), and in a
second step we define the interpretation of negative formulae simply
as the orthogonal of the interpretation of their dual formula. The
defininition for negative formulae does not even mention the syntax of
their inhabitants (such as $\lambda x.M$), but if we have a syntax for
them, they ``happen to live'' (somewhat miraculously) in the
interpretation.

We now formalise a way to express this \emph{syntactically}, as a
proof-term calculus for big-step \LKF.

\renewcommand\DerOSPos[3][]{#2\Seq[#1]{}{\Downarrow #3}}
\renewcommand\DerOSNeg[3][]{#2\Seq[#1]{}{\Uparrow #3}}

\subsection{Functional interpretation as pattern-matching}
\label{sec:patmatch}

Earlier we wrote that ``the proofs of negatives \emph{must interact
  well with} the proofs of the positive dual''. The intuition we
formalise is that 
\begin{itemize}
\item
  the proofs of a positive connective (\ie of some $\DerOSPos
  {\Gamma_0} {P}$) are some data that can be \emph{pattern-matched};
\item
  the proofs of a negative connective (\ie of some $\DerOSNeg
  {\Gamma_0} {\neg P}$) are \emph{functions} that consume data by
  \emph{pattern-matching}.
\end{itemize}

The fact that the proofs of a negative are determined by duality from
the proofs of the positive dual, is reflected by the fact that the
shape of a pattern-matching function is indeed completely determined
by the data-type of its argument.

So the Curry-Howard interpretation of big-step \LKF\ is an abstract
system of pattern-matching.
  
The ``proof-terms'' for the decomposition of (positive) connectives
are \Index[pattern]{patterns}. For instance for the connectives
$\andP,\orP$:

\begin{definition}[Patterns for $\andP,\orP$]\label{def:patforAndOr}
  Patterns are defined by the following syntax:
  \[p\recdef x^+\sep x^- \sep \paire{p_1}{p_2}\sep \inj i p\]

  Their typing rules are presented in Fig.~\ref{fig:DecompWPat}.
\end{definition}
\begin{bfigure}[!h]
    \[\begin{array}c
    \infer{\DerDec{}{x^-\col\neg P} {{x^-}\col\neg P}}{}
    \qquad
    \infer{\DerDec{}{x^+\col a} {{x^+}\col a}}{}\\\\
    \infer{\DerDec{}{\Gamma_1,\Gamma_2}{{\paire {p_1}{p_2}} \col A_1\andP A_2}}
    {\DerDec{}{\Gamma_1}{{p_1} \col A_1}
      \qquad
      \DerDec{}{\Gamma_2}{{p_2} \col A_2}
    }
    \qquad
    \infer{\DerDec{}{\Gamma}{{\inj i{p}}\col A_1\orP A_2}}
    {\DerDec{}{\Gamma}{{p}\col A_i}}
  \end{array}
    \]
  \caption{Decomposition with patterns}
  \label{fig:DecompWPat}
\end{bfigure}

We now give the proof-terms for big-step \LKF:

\begin{definition}[Pattern-matching calculus]

  Let $\Data$ be a set of elements called \Index[pattern]{patterns}, and denoted $p$, $p'$, \ldots

  The syntax of proof-terms is given by the following grammar:
  \[\begin{array}{llll}
  \mbox{\bf Positive terms}&t^+&\recdef& p.\sigma\\
  \mbox{\bf Negative terms}&t^-&\recdef&  f\\
  \mbox{\bf Commands}&c&\recdef& \cutc {x^-} {t^+}\sep {\cutc f {t^+}}
  \end{array}
  \]
  where
  \begin{itemize}
  \item $\sigma$ is a substitution from negative variables such as $x^-$ to negative terms, and from positive variables such as $x^+$ to positive terms;
  \item $f$ is a function from patterns to commands.
  \end{itemize}

  Let $\Atms$ and $\Moles$ be two sets of elements called
  \Index[atom]{atoms} and \Index[molecule]{molecules} and denoted 
  $p$ and $P$, respectively.

  Let \Index[typing context]{typing contexts} be functions mapping
  negative variables to molecules (written $x^-\col\neg P$) and
  positive variables to atoms (written $x^+\col a$).

  Let $\DerDec{}{\Gamma}{\XcolY p P}$ be a typing relation where
  $p$ is a pattern, $P$ is a molecule, and $\Gamma$ is a typing context.

  The typing rules for proof-terms are presented in Fig.~\ref{fig:PatMatchTyping}.

  There is just one cut-elimination rule:
    \[
    \rulenamed{\textsf{pat-match}}\qquad\cutc f {p.\sigma}\Rew{} (f(p))\sigma
    \]
    where $c\sigma$ denotes the application of substitution $\sigma$ to the command $c$.
\end{definition}
\begin{bfigure}[!h]
  \[
  \begin{array}{c}
    \infer{\DerOSPos{{\Gamma_0}}{{p.\sigma}\col P}}
          {\DerDec{}{\Gamma}{{p}\col P}
            \qquad\forall (x^-\col \neg P)\in{\Gamma},\quad \DerOSNeg{{\Gamma_0}}{{\sigma(x^-)}\col {\neg P}}
            \qquad\forall (x^+\col a)\in{\Gamma},\quad (\sigma(x^+)\col a)\in {\Gamma_0}
          }
          \\\\
          \infer{\DerOSNeg {\Gamma_0} {{f}\col\neg P}}
                {\forall \Gamma, (\DerDec{}{\Gamma}{{p}\col P})
                  \Rightarrow
                  {f(p)}\col(\DerOSNeg {{\Gamma_0},\Gamma} {})}\\\\
                \infer{{\cutc {x^-} {p.\sigma}}\col(\DerOSNeg {{\Gamma_0},x^-\col\neg  P} {})}{\DerOSPos {{\Gamma_0},x^-\col\neg P} {{p.\sigma}\col P}}
                \qquad\qquad
                \infer{{\cutc {f} {t^+}}\col(\DerOSNeg {{\Gamma_0}} {})}{
                  \DerOSNeg {{\Gamma_0}} {{f}\col\neg P}\qquad\DerOSPos {{\Gamma_0}} {{t^+}\col P}
                }\\\\
  \end{array}
  \]
  where $\Gamma_0$, $\Gamma$, \ldots are typing contexts.
  \caption{Typing for the pattern-matching calculus}
  \label{fig:PatMatchTyping}
\end{bfigure}

The one cut-elimination rule is the very standard mechanism of pattern-matching, with the command
$\cutc f {t^+}$ representing what we could informally write as:\\\nobreak
``\texttt{match $t^+$ with $\underbrace{\ldots \mapsto \ldots}_{f}$}''

\begin{remark}
  \begin{enumerate}
  \item Notice how negative terms are not really terms, but functions
    of the meta-level (or meta-level functions that are reified in the
    term syntax); this is a higher-order definition, and we do not
    give any concret syntax for such functions.  

    Strictly speaking, our definition depends on the notion of
    function space that we take for the definition of negative terms
    (We could for instance restrict it to computable functions, but so
    far we do not specify such things).

    Also, with such a definition, it may not be clear exactly what the
    contextual closure of the rule $\rulenamed{\textsf{pat-match}}$
    is. By $\Rew{\rulenamed{\sf pat-match}}$ we therefore denote the
    reduction relation where $\rulenamed{\textsf{pat-match}}$ is
    applied at the top-level of a given command (no contextual
    closure).

  \item
    Also notice how we emphasised that the definition of the proof-term
    calculus is \emph{independent} from the syntax of patterns and the
    typing system for them.

    Definition~\ref{def:patforAndOr} and Fig.~\ref{fig:DecompWPat} give
    one example (where atoms are positive atomic formulae and molecules
    are positive formulae).

    But the construction of the Curry-Howard interpretation for big-step
    focussing is modular in those notions.

    This is a gain of genericity / abstraction that we will further
    develop in the next Chapters.
 \end{enumerate}
 \end{remark}

\begin{theorem}[Subject Reduction]

  If $c\col(\DerOSNeg {\Gamma} {})$ and $c\Rew{\rulenamed{\sf pat-match}}c'$ then $c'\col(\DerOSNeg {\Gamma} {})$.
\end{theorem}

Reviewing the various properties that are desirable for an instance of
the Curry-Howard correspondence, we find that Progress (in this case,
cut-elimination), depends on how we may reduce functions (so far,
$\Rew{\rulenamed{\sf pat-match}}$ only applies at the root);
Confluence does not make sense here because, until we define how to
reduce functions, there is at most one redex to reduce; and for
Normalisation, we need to explore models (\eg orthogonality models) of
such a calculus.

\sectionno{Conclusion}

The study of orthogonality models for the pattern-matching
calculus should be particularly interesting: 

In~\cite{MunchCSL09}, Munch-Maccagnoni already explored the
construction of orthogonality models for polarised System~\SysL\ with
an emphasis on focussing properties. Big-step \LKF\ and its underlying
pattern-matching calculus seems to be an even more appropriate
framework to look at the connection between focussing and
orthogonality models, since this framework reflects at the syntactical
level what orthogonality models describe at the semantical level,
namely the fact that we first declare what the ``inhabitants of
positive formulae'' are, and then we define the ``inhabitants of
negative formulae'' by duality as those inhabitants that ``interact
well with'' the inhabitants of the dual (positive) formula. In case of
an orthogonality model, to ``interact well with'' means to ``be
orthogonal to''; in the case of pattern-matching, it means to ``be
able to consume''. To be more precise:
\begin{itemize}
\item Inhabitants of positive types have \emph{structure}: in an
  orthogonality model we need an algebraic structure to interpret
  positive constructs such as $\cons\_\_$ or $\inj\_\_$; in big-step
  \LKF, these inhabitants come as the combination of a pattern
  (\eg $\cons\_\_$ or $\inj\_\_$) and a substitution that fills its
  holes.
\item Inhabitants of negative formulae may lack any structure, but
  they come with a \emph{behaviour}: in an orthogonality model, they
  can range over any abstract set for which the orthogonality relation
  with positive inhabitants is defined; in big-step \LKF, they range
  over any abstract set of functions (we do not specify which) that
  can consume patterns.
\end{itemize}

So, in order to formalise the connections that are informally
described above, the second part of this dissertation explores
orthogonality models for big-step focussing systems. We shall strip
anything that is not essential off the constructions we make,
systematically seeking the greatest generality, and aiming at the cores
of orthogonality models and focussing systems. Doing so reveals the
essential difference between realisability and typing:
\begin{itemize}
\item in realisability, checking whether a given negative inhabitant
  ``interacts well with'' an arbitrary inhabitant of a positive
  formula, requires the computation of an interaction that
  explores the positive inhabitant's structure to an {\bf arbitrary
    depth} (as nothing restricts the criterion given by orthogonality);
\item in typing, checking whether a given negative inhabitant
  ``interacts well with'' an arbitrary inhabitant of a positive
  formula, only requires the computation of an interaction that
  explores the positive inhabitant's structure to a {\bf bounded
    depth} (as the negative inhabitant is a function that performs a
  case analysis on the positive inhabitant's top-level pattern and the
  interaction has to uniformly treat the rest of the inhabitant's
  structure).
\end{itemize}
In case each positive formula comes with a finite number of patterns
for it, the above distinction is what makes typing decidable and
realisability undecidable (in general).

\ignore{
  \Slide[A term calculus for this (Munch-Maccagnoni)]{
    
    {\[
      \begin{array}{c}
        \infer{\DerOSPost{\Theta}{A\andP B}{\paire t u}}
        {\DerOSPost{\Theta}{A}t\qquad \DerOSPost{\Theta}{B}u}
        \qquad
        \infer{\DerOSPost{\Theta}{A_1\orP A_2}{\inj[i]t}}
        {\DerOSPost{\Theta}{A_i}t}
        \qquad
        \infer{\DerOSPost{\Theta}{\EX \alpha A}{\appl B t}}
        {\DerOSPost{\Theta}{\subst A \alpha B}t}
        \\\\
        \infer[\begin{array}l P \mbox{ positive formula}\end{array}]{\DerOSPost{\Theta,x^+\col\non P}{P}{x^+}}{}
        \qquad
        \\\\
        \infer{\DerOSPost{\Theta}{A\andN B}{\muet x c {x'} {c'}}}
        {\DerOSNegt{\Theta,x\col A}{c}\qquad \DerOSNegt{\Theta,x'\col B}{c'}}
        \qquad
        \infer{\DerOSPost{\Theta}{A_1\orN A_2}{\muou x y c}}
        {\DerOSNegt{\Theta,x\col A_1,y\col A_2}{c}}
        \\\\
        \infer[\alpha\notin\FV{\Theta}]{\DerOSPost{\Theta}{\FA \alpha A}{\mufa x c}}
        {\DerOSNegt{\Theta,x\col A}c}
        \\\\
        \infer[N\mbox{ negative formula}]{\DerOSPost \Theta {N}{\m {x^+}c}}{\DerOSNegt {\Theta,x^+\col N} {c}}
        \qquad
        \infer[A\mbox{ {any formula}}]{\DerOSNegt {\Theta, {x\col A}} {\cutc x t}}{\DerOSPost {\Theta,x\col {A}} {A}t}
      \end{array}
      \]}
  }

  \Slide[The term syntax]{
    {\[
      \begin{array}{ll}
        t^+,u^+,\ldots&\recdef x^+\sep \paire t u\sep \inj[1]t\sep\inj[2]t\sep\appl B t\\
        t^-,u^-,\ldots&\recdef \muet x {x'} c {c'} \sep \muou x y c\sep\mufa x c\sep\m{x^+}c\\
        t,u,\ldots&\recdef t^+\sep t^-\\
        c,\ldots&\recdef \cutc {x^+} {t^-} \sep\cutc {t^+}{x^-} 
      \end{array}
      \]}
    {Only proofs that can be reduced: $\cutc {x^+} {\m{y^+}c}\Rew{}\subst c {y^+}{x^+}$}

    {So far, no real cuts}
  }

  \Slide[Cuts and unfocused proofs]{
    
    {\[\begin{array}{ll}
        t^+,u^+,\ldots&\recdef x^+\sep \paire t u\sep \inj[1]t\sep\inj[2]t\sep\appl B t\sep{\m{x^-}c}\\
        t^-,u^-,\ldots&\recdef {x^-}\sep \muet x c {x'} {c'} \sep \muou x y c\sep\mufa x c\sep\m{x^+}c\\
        t,u,\ldots&\recdef t^+\sep t^-\\
        c,\ldots&\recdef {\cutc {t^+} {t^-}}\end{array}\]}
    {\[\begin{array}{c}
        \infer[P\mbox{ positive formula}]{\DerOSPost \Theta {P}{\m {x^-}c}}{\DerOSNegt {\Theta,x^-\col P} {c}}
        \qquad
        \infer[\begin{array}l N \mbox{ negative formula}\end{array}]{\DerOSPost{\Theta,x^-\col\non N}{N}{x^-}}{}
        \\\\
        \infer{\DerOSNegt {\Theta} {\cutc {t^+}{t^-}}}{\DerOSPost {\Theta} {P}{t^+}\qquad \DerOSPost {\Theta} {N}{t^-}}
      \end{array}\]}
    {Values:\[\begin{array}{ll}
        v^+\ldots&\recdef x^+\sep \paire v {v'}\sep \inj[1]v\sep\inj[2]v\sep\appl B v\\
        v,\ldots&\recdef v^+\sep t^-\\
      \end{array}\]
      Values are proof-terms for focused proofs}
  }

  \Slide[Cut-reduction rules and re-focusing]{
    
    {Logical cut-reduction rules\[\begin{array}{ll}
        \cutc{\paire{v}{v'}}{\muou x y c}&\Rew{}\subst c{x,y}{v,v'}\\
        \cutc{\inj[i] v}{\muet {x_1} {c_1}{x_2}  {c_2}}&\Rew{}\subst {c_i}{x_i}{v}\\
        \cutc{\appl B v}{\mufa {x}{c}}&\Rew{}\subst {c}{x}{v}\end{array}\]}
    {Structural reduction rules\[\begin{array}{ll}
        \cutc {v^+} {\m{y^+}c}&\Rew{}\subst c {y^+}{v^+}\\
        \cutc {\m{y^-}c} {t^-}&\Rew{}\subst c {y^-}{t^-}
      \end{array}\]}
    {Focusing reduction rules
      \[\begin{array}{lll}
        \cutc{\paire{t}{t'}}{u^-}&\Rew{}\cutc{t}{\m x{\cutc{t'}{\m{x'}{\cutc{\paire{x}{x'}}{u^-}}}}}
        &\mbox{if $t$ or $t'$ is not a value}\\
        \cutc{\inj[i] t}{u^-}&\Rew{}\cutc t {\m x{\cutc{\inj[i] x}{u^-}}}      
        &\mbox{if $t$ is not a value}\\
        \cutc{\appl B t}{u}&\Rew{}\cutc t {\m x{\cutc{\appl B x}{u}}}
        &\mbox{if $t$ is not a value}\\
      \end{array}\]
      where $\cutc{t^-}{t^+}$ is another notation for $\cutc{t^+}{t^-}$}
    
  }
}

\part{Abstract focussing}
\label{partII}

\chapter*{Introduction}

The second part of this dissertation presents unpublished material, on
the theme of \Index{abstract focussing}. 

In the previous chapters we have seen the use of polarities and
focussing in the proof theory of classical logic, where a focussed
proof is a tree that alternates \Index[synchronous]{synchronous
  phases} with \Index[asynchronous]{asynchronous phases}.

A level of abstraction is reached by \Index{big-step focussing}, which
compacts each phase into one inference step and thus allows the inner
details of phases to be ``forgotten''. As revealed by Zeilberger's
formulation of big-step
focussing~\cite{ZeilbergerPOPL08,Zeilberger08}, the computational
interpretation of this is pattern-matching.

In parallel to this, Munch-Maccagnoni~\cite{MunchCSL09} formalised the
connection between focussing and the orthogonality techniques, which
were presented in Chapter~\ref{ch:ortho} for strong normalisation
proofs and witness extraction.

The origin of the material presented in this second part of this
dissertation is the idea that this deep connection could be revealed
at a more abstract level if a Zeilberger-style system was used: For
instance, the fact that, in such a system, the inhabitants of negative
formulae live in an abstract function space and are not made of any
syntax reflects the fact that, in orthogonality models, inhabitants of
negative formulae can range over an abstract set and have no algebraic
structure.
The second part of this dissertation therefore started as a
formalisation, in the proof-assistant \Coq~\cite{Coq}, of orthogonality
models for a Zeilberger-style system, culminating with the Adequacy
Lemma that connects the big-step focussing proof system with the
orthogonality approach.

Doing this formalises the connection at a level of abstraction that
forgets about the syntax or structure not only of the inhabitants of
negative formulae (as suggested above) but also of positive formulae,
abstracting over the logical connectives and the very syntax of
formulae.

In the abstract framework that we present here, called \LAF, an extra
step of abstraction is also reached (compared
to~\cite{ZeilbergerPOPL08,Zeilberger08}) over the construction of
(typing) contexts, which allows the same framework to capture both
classical and intuitionistic systems. More substantially, the
treatment of quantifiers is also new.

As the material developed and expanded, it also appeared that our
framework, together with its machine-checked formalisation, could be
directly implemented and serve as the theoretical foundations for a
new version of the \Psyche\ system, discussed in Part~\ref{partIII} of
this dissertation. It could perhaps even serve as the basis for a
formal proof of the system's correctness. Thinking along those lines
oriented the design of the \LAF\ framework with implementation issues
in mind (\eg using De Bruijn's indices or De Bruijn's levels), and
resulted in the formalisation of mathematical structures behind which
the OCaml modules can clearly be seen.

This \Coq\ formalisation and the implementability concern also
resulted in a presentation of the material that is admittedly
technical, with \eg numerous parameters and long specifications, which
was also fuelled by the desire to identify the connection between
focussing and orthogonality at the ``purest'' level: every design
choice or ingredient of the framework that was not essential to
establishing the connection was systematically turned into a parameter
of the framework, with an axiomatisation for it that we sought to be as
weak as possible for the theory to hold.

Chapter~\ref{ch:LAFwoQ} presents a description of the proof-term
system for big-step focussing that is more formal than that with which
we concluded Part~\ref{partI} of this dissertation. This
formalisation, called \LAF, is essentially a reformulation of the
ideas in~\cite{ZeilbergerPOPL08,Zeilberger08}, with no substantial
difference but the modular description of typing contexts. This allows
classical and intuitionistic systems to be instances of the same
parameterised system \LAF, as we describe at the end of the chapter.

Chapter~\ref{ch:LAFwQ}, on the other hand, presents a substantial
extension: the \LAF\ system with quantifiers. It therefore subsumes
Chapter~\ref{ch:LAFwoQ}, but giving the version of \LAF\ with
quantifiers straight away would be a bit harsh on the reader.

Chapter~\ref{ch:real} explores realisability models for \LAF, based on
orthogonality, and its contents was the original motivation for the
development of this material, as the Adequacy Lemma connects big-step
focussing with orthogonality, typing with realisability, syntax with
semantics. We apply this methodology to derive the consistency of
\LAF\ systems.

Chapter~\ref{ch:trans} then investigates the operational semantics of
\LAF, which interprets the proof-terms for big-step focussing as a
pattern-matching calculus.  We first present a small-step semantics by
means of an abstract machine for head-reduction. Adapting the
methodology of Chapter~\ref{ch:ortho}, we apply the orthogonality
models of Chapter~\ref{ch:real} to prove the normalisation of typed
terms with respect to this abstract machine.  Then we develop the
abstract machine into a big-step operational semantics, for which a
new application of orthogonality models provides the cut-elimination
result for \LAF.

\cchapter[Abstract focussed sequent calculus w/o quantifiers]{An abstract focussed sequent calculus - without quantifiers}
\label{ch:LAFwoQ}

\minitoc 

In this chapter, we show how Zeilberger's
ideas~\cite{ZeilbergerPOPL08,Zeilberger08}, as presented in
Chapter~\ref{ch:polarfocus}, can be developed into an \Index{abstract
  focussed sequent calculus} called \LAF, and whose instances express
the big-step versions of standard focussed sequent calculi.

The system of Chapter~\ref{ch:polarfocus} is already abstract in the
relation $\decf{}$ that decomposes a positive formula into a
collection of positive atoms and negative formulae. Correspondingly,
it is also abstract in the notion of \emph{pattern} whose typing
judgement is given by the relation $\decf{}$.

We push this abstraction further:
\begin{itemize}
\item Since this decomposition relation $\decf{}$ was the only ingredient of the
  system that used the syntax of formulae, we do not even have to
  assume that formulae are syntax, \ie have an inductive structure,
  nor do we have to assume that ``positive atoms'' are particular
  kinds of formulae; positive atoms and formulae could literally be
  two arbitrary sets. We shall now respectively call them
  \Index[atom]{atoms} and \Index[molecule]{molecules}.
\item Moreover, a typing context $\Gamma$ could be extended in an
  asynchronous step into $\Gamma,\Delta$, where $\Delta$ is the result
  of decomposing some positive formula according to some pattern $p$
  and the decomposition relation $\decf{}$. We have in fact no reason
  to assume that $\Gamma$ and $\Delta$ are of the same nature and that
  $\Gamma,\Delta$ corresponds to set union (or whatever standard
  combination of typing contexts one usually considers). Therefore,
  ``typing contexts'' such as $\Gamma$ will form an abstract notion,
  namely an algebra equipped with specific functions among which an
  arbitrary asymmetric construction $\Gamma\Cextend\Delta$ that
  replaces the above.\footnote{Following Zeilberger's style, $\Delta$
    itself will not be a typing context but will have a tree structure
    that may reflect the way a positive formula is decomposed into it.}
\end{itemize}

Something that is difficult to treat formally at this abstract level
is the use of a non-deterministic way of naming variables, and then
having to deal with $\alpha$-conversion, in particular when we
formalise our framework \LAF\ and its meta-theory in the
proof-assistant \Coq.  Therefore we adopt a deterministic way of
naming variables (now called \Index[label]{labels} since they are not
subject to $\alpha$-conversion), but we remain abstract in the exact
system that we use for naming them: this approach will capture for
instance De Bruijn's indices as well as De Bruijn's levels.

Section~\ref{sec:LAFDefQF} presents \LAF. Section~\ref{sec:LAFexprop}
describes how to tune (\ie instantiate) the abstract parameters so as
to capture different logics (or logical systems).
Section~\ref{sec:LAFlabels} provide instances illustrating different
implementations of labels corresponding to De Bruijn's indices
and De Bruijn's levels.

\section{Presentation of the system}
\label{sec:LAFDefQF}

This section presents the quantifier-free version of system \LAF, a
highly modular / parameterised sequent calculus for big-step
focussing.

An instance of \LAF\ is given by a tuple of parameters
\[(\Atms,\Moles,\var[+],\var[-],\TContexts,\Data,\decf{})\]
where each parameter is described below.

\subsection{Atoms, molecules, typing decompositions and typing contexts}

The first group of parameters \((\Atms,\Moles)\) specifies what the
instance of \LAF, as a logical system, talks about. A typical example
is when \(\Atms\) and \(\Moles\) are respectively the sets of
(positive) atoms and the set of formulae from a polarised logic. We
will see in the next sections how our level of abstraction allows for
some interesting variants. In the Curry-Howard view, \(\Atms\) and
\(\Moles\) are our sets of types.

\begin{definition}[Atoms \& molecules]\strut

  \LAF\ is parameterised by two sets $\Atms$ and $\Moles$, whose
  elements are respectively called \Index[atom]{atoms} (denoted $a$,
  $a'$,\ldots), and \Index[molecule]{molecules} (denoted $M$,
  $M'$,\ldots).
\end{definition}

We then aim at defining \emph{typing contexts}, those structures
denoted $\Gamma$ in a typing judgement of the form $\Der\Gamma\ldots{}$.

Intuitively, we expect $\Gamma$ to ``contain'' atoms and molecules, or
more precisely to declare some variables as having atoms and molecules
as their types.

For this it will be useful (\eg to build models of \LAF) to define
contexts more generically, mapping variables to elements of two sets
$\mathcal A$ and $\mathcal B$.

Contexts will be extendable (in the case of typing contexts, we may
want to extend $\Gamma$ with a new type declaration for a fresh
variable), and the following data-structure formalises what generic
contexts will be extended with.

\begin{definition}[Generic decomposition algebras]

  Given two sets $\mathcal A$ and $\mathcal B$, the $(\mathcal
  A,\mathcal B)$-\Index{decomposition algebra} $\DecompType[\mathcal
    A,\mathcal B]$, whose elements are called $(\mathcal A,\mathcal
  B)$-\Index[decomposition]{decompositions}, is the free algebra
  defined by the following grammar:
  \[\Delta,\Delta_1,\ldots\recdef a\sep \Drefute b\sep\Dunit\sep\Delta_1\Dand\Delta_2\]
  where $a$ (\resp $b$) ranges over $\mathcal A$ (\resp $\mathcal B$).

  Let $\Dstruct$ abbreviate
  $\DecompType[\unitt,\unitt]$, whose elements we call
  \Index[decomposition structure]{decomposition structures}. 

  The \Index[structure]{(decomposition) structure} of an $(\mathcal
  A,\mathcal B)$-decomposition $\Delta$, denoted $\abs\Delta$, is its
  obvious homomorphic projection in $\Dstruct$.
\end{definition}

Intuitively, a $(\mathcal A,\mathcal B)$-decomposition $\Delta$ is
simply the packaging of elements of $\mathcal A$ and elements of
$\mathcal B$; we could flatten this packaging by seeing $\Dunit$ as
the empty set (or multiset), and $\Delta_1\Dand\Delta_2$ as the union
of the two sets (or multisets) $\Delta_1$ and $\Delta_2$.

Note that the coercion from $\mathcal B$ into $\DecompType[\mathcal
  A,\mathcal B]$ is denoted with $\Drefute{}$. It helps
distinguishing it from the coercion from $\mathcal A$ (\eg when
$\mathcal A$ and $\mathcal B$ intersect each other), and in many
instances of \LAF\ it will remind us of the presence of an otherwise
implicit negation. But so far it has no logical meaning, and in
particular $\mathcal B$ is not equipped with an operator $\Drefute{}$
of syntactical or semantical nature.

\begin{definition}[Generic contexts]

  \LAF\ is parameterised by two sets $\var[+]$ and $\var[-]$, of
  elements called \Index[positive label]{positive labels} and
  \Index[negative label]{negative labels}, respectively.

  Given two sets $\mathcal A$ and $\mathcal B$, an \Index[context
    algebra]{$(\mathcal A,\mathcal B)$-context algebra} is an algebra
  of the form
  \[\left(
  \Contexts,
  {\footnotesize
    \left(\begin{array}{c@{}l@{}l}
      \Contexts\times\var[+]&\pfunspace&\mathcal A\\
      (\Gam,x^+)&\mapsto&\varRead[x^+]\Gam
    \end{array}\right),
    \left(\begin{array}{c@{}l@{}l}
      \Contexts\times\var[-]&\pfunspace&\mathcal B\\
      (\Gam,x^-)&\mapsto&\varRead[x^-]\Gam
    \end{array}\right),
    \left(\begin{array}{c@{}l@{}l}
      \Contexts\times\DecompType[\mathcal A,\mathcal B]&\rightarrow&\Contexts\\
      (\Gam,\Delta)&\mapsto&\Gam\Cextend\Delta
    \end{array}\right)
  }
  \right)
  \]
   whose elements are called $(\mathcal A,\mathcal B)$-\Index[context]{contexts}.

   As $(\Gam,x^+)\mapsto\varRead[x^+]\Gam$ and $(\Gam,x^-)\mapsto\varRead[x^-]\Gam$ are partial functions, we denote by $\domP\Gamma$ (\resp $\domN\Gamma$) the subset of $\var[+]$ (\resp $\var[-]$) where $\varRead[x^+]\Gam$ (\resp $\varRead[x^-]\Gam$) is defined.
\end{definition}


We choose to call elements of $\var[+]$ and $\var[-]$ ``labels'', rather than ``variables'', because ``variable'' suggests an object identified by a name that ``does not matter'' and somewhere subject to $\alpha$-conversion. For instance in the following typing rule for the (simply-typed) $\lambda$-calculus
\[\infer{\Der \Gamma {\XcolY{\lambda x.t}{A\arr B}}}
        {\Der{\Gamma,\XcolY x A}{\XcolY t B}}\]
the $\alpha$-convertibility of the variable $x$ bound in $\lambda x.t$ relates to a non-deterministic choice of name for the variable used to extend the context $\Gamma$ into $\Gamma,\XcolY x A$.\footnote{The fact that the non-deterministic choice does not matter, \aka \emph{equivariance}, is covered at length in nominal logic~\cite{PittsAM:nomlfo-jv} and other works formalising binding.} It turns out that such non-determinism in context extension is quite tricky to adapt (though probably not impossible) to the level of abstraction of \LAF, and in practice would not be used in an implementation of proof-search, where a deterministic choice of name would be performed (``first fresh name'' picking, etc). 

Therefore, we decide to present \LAF\ without the non-determinism related to $\alpha$-conversion, yet without committing to using De Bruijn's indices or De Bruijn's levels. Hence the use of ``labels'', that will accommodate both systems (and others, as long as the concept of context extension $\Gamma\Cextend \Delta$ is a proper function, \ie remains deterministic).


\begin{definition}[Typing decompositions and typing contexts]\strut

  The \Index[typing decomposition algebra]{typing decomposition
    algebra}, denoted $\DecompType$, whose elements are called
  \Index{typing decompositions}, is the $(\Atms,\Moles)$-decomposition
  algebra.

  \LAF\ is then parameterised by an $(\Atms,\Moles)$-context algebra $\TContexts$,
  whose elements are called \Index[typing context]{typing contexts}.
\end{definition}

\subsection{Logical connectives}

Finally, the last group of parameters \((\Data,\decf{})\) specifies
the structure of molecules. If \(\Moles\) is a set of formulae
featuring logical connectives, those parameters specify the
introduction rules for the connectives.

\begin{definition}[Patterns \& decomposition relation]\strut

  \LAF\ is parameterised by a \Index{pattern algebra}, an algebra of the form
  \[\left(\Data,
      {\footnotesize
        \left(\begin{array}{c@{}l@{}l}
            \Data&\rightarrow&\Dstruct\\
            p&\mapsto&\Datast p
          \end{array}\right)
      }
    \right)
  \] 
  whose elements are called \Index[pattern]{patterns},
  and by a \Index{decomposition relation}, \ie a set of elements
  \[(\DerDec{}{\_}{\XcolY \_{\_}}):(\DecompType\times\Data\times\Moles)\]
  such that if $\DerDec{}{\Delta}{\XcolY p{M}}$ then the structure of
  $\Delta$ is $\Datast p$.
\end{definition}

The intuition behind the terminology is that the decomposition
relation $\decf{}$ decomposes a molecule, according to a pattern, into
a typing decomposition which, as a first approximation, can be seen as
a ``collection of atoms and (hopefully smaller) molecules''.

\subsection{Definition of the system}

\begin{definition}[Proof-Terms]\strut

  Proof-terms are defined by the following syntax:
  \[
  \begin{array}{lll@{\recdef}l}
    \mbox{Positive terms }&\PTerms^+&t^+&pd\\
    \mbox{Decomposition terms }&\Decomp&d&x^+ \sep \THO f\sep\Tunit\sep d_1\Tand d_2 \\
    \mbox{Commands}&\PTerms&c& \cutc{x^-}{t^+} \sep \cutc{f}{t^+}
  \end{array}
  \]
  where $p$ ranges over $\Data$, $x^+$ ranges over $\var[+]$, $x^-$ ranges over $\var[-]$, and $f$ ranges over the partial function space $\Data\pfunspace\PTerms$.
\end{definition}

We can finally present the typing system \LAF:

\begin{definition}[\LAF]
  \LAF\ is the inference system of Fig.~\ref{def:LAFqf} defining the derivability of three kinds of sequents
  \[
  \begin{array}{l@{\quad:\quad}l}
    (\DerF\_ {\XcolY {\_} \_}{}) & (\TContexts\times \PTerms^+\times\Moles)\\
    (\Der\_ {\XcolY \_ \_}) & (\TContexts\times \Decomp\times\DecompType)\\
    (\Der\_ \_) & (\TContexts\times \PTerms)
  \end{array}
  \]
  We further impose in rule \textsf{async} that the domain of function $f$ be exactly those patterns that can decompose $M$ (if $p\in\dom f$ then there exists $\Delta$ such that $\DerDec{}{\Delta}{\XcolY p{M}}$).

  \LAFcf\ is the inference system \LAF\ without the \cut-rule.
\end{definition}
  \begin{bfigure}[!h]
    \[
    \begin{array}{c}
      \infer[\textsf{sync}]
            {\DerF\Gam {\XcolY {p d} M}{}}
            {\DerDec{}{\Delta}{\XcolY p{M}}
              \quad
              \Der{\Gamma}{\XcolY  d{\Delta}}}
            \\
            \midline
            \\
            \infer{\Der\Gam{\XcolY{\Tunit}{\Dunit}}}{\strut}
            \qquad
            \infer{\Der\Gam{\XcolY{ d_1\Tand d_2}{\Del_1\Dand\Del_2}}}
                  {\Der\Gam{\XcolY{ d_1}{\Del_1}}
                    \quad
                    \Der\Gam{\XcolY{ d_2}{\Del_2}}}
                        \\\\
                        \infer[\textsf{init}]
                              {\Der{\Gam}{\XcolY{x^+}{a}}}{\varRead[x^+]\Gam = a}
                              \qquad
                              \infer[\textsf{async}]
                                    {\Der\Gam {\XcolY {\THO f} {\Drefute M}}}
                                    {\forall p,\forall\Delta,\quad
                                      \DerDec{}{\Delta}{\XcolY p{M}}\quad\imp\quad\Der{\Gamma\Cextend[x]\Delta}{f(p)}}
                                    \\
                                    \midline
                                    \\
                                    \infer[\textsf{select}]{\Der\Gam{\cutc{x^-}{t^+}}}
                                          {\quad\DerF\Gam {\XcolY{t^+}{\varRead[x^-]\Gam}} {}}
                                          \qquad
                                          \infer[\cut]{\Der\Gam{\cutc{f}{t^+}}}
                                                {\Der\Gam {\XcolY {f}{\Drefute M}}\qquad
                                                  \DerF\Gam {\XcolY{t^+}M} {}}
    \end{array}
    \]
    \caption{\LAF}
    \label{def:LAFqf}
  \end{bfigure}

An intuition of \LAF\ can be given in terms of proof-search:

When we want to ``prove'' a molecule, we first need to decompose it
into a collection of atoms and (refutations of) molecules (rule
$\textsf{sync}$). Each of those atoms must be found in the current
typing context (rule $\textsf{init}$).  Each of those molecules must
be refuted, and the way to do this is to consider all the possible
ways that this molecule could be decomposed, and for each of those
decompositions, prove the inconsistency of the current typing context
extended with the decomposition (rule $\textsf{async}$). This can be
done by proving one of the molecules refuted in the typing context
(rule $\textsf{select}$) or refuted by a complex proof (rule
$\textsf{cut}$).  Then a new cycle begins again.

Typing decompositions and decomposition terms organise the packaging
of the proofs of atoms and (refuted) molecules decomposed by rule
$\textsf{sync}$. Typing decompositions could here be taken to be a
multiset of atoms and (refuted) molecules, but keeping a dedicated
structure for the packaging will be more convenient when we add
quantifiers: giving decompositions an inductive structure allows a
lossless modelling of quantifiers' scopes.

\section{Capturing existing systems}

The above intuitions may become clearer when we instantiate the
parameters of \LAF\ with actual literals, formulae, etc in order to
capture existing systems:

In the rest of this chapter we illustrate system \LAF\ by specifying
different instances, providing each time the long list of parameters,
that capture different focussed sequent calculus systems.

By ``capture'', we mean of course a stronger result than just the
equivalence between the notions of provability. In order to strengthen
such a weak property between two systems, it is relevant to consider
the notions of \Index{adequacy} as defined
in~\cite{nigam09phd,nigam10jar}:

The shallowest level of adequacy, \Index{relative completeness}, or
adequacy of level -1, requires that a sequent is provable in one
system \iff\ the sequent to which it is mapped is provable in the
other system. Level -2 of adequacy, \Index{full completeness of
  proofs}, requires that there be a one-to-one correspondence between
their (complete) proofs. Level -3 of adequacy, \Index{full
  completeness of derivations} (a word used
in~\cite{nigam09phd,nigam10jar} for incomplete proofs), requires a
one-to-one correspondence between the derivations in one system and
those of the other system.

Strictly speaking, level -2 adequacy does not say more than level -1
as soon as the sequent has infinitely and denumerably many
proofs. With level -3 adequacy, we aim at capturing much more. The
simplest way to formalise its informal description above, for a
function $\phi$ that maps the sequents of system $\mathcal A$ into the
sequents of system $\mathcal B$, is probably as follows:

\emph{For every sequent $\mathcal S$ and multiset $\multiset{\mathcal
  S_1,\ldots,\mathcal S_n}$ of sequents in $\mathcal A$, there is a
one-to-one correspondence $\phi_{\mathcal S,\multiset{\mathcal
    S_1,\ldots,\mathcal S_n}}$ between
\begin{itemize}
\item the partial proofs in $\mathcal A$ whose conclusion is $\mathcal
  S$ and whose multiset of open leaves is $\multiset{\mathcal
    S_1,\ldots,\mathcal S_n}$
\item the partial proofs in $\mathcal B$ whose conclusion is
  $\phi(\mathcal S)$ and whose multiset of open leaves is
  $\multiset{\phi(\mathcal S_1),\ldots,\phi(\mathcal S_n)}$
\end{itemize}}
The above is a symmetric property when $\phi$ is itself a one-to-one
correspondence between sequents, but can also make sense if it is not.
However, the above property needs to be adapted
\begin{itemize}
\item when in either of the two systems, we are interested not in each
  individual application of the inference rules but rather in
  groupings of rules: for instance in a focussed calculus, we may want
  to consider the grouping of a synchronous phase followed by a
  asynchronous phase (a.k.a.\ a \Index{macro-rule} decomposing a
  \Index{synthetic connective}) as a single step whose internal
  details should be ignored by the correspondence (this is what
  happens in~\cite{nigam09phd,nigam10jar});
\item when either of the two systems features proof-terms, as the
  notion of incomplete proof is polluted by the presence, in the
  sequent, of a proof-term denoting a complete proof (unless we start
  considering incomplete proof-terms as well).
\end{itemize}
Both situations jeopardise the bijective aspect of each
$\phi_{\mathcal S,\multiset{\mathcal S_1,\ldots,\mathcal S_n}}$: in
the former situation, we probably want to quotient proofs in some way
so that the internal details of a rule grouping do not lead to
multiple proofs that are not reflected in the other system
(\cite{nigam10jar} mentions for instance ``up to the permutation of
asynchronous rules''); in the latter situation, proof-term annotations
would provide for instance two proofs of $\Der{\XcolY x A,\XcolY y
  A}{\XcolY ? A}$ while we only count one proof of $\Der{A,A}{A}$
(whether $A,A$ denotes a set or a multiset).

Another issue with the above notion of adequacy is that it fails to
impose any notion of compositionality (when derivations are ``plugged
into'' the open leaves of another derivation) about the family
$(\phi_{\mathcal S,\multiset{\mathcal S_1,\ldots,\mathcal
    S_n}})_{\mathcal S,\multiset{\mathcal S_1,\ldots,\mathcal S_n}}$,
something which we may have in mind when thinking about the ``deepest
level of adequacy''.

System \LAF\ features both focussing and proof-terms. Rather than
trying to adapt to these concepts, and strengthen with
compositionality, the above formalisation of level -3 adequacy, we opt
for a version of level -3 adequacy that drops the use of bijections
and the quantitative aspects that they provide (we no longer try to
count proofs). On the other hand, we retain from level -3 adequacy,
and formalise, the fact that the \textbf{structure} of proofs in one
system matches the structure of proofs in the other system.

\begin{definition}[Structural adequacy]\strut
  \begin{itemize}
  \item
    Let $\mathcal A$ be an inference system providing a notion of
    proof-trees for elements called ``sequents'', and let $\mathcal P$ be
    a set of sequents.

    Given a proof-tree $\pi$ in $\mathcal A$, the multiset of
    \Index{$\mathcal P$-immediate sequents} of $\pi$ is defined
    recursively on $\pi$: it contains the conclusions that are in
    $\mathcal P$ of the direct sub-trees of $\pi$, as well as the
    $\mathcal P$-immediate sequents of the direct sub-trees of $\pi$
    whose conclusions are not in $\mathcal P$.

  \item
    Let $\mathcal A$ and $\mathcal B$ be two inference systems as in
    the previous point, and $\mathcal R$ be a relation between the
    sequents of $\mathcal A$ and the sequents of $\mathcal B$, with
    domain $\mathcal D$ and co-domain $\mathcal C$.

    $\mathcal R$ satisfies \Index{structural adequacy} if, whenever
    $\mathcal S\mathcal R\mathcal S', \mathcal S_1\mathcal R\mathcal
    S'_1\ldots,\mathcal S_n\mathcal R\mathcal S'_n$,
    \begin{center}
      there is in $\mathcal A$ a proof of $\mathcal S$ with $\mathcal D$-immediate sequents $\multiset {S_1,\ldots,S_n}$\\
      \iff\\
      there is in $\mathcal B$ a proof of $\mathcal S'$ with $\mathcal C$-immediate sequents $\multiset {S'_1,\ldots,S'_n}$
    \end{center}
  \end{itemize}
\end{definition}

Structural adequacy clearly entails level -1 adequacy (by induction on
a proof in $\mathcal A$, recursively finding its $\mathcal
D$-immediate sequents, we recompose a proof in $\mathcal B$), but
implies neither level -2 nor level -3 since we are not counting
proofs. Also notice that we do not require anything about incomplete
proofs that cannot be completed.

Every instance below relates to a traditional system, as we define an
encoding satisfying structural adequacy.  While \LAF\ is defined as a
typing system (in other words with proof-terms decorating proofs in
the view of the Curry-Howard correspondence), most traditional systems
that we capture below are purely logical, with no proof-term
decorations.  The encoding therefore needs to erase proof-term
annotation, and for this it is useful to project the notion of typing
context as follows:

\begin{definition}[Referable atoms and molecules]\strut

  Let $\Acc[+]\Gamma$ (\resp $\Acc[-]\Gamma$) be the image set of
  $x^+\mapsto\varRead[x^+]\Gamma$ (\resp
  $x^+\mapsto\varRead[x^+]\Gamma$), \ie the set of atoms (\resp
  molecules) that can be refered to, in $\Gamma$, by the use of a
  positive (\resp negative) label.
\end{definition}

\section{Examples in propositional logic}\label{sec:LAFexprop}

The parameters of \LAF\ will be specified so as to capture: the
one-sided version of \LKF~\cite{liang09tcs,liang11apal}, its two-sided
version, and \LJF~\cite{liang09tcs}.

\subsection{Polarised classical logic - one-sided}
\label{sec:LAFK1qf}

In this sub-section we define the instance \LAF[K1] corresponding to
the one-sided focused sequent calculus \LKF\ for polarised classical
logic~\cite{liang09tcs,liang11apal}.

\begin{definition}[Literals, formulae, patterns, decomposition]
\label{def:classpolsyntax}

  Let $\Lit$ be a set of elements called \Index[literal]{literals}, equipped with an involutive function called \Index{negation} mapping every literal $l$ to a literal $\non l$.

  Let $\Atms$ be a \Index{polarisation set}, \ie a subset of $\Lit$ such that $l\in\Atms$ \iff\ $\non l\notin\Atms$. Elements of $\Atms$ will be ranged over by $a, a',\ldots$.

  Let $\Moles$ be the set defined by the first line of the following grammar for (polarised) formulae of classical logic:
  \[ 
  \begin{array}{llll}
    \mbox{Positive formulae}&  P,\ldots&\recdef a\sep\trueP \sep \falseP \sep A\andP B\sep A \orP B\\
    \mbox{Negative formulae}& N,\ldots&\recdef \non a\sep\trueN \sep \falseN \sep A\andN B\sep A \orN B\\
    \mbox{Unspecified formulae}&A&\recdef P \sep N
  \end{array}
  \]

  Negation is extended to formulae as follows:
  \[ \begin{array}{llll}
    \non{\trueP}&\eqdef\falseN&\non{\trueN}&\eqdef\falseP\\
    \non{\falseP}&\eqdef\trueN&\non{\falseN}&\eqdef\trueP\\
    \non{(A\andP B)}&\eqdef\non A\orN \non B&\non{(A\andN B)}&\eqdef\non A\orP \non B\\
    \non{(A\orP B)}&\eqdef\non A\andN \non B&\non{(A\orN B)}&\eqdef\non A\andP \non B
  \end{array}
  \]
  and we extend it to sets or multisets of formulae pointwise.

  The set $\Data$ of \Index[pattern]{pattern} is defined by the following grammar:
  \[p,p_1,p_2,\ldots\recdef \Ppos \sep \Pneg \sep \Ptrue\sep\paire {p_1}{p_2}\sep\inj i p\]

  The decomposition relation $(\DerDec{}{\_}{\XcolY
    \_{\_}}):(\DecompType\times\Data\times\Moles)$ is the restriction
  to molecules of the relation of $\DecompType\times\Data\times\Forms$
  defined inductively for all formulae by the inference system of
  Fig.~\ref{fig:decompK1}.

  The map $p\mapsto \Datast p$ can be inferred from the decomposition relation.
\end{definition}
\begin{bfigure}
    \[
    \begin{array}c
      \infer{\DerDec{}{\Dunit}{\XcolY{\Ptrue}\trueP}}{}
      \qquad
      \infer{\DerDec{}{\Drefute {\non N}}{\XcolY{\Pneg}{N}}}{}
      \qquad
      \infer{\DerDec{}{a}{\XcolY{\Ppos}{a}}}{}\\\\
      \infer{\DerDec{}{\Delta_1,\Delta_2}{\XcolY{\paire{p_1}{p_2}}A_1\andP A_2}}{
        \DerDec{}{\Delta_1}{\XcolY{p_1}A_1}
        \quad
        \DerDec{}{\Delta_2}{\XcolY{p_2}A_2}
      }
      \qquad
      \infer{\DerDec{}{\Delta}{\XcolY{\inj i p}A_1\orP A_2}}{\DerDec{}{\Delta}{\XcolY{p}A_i}}  
    \end{array}
    \]
  \caption{Decomposition relation for \LAF[K1]}
  \label{fig:decompK1}
\end{bfigure}

Keeping the \textsf{sync} rule of \LAF[K1] in mind, we can already see
in Fig.~\ref{fig:decompK1} the traditional introduction rules of
positive connectives in polarised classical logic. The rest of this
sub-section formalises that intuition and explains how \LAF[K1]
manages the introduction of negative connectives, etc.

But in order to finish the instantiation of \LAF\ for propositional
polarised classical logic (1-sided), we need to define typing
contexts, \ie give $\var[+]$, $\var[-]$, and $\TContexts$. In
particular, we have to decide how to refer to elements of the typing
context. To avoid getting into aspects that may be considered as
implementation details, we will stay rather generic and only assume
the following property:

\begin{definition}[Typing contexts]\label{def:classicalcontext} We assume
  \[
  \begin{array}{ll@{\qquad}ll}
    \Acc[+]{\Gamma\Cextend a}&=\Acc[+]{\Gamma}\cup \{a\}&  \Acc[-]{\Gamma\Cextend a}&=\Acc[-]{\Gamma}\\
    \Acc[+]{\Gamma\Cextend \Drefute M}&=\Acc[+]{\Gamma}&  \Acc[-]{\Gamma\Cextend \Drefute M}&=\Acc[-]{\Gamma}\cup \{M\}\\
    \Acc[\pm]{\Gamma\Cextend\Dunit}&=\Acc[\pm]{\Gamma}&
    \Acc[\pm]{\Gamma\Cextend(\Delta_1\Dand\Delta_2)}&=\Acc[\pm]{\Gamma\Cextend\Delta_1\Cextend\Delta_2}
  \end{array}
  \]
  where $\pm$ stands for either $+$ or $-$.
\end{definition}

In section~\ref{sec:LAFlabels} we present several implementations satisfying the
above.

We now relate (cut-free) \LAFcf[K1] and the \LKF\ system
of~\cite{liang09tcs,liang11apal}.

\begin{definition}[Flattening typing decompositions]
  Let $\flatten\Delta$ be the flattening of a typing decomposition as a multiset of positive literals and negative formulae, \ie
  \[\begin{array}{ll@{\qquad}ll}
  \flatten a&\eqdef \multiset a&\flatten {\Drefute P}&\eqdef \multiset{\non P}\\
  \flatten \Dunit&\eqdef \emptyset&\flatten {\Delta_1\Dand \Delta_2}&\eqdef \flatten {\Delta_1}\cup\flatten {\Delta_2}
  \end{array}
  \]
\end{definition}

\begin{remark}\strut\label{rem:ImFlat}
  \begin{itemize}
  \item
    Notice that, for all formulae $A$ and typing decomposition $\Delta$, 
    there exists $p\in\Data$ such that $\DerDec{}{\Delta}{\XcolY p{A}}$
    \iff\ $A\downarrow\flatten\Delta$ as defined in~\cite{liang11apal}.
  \item
    Our assumption about typing contexts implies that, for all $\Gamma$ and $\Delta$,
    \[
    \Acc[+]{\Gamma\Cextend\Delta}\cup\non{\Acc[-]{\Gamma\Cextend\Delta}} = \Acc[+]{\Gamma}\cup\non{\Acc[-]{\Gamma}}\cup\flatten\Delta
    \]
  \end{itemize}
\end{remark}

\begin{definition}[Mapping sequents]

  We encode the sequents of \LAF[K1] (regardless of derivability) to those of \LKF\ as follows:
  \[\begin{array}{l@{\qquad\eqdef\qquad}l}
  \phi(\Der\Gamma {c})&\DerLKF {\non{\Acc[+]\Gamma},{\Acc[-]\Gamma}}{}\\
  \phi(\Der\Gamma {\XcolY {x^+} {a}})&\DerFLKF{\non{\Acc[+]\Gamma},{\Acc[-]\Gamma}} {a}\\
  \phi(\Der\Gamma {\XcolY {f} {\Drefute P}})&\DerFLKF {\non{\Acc[+]\Gamma},{\Acc[-]\Gamma}}{\non P}\\
  \phi(\DerF\Gamma {\XcolY {t^+} P}{})&\DerFLKF{\non{\Acc[+]\Gamma},{\Acc[-]\Gamma}} {P}
  \end{array}
  \]
\end{definition}

\begin{theorem}[Adequacy between {\LAFcf[K1]}\ and \LKF]\label{th:adequacyK1}

  $\phi$ satisfies structural adequacy between \LAFcf[K1]\ and \LKF.
\end{theorem}
\begin{proof}
  The Lemmata 2 and 3 of~\cite{liang11apal} (for the particular case of \LKF) provide the correspondence with the big-step rules of \LAFcf[K1]:
  \begin{itemize}
  \item[$\textsf{async}$]
    Clearly, a derivation in \LAFcf[K1]\ concludes $\Der\Gamma {\XcolY {f} {\Drefute P}}$ for some term $f$ \iff\ it is of the form
    \[
    \infer{\Der\Gamma {\XcolY {f} {\Drefute P}}}
          {
            \Der{\Gamma\Cextend\Delta_1} {c_1}
            \quad
            \ldots
            \quad
            \Der{\Gamma\Cextend\Delta_n} {c_n}
          }
          \]
          for some commands $\{c_1,\ldots,c_n\}$, and where $\{\Delta_1,\ldots,\Delta_n\} = \{ \Delta \mid \exists p, \DerDec{}{\Delta}{\XcolY p P}\}$.

          Correspondingly, Lemma 2 of~\cite{liang11apal}\footnote{slightly reworded using its Lemma 4 as well} entails that a derivation in \LKF\ concludes $\DerFLKF{\non{\Acc[+]\Gamma},{\Acc[-]\Gamma}} {\non P}$
          \iff\ it is of the form
          \[
          \infer{\DerFLKF{\non{\Acc[+]\Gamma},{\Acc[-]\Gamma}} {\non P}}
                { 
                  \infer{\DerLKF{\non{\Acc[+]\Gamma},{\Acc[-]\Gamma}} {\non P}}{
                    \infer{\vdots}{\DerLKF{\non{\Acc[+]\Gamma},\Acc[-]\Gamma,\non{\Phi_1}} {}} 
                    \infer{\vdots}{\DerLKF{\non{\Acc[+]\Gamma},\Acc[-]\Gamma,\non{\Phi_n}} {}} 
                  }
                }
                \]
                where $\{\Phi_1,\ldots,\Phi_n\} = \{ \Phi \mid P\downarrow\Phi\}$.

                Writing $\phi$ for the bijection from $1,\ldots,n$ to itself such that $\flatten{\Delta_i}=\Phi_{\phi(i)}$, we notice that every sequent $\Der{\Gamma\Cextend\Delta_i} {c_i}$ is mapped to the sequent $\DerLKF{\non{\Acc[+]\Gamma},\Acc[-]\Gamma,\non{\Phi_{\phi(i)}}} {}$.
                Indeed, Remark~\ref{rem:ImFlat}.2 entails that 
                \[
                \non{\Acc[+]{\Gamma\Cextend\Delta_i}}\cup{\Acc[-]{\Gamma\Cextend\Delta_i}} = \non{\Acc[+]{\Gamma}}\cup{\Acc[-]{\Gamma}}\cup\non{\Phi_{\phi(i)}}
                \]

              \item[$\textsf{sync}$]

                Clearly, a derivation in \LAFcf[K1]\ concludes $\DerF\Gamma {\XcolY {t^+} P}{}$ for some term $t^+$ \iff\ it is of the form
                \[
                \infer{\DerF\Gamma {\XcolY {p\sigma} P}{}}{
                  \DerDec{}{\Delta}{\XcolY p P}
                  \infer{\Der\Gamma{\XcolY\sigma\Delta}}{
                    \infer{\vdots}{\Der\Gamma {\XcolY {t^-_1} {u_1}}} 
                    \infer{\vdots}{\Der\Gamma {\XcolY {t^-_n} {u_n}}} 
                  }
                }
                \]
                for some $\Delta$, $p$, $\sigma$, $t^-_1,\ldots,t^-_n$, and where $\flatten \Delta =
                \{u_1,\ldots,u_n\}$.

                Correspondingly, Lemma 3 of~\cite{liang11apal} entails that a derivation in \LKF\ concludes $\DerFLKF{\non{\Acc[+]\Gamma},{\Acc[-]\Gamma}} {P}$
                \iff\ it is of the form
                \[
                \infer{\DerFLKF{\non{\Acc[+]\Gamma},{\Acc[-]\Gamma}} {P}}{
                  \infer{\vdots}{\DerFLKF{\non{\Acc[+]\Gamma},{\Acc[-]\Gamma}} {u_1}} 
                  \infer{\vdots}{\DerFLKF{\non{\Acc[+]\Gamma},{\Acc[-]\Gamma}} {u_n}} 
                }
                \]
                for some $P\downarrow \{u_1,\ldots,u_n\}$.

              \item[$\textsf{init}$]
                Clearly, a derivation in \LAFcf[K1]\ concludes $\Der\Gamma {\XcolY {x^+} a}{}$ for some positive label $x^+$ \iff\ it is of the form
                \[
                \infer{\Der\Gamma {\XcolY {x^+} a}{}}{}
                \]
                with $a\in\Acc[+]\Gamma$.

                Correspondingly, a derivation in \LKF\ concludes $\DerFLKF{\non{\Acc[+]\Gamma},{\Acc[-]\Gamma}} {a}$
                \iff\ it is of the form
                \[
                \infer{\DerFLKF{\non{\Acc[+]\Gamma},{\Acc[-]\Gamma}} {a}}{}
                \]
                with $a\in\Acc[+]\Gamma$.

              \item[$\textsf{select}$]
                Clearly, a derivation in \LAFcf[K1]\ concludes $\Der\Gamma {c}{}$ for some command $c$ \iff\ it is of the form
                \[
                \infer{\Der\Gamma {\cutc{x^-}{t^+}}{}}{
                  \DerF\Gamma {\XcolY {t^+} P}{}
                }
                \]
                with $P\in\Acc[-]\Gamma$.

                Correspondingly, a derivation in \LKF\ concludes $\DerLKF{\non{\Acc[+]\Gamma},{\Acc[-]\Gamma}} {}$
                \iff\ it is of the form
                \[
                \infer{\DerLKF{\non{\Acc[+]\Gamma},{\Acc[-]\Gamma}} {}}{
                  \DerFLKF{\non{\Acc[+]\Gamma},{\Acc[-]\Gamma}} {P}
                }
                \]
                with $P\in\Acc[-]\Gamma$.
  \end{itemize}
\end{proof}

\begin{corollary}[Equivalence of provability]

  The provability of a sequent in \LAFcf[K1] is the same as that of its
  encoding in \LKF.
\end{corollary}

The proof may raise the question of why, in the definition of \LAF, we
gave a structure to typing decompositions, instead of directly using a
flattened version (\eg multiset). The reason is to allow the
parametrisation of the system so as to capture logics for which the
structure of typing decomposition may be important; if only for
first-order logic, the scope of eigenvariables is more easily managed
with a structure; this is even more true in higher-order logic.

\subsection{Polarised classical logic - two-sided}

Having seen how an instance of \LAF\ captures a one-sided sequent
calculus, we could see \LAF\ itself as a sequent calculus that is
intrinsically one-sided, considering as a notational idiosyncrasy our
writing the typing environments on the left of the turnstyle.

Here, we show that, by enriching the atoms and molecules with a
``side information'', we can also capture a two-sided version of \LKF.

\begin{definition}[Literals, formulae, patterns, decomposition]\strut

  Let $\Lit^+$ (\resp $\Lit^-$) be a set of elements called positive (\resp negative) literals, and ranged over by $l^+,l_1^+,l_2^+,\ldots$ (\resp $l^-,l^-_1,l^-_2,\ldots$).

  Formulae are defined by the following grammar:
  \[ 
  \begin{array}{llll}
    \mbox{Positive formulae}&  P,\ldots&\recdef l^+\sep\trueP \sep \falseP \sep A\andP B\sep A \orP B\sep \negP A\\
    \mbox{Negative formulae}& N,\ldots&\recdef l^-\sep\trueN \sep \falseN \sep A\andN B\sep A \orN B\sep \negN A\\
    \mbox{Unspecified formulae}&A&\recdef P \sep N
  \end{array}
  \]

  We \emph{position} a literal or a formula on the left-hand side or the right-hand side of a sequent by combining it with an element, called \Index{side information}, of the set $\{\lefths,\righths\}$: we define
  \[\begin{array}{ll}
  \Atms&\eqdef\{(l^+,\righths)\sep l^+\mbox{ positive literal}\}\cup\{(l^-,\lefths)\sep l^-\mbox{ negative literal}\}\\
  \Moles&\eqdef\{(P,\righths)\sep P\mbox{ positive formula}\}\cup\{(N,\lefths)\sep N\mbox{ negative formula}\}
  \end{array}
  \]

  The set $\Data$ of \Index[pattern]{patterns} is defined by the following grammar:
  \[\begin{array}{lrl}
  p,p_1,p_2,\ldots&\recdef& \Ppos_\righths \sep \Pneg_\righths \sep \Ptrue_\righths\sep\paire {p_1}{p_2}\sep\inj i p\sep\switchl p\\
  &\sep&\Ppos_\lefths \sep \Pneg_\lefths \sep \Ptrue_\lefths\sep\caseanal {p_1}{p_2}\sep\project i p\sep\switchr p
  \end{array}
  \]

  The decomposition relation $(\DerDec{}{\_}{\XcolY
    \_{\_}}):(\DecompType\times\Data\times\Moles)$ is the restriction
  to molecules of the relation of
  $\DecompType\times\Data\times(\Forms\times\{\lefths,\righths\})$
  defined inductively for all positioned formulae by the inference
  system of Fig.~\ref{fig:decompK2}.

  Again, since we want to capture classical logic, we assume the same
  property about $(\var[+],\var[-],\TContexts)$ as we did in
  Definition~\ref{def:classicalcontext}.
\end{definition}
\begin{bfigure}[!h]
    \[
    \begin{array}c
      \infer{\DerDec{}{\Drefute {(N,\lefths)}}{\XcolY{\Pneg_\righths}{(N,\righths)}}}{\strut}
      \qquad
      \infer{\DerDec{}{(l^+,\righths)}{\XcolY{\Ppos_\righths}{(l^+,\righths)}}}{\strut}
      \\\\
      \infer{\DerDec{}{\Dunit}{\XcolY{\Ptrue_\righths}{(\trueP,\righths)}}}{\strut}
      \qquad
      \infer{\DerDec{}{\Delta}{\XcolY{\switchl p}{(\negP A,\righths)}}}
            {\DerDec{}{\Delta}{\XcolY{p}{(A,\lefths)}}}
            \\\\
            \infer{\DerDec{}{\Delta_1,\Delta_2}{\XcolY{\paire{p_1}{p_2}}{(A_1\andP A_2,\righths)}}}{
              \DerDec{}{\Delta_1}{\XcolY{p_1}{(A_1,\righths)}}
              \quad
              \DerDec{}{\Delta_2}{\XcolY{p_2}{(A_2,\righths)}}
            }
            \qquad
            \infer{\DerDec{}{\Delta}{\XcolY{\inj i p}{(A_1\orP A_2,\righths)}}}{\DerDec{}{\Delta}{\XcolY{p}{(A_i,\righths)}}}\\\\
            \infer{\DerDec{}{\Drefute {(P,\righths)}}{\XcolY{\Pneg_\lefths}{(P,\lefths)}}}{\strut}
            \qquad
            \infer{\DerDec{}{(l^-,\lefths)}{\XcolY{\Ppos_\lefths}{(l^-,\lefths)}}}{\strut}
            \\\\
            \infer{\DerDec{}{\Dunit}{\XcolY{\Ptrue_\lefths}{(\falseN,\lefths)}}}{\strut}
            \qquad
            \infer{\DerDec{}{\Delta}{\XcolY{\switchr p}{(\negN A,\lefths)}}}
                  {\DerDec{}{\Delta}{\XcolY{p}{(A,\righths)}}}
                  \\\\
                  \infer{\DerDec{}{\Delta_1,\Delta_2}{\XcolY{\caseanal{p_1}{p_2}}{(A_1\orN A_2,\lefths)}}}{
                    \DerDec{}{\Delta_1}{\XcolY{p_1}{(A_1,\lefths)}}
                    \quad
                    \DerDec{}{\Delta_2}{\XcolY{p_2}{(A_2,\lefths)}}
                  }
                  \qquad
                  \infer{\DerDec{}{\Delta}{\XcolY{\project i p}{(A_1\andN A_2,\lefths)}}}{\DerDec{}{\Delta}{\XcolY{p}{(A_i,\lefths)}}}
    \end{array}
    \]
  \caption{Decomposition relation for \LAF[K2]}
  \label{fig:decompK2}
\end{bfigure}

Keeping the \textsf{sync} rule of \LAF[K2] in mind, we see in
Fig.~\ref{fig:decompK2} the traditional right-introduction rules of
positive connectives and left-introduction rules of negative
connectives.

A deeper intuition can be given by encoding \LAF[K2] sequents as two-sided sequents, just like we encoded \LAF[K1] sequents as one-sided \LKF\ sequents:
\begin{definition}[{\LAF[K2]} sequents as two-sided sequents]\strut\label{def:K2asLKF}
  \begin{enumerate}
  \item First, when $\pm$ is either $+$ or $-$, we define
    \[\begin{array}l
    \Acc[\pm\righths]\Gamma\eqdef\{A\mid(A,\righths)\in\Acc[\pm]\Gamma\}\\
    \Acc[\pm\lefths]\Gamma\eqdef\{A\mid(A,\lefths)\in\Acc[\pm]\Gamma\}
    \end{array}
    \]
  \item Then we define the encoding:
    \[    \phi(\Der\Gamma {c})\qquad\eqdef\qquad\Der{\Acc[+\righths]\Gamma,\Acc[-\lefths]\Gamma} {\Acc[+\lefths]\Gamma,\Acc[-\righths]\Gamma}   \]
  \end{enumerate}
\end{definition}

Keeping the above interpretation of sequents in mind, we should now see how to develop the details of the correspondence (similar to that expressed in Theorem~\ref{th:adequacyK1}) between \LAFcf[K2] and the two-sided version of \LKF\ (which may actually not be written down in the literature).

As we can see, the decomposition relation, and the whole inference system described by \LAF[K2], is completely symmetric.

\subsection{Polarised intuitionistic logic}

\begin{definition}[Literals, formulae, patterns, decomposition]\strut

  Let $\Lit^+$ (\resp $\Lit^-$) be a set of elements called positive (\resp negative) literals, and ranged over by $l^+,l_1^+,l_2^+,\ldots$ (\resp $l^-,l^-_1,l^-_2,\ldots$).

  Formulae are defined by the following grammar:
  \[ 
  \begin{array}{llll}
    \mbox{Positive formulae}&  P,\ldots&\recdef l^+\sep\trueP \sep \falseP \sep A\andP B\sep A \ou B\\
    \mbox{Negative formulae}& N,\ldots&\recdef l^-\sep\trueN \sep \falseN \sep A\andN B\sep A \imp B\sep \neg A\\
    \mbox{Unspecified formulae}&A&\recdef P \sep N
  \end{array}
  \]

  We \emph{position} a literal or a formula on the left-hand side or the right-hand side of a sequent by combining it with an element, called \Index{side information}, of the set $\{\lefths,\righths\}$: we define
  \[\begin{array}{ll}
  \Atms&\eqdef\{(l^+,\righths)\sep l^+\mbox{ positive literal}\}\cup\{(l^-,\lefths)\sep l^-\mbox{ negative literal}\}\cup\{(\falseN,\lefths)\}\\
  \Moles&\eqdef\{(P,\righths)\sep P\mbox{ positive formula}\}\cup\{(N,\lefths)\sep N\mbox{ negative formula}\}
  \end{array}
  \]
  In the rest of this sub-section $v$ stands for either a negative literal $l^-$ or $\falseN$.

  The set $\Data$ of \Index[pattern]{pattern} is defined by the following grammar:
  \[\begin{array}{lrl}
  p,p_1,p_2,\ldots&\recdef& \Ppos_\righths \sep \Pneg_\righths \sep \Ptrue_\righths\sep\paire {p_1}{p_2}\sep\inj i p\\
  &\sep&\Ppos_\lefths \sep \Pneg_\lefths \sep \Ptrue_\lefths\sep\cons {p_1}{p_2}\sep\project i p\sep\switchr p
  \end{array}
  \]

  The decomposition relation $(\DerDec{}{\_}{\XcolY
    \_{\_}}):(\DecompType\times\Data\times\Moles)$ is the restriction
  (to molecules) of the relation of
  $\DecompType\times\Data\times(\Forms\times\{\lefths,\righths\})$ defined inductively for all
  positioned formulae by the inference system of Fig.~\ref{fig:decompJ}.
\end{definition}
\begin{bfigure}[!h]
    \[
    \begin{array}c
      \infer{\DerDec{}{\Drefute {(N,\lefths)}}{\XcolY{\Pneg_\righths}{(N,\righths)}}}{\strut}
      \qquad
      \infer{\DerDec{}{(l^+,\righths)}{\XcolY{\Ppos_\righths}{(l^+,\righths)}}}{\strut}
      \\\\
      \infer{\DerDec{}{\Dunit}{\XcolY{\Ptrue_\righths}{(\trueP,\righths)}}}{\strut}
            \\\\
            \infer{\DerDec{}{\Delta_1,\Delta_2}{\XcolY{\paire{p_1}{p_2}}{(A_1\andP A_2,\righths)}}}{
              \DerDec{}{\Delta_1}{\XcolY{p_1}{(A_1,\righths)}}
              \quad
              \DerDec{}{\Delta_2}{\XcolY{p_2}{(A_2,\righths)}}
            }
            \qquad
            \infer{\DerDec{}{\Delta}{\XcolY{\inj i p}{(A_1\ou A_2,\righths)}}}{\DerDec{}{\Delta}{\XcolY{p}{(A_i,\righths)}}}\\\\
            \infer{\DerDec{}{\Drefute {(P,\righths)}}{\XcolY{\Pneg_\lefths}{(P,\lefths)}}}{\strut}
            \qquad
            \infer{\DerDec{}{(l^-,\lefths)}{\XcolY{\Ppos_\lefths}{(l^-,\lefths)}}}{\strut}
            \\\\
            \infer{\DerDec{}{(\falseN,\lefths)}{\XcolY{\Ptrue_\lefths}{(\falseN,\lefths)}}}{\strut}
            \qquad
            \infer
                {\DerDec{}{\Delta\Dand{(\falseN,\lefths)}}{\XcolY{\switchr p}{(\neg A,\lefths)}}}
                {\DerDec{}{\Delta}{\XcolY{p}{(A,\righths)}}}
                \\\\
                  \infer{\DerDec{}{\Delta_1,\Delta_2}{\XcolY{\cons{p_1}{p_2}}{(A_1\imp A_2,\lefths)}}}{
                    \DerDec{}{\Delta_1}{\XcolY{p_1}{(A_1,\righths)}}
                    \quad
                    \DerDec{}{\Delta_2}{\XcolY{p_2}{(A_2,\lefths)}}
                  }
                  \qquad
                  \infer{\DerDec{}{\Delta}{\XcolY{\project i p}{(A_1\andN A_2,\lefths)}}}{\DerDec{}{\Delta}{\XcolY{p}{(A_i,\lefths)}}}
    \end{array}
    \]
  \caption{Decomposition relation for \LAF[J]}
  \label{fig:decompJ}
\end{bfigure}

Again, we can already see in Fig.~\ref{fig:decompJ} the traditional right-introduction rules of positive connectives and left-introduction rules of negative connectives.

A few words about the connectives: compared to \LAF[K2], we have dropped the positive negation and we have replaced the negative disjunction by the implication, also negative (the negative negation and the positive disjunction are consequently written $\neg$ and $\ou$, respectively).
Since in (polarised) classical logic, $A\imp B$ can be seen as an abbreviation for $(\negN A)\orN B$, the decomposition rule for $(A\imp B,\lefths)$ is simply the combination of the $K2$ rules for $\negN$ and $\orN$.

With implication as a primitive connective, we could actually remove the (negative) negation from the system, since it can in turn be seen as the combination of implication and absurdity ($\neg A$ can be seen as the abbreviation for $A\imp\falseN$) and its decomposition rule reflects this.
Notice that the decomposition rule for $\falseN$ (and therefore that of $\neg$) are slightly modified compared to $K2$. To understand this, we should start by making the following remark:

\begin{remark}\strut
\label{rem:exactlyone}
  \begin{enumerate}
  \item
    Whenever $\DerDec {}\Delta {\XcolY p{(A,\righths)}}$, $\Delta$ contains no items of the form $\Drefute (P,\righths)$ or $(v,\lefths)$.
  \item
    Whenever $\DerDec {}\Delta {\XcolY p{(A,\lefths)}}$, $\Delta$ contains \textbf{exactly one} item of the form $\Drefute (P,\righths)$ or $(v,\lefths)$.
  \end{enumerate}
($v$ stands for either a negative literal $l^-$ or $\falseN$).
\end{remark}

The first point corresponds to the fact that, when we have a right-hand side focus in intuitionistic logic, the focus never switches to the left-hand side when looking at a proof-tree bottom-up.
Notice that this would be false in presence of the positive negation, which would precisely switch the focus to the left-hand side as in $K2$.

Now the second point would not hold if we kept the negative disjunction from $K2$, since its decomposition rule would create a branching with two premisses of the form $(v,\lefths)$. Hence its replacement with implication, whose decomposition rule has only one premiss of that form, so that, in every derivation of the above inference system, at most one branch keeps decomposing formulae on the left. And that would be true with the $K2$ rules for $\falseN$ and $\negN$.
The reason to tweak them is to get point 2 with \emph{exactly one} rather than \emph{at most one}, and it is for this tweak that we added $(\falseN,\lefths)$ to $\Atms$ (compared to the $K2$ version).

To see why Remark~\ref{rem:exactlyone}.2 is so important for intuitionistic logic, we should now interpret \LAF[K2] sequents as intuitionistic sequents (from \eg \LJF~\cite{liang09tcs}):
\begin{definition}[{\LAF[J]} sequents as two-sided \LJF\ sequents]\strut
  \begin{enumerate}
  \item First, when $\pm$ is either $+$ or $-$, we define
    \[
    \begin{array}l
      \Acc[\pm\righths]\Gamma\eqdef\{A\mid(A,\righths)\in\Acc[\pm]\Gamma\}\\
      \Acc[+\lefths]\Gamma\eqdef\{l^-\mid(l^-,\lefths)\in\Acc[+]\Gamma\}\\
      \Acc[-\lefths]\Gamma\eqdef\{N\mid(N,\lefths)\in\Acc[-]\Gamma\}
    \end{array}
    \]
  \item Then we define the encoding:
  \[\begin{array}{l@{\qquad\eqdef\qquad}l}
  \phi(\Der\Gamma {c})&\DerLJF {\Acc[+\righths]\Gamma,\Acc[-\lefths]\Gamma}{}{\Acc[+\lefths]\Gamma,\Acc[-\righths]\Gamma}\\
  \phi(\Der\Gamma {\XcolY {x^+} {(l^-,\lefths)}})
  &\DerFlLJF{\Acc[+\righths]\Gamma,\Acc[-\lefths]\Gamma} {l^-}{\Acc[+\lefths]\Gamma,\Acc[-\righths]\Gamma}\\
  \phi(\Der\Gamma {\XcolY {f} {\Drefute {(P,\righths)}}})
  &\DerFlLJF {\Acc[+\righths]\Gamma,\Acc[-\lefths]\Gamma}{P}{\Acc[+\lefths]\Gamma,\Acc[-\righths]\Gamma}\\
  \phi(\DerF\Gamma {\XcolY {t^+} {(N,\lefths)}}{})
  &\DerFlLJF{\Acc[+\righths]\Gamma,\Acc[-\lefths]\Gamma} {N}{\Acc[+\lefths]\Gamma,\Acc[-\righths]\Gamma}\\
  \phi(\Der\Gamma {\XcolY {x^+} {(l^+,\righths)}})
  &\DerFrLJF{\Acc[+\righths]\Gamma,\Acc[-\lefths]\Gamma} {l^+}\\
  \phi(\Der\Gamma {\XcolY {f} {\Drefute {(N,\lefths)}}})
  &\DerFrLJF {\Acc[+\righths]\Gamma,\Acc[-\lefths]\Gamma}{N}\\
  \phi(\DerF\Gamma {\XcolY {t^+} {(P,\righths)}}{})
  &\DerFrLJF{\Acc[+\righths]\Gamma,\Acc[-\lefths]\Gamma} {P}
  \end{array}
  \]
  \end{enumerate}
In the first four cases, we require $\Acc[+\lefths]\Gamma,\Acc[-\righths]\Gamma$ to be a singleton (or be empty).
\end{definition}

The first line of the encoding is the same as for \LAF[K2]
(Definition~\ref{def:K2asLKF}), but for the fact that we require
$\Acc[+\lefths]\Gamma,\Acc[-\righths]\Gamma$ to be a singleton (or be
empty), since we are to capture an intuitionistic system such as
\LJF.  We also see in the last three cases (when there is a right-hand
side focus), that the encoding forgets
$\Acc[+\lefths]\Gamma,\Acc[-\righths]\Gamma$ altogether. If it is not
empty, then it should definitely play no further role in the proof of
the \LAF[J] sequent.

The issue arises in particular when analysing the \textsf{select} rule:

In \LJF, placing the focus on a formula on the left-hand side does not
affect the formula stored on the right-hand side; on the contrary,
placing the focus on the right-hand side formula removes it from
the right-hand side (no backup copy is made).

This is an important feature of intuitionistic logic: if a backup copy
of the formula was kept, we could place again the focus on it further
up in the proof, and we could thus prove formulae such as $A\ou \neg
A$ or the drinker's theorem; indeed this amounts to authorising
contraction on the right.

Now looking at the \textsf{select} rule of \LAF[J], notice that the
typing context $\Gamma$ is \textbf{unchanged} by the rule: placing the
focus on a right-hand side formula (\ie a formula from
$\Acc[+\lefths]\Gamma,\Acc[-\righths]\Gamma$) does not remove it from
the typing context.

We could therefore fear that, because of this feature, \LAF\ forces
the presence of contraction (left and right) and is therefore
intrinsically classical. Fortunately, this is not the case:

After selecting a right-hand side formula for focus, it is decomposed
according to the rules of the decomposition relation. As mentioned in
Remark~\ref{rem:exactlyone}.1, the focus never switches to the
left-hand side and we are therefore left to prove a collection of
sequents of the form $\Der\Gamma {\XcolY {x^+} {(v,\righths)}}$ or
$\Der\Gamma {\XcolY {f} {\Drefute {(N,\lefths)}}}$ for some $x^+$ or
$f$ to be found. In the former case, the part of $\Gamma$ that stores
the unfortunate backup copy of the right-hand side formula that was
selected for focus, does not affect whether
$(v,\righths)\in\Acc[+]{\Gamma}$. In the latter case, only rule
\textsf{async} can be applied and a sequent of the form
$\Der{\Gamma\Cextend\Delta} {c}$ is left to be proved (for some $c$ to
be found) for every $\Delta$ that can decompose $(N,\lefths)$.  For
the adequacy with intuitionistic logic to work, it suffices that for
every such $\Delta$, the operation $\Gamma\Cextend\Delta$ erases from
$\Gamma$ the unfortunate backup copy of the right-hand side formula
that was selected for focus.  According to
Remark~\ref{rem:exactlyone}.2, every such $\Delta$ contains
\textbf{exactly one} item of the form $(v,\lefths)$ or
$\Drefute{(P,\righths)}$, \ie a new right-hand side formula which can overwrite
the old one.  At least, provided that $(\var[+],\var[-],\TContexts)$
are defined to do that job.

Having tweaked the decomposition rule for $(\falseN,\lefths)$ to
guarantee Remark~\ref{rem:exactlyone}.2,\linebreak $(\var[+],\var[-],\TContexts)$
should also make sure that, for any $\Gamma$, the (focussed) sequent
$\DerF{\Gamma}{\XcolY{t^+}{(\falseN,\lefths)}}{}$ can still be proved
for some $t^+$ to be found, \ie the sequent
$\Der{\Gamma}{\XcolY{x^+}{(\falseN,\lefths)}}$ can be proved for some
$x^+$ to be found, \ie $(\falseN,\lefths)\in\Acc[+]\Gamma$ (even when
$\Gamma$ is interpreted as something completely empty).  This is easy
to do, by having a permanent and special label
$x^+_{(\falseN,\lefths)}\in\var[+]$ mapped to $(\falseN,\lefths)$ in
every $\Gamma$. This is the same as permanently adding $\falseN$ on
the right-hand side of intuitionistic sequents (as some kind of
multi-conclusion), lest that right-hand side ever gets empty: it is
harmless to both the intuitionistic provability and the structural
theory of proofs (none are added, non are removed).

\begin{definition}[Typing contexts]\label{def:intuitionisticcontext}\strut

 We assume that we always have $(\falseN,\lefths)\in\Acc[+]{\Gamma}$ and that
 \[
 \begin{array}c
   \begin{array}{ll@{\qquad}ll}
    \Acc[+]{\Gamma\Cextend (l^+,\righths)}&=\Acc[+]{\Gamma}\cup \{(l^+,\righths)\}
    &
    \Acc[-]{\Gamma\Cextend a}&=\Acc[-]{\Gamma}\\
    \Acc[+]{\Gamma\Cextend \Drefute M}&=\Acc[+]{\Gamma}
    &
    \Acc[-]{\Gamma\Cextend \Drefute{(N,\lefths)}}&=\Acc[-]{\Gamma}\cup \{(N,\lefths)\}\\
    \Acc[\pm]{\Gamma\Cextend\Dunit}&=\Acc[\pm]{\Gamma}
    &
   \Acc[\pm]{\Gamma\Cextend(\Delta_1\Dand\Delta_2)}&=\Acc[\pm]{\Gamma\Cextend\Delta_1\Cextend\Delta_2}
  \end{array}\\
   \begin{array}{ll}
    \Acc[+]{\Gamma\Cextend (v,\lefths)}
    &=\{(l^+,\righths)\mid (l^+,\righths)\in\Acc[+]{\Gamma}\}\cup\{(v,\lefths),(\falseN,\lefths)\}\\
    \Acc[-]{\Gamma\Cextend \Drefute{(P,\righths)}}
    &=\{(N,\lefths)\mid (N,\lefths)\in\Acc[-]{\Gamma}\}\cup\{(P,\righths)\}\\
  \end{array}
  \end{array}
  \]
  where again $\pm$ stands for either $+$ or $-$ and $v$ stands for either a negative literal $l^-$ or $\falseN$.
\end{definition}

The first three lines are the same as those assumed for $K1$ and $K2$, except it is restricted to those cases where we do not try to add to $\Gamma$ an atom or a molecule that is interpreted as going to the right-hand side of a sequent. When we want to do that, this atom or molecule should overwrite the previous atom(s) or molecule(s) that was (were) interpreted as being on the right-hand side; this is done in the last two lines, where $\Acc[+\lefths]{\Gamma},\Acc[-\righths]{\Gamma}$ is completely erased.

\begin{theorem}[Adequacy between {\LAFcf[J]}\ and \LJF]\label{th:adequacyJ}

  $\phi$ satisfies structural adequacy between \LAFcf[J]\ and \LJF.
\end{theorem}
\begin{proof}
The details are similar to those of Theorem~\ref{th:adequacyK1}, relying again on the \LJF\ properties expressed in~\cite{liang09tcs,liang11apal} and following the series of remarks and design decisions that were made above.
\end{proof}

\ssection[Examples of labels implementation]{Examples of labels implementation: De Bruijn's indices and levels}
\label{sec:LAFlabels}

In this section we give some concrete implementations of labels to
completely specify the typing context algebras used in the examples of
the previous section.

\subsection{Labels for classical logic}
\label{sec:LAFlabelsK}

In the instances \LAF[K1] and \LAF[K2], we have simply made some
assumptions on the typing context algebra (in
Definition~\ref{def:classicalcontext}). We now give it a full
definition satisfying these assumptions and using of De Bruijn's
indices.

In fact, we generically build an $(\mathcal A,\mathcal B)$-context for
each pair of sets $\mathcal A$ and $\mathcal B$, and the typing
context algebra will simply be the instance where $\mathcal A=\Atms$
and $\mathcal B=\Moles$.

\begin{definition}[Generic context algebras with De Bruijn's indices - classical]\label{def:DBlabels}\nopagesplit

  Given two sets $\mathcal A$ and $\mathcal B$, we define an
  $(\mathcal A,\mathcal B)$-context algebra $\TContexts[\mathcal
    A,\mathcal B]$ as follows:

  An $(\mathcal A,\mathcal B)$-context $\Gamma$ is a pair
  $(\Gamma^+,\Gamma^-)$ where $\Gamma^+$ is a list of elements of
  $\mathcal A$ and $\Gamma^-$ is a list of elements of $\mathcal B$.

  Extensions are defined as follows:
  \[
  \begin{array}{ll@{\qquad}ll}
    (\Gamma^+,\Gamma^-)\Cextend a&\eqdef(\cons a {\Gamma^+},\Gamma^-)&  
    (\Gamma^+,\Gamma^-)\Cextend \Drefute b &\eqdef(\Gamma^+,\cons b {\Gamma^-})\\
    (\Gamma^+,\Gamma^-)\Cextend \Dunit &\eqdef(\Gamma^+,\Gamma^-)
    &(\Gamma^+,\Gamma^-)\Cextend(\Delta_1\Dand\Delta_2)
    &\eqdef(\Gamma^+,\Gamma^-)\Cextend\Delta_1\Cextend\Delta_2
  \end{array}
  \]

  Positive labels and negative labels are two disjoint copies of the set of integers, with elements denoted $n^+$ and $n^-$, and we define\\
  $\varRead[n^+]{(\Gamma^+,\Gamma^-)}$ as the $(n+1)^{th}$ element of $\Gamma^+$\\
  $\varRead[n^-]{(\Gamma^+,\Gamma^-)}$ as the $(n+1)^{th}$ element of $\Gamma^-$.
\end{definition}

These are indeed De Bruijn's indices, since the element accessed by
label $0^+$ (\resp $0^-$) is the head of the list $\Gamma^+$ (\resp
$\Gamma^-$), \ie the element of the list that has last been added.

Alternatives using De Bruijn's indices are possible:
\begin{itemize}
\item the choice we made, when extending a context with
  $(\Delta_1\Dand\Delta_2)$, of first extending the context with
  $\Delta_1$ and then extending the result with $\Delta_2$, was
  completely arbitrary, we could have defined
  \[(\Gamma^+,\Gamma^-)\Cextend(\Delta_1\Dand\Delta_2)
  \eqdef(\Gamma^+,\Gamma^-)\Cextend\Delta_2\Cextend\Delta_1\]
\item we could have defined an $(\mathcal A,\mathcal B)$-context
  $\Gamma$ as one single list of atoms and molecules, with
  $\var[+]=\var[-]=\mathbb N$ and $\varRead[n^+]{\Gamma}$ (\resp
  $\varRead[n^-]{\Gamma}$) being defined only on those integers
  mapped to atoms (\resp molecules);
\item we could have defined an $(\mathcal A,\mathcal B)$-context
  $\Gamma$ as a list of $(\mathcal A,\mathcal B)$-decompositions,
  with \[\Gamma\Cextend\Delta\eqdef \cons{\Delta}\Gamma\] and then a
  positive or negative label would be a pair $(n,i)$, where the
  integer $n$ identifies the $n^{th}$ element $\Delta$ of the list
  and $i$ is a string of $0$ and $1$ representing the path from the
  root of $\Delta$ (seen as a tree) to one of its leaves.
\end{itemize}

But we can also use De Bruijn's levels, rather than indices.

One of the drawbacks of the implementation with De Bruijn's indices,
is that the name of a label, declared with a type in a typing context
$\Gamma$, ``changes'' when $\Gamma$ is extended with some typing
decomposition $\Delta$. For instance if $\varRead[0^+]{\Gamma}$ is an
atom $a$ because $a$ is the head of the list $\Gamma^+$, then in
$\Gamma\Cextend\Delta$, $a$ may no longer be the head of
$(\Gamma\Cextend\Delta)^+$ and it will be refered to with an updated
label name.

Depending on how the computations of $\varRead[x^+]{\Gamma}$ and
$\varRead[x^-]{\Gamma}$ are implemented (imagine we have a HashMap for
this), it could be problematic to have to update all the label names
at every extension. We could do this update lazily, or we could also
go for De Bruijn's levels: once it has been introduced in a typing
context, a label will remain unchanged by the subsequent extensions of
the context.

\begin{definition}[Context algebras with De Bruijn's levels - classical]
  \label{def:DBlabelslevels}

  Positive labels and negative labels are two disjoint copies of the
  set of integers, with elements denoted $n^+$ and $n^-$, and we
  define\\
  $\varRead[n^+]{(\Gamma^+,\Gamma^-)}$ as the
  $(\abs{\Gamma^+}-n)^{th}$ element of
  $\Gamma^+$\\ $\varRead[n^-]{(\Gamma^+,\Gamma^-)}$ as the
  $(\abs{\Gamma^-}-n)^{th}$ element of $\Gamma^-$.
\end{definition}

In other words, the difference between De Bruijn's indices and De
Bruijn's levels is that we are counting from the bottom of the list
rather than from the head.

All of the above alternatives work for \LAF[K1] and \LAF[K2], in that
the assumptions of Definition~\ref{def:classicalcontext} are clearly
satisfied.

Choosing between them is really a question of implementation, with no
theoretical impact.

\subsection{Labels for intuitionistic logic}

In the instance \LAF[J], we have made some different assumptions on
the typing context algebra (in
Definition~\ref{def:intuitionisticcontext}). We adapt our definition
of the typing context algebra accordingly.

This time, we directly define it rather than go through the generic
definition of an $(\mathcal A,\mathcal B)$-context algebra for each
$\mathcal A$ and $\mathcal B$, since the assumptions in
Definition~\ref{def:intuitionisticcontext} (unlike those in
Definition~\ref{def:classicalcontext}) make a case analysis on the
kind of atom (\resp on the kind of molecule) that is added to the
typing context. That case analysis would not make sense for arbitrary 
sets $\mathcal A$ and $\mathcal B$.

\begin{definition}[Context algebras with De Bruijn's indices - intuitionistic]

  The typing context algebra $\TContexts$ is defined as follows:

  A typing context $\Gamma$ is a triple $(\Gamma^+,\Gamma^-,R)$ where
  $\Gamma^+$ is a list of atoms, $\Gamma^-$ is a list of molecules,
  and $R$ is either an atom of the form $(v,\lefths)$ or a molecule of the form $(P,\righths)$.\footnote{Intuitively, $R$ represents the right-hand side of the \LJF\ sequent.}

  Extensions are defined as follows:
  \[
  \begin{array}{ll@{\quad}ll}
    (\Gamma^+,\Gamma^-,R)\Cextend (l^+,\righths)&\eqdef(\cons {(l^+,\righths)} {\Gamma^+},\Gamma^-,R)&  
    (\Gamma^+,\Gamma^-,R)\Cextend \Drefute {(N,\lefths)} &\eqdef(\Gamma^+,\cons {(N,\lefths)} {\Gamma^-},R)\\
    (\Gamma^+,\Gamma^-,R)\Cextend (v,\lefths)&\eqdef(\Gamma^+,\Gamma^-,(v,\lefths))&  
    (\Gamma^+,\Gamma^-,R)\Cextend \Drefute {(P,\righths)} &\eqdef(\Gamma^+,\Gamma^-,(P,\righths))\\
    (\Gamma^+,\Gamma^-,R)\Cextend \Dunit &\eqdef(\Gamma^+,\Gamma^-)
    &(\Gamma^+,\Gamma^-,R)\Cextend(\Delta_1\Dand\Delta_2)
    &\eqdef(\Gamma^+,\Gamma^-,R)\Cextend\Delta_1\Cextend\Delta_2
  \end{array}
  \]

  Again, we use for labels two disjoint copies $\mathbb N^+$ and $\mathbb N^-$ of the set of integers:

  A positive label is either some $n^+\in\mathbb N^+$ or one of two special labels $\star^+$ and $x^+_{(\falseN,\lefths)}$.\\
  A negative label is either some $n^-\in\mathbb N^-$ or the special label $\star^-$.\\
  And we define\\
  $\varRead[n^+]{(\Gamma^+,\Gamma^-,R)}$ as the $(n+1)^{th}$ element of $\Gamma^+$\\
  $\varRead[\star^+]{(\Gamma^+,\Gamma^-,R)}$ as $R$ if it is of the form $(v,\lefths)$ (undefined if not)\\
  $\varRead[x^+_{(\falseN,\lefths)}]{(\Gamma^+,\Gamma^-,R)}$ as $(\falseN,\lefths)$\\
  $\varRead[n^-]{(\Gamma^+,\Gamma^-,R)}$ as the $(n+1)^{th}$ element of $\Gamma^-$\\
  $\varRead[\star^-]{(\Gamma^+,\Gamma^-,R)}$ as $R$ if it is of the form $(P,\righths)$ (undefined if not).
\end{definition}

Clearly, the above definition of the typing context algebra satisfies
the assumptions in Definition~\ref{def:intuitionisticcontext}.

And again, there are many alternatives for the above definition,
including the use of De Bruijn's levels, etc. Choosing between them
would again simply be a question of implementation.

In brief, this section (as well as other parts of this dissertation)
shows that the theory is able to handle diverse implementations,
instead of having the theory commit to a particular choice of
formalisation, and then having an implementation depart from it. Here,
we can directly see the OCaml modules and module signatures that we
can or should implement.

\cchapter[Abstract focussed sequent calculus w.\ quantifiers]{An abstract focussed sequent calculus - with quantifiers}
\label{ch:LAFwQ}

\minitoc 

In this chapter we extend the \LAF\ sequent calculus to handle
quantifiers.

First, we should notice that the calculus we presented in
Chapter~\ref{ch:LAFwoQ} can already ``handle quantifiers'', in the way
the $\omega$-rule does~\cite{hilbert31,schuette50}. Indeed, we can
adapt and extend system \LAF[K1] with an extra rule for the
decomposition relation such as
\[
\infer
    {\DerDec{}{\Delta}{\XcolY{\paire r p}{\exists x A}}}
    {\DerDec{}{\Delta}{\XcolY{p}{\subst A x r}}}
\]
capturing the positive behaviour of the existential quantifier in the
synchronous rule.

But this will also determine the asynchronous treatment of the
universal quantifier: Ignoring proof-terms for the moment, proving the
refutation $\Der\Gamma {{\Drefute {\exists x {N}}}}$ (\ie intuitively
proving $\forall x {\non N}$) requires the use of rule \textsf{async},
with sub-proofs for each of the sequents \[\Der{\Gamma,\Drefute{\subst
    {\non N} x t}} {}\] where $t$ ranges over all potential witnesses
for $x$, which is the behaviour of the $\omega$-rule.

In particular if $N$ is of the form $\forall y\ P$, each of those
premisses can then be derived by a proof of the form
\[\infer{\Der{\Gamma,\Drefute{\exists y \subst {\non P} x t}} {}}
        {\infer{\DerF\Gam {\exists y \subst {\non P} x t} {}}
          {\Der\Gam {\Drefute{\subst {P} {x,y} {t,t'}}} {}}}
\]
where $t'$ is witness for $y$ whose choice may depend (possibly
in a non-uniform way) on the instance $t$ of $x$.

So, in order to recover a standard rule for $\forall$-introduction,
which uses something like an eigenvariable, we need to enrich \LAF,
which will now be given by a tuple of parameters
\[(\Sorts,\Terms,\SoContexts,\decs{},\Atms,\Moles,\atmEq{}{},\var[+],\var[-],\TContexts,\Data,\decf{})\]
where each parameter is described in Section~\ref{sec:LAFDefwQ}.

Section~\ref{sec:LAFexbind} then provides an instance illustrating
first-order quantification.

\section{Presentation of the system}
\label{sec:LAFDefwQ}

\subsection{Quantifying structure}

The first group of parameters \((\Sorts,\Terms,\SoContexts,\decs{})\) specifies the
objects that \LAF\ quantifies over. For logics with quantifiers, the
following definition provides a rather general setting: the terms that
can be provided as witnesses are multi-sorted, and the sorting may
depend on a local sorting context (as we would need for higher-order
logic, dependent types, etc).

\begin{definition}[Quantifying structure]\label{def:quantstruct}

  \LAF\ is parameterised by a \Index{quantifying structure} \((\Sorts,\Terms,\SoContexts,\decs{})\), made of 
  \begin{itemize}
  \item
    a fixed set $\Sorts$ of elements called \Index[sort]{sorts}, denoted $s$, $s_1$, etc.
  \item
    a fixed set $\Terms$ of elements called \Index[term]{terms}, denoted $r$, $r_1$, etc,
  \item
    a fixed set $\SoContexts$ of elements called \Index[sorting context]{sorting contexts}, denoted $\Sigma$, $\Sigma_1$, etc,
  \item a \Index{sorting relation}, \ie a set of elements $(\Ders{\_}{\XcolY \_{\_}}):(\SoContexts\times\Terms\times\Sorts)$
  \end{itemize}
\end{definition}

\subsection{Atoms and Molecules}

The next group of parameters $(\Atms,\Moles,\atmEq{}{})$ adapts the
notions of atoms and molecules to the presence of quantifiers.

Atoms and molecules now need more structure than in the propositional
case, because intuitively, we would like atoms and molecules to refer
to terms. Whether it is in the decomposition relation or elsewhere in
the \LAF\ inference system, we will have a rule where witnesses for
existential variables are picked. This usually involves substituting
the witness for the existential variable in the premiss of the rule.

Two reasons suggest to go for a different approach: 

First, this requires us to specify how substitutions affect atoms and
molecules; which then requires us to specify what variables and terms
are for the abstract notions of atoms and molecules; then we would
probably need to specify how substitutions affect the decomposition
relation. All of which are rather heavy in our abstract setting.

Second, an implementation of proof-search would probably depart from
such a rule anyway, as it could be costly to traverse the whole
sequent, or even just some of its atoms and molecules, to compute the
substitution every time a witness is picked. The substitution would
more efficiently be done lazily, keeping the fact that the existential
variable ``is in fact the witness'' in a separate data-structure to be
looked up when we finally need the information.

Hence, we will formalise what would be actually closer to
implementation, expressing an atom (it will be the same for a
molecule) as a pair $(a,\lv r)$ where $a$ is a structure not
(explicitly) refering to terms and $\lv r$ is a list of terms: the
former is a \emph{parameterised atoms} and the latter is its list of
parameters. The list of parameters $\lv r$ can be seen as a delayed
substitution, in the view that $a$ refers to its parameters by either
using something like De Bruijn's indices, or by having a series of
$\lambda$-abstractions at its top-level.

For instance, to represent the atoms of first-order logic, we could
use a pair
\[(p(4,\#1,\#2) , \cons x{\cons 5 \emptylist})\]
(where $p$ is a ternary predicate symbol and $x$ is an eigenvariable) to
represent the atom usually written as $p(4,x,5)$.

A parameterised atom (such as $p(4,\#1,\#2)$) comes with a notion of
\emph{arity}: a list of sorts $l$ describing the sorts of its
parameters numbered from 1 to $\abs l$ (the arity of $p(4,\#1,\#2)$
would here be a list of length at least $2$).

This leads to the following definition.

\begin{definition}[Atoms \& molecules]\label{def:atommolecules1o}

  \LAF\ is parameterised by two sets $\Atms$ and $\Moles$, whose
  elements are respectively called \Index[atom]{(parameterised) atoms}
  (denoted $a$, $a'$,\ldots) and \Index[molecule]{(parameterised)
    molecules} (denoted $M$, $M'$,\ldots), each of which is equipped
  with a function that maps every atom $a$ (\resp molecule $M$) to a
  list of sorts denoted $\abs a$ (\resp $\abs M$) and called its
  \Index{arity}.

  The set $\IAtms$ (\resp $\IMoles$) of \Index[instanciated
    atom]{instanciated atoms} (\resp \Index[instanciated
    molecule]{instanciated molecules}) is the set of pairs of the form
  $(a,\lv r)$ (\resp $(M,\lv r)$), where $a$ is an atom (\resp $M$ is a
  molecule) of arity $l$ and $\lv r$ is a list of terms of length
  $\abs l$.\footnote{We do not relate the sorts specified in $l$ to
    the sorts of the terms, which only make sense in a specific
    sorting context.}

  \LAF\ is also parameterised by an equivalence relation
  $\atmEq{}{}$ over $\IAtms$, which we call \Index{equality}.
\end{definition}

The equality relation is what replaces the computation of
substitutions: using the previous example, if $(p(4,\#1,\#2) , \cons
x{\cons 5 \emptylist})$ ``represents'' the atom usually written as
$p(4,x,5)$, so do the pairs $(p(4,\#2,\#1) , \cons 5{\cons x
  \emptylist})$, $(p(4,\#1) , \cons x{\emptylist})$ and
$(p(\#3,\#1,\#2) , \cons x{\cons 5 {\cons 4\emptylist}})$. The
equality relation on instantiated atoms is then used to declare all of
these pairs be equal.

Interestingly enough, this equality relation will only be used at the
leaves of proof-trees when proof-search has to compare two
instantiated atoms to close the branch. More surprisingly, there is no
need to have a similar equality relation between molecules; they are
never compared during proof-search.

\subsection{Typing decompositions and typing contexts}\label{sec:tdtc}

As we have seen, if the choice of witnesses for existential variables
is made in the decomposition relation, the asynchronous phase treats
universal variables in the style of the $\omega$-rule.  To avoid this,
the choice of witnesses cannot be made in the decomposition relation;
instead, we ``leave a hole'' and delay its filling until we inhabit
typing decompositions.

Therefore, the notion of typing decomposition itself needs to be
enriched with a new construct, denoted $\Dex[s]{}{\Delta}$, that we use
to mark a place where an existential variable of sort $s$ was found while
decomposing a molecule: For instance we can use the construct in the
decomposition relation with a rule (again, forgetting proof-terms) such as
\[
\infer{\DerDec{}{\Dex[s]{}{\Delta}}{{\exists {x^s} A}}}
      {\DerDec{}{\Delta}{{\subst A x {\#1}}}}
\]
where $\#1$ is a temporary name for the hole (you may think of it as a
De Bruijn's index), or with the equivalent rule
\[
\infer{\DerDec{}{\Dex[s]{}{\Delta}}{{\exists^s A}}}
      {\DerDec{}{\Delta}{{A}}}
\]
if De Bruijn's indices are already used in formulae to represent bound variables.

The choice of witness will then be made when proving/inhabiting $\Dex[s]{}{\Delta}$.

Correspondingly, we extend the notion of typing decompositions as
follows:

\begin{definition}[Typing decompositions]\label{def:typingdecompositions}

  The \Index[typing decomposition algebra]{typing decomposition
    algebra}, denoted $\TDecs$, is the family of sets
  $(\TDecs[l])_l$ defined by the following grammar:
  \[\Delta^l,\Delta^l_1,\ldots\recdef a^l\sep \Drefute M^l\sep\Dunit\sep\Delta^l_1\Dand\Delta^l_2\sep\Dex[s]{}{\Delta^{\cons s l}}\]
  where $\Delta^l,\Delta^l_1,\ldots$ range over $\TDecs[l]$, $a^l$
  ranges over parameterised atoms of arity $l$ and $M^l$ ranges over
  parameterised molecules of arity $l$ (and $s$ still ranges over
  $\Sorts$).

  Elements of $\TDecs[l]$ are called \Index[typing
    decomposition]{typing decompositions} of arity $l$.

  The set $\ITDecs$ of \Index[instanciated typing
    decomposition]{instanciated typing decompositions} is the set of
  pairs of the form $(\Delta^l,\lv r)$ where $\Delta^l$ is a typing
  decomposition of arity $l$ and $\lv r$ is a list of terms of length
  $\abs l$.
\end{definition}

Here, the construct $\Dex[s]{}{\Delta^{\cons s l}}$ declares a
new ``hole'' of sort $s$ so that the atoms and molecules in
$\Delta^{\cons s l}$ may now depend on this extra parameter.

But notice that typing decompositions (unless instantiated) never
mention terms; they will be used to decompose \emph{parameterised}
molecules, rather than \emph{instantiated} molecules: this is because,
intuitively, the decomposition of a molecule into a typing
decomposition only concerns the logical structure of the molecule, not
the terms that it contains.\footnote{This requires the distinction
  between the two to be clear, which will prevent us from modelling
  higher-order logic.}

Now, as in the quantifier-free case, typing decompositions will be
used to extend typing contexts, but we do expect the types, in the
typing declarations of a typing context, to be \emph{instantiated}
atoms and molecules.

This raises two questions when extending a typing context $\Gamma$
with a typing decomposition $\Delta^l$:
\begin{itemize}
\item how do the parameterised atoms and molecules of $\Delta^l$ turn
  into instantiated atoms and molecules in the extended environment?
\item how should the new construct $\Dex[s]{}{\Delta^{\cons s l}}$
  impact the extension?
\end{itemize}

To answer these questions, we anticipate that, as in the
quantifier-free case, the typing context $\Gamma$ gets extended in the
$\textsf{async}$ rule. In our setting with quantifiers, that rule will
be used to refute an \emph{instantiated} molecule. This extension thus
takes place in presence of the list of parameters ${\bf r}$ of the
molecule we are refuting, and the length of this list will match the
arity $l$ of $\Delta^l$.

Therefore, the operation that we need to extend environments is of the
form $\Gam\Textend[{\bf r}]{\Delta^l}$, hence the notion of
instanciated typing decomposition.

\begin{definition}[Typing contexts]\label{def:typingcontexts}

  \LAF\ is parameterised by two sets $\var[+]$ and $\var[-]$,
  of elements called \Index[positive label]{positive labels} and
  \Index[negative label]{negative labels}, respectively.

  \LAF\ is then parameterised by an algebra $\TContexts$ of the form
  \[
  \left(\TContexts,
       {\footnotesize
         \left(\begin{array}{c@{}l@{}l}
           \TContexts\times\var[+]&\pfunspace&\IAtms\\
           (\Gam,x^+)&\mapsto&\varRead[x^+]\Gam
         \end{array}\right),
         \left(\begin{array}{c@{}l@{}l}
           \TContexts\times\var[-]&\pfunspace&\IMoles\\
           (\Gam,x^-)&\mapsto&\varRead[x^-]\Gam
         \end{array}\right),
         \left(\begin{array}{c@{}l@{}l}
           \TContexts&\rightarrow&\SoContexts\\
           \Gam&\mapsto&\Cfune\Gam
         \end{array}\right),
         \left(\begin{array}{c@{}l@{}l}
           \TContexts\times\ITDecs&\rightarrow&\TContexts\\
           (\Gam,(\Delta^l,{\bf r}))&\mapsto&\Gam\Textend[{\bf r}]{\Delta^l}
         \end{array}\right)
       }\right)
  \]
  whose elements are called \Index[typing context]{typing contexts}.

   We denote by $\domP\Gamma$ (\resp $\domN\Gamma$) the subset of
   $\var[+]$ (\resp $\var[-]$) where $\varRead[x^+]\Gam$ (\resp
   $\varRead[x^-]\Gam$) is defined.
\end{definition}

Of course, nothing in the above definition specifies the behaviour of
the operation $\Gam\Textend[{\bf r}]{\Delta^l}$.

This will not be problematic to define the \LAF\ system, nor to define
its realisability models; but in order to \emph{build} those, it will
be easier if we also know that the typing context algebra satisfies
more specific properties. In Section~\ref{sec:LAFexbind} we give an
example of \LAF\ instance where the behaviour of $\Gam\Textend[{\bf
    r}]{\Delta^l}$ is specified.

Also, note the presence of the operation $\Cfune\Gam$ that extracts
from $\Gamma$ a sorting context, which will be used in the
\LAF\ system to constrain the pick of witnesses.

Notice in the above two definitions (\ref{def:typingdecompositions}
and~\ref{def:typingcontexts}) that, in contrast to what we did in the
quantifier-free case, we have here directly defined typing
decompositions and typing contexts instead of defining them as
particular instances of generic decompositions and generic contexts.
This is due to the need of taking into account, in those data
structures, the parameters ${\bf r}$ that are specific to atoms and
molecules and non-existant for arbitrary sets $\mathcal A$ ad
$\mathcal B$.
However, we shall still define generic decompositions and generic
contexts, as these will be used for instance to build models of \LAF,
and also more immediately to define the \emph{structure} of a typing
derivation (as we did in the quantifier-free case).

\begin{definition}[Generic decomposition algebras and decomposition structures]

  Given three set $\mathcal A$, $\mathcal B$, and $\mathcal C$, the
  $(\mathcal A,\mathcal B,\mathcal C)$-\Index{decomposition algebra}
  $\DecompType[\mathcal A,\mathcal B,\mathcal C]$, whose elements are
  called $(\mathcal A,\mathcal B,\mathcal
  C)$-\Index[decomposition]{decompositions}, is the free
  algebra defined by the following grammar:
  \[\Delta,\Delta_1,\ldots\recdef a\sep \Drefute b\sep\Dunit\sep\Delta_1\Dand\Delta_2\sep\Dex[c]{}{\Delta}\]
where $a$ (\resp $b$, $c$) ranges over $\mathcal A$ (\resp $\mathcal B$, $\mathcal C$).

  Let $\Dstruct$ abbreviate
  $\DecompType[\unitt,\unitt,\unitt]$, whose
  elements we call \Index[decomposition structure]{decomposition
    structures}.  

  The \Index[structure]{(decomposition) structure} of an $(\mathcal
  A,\mathcal B, \mathcal C)$-decomposition $\Delta$, denoted
  $\abs\Delta$, is its obvious homomorphic projection in $\Dstruct$.

  The \Index[structure]{(decomposition) structure} of a typing
  decomposition $\Delta^l$, denoted $\abs{\Delta^l}$, is defined as
  follows:
  \[\begin{array}{llll}
  \abs {a^l}&\eqdef\uniti& \abs {\Drefute {M^l}}&\eqdef\uniti\\
  \abs\Dunit&\eqdef\uniti&\abs {{\Delta^l_1}\Dand {\Delta^l_2}}&\eqdef{\abs{\Delta^l_1}}\Dand{\abs{\Delta^l_2}}\\
  \abs{\Dex[s]{}{\Delta^{\cons s l}}}&\eqdef\Dex[()]{}{\abs{\Delta^{\cons s l}}}
  \end{array}
  \]
\end{definition}

Here, we see that the typing decomposition algebra is more subtle than
the $(\Atms,\Moles,\Sorts)$-decomposition algebra, because arities are taken
into account.

Similarly, we here define generic contexts, which will be used in the next chapters.
\begin{definition}[Generic contexts]

  Given four sets $\mathcal A$, $\mathcal B$, $\mathcal C$ and $\mathcal D$, an
  \Index[context algebra]{$(\mathcal A,\mathcal B,\mathcal C,\mathcal D)$-context algebra}
  is an algebra of the form
  \[
  \left(\Contexts,
       {\footnotesize
         \left(\begin{array}{c@{}l@{}l}
           \Contexts\times\var[+]&\pfunspace&\mathcal A\\
           (\Gam,x^+)&\mapsto&\varRead[x^+]\Gam
         \end{array}\right),
         \left(\begin{array}{c@{}l@{}l}
           \Contexts\times\var[-]&\pfunspace&\mathcal B\\
           (\Gam,x^-)&\mapsto&\varRead[x^-]\Gam
         \end{array}\right),
         \left(\begin{array}{c@{}l@{}l}
           \Contexts&\rightarrow&\mathcal D\\
           \Gam&\mapsto&\Cfune\Gam
         \end{array}\right),
         \left(\begin{array}{c@{}l@{}l}
           \Contexts\times\DecompType[\mathcal A,\mathcal B,\mathcal C]&\rightarrow&\Contexts\\
           (\Gam,\Delta)&\mapsto&\Gam\Cextend\Delta
         \end{array}\right)
       }\right)
  \]
  whose elements are called $(\mathcal A,\mathcal B,\mathcal C,\mathcal D)$-\Index[context]{contexts}.

   Again, we denote by $\domP\Gamma$ (\resp $\domN\Gamma$) the subset
   of $\var[+]$ (\resp $\var[-]$) where $\varRead[x^+]\Gam$ (\resp
   $\varRead[x^-]\Gam$) is defined.
\end{definition}

\subsection{Logical connectives}

The concepts of patterns and decomposition relations are unchanged, except they rely on the enriched concepts of atoms, molecules and typing decompositions.

\begin{definition}[Patterns \& decomposition relation]\label{def:patterns1o}\nopagesplit

  \LAF\ is parameterised by a \Index[pattern]{pattern algebra}, an algebra of the form
  \[\left(\Data,
      {\footnotesize
        \left(\begin{array}{c@{}l@{}l}
            \Data&\rightarrow&\Dstruct\\
            p&\mapsto&\Datast p
          \end{array}\right)
      }
    \right)
  \] 
  whose elements are called \Index[pattern]{patterns}, and by a
  \Index{decomposition relation} (for every $l$), \ie a set of elements
  \[(\DerDec{}{\_}{\XcolY \_{\_}}):(\TDecs[l]\times\Data\times\Moles[l])\]
  such that if $\DerDec{}{\Delta}{\XcolY p{M}}$ then the structure of
  $\Delta$ is $\Datast p$.
\end{definition}

\subsection{Definition of the system}

\begin{definition}[Proof-Terms] Proof-terms are defined by the following grammar:
  \[
  \begin{array}{lll@{\recdef}l}
    \mbox{Positive terms }&\PTerms^+&t^+&pd\\
    \mbox{Decomposition terms }&\Decomp&d&x^+ \sep \THO f\sep\Tunit\sep d_1\Tand d_2 \sep \Tex r  d \\
    \mbox{Commands}&\PTerms&c& \cutc{x^-}{t^+} \sep \cutc{f}{t^+}
  \end{array}
  \]
  where $p$ ranges over $\Data$, $x^+$ ranges over $\var[+]$, $x^-$ ranges over $\var[-]$, and $f$ ranges over $\Data\pfunspace\PTerms$.
\end{definition}

We can finally present the typing system \LAF:

\begin{definition}[\LAF]

  \LAF\ is the inference system of Fig.~\ref{def:LAF} defining the derivability of three kinds of sequents
  \[
  \begin{array}{l@{\quad:\quad}l}
    (\DerF\_ {\XcolY {\_} \_}{}) & (\TContexts\times \PTerms^+\times\IMoles)\\
    (\Der\_ {\XcolY \_ \_}) & (\TContexts\times \Decomp\times\ITDecs)\\
    (\Der\_ \_) & (\TContexts\times \PTerms)
  \end{array}
  \]
  We further impose in rule \textsf{async} that the domain of function $f$ be exactly those patterns that can decompose $M$ (if $p\in\dom f$ then there exists $\Delta$ such that $\DerDec{}{\Delta}{\XcolY p{M}}$).

  \LAFcf\ is the inference system \LAF\ without the \cut-rule.
\end{definition}

\begin{bfigure}[!h]
  \[
  \begin{array}{c}
    \infer[\textsf{sync}]{\DerF\Gam {\XcolY {p d} {(M,\lv r)}}{}}{
      \DerDec{}{\Delta}{\XcolY p{M}}
      \quad
      \Der{\Gamma}{\XcolY  d{(\Delta,\lv r)}}
    }\\
    \midline
    \\
    \infer{\Der\Gam{\XcolY{\Tunit}{(\Dunit,\lv r)}}}{\strut}
    \qquad
    \infer{\Der\Gam{\XcolY{ d_1\Tand d_2}{((\Del_1\Dand\Del_2),\lv r)}}}{
      \Der\Gam{\XcolY{ d_1}{(\Del_1,\lv r)}}
      \quad
      \Der\Gam{\XcolY{ d_2}{(\Del_2,\lv r)}}
    }
    \qquad
    \infer{\Der\Gam{\XcolY{\Tex {r'} d}{\Dex[s]{} {(\Del,\lv r)}}}}{
      \Ders{\Cfune\Gam}{\XcolY {r'} s}
      \quad
      \Der\Gam{\XcolY{ d}{(\Del,\cons {r'}{\lv r})}}
    }
    \\\\
    \infer[\Init]{\Der{\Gam}{\XcolY{x^+}{(a,\lv r)}}}{
      \atmEq{\varRead[x^+]\Gam} {(a,\lv r)}
    }
    \qquad
    \infer[\textsf{async}]{\Der\Gam {\XcolY {\THO f} {(\Drefute M,\lv r)}}}{
      \forall p,\forall\Delta,\quad
      \DerDec{}{\Delta}{\XcolY p{M}}\quad\imp\quad\Der{\Gamma\Textend[{\bf r}]{\Delta}}{f(p)}
    }
    \\
    \midline
    \\
    \infer[\Select]{\Der\Gam{\cutc{x^-}{t^+}}}{
      \DerF\Gam {\XcolY{t^+}{\varRead[x^-]\Gam}} {}}
    \qquad
    \infer[\cut]{\Der\Gam{\cutc{f}{t^+}}}{
      \Der\Gam {\XcolY {f}{(\Drefute M,\lv r)}}
      \qquad
      \DerF\Gam {\XcolY{t^+}{(M,\lv r)}} {}
    }
  \end{array}
  \]
  \caption{\LAF}
  \label{def:LAF}
\end{bfigure}

\section{Extending \LAF[K1] with quantifiers}
\label{sec:LAFexbind}

In this section we extend to multi-sorted first-order logic the
example of polarised classical logic (one-sided version \LAF[K1]) in
Section~\ref{sec:LAFexprop}. Such a first-order extension could also
be done for the two-sided versions of polarised classical logic or
intuitionistic logic.

To handle quantifiers, we make a clear separation between \emph{bound
  variables} and \emph{eigenvariables}: the intuition being that in
order to prove $\forall x\ p(x)$ we prove $p(\frak x)$ for ``an
arbitrary $\frak x$'', using an eigenvariable $\frak x$.

Actually, the reasons why we used ``labels'' instead of ``variables''
in the quantifier-free system also apply to eigenvariables: both in
the perspective of an implementation and for the formalisation of such
an abstract system as \LAF, it will be convenient to have a
deterministic way to name eigenvariables with no notion of
$\alpha$-conversion or equivariance. We will therefore call them
\emph{eigenlabels}.

\LAF\ is more flexible regarding bound variables, which could be named
and subject to $\alpha$-conversion. However, already using De Bruijn's
indices to represent binding in the syntax of formulae will be
convenient since, as already mentioned in Section~\ref{sec:tdtc}, we
can simply write
\[
\infer{\DerDec{}{\Dex[s]{}{\Delta}}{{\exists^s A}}}
      {\DerDec{}{\Delta}{{A}}}
\]
instead of 
\[
\infer{\DerDec{}{\Dex[s]{}{\Delta}}{{\exists {x^s} A}}}
      {\DerDec{}{\Delta}{{\subst A x {\#1}}}}
\]
saving us the trouble of defining the substitution operation on
formulae.

\begin{definition}[Literals, formulae, patterns, decomposition]\label{def:K1struct}\nopagesplit

  Let $\Sorts$ be a set of \Index[sort]{sorts} and $\Xi$ be an $\Sorts$-signature
  in the sense of multi-sorted first-order logic.

  Predicate arities are represented as lists of sorts, denoted $l,l',\ldots$.

  Given such an arity $l$, the set of
  \Index[$l$-literal]{$l$-literals} is the set of literals over
  $\Xi$ (well-sorted atomic propositions and their negations) whose
  free variables are among $\# 1,\ldots,\#\abs l$ with (respective)
  sorts given by $l$.

  Consider a subset of the set of predicate symbols, whose elements are
  called \Index[positive predicate symbol]{positive predicate
    symbols}; predicate symbols that are not in that set are called
  \Index[negative predicate symbol]{negative}.

  Let the set $\Atms[l]$ of \Index[parameterised atom]{parameterised
    atoms of arity $l$} be the set of $l$-literals that are either of
  the form $p(t_1,\ldots,t_n)$, with $p$ being a positive predicate
  symbol, or of the form $\neg p(t_1,\ldots,t_n)$, with $p$ being a
  negative predicate symbol.

  Similarly to Definition~\ref{def:classpolsyntax}, let the set
  $\Moles[l]$ of \Index[parameterised molecule]{parameterised
    molecules of arity $l$} be the set defined by the first line of
  the following grammar for (polarised) formulae of classical logic:
  \[ 
  \begin{array}{llll}
    \mbox{Positive $l$-formulae}&  P^l,\ldots&\recdef a^l\sep\trueP \sep \falseP \sep A^l\andP B^l\sep A^l \orP B^l\sep \exists^s A^{\cons s l}\\
    \mbox{Negative $l$-formulae}& N^l,\ldots&\recdef \non{a^l}\sep\trueN \sep \falseN \sep A^l\andN B^l\sep A^l \orN B^l\sep \forall^s A^{\cons s l}\\
    \mbox{Unspecified $l$-formulae}&A^l&\recdef P^l \sep N^l
  \end{array}
  \]
  with $a^l$ ranging over $\Atms[l]$ and $\non{a^l}$ ranging over
  $l$-literals that are not in $\Atms[l]$.
  
  Similarly to Definition~\ref{def:classpolsyntax}, let negation be the
  involutive function defined as follows:
  \[ \begin{array}{c}
    \begin{array}{ll}
      \non {(p(t_1,\ldots,t_n))}&\eqdef \neg p(t_1,\ldots,t_n)\\
      \non {(\neg p(t_1,\ldots,t_n))}&\eqdef p(t_1,\ldots,t_n)
    \end{array}\\
    \begin{array}{ll@{\qquad}ll}
      \non{\trueP}&\eqdef\falseN&\non{\trueN}&\eqdef\falseP\\
      \non{\falseP}&\eqdef\trueN&\non{\falseN}&\eqdef\trueP\\
      \non{(A\andP B)}&\eqdef\non A\orN \non B&\non{(A\andN B)}&\eqdef\non A\orP \non B\\
      \non{(A\orP B)}&\eqdef\non A\andN \non B&\non{(A\orN B)}&\eqdef\non A\andP \non B
    \end{array}
  \end{array}
  \]
  and we extend it to sets or multisets of formulae pointwise.

  The set $\Data$ of \Index{patterns} extends that of
  Definition~\ref{def:classpolsyntax} according to the following
  grammar:
  \[p,p_1,p_2,\ldots\recdef \Ppos \sep \Pneg \sep \Ptrue\sep\paire {p_1}{p_2}\sep\inj i p\sep \pexists p\]

  The decomposition relation $(\DerDec{}{\_}{\XcolY
    \_{\_}}):(\DecompType\times\Data\times\Moles)$ is the extension of
  that of Definition~\ref{def:classpolsyntax}, as shown in
  Fig.~\ref{fig:decompK1wQ}.
\end{definition}
\begin{bfigure}[!h]
    \[
    \begin{array}c
      \infer{\DerDec{}{\Dunit}{\XcolY{\Ptrue}\trueP}}{}
      \qquad
      \infer{\DerDec{}{\Drefute {\non N}}{\XcolY{\Pneg}{N}}}{}
      \qquad
      \infer{\DerDec{}{a}{\XcolY{\Ppos}{a}}}{}
      \\\\
      \infer{\DerDec{}{\Delta_1,\Delta_2}{\XcolY{\paire{p_1}{p_2}}A_1\andP A_2}}{
        \DerDec{}{\Delta_1}{\XcolY{p_1}A_1}
        \quad
        \DerDec{}{\Delta_2}{\XcolY{p_2}A_2}
      }
      \qquad
      \infer{\DerDec{}{\Delta}{\XcolY{\inj i p}A_1\orP A_2}}{\DerDec{}{\Delta}{\XcolY{p}A_i}}  
      \qquad
      \infer{\DerDec{}{\Dex[s]{}\Delta}{\XcolY{\pexists p}{\exists^s A}}}
            {\DerDec{}{\Delta}{\XcolY{p}{A}}}
    \end{array}
    \]
  \caption{Decomposition relation for \LAF[K1]}
  \label{fig:decompK1wQ}
\end{bfigure}

Several concepts are still missing to define an instance of \LAF: we
need to define the set $\Terms$ of terms, the set $\SoContexts$ of
sorting contexts, the sorting relation $\decs{}$, the equality
relation $\atmEq{}{}$ on instantiated atoms, and the typing context
algebra $\TContexts$.

\begin{definition}[Terms, sorting and equality]\label{def:K1terms}\nopagesplit

  Let $\vare$ be a copy of the set of natural numbers, whose elements
  are called \Index[eigenlabel]{eigenlabels} and denoted $n^e$,
  $n^e_1$, \ldots

  The set $\Terms$ of \Index[term]{terms}, denoted $r,r',\ldots$, is
  defined as the set of first-order terms whose variables are
  eigenlabels and whose function symbols are those declared in the
  signature $\Xi$.

  The set $\SoContexts$ of \Index[sorting context]{sorting contexts},
  denoted $\Sigma,\Sigma',\ldots$, is $\vare\pfunspace\Sorts$.

  We write $\Ders{\Sigma}{\XcolY r s}$ when the term $r$ is of sort
  $s$ in the sorting context $\Sigma$, according to the signature
  $\Xi$.

  We define the equality relation as follows: $\atmEq{(a^l,{\bf
      r})}{(a'^{l'},{\bf r'})}$ if the literal
  $\subst{a^l}{\#1,\ldots,\#\abs{l}}{{\bf r}}$\ \footnote{\ie the
    substitution of ${\bf r}$ for $\#1,\ldots,\#\abs{l}$ in $a^l$} is
  syntactically equal to the literal
  $\subst{a'^{l'}}{\#1,\ldots,\#\abs{l'}}{{\bf r'}}$.\footnote{\ie the
    substitution of ${\bf r'}$ for $\#1,\ldots,\#\abs{l'}$ in
    $a'^{l'}$}
\end{definition}

The last task is to define the typing context algebra $\TContexts$. We
do this by adapting Definitions~\ref{def:DBlabels}
and~\ref{def:DBlabelslevels}. We will use De Bruijn's levels for
eigenlabels, because as explained in Section~\ref{sec:LAFlabels}, De
Bruijn's levels do not need to be updated once they are introduced (in
contrast to De Bruijn's indices).

\begin{definition}[Typing context algebra]\label{def:TcontextsK1wQ}

  We define the support set of $\TContexts$ as the set of triples
  $(\Gamma^+,\Gamma^-,\Gamma^e)$ where $\Gamma^+$ is a list of
  elements of $\IAtms$, $\Gamma^-$ is a list of elements of $\IMoles$,
  and $\Gamma^e$ is a list of elements of $\Terms$.

  As in Definition~\ref{def:DBlabels}, two disjoint copies $\var[+]$
  and $\var[-]$ of the set of natural numbers are used for positive
  labels and negative labels, respectively denoted $n^+$ and
  $n^-$, and we define\\ 
  $\varRead[n^+]{(\Gamma^+,\Gamma^-,\Gamma^e)}$ as the $(\abs{\Gamma^+}-n)^{th}$
  element of $\Gamma^+$\\ 
  $\varRead[n^-]{(\Gamma^+,\Gamma^-,\Gamma^e)}$ as the $(\abs{\Gamma^-}-n)^{th}$
  element of $\Gamma^-$.

  We now also define $\varRead[n^e]{(\Gamma^+,\Gamma^-,\Gamma^e)}$ as the
  $(\abs{\Gamma^e}-n)^{th}$ element of $\Gamma^e$, for an eigenlabel $n^e$.

  We turn the resulting structure
  \[
  \left(\TContexts,
       {\footnotesize
         \left(\begin{array}{c@{}l@{}l}
           \TContexts\times\var[+]&\pfunspace&\IAtms\\
           (\Gam,n^+)&\mapsto&\varRead[n^+]\Gam
         \end{array}\right),
         \left(\begin{array}{c@{}l@{}l}
           \TContexts\times\var[-]&\pfunspace&\IMoles\\
           (\Gam,n^-)&\mapsto&\varRead[n^-]\Gam
         \end{array}\right),
         \left(\begin{array}{c@{}l@{}l}
           \TContexts&\rightarrow&(\vare\pfunspace\Terms)\\
           \Gam&\mapsto&(n^e\mapsto\varRead[n^e]\Gam)
         \end{array}\right)
       }\right)
       \]
       into a typing context algebra, by adding a notion of typing context extension
       \[
       \left(\begin{array}{c@{}l@{}l}
         \TContexts\times\ITDecs&\rightarrow&\TContexts\\
         (\Gam,(\Delta^l,{\bf r}))&\mapsto&\Gam\Textend[{\bf r}]{\Delta^l}
       \end{array}\right)
       \]
       which we define as follows:
  \[
  \begin{array}{llll}
    (\Gamma^+,\Gamma^-,\Gamma^e)\Cextend {(a^l,{\bf r})}
    &\eqdef(\cons {(a^l,{\bf r})} {\Gamma^+},\Gamma^-,\Gamma^e)\\  
    (\Gamma^+,\Gamma^-,\Gamma^e)\Cextend {(\Drefute {M^l},{\bf r})} 
    &\eqdef(\Gamma^+,\cons {(M^l,{\bf r})} {\Gamma^-},\Gamma^e)\\
    (\Gamma^+,\Gamma^-,\Gamma^e)\Cextend \Dunit &\eqdef(\Gamma^+,\Gamma^-,\Gamma^e)\\
    (\Gamma^+,\Gamma^-,\Gamma^e)\Cextend((\Delta^l_1\Dand\Delta^l_2),{\bf r})
    &\eqdef(\Gamma^+,\Gamma^-,\Gamma^e)\Cextend{(\Delta^l_1,{\bf r})}\Cextend{(\Delta^l_2,{\bf r})}\\
    (\Gamma^+,\Gamma^-,\Gamma^e)\Cextend {(\Dex[s]{}{\Delta^{\cons s l}},{\bf r})} 
    &\eqdef(\Gamma^+,\Gamma^-,\cons s {\Gamma^e})\Cextend{(\Delta^{\cons s l},\cons{\abs{\Gamma^e}^e}{\bf r})}
  \end{array}
  \]
\end{definition}

The operation of typing context extension adapts to the presence of
quantifiers the operation defined in Definition~\ref{def:DBlabels} for
the quantifer-free case.

The only difference is the third component $\Gamma^e$ of the typing
context, which records the declared eigenlabels together with their
sorts. This sorting context is extended whenever the typing context is
extended with an instantiated typing decomposition of the form
$(\Dex[s]{}{\Delta^{\cons s l}},{\bf r})$, which creates a new
eigenlabel of sort $s$, which becomes the new head of the sorting
context. As we use De Bruijn's levels rather than De Brujn's indices,
the new eigenlabel is therefore $\abs{\Gamma^e}^e$ (and it gets added
to the current list of terms). This can be seen as picking the ``first
available name'' for the eigenlabel to be created, a process that is
often used in implementations of such systems.

With De Bruijn's indices rather than De Bruijn's levels, the new
eigenlabel would be $0^e$, but the price to pay for this is that the
previously declared eigenlabels have ``changed names'', \ie would be
referred to as $(n+1)^e$ instead of $n^e$. Every structure referring
to those previously declared eigenlabels (namely, the instanciated
atoms and molecules in $\Gamma^+$ and $\Gamma^-$, as well as ${\bf r}$
itself) would then need to be updated with the name change.

\bigskip

We could easily define variants of the above system to quantify over
other objects than first-order terms, as most of the definitions are
rather modular: For example we could quantify over simply-typed
$\lambda$-terms to design a \LAF\ instance similar to the type theory
$\lambda\Pi$ (see \eg\cite{Bar:intgts}), except our proof-terms do not
have the same syntax and and typing properties as the $\lambda$-terms
we quantify over.

For this we take the same definitions as for multi-sorted first-order
logic, except in Definition~\ref{def:K1struct} we take $\Sorts$ be the
set of \Index[simple type]{simple types}, and in
Definition~\ref{def:K1terms} we take terms to be $\lambda$-terms, we
take $\Ders{\Sigma}{\XcolY r s}$ to be the typing relation of the
simply-typed $\lambda$-calculus, and we define atom equality with
$\beta$- (or $\beta\eta$-) conversion instead of syntactic equality:
$\atmEq{(a^l,{\bf r})}{(a'^{l'},{\bf r'})}$ if $(\lambda\#1\ldots
\#{\abs{l}}.a^l)\ {\bf r}\quad \Rewsn{\beta}\quad (\l \#1\ldots
\#{\abs{l'}}.b'^{l'})\ {\bf r'}$ (or similarly with $\beta\eta$).

All of the other definitions are the same.

\chapter{Realisability models of abstract focussing}
\label{ch:real}

\minitoc 

In this chapter we investigate the semantics of \LAF. More precisely,
we investigate models of proofs / typing derivations with the
\emph{Adequacy Lemma} as the main objective: In very generic terms, if
$t$ is of type $A$ then in the model we want the interpretation of $t$
to be in the interpretation of $A$ (if that is a set, or we want the
interpretation of $t$ to satisfy the interpretation of $A$, if that is
a predicate).

Of course there are many models satisfying the above, starting with
the uninformative ones where everything is
collapsed.\footnote{Interpret every type by the same singleton set and
  every inhabitant of that type by the inhabitant of the singleton
  set, and the Adequacy Lemma trivially holds but does not provide any
  useful information.} So we investigate here a class of models, as
large as possible, and prove the Adequacy Lemma generically for that
class; then we will show interesting models in that class for which
the Adequacy Lemma (that we now have for free) is informative (\eg
concludes the consistency of \LAF, despite the presence of cuts and
without proving cut-elimination).

This class of models is that of \emph{abstract realisability
  algebras}; the specifications that we require of such an algebra do
depend on the instance of \LAF\ that we want to model - they will be
different if we are to model for instance \LAF[K1], \LAF[K2], or
\LAF[J]; but we can give those specifications parametrically and prove
the Adequacy Lemma generically.

Hence in this chapter we start by assuming we are given an instance of \LAF
\[(\Sorts,\Terms,\SoContexts,\decs{},\Atms,\Moles,\atmEq{}{},\var[+],\var[-],\TContexts,\Data,\decf{})\]

Section~\ref{sec:RealTerms} gives the specifications needed to
interpret terms, Section~\ref{sec:RealTypes} gives the specifications
needed to interpret types and proves the Adequacy Lemma. Finally,
Section~\ref{sec:RealConsistency} exhibits a simple model from which
we derive the consistency of \LAF.

\section{Model structures and the interpretation of proof-terms}
\label{sec:RealTerms}

In this section we interpret the proof-terms of \LAF\ in a
realisability algebra, and for this we introduce the notion of \emph{model structure}.

\begin{definition}[Model structure]

  A \Index{model structure} is
  an algebra of the form
  \[
  \begin{array}l
  \left(
  \STerms,\SSoContexts,\SPrim,\SPos,\SNeg, \orth \ \ , \SContexts,\right.\\
       \hspace{60pt}
       \left.
       {\footnotesize
         \left(\begin{array}{c@{}l@{}l}
           \Data&\rightarrow&(\DecompType[\SPrim,\SNeg,\STerms]\rightarrow\SPos)\\
           p&\mapsto&\spat p
         \end{array}\right)
       },
       {\footnotesize
         \left(\begin{array}{c@{}l@{}l}
           \Terms\times\SSoContexts&\pfunspace&\STerms\\
           (r,\sigma)&\mapsto&\sem[\sigma] r
         \end{array}\right)
       },

       {\footnotesize
         \left(\begin{array}{c@{}l@{}l}
           (\Data\pfunspace\PTerms)\times\SContexts&\pfunspace&\SNeg\\
           (f,\rho)&\mapsto&\sem[\rho] f
         \end{array}\right)
       }       
  \right)
  \end{array}
  \]
where
\begin{itemize}
\item $\STerms,\SSoContexts,\SPrim,\SPos,\SNeg$ are five arbitrary sets of elements called
  \Index[term denotation]{term denotations},
  \Index[valuation]{valuations},
  \Index[label denotation]{label denotations},
  \Index[positive denotation]{positive denotations},
  \Index[negative denotation]{negative denotations},
  respectively;
\item $\orth \ \ $ is a relation between negative and positive
  denotations ($\orth \ \ \subseteq \SNeg\times\SPos$), called the
  \Index{orthogonality relation};
\item $\SContexts$ is a $(\SPrim,\SNeg,\STerms,\SSoContexts)$-context algebra,
  whose elements, denoted $\rho,\rho',\ldots$, are called
  \Index[semantic context]{semantic contexts}.
\end{itemize}

We extend the notation $\sem[\sigma] r$ to apply to a list of terms ${\bf r}$: $\sem[\sigma] {\bf r}$, using the standard map function on lists.

The $(\SPrim,\SNeg,\STerms)$-decomposition algebra $\DecompType[\SPrim,\SNeg,\STerms]$ is abbreviated $\SDecs$; its elements, denoted $\frak\Delta$, $\frak\Delta'$\ldots, are called \Index[semantic decomposition]{semantic decompositions}.
\end{definition}

Given a model structure, we can define the interpretation of
proof-terms. The model structure already gives an interpretation for
the partial functions $f$ from patterns to commands. We extend it to all proof-terms as follows

\begin{definition}[Interpretation of proof-terms]

  Positive terms (in $\PTerms^+$) are interpreted as positive
  denotations (in $\SPos$),\\ decomposition terms (in $\Decomp$) are
  interpreted as semantic decompositions (in
  $\SDecs$),\\ and commands (in $\PTerms$)
  are interpreted as pairs in $\SNeg\times\SPos$ (that may or may not
  be orthogonal), according to the following definition:

  \[
  \begin{array}{ll@{\qquad}ll@{\qquad}ll}
    \sem[\rho]{pd}&\eqdef\spat p(\sem[\rho]d)
    &\sem[\rho]{x^+}&\eqdef\varRead[x^+]\rho
    &\sem[\rho]{\cutc{x^-}{t^+}}&\eqdef(\varRead[x^-]\rho,\sem[\rho]{t^+})\\
    &&\sem[\rho]{\THO f}&\eqdef\sem[\rho]{\THO f}\footnote{as given by the model structure}
    &\sem[\rho]{\cutc{f}{t^+}}&\eqdef(\sem[\rho]{\THO f},\sem[\rho]{t^+})\\
    &&\sem[\rho]\Tunit&\eqdef\Dunit\\
    &&\sem[\rho]{d_1\Tand d_2}&\eqdef\sem[\rho]{d_1}\Dand\sem[\rho]{d_2}\\
    &&\sem[\rho]{\Tex r  d}&\eqdef\Dex[{\sem[\Cfune\rho]r}]{}{\sem[\rho]d}
  \end{array}
  \]
\end{definition}

\ssection{Realisability algebras, interpretation of types \& Adequacy}
\label{sec:RealTypes}
\label{sec:RealAdequacy}

Again, let us keep in mind the Adequacy Lemma: if $t$ is of type $A$
then the interpretation of $t$ to be in the interpretation of $A$.  We
have already defined the interpretation of proof-terms in a model
structure.  We now proceed to define the interpretation of types.

In system \LAF, there are three concepts of ``type inhabitation'' for
atoms and molecules:
\begin{itemize}
\item ``proving'' an atom by finding a suitable positive label in the
  typing context (the inhabitant is in $\var[+]$);
\item ``proving'' a molecule by choosing a way to decompose it into a
  typing decomposition (the inhabitant is in $\PTerms^+$);
\item ``refuting'' a molecule by case analysing all the possible ways
  of decomposing it into a typing decomposition (the inhabitant is in
  $\Data\pfunspace\PTerms$).
\end{itemize}

As positive labels are interpreted in $\SPrim$, positive proof-terms
are interpreted in $\SPos$ and functions in $\Data\pfunspace\PTerms$
are interpreted in $\SNeg$, we will correspondingly
\begin{itemize}
\item have an interpretation of every atom as a particular subset of $\SPrim$;
\item have a \emph{positive} interpretation of every molecule as a particular subset of $\SPos$;
\item have a \emph{negative} interpretation of every molecule as a particular subset of $\SNegV$.
\end{itemize}

To make sure that we capture, in our notion of abstract realisability
algebra, a wide class of models, the first of the three above
interpretations will be left as a parameter; we barely impose any
specification on this parameter.  The other two, however, will be
\emph{defined} notions.

Also, we have in \LAF\ a notion of \emph{sorting} for terms, whose
counter-part in an abstract realisability algebra is also left as a
parameter to be fixed \emph{ad libitum}. This leads to the following
definition:

\begin{definition}[Realisability algebra]

  A \Index{realisability algebra} is
  \begin{itemize}
  \item a model structure
  \item together with three functions
    \[
      {\footnotesize
        \left(\begin{array}{c@{}l@{}l}
          \Sorts&\rightarrow&\powerset{\STerms}\\
          s&\mapsto&\sem{s}
        \end{array}\right)
      },
      \qquad
      {\footnotesize
        \left(\begin{array}{c@{}l@{}l}
          \SoContexts&\rightarrow&\powerset{\SSoContexts}\\
          \Sigma&\mapsto&\sem{\Sigma}
        \end{array}\right)
      },
      \qquad
      {\footnotesize
        \left(\begin{array}{c@{}l@{}l}
          \Atms[l]&\rightarrow&(\STerms^{\abs l}\rightarrow\powerset{\SPrim})\\
          a^l&\mapsto&\sem{a^l}
        \end{array}\right)
      }
      \]
      satisfying
      \begin{itemize}
      \item if $\Ders{\Sigma}{\XcolY r s}$ and $\sigma\in\sem\Sigma$ then ($\sem[\sigma]r$ is defined and) $\sem[\sigma]r\in\sem s$;
        \item if $\atmEq{(a,{\bf r})}{(a',{\bf r'})}$
      then for all $\sigma:\SSoContexts$ we have $\sem
      a(\sem[\sigma]{\bf r})=\sem{a'}(\sem[\sigma]{\bf
        {r'}})$.\footnote{if both sides are defined}
  \end{itemize}
  \end{itemize}
\end{definition}

Now notice in \LAF\ that the derivability of the typing judgements for
atoms and molecules is defined inductively together with the
derivability of a typing judgement for typing decompositions; inhabitants
of those are decomposition terms.

Therefore, we will also define an interpretation for typing
decompositions, as subsets of $\DecompType[\SPrim,\SNeg,\STerms]$. For
this we need to specify how to ``lift relations to typing decompositions'':

\begin{definition}[Lifting relations]
  Given
  \begin{itemize}
  \item two relations $\mathcal
    R_1\subseteq\Atms[l]\times\STerms^{\abs l}\times\SPrim$ and $\mathcal
    R_2\subseteq\Moles[l]\times\STerms^{\abs l}\times\SNeg$ (for every
    arity $l$)
  \item a relation $\mathcal
    R_3\subseteq\Sorts\times\STerms$,
  \item an arity $l$ and a list of term denotations $\frak{rl}$ of length $\abs l$,
  \item a typing decomposition $\Delta^l$ of arity $l$ and a semantic decomposition $\mathfrak\Delta$
  \end{itemize}
  we say that \Index[relates\ldots
    according\ldots]{$\Delta^l$ ${\mathfrak{rl}}$-relates to $\mathfrak\Delta$
    according to $\mathcal R_1$, $\mathcal R_2$ and $\mathcal R_3$} if
  the relation $(\Delta^l,{\mathfrak{rl}})\ \mathcal R\ \mathfrak\Delta$ can be
  derived by the following rules:

  \[\begin{array}c
  \infer{(a^l,{\mathfrak{rl}})\ \mathcal R\ \mathfrak l}{(a^l,\mathfrak{rl})\ \mathcal R_1\ \mathfrak l}
  \qquad
  \infer{(\Drefute M^l,{\mathfrak{rl}})\ \mathcal R\ \mathfrak n}{(M^l,\mathfrak{rl})\ \mathcal R_2\ \mathfrak n}
  \qquad
  \infer{(\Dunit,{\mathfrak{rl}})\ \mathcal R\ \Dunit}{\strut}\\\\
  \infer{((\Delta^l_1\Dand\Delta^l_2),{\mathfrak{rl}})\ \mathcal R\ \mathfrak\Delta_1\Dand\mathfrak\Delta_2}
        {(\Delta^l_1,{\mathfrak{rl}})\ \mathcal R\ \mathfrak\Delta_1\qquad(\Delta^l_2,{\mathfrak{rl}})\ \mathcal R\ \mathfrak\Delta_2}
  \qquad
  \infer{(\Dex[s]{}{\Delta^{\cons s l}},{\mathfrak{rl}})\ \mathcal R\ \Dex[\mathfrak r]{}{\Delta'}}
        {s\ \mathcal R_3\ \mathfrak r\qquad(\Delta^{\cons s l},\cons{\mathfrak r}{\mathfrak{rl}})\ \mathcal R\ \mathfrak{\Delta}}
  \end{array}
        \]
\end{definition}
Obviously in that case $\Delta^l$ and $\frak\Delta$ have the same decomposition structure.

The interpretation of types will be defined by simultaneous induction on molecules and typing decompositions. This induction needs to follow a well-founded relation:

\begin{definition}[Well-founded \LAF\ instance]

  We write $M^{l'}\MolesIneq M^l$ if there are $\Delta^l$ and $p$
  such that $\DerDec{}{\Delta^l} {\XcolY p {M^l}}$ and $M^{l'}$ is a
  leaf of $\Delta^l$.

  The \LAF\ instance is \Index[well-founded
    \LAF\ instance]{well-founded} if $\MolesIneq$ is well-founded.
\end{definition}

It could be the case that the \LAF\ instance is not well-founded, \eg
if molecules contain fixpoints.

\begin{warning}
  In the rest of this chapter, we will assume \LAF\ instances to be well-founded.
\end{warning}

Under this assumption, the following interpretations of types are well-defined:

\begin{definition}[Interpretation of types and typing contexts]
  A realisability algebra already provides
  the interpretation of a parameterised atom of arity $l$ in $(\STerms^{\abs l}\rightarrow\powerset{\SPrim})$.

  We now define\\ 
  the positive interpretation of a parameterised molecule of arity $l$ in $(\STerms^{\abs l}\rightarrow\powerset{\SPos})$;\\ 
  the negative interpretation of a parameterised molecule of arity $l$ in $(\STerms^{\abs l}\rightarrow\powerset{\SNeg})$;\\ 
  the interpretation of a typing decomposition of arity $l$ is in $(\STerms^{\abs l}\rightarrow\powerset{\DecompType[\SPrim,\SNeg,\STerms]})$:
  \[\begin{array}{llll}
  \SemTyP {M^l} (\frak{rl}) &\eqdef \{ \spat p (\frak \Delta)\in\SPos
  &\mid \frak\Delta\in\SemTy{\Delta^l}(\frak{rl}),\mbox{ and } \DerDec{}{\Delta^l} {\XcolY p {M^l}}\}\\
  \SemTyN {M^l} (\frak{rl}) &\eqdef \{ \frak n\in \SNeg
  &\mid \forall \frak p \in\SemTyP M (\frak{rl}), \orth{\frak n}{\frak p}\}\\
  \SemTy {\Delta^l} (\frak{rl})&\eqdef\{\frak \Delta\in\SDecs
  &\mid \mbox{$\Delta^l$ ${\mathfrak{rl}}$-relates to $\mathfrak\Delta$
    according to $\{(a^l,\frak {rl},\frak l)\mid\frak l\in\SemTy{a^l}(\frak{rl})\}$}\\
    &&\hfill \{(M^l,\frak {rl},\frak n)\mid\frak n\in\SemTyN{M^l}(\frak{rl})\}\\
  &&\hfill\mbox{and }\{(s,\frak r)\mid\frak r\in\sem s\}&\}
  \end{array}
  \]

  We now define the semantics of instanciated atoms, molecules and typing decompositions:
  \[\begin{array}{ll@{\qquad}ll}
  \SemTy[\sigma]{(a^l,{\bf r})}&\eqdef\SemTy {a^l}(\sem[\sigma]{\bf r})
  &\SemTyP[\sigma]{(M^l,{\bf r})}&\eqdef\SemTyP {M^l}(\sem[\sigma]{\bf r})\\
  \SemTy[\sigma]{(\Delta^l,{\bf r})}&\eqdef\SemTy {\Delta^l}(\sem[\sigma]{\bf r})
  &\SemTyN[\sigma]{(M^l,{\bf r})}&\eqdef\SemTyN {M^l}(\sem[\sigma]{\bf r})
  \end{array}\]

  We finally define the interpretation of a typing
  context:\footnote{In this definition we implicitly require that
    $\domP\rho=\domP\Gamma$, $\domN\rho=\domN\Gamma$ and for all
    $x^+\in\domP\rho$ (\resp $x^-\in\domN\rho$)
    $\SemTy[\Cfune\rho]{\varRead[x^+]\Gamma}$ (\resp
    $\SemTyN[\Cfune\rho]{\varRead[x^-]\Gamma}$) is defined.}
  \[\begin{array}{llll}
  \SemTy\Gamma\eqdef\{\rho\in\SContexts\mid
  &\Cfune\rho\in\sem{\Cfune\Gamma}\\
  &\forall x^+\in\domP\rho,\ \varRead[x^+]\rho\in\SemTy[\Cfune\rho]{\varRead[x^+]\Gamma}\\
  &\forall x^-\in\domN\rho,\ \varRead[x^-]\rho\in\SemTyN[\Cfune\rho]{\varRead[x^-]\Gamma}
  &\}
  \end{array}\]
\end{definition}

Now that we have defined the interpretation of terms and the interpretation of types, we prove the Adequacy Lemma.

\begin{lemma}[Adequacy for \LAF]\label{lem:adequacy}
  We assume the following hypotheses:
  \begin{enumerate}
  \item[Well-foundedness:]\\
    The \LAF\ instance is well-founded.
  \item[Typing correlation:]\\
    If $\rho\in\SemTy\Gamma$ and $\frak \Delta\in\SemTy[\Cfune\rho]{(\Delta^l,{\bf r})}$ then $(\rho\Cextend\frak\Delta)\in\SemTy{\Gamma\Textend{\Delta^l}}$.
  \item[Stability:]\\
    If $\frak d\in\SemTy[\sigma]{(\Delta^l,{\bf r})}$ for some $\Delta^l,\sigma,{\bf r}$ and $\SemTe{f(p)}{\rho\Cextend \frak d}\in\orth{}{}$, then $\orth{\sem[\rho] f}{\spat p(\frak d)}$.
  \end{enumerate}

  We conclude that, for all $\rho\in\SemTy\Gamma$,
  \begin{enumerate}
  \item if $\DerF\Gamma {\XcolY {t^+} {(M^l,{\bf r})}}{}$ then $\SemTe{t^+}\rho\in\SemTyP {(M^l,{\bf r})}$; 
  \item if $\Der\Gamma {\XcolY {d} {(\Delta^l,{\bf r})}}{}$ then $\SemTe{d}\rho\in\SemTy {(\Delta^l,{\bf r})}$; 
  \item if $\Der\Gamma {{t} {}}{}$ then $\SemTe{t}\rho\in\orth{}{}$.
  \end{enumerate}
\end{lemma}
\begin{proof}See the proof in \Coq~\cite{LengrandHDRCoq}.
\end{proof}

\section{A more concrete class of \LAF\ instances}
\label{sec:RealConcrete}

Looking at the Adequacy Lemma, the stability condition is traditional:
it is the generalisation, to that level of abstraction, of the usual
condition on the orthogonality relation in orthogonality models (those
realisability models that are defined in terms of orthogonality,
usually to model classical
proofs~\cite{girard-ll,DanosKrivine00,Krivine01,MunchCSL09,LM:APAL07}):
orthogonality is ``closed under anti-reduction''. Here, we have not
yet defined a notion of reduction on \LAF\ proof-terms, but
intuitively, we would expect, in order to reduce cuts, to rewrite
$\cutc{\THO f}{pd}$ to $f(p)$ ``substituted by $d$''.

On the other hand, the typing correlation property is new, and is due
to the level of abstraction we operate at: there is no reason why our
data structure for typing contexts would relate to our data structure
for semantic contexts, and the extension operation, in both of them,
has so far been completely unspecified. Hence, we clearly need such an
assumption to relate the two.

However, one may wonder when and why the typing correlation property
should be satisfied. Looking at the example of \LAF[K1], one may
anticipate how typing correlation could hold for this instance of
\LAF: at least in the quantifier-free case (Sections~\ref{sec:LAFK1qf}
and~\ref{sec:LAFlabelsK}), we have a generic definition of $(\mathcal
A,\mathcal B)$-contexts with a parametric operation of extension,
which we can use for both typing contexts and semantic contexts.

In this section we generalise this example to a class of \LAF\ systems
(and later we identify a corresponding subclass of realisability
algebras), where Adequacy holds under a hypothesis that simplifies
(and entails) typing correlation, and that is satisfied in particular
when the extension of contexts is defined parametrically.

\subsection{\LAF\ instances with eigenlabels}

In this class, the extension of a typing context
$\Gamma\Textend{\Delta^l}$ is expressed in terms of the extension
operation $\Gamma'\Cextend\Delta'$ of an
$(\IAtms,\IMoles,\Sorts,\SoContexts)$-context algebra.

This is done along the same lines as in the example of \LAF[K1] in
Section~\ref{sec:LAFexbind}, \ie with a notion of eigenlabel and the
understanding of sorting contexts (in $\SoContexts$) a functions
mapping eigenlabels to sorts. Hence, we update and refine previous
definitions (and introduce new ones) with this understanding of
sorting contexts.

\begin{definition}[Three-parameter contexts]

  Assume we have three disjoint sets $\var[+]$, $\var[-]$ and $\vare$,
  the union of which ($\var[+]\cup\var[-]\cup\vare$) we denote $\var$.

  Given three sets $\mathcal A$, $\mathcal B$, $\mathcal C$, we
  abreviate the terminology $(\mathcal A,\mathcal B,\mathcal
  C,\vare\pfunspace\mathcal C)$-context into $(\mathcal A,\mathcal
  B,\mathcal C)$-context.

  We also abbreviate $\Cfune \Gamma(x)$ as $\varRead[x]\Gamma$, for an
  $(\mathcal A,\mathcal B,\mathcal C)$-context $\Gamma$ and an
  element $x\in\vare$, writing $\domE\Gamma$ for $\dom{\Cfune\Gamma}$.
  Finally, we abbreviate $\domP\Gamma\cup\domN\Gamma\cup\domE\Gamma$
  as $\domAll\Gamma$, and we say that $\Gamma$ is \Index[empty
    context]{empty} if $\domAll{\Gamma}=\emptyset$.
\end{definition}

We now introduce the lifting of relations to generic decompositions
and contexts:

\begin{definition}[Lifting relations]

  Assume we are given three relations $\mathcal R_1\subseteq\mathcal
  A\times\mathcal A'$, $\mathcal R_2\subseteq\mathcal B\times\mathcal
  B'$ and $\mathcal R_3\subseteq\mathcal C\times\mathcal C'$.

  We say that \Index[relates\ldots according\ldots]{an $(\mathcal
    A,\mathcal B,\mathcal C)$-decomposition $\Delta$ relates to an
    $(\mathcal A',\mathcal B',\mathcal C')$-decomposition $\Delta'$
    according to $\mathcal R_1$, $\mathcal R_2$ and $\mathcal R_3$} if
  the relation $\Delta\ \mathcal R\ \Delta'$ can be derived by the
  following rules:
  \[
  \infer{a\ \mathcal R\ a'}{a\ \mathcal R_1\ a'}
  \quad
  \infer{b\ \mathcal R\ b'}{b\ \mathcal R_2\ b'}
  \quad
  \infer{\Dunit\ \mathcal R\ \Dunit}{\strut}
  \quad
  \infer{\Delta_1\Dand\Delta_2\ \mathcal R\ \Delta'_1\Dand\Delta'_2}
        {\Delta_1\ \mathcal R\ \Delta'_1\qquad\Delta_2\ \mathcal R\ \Delta'_2}
        \quad
        \infer{\Dex[c]{}{\Delta}\ \mathcal R\ \Dex[c']{}{\Delta'}}
              {c\ \mathcal R_3\ c'\qquad\Delta\ \mathcal R\ \Delta'}
  \]

  We say that \Index[relates\ldots according\ldots]{an $(\mathcal A,\mathcal B,\mathcal C)$-context $\Gamma$ relates
    to an $(\mathcal A',\mathcal B',\mathcal C')$-context $\Gamma'$ according to $\mathcal R_1$, $\mathcal R_2$ and
    $\mathcal R_3$} if\footnote{By writing the three conditions we
    implicitly request $\domP{\Gamma}=\domP{\Gamma'}$, $\domN{\Gamma}=\domN{\Gamma'}$ and $\domE{\Gamma}=\domE{\Gamma'}$.}
  \[\begin{array}{lc}
  \forall x^+\in\var[+],& \varRead[x^+]{\Gamma}\ \mathcal R_1\  \varRead[x^+]{\Gamma'}\\
  \forall x^-\in\var[-],& \varRead[x^-]{\Gamma}\ \mathcal R_2\  \varRead[x^-]{\Gamma'}\\
  \forall x\in\vare,& \varRead[x]{\Gamma}\ \mathcal R_3\  \varRead[x]{\Gamma'}
  \end{array}\]

  Assume we are now given three functions $f_1:\mathcal
  A\rightarrow \mathcal A'$, $f_2:\mathcal
  B\rightarrow \mathcal B'$ and $f_3:\mathcal
  C\rightarrow \mathcal C'$.\\
  we say that \Index[map]{$\Gamma'$ is a map of $\Gamma$ according to $f_1$, $f_2$ and
    $f_3$} if it relates to $\Gamma$ according to the relations
  $\{(f_1(a),a)\mid a\in\mathcal A\}$, $\{(f_2(b),b)\mid b\in\mathcal B\}$ and $\{(f_3(c),c)\mid c\in\mathcal C\}$.
\end{definition}

Using the above two definition, we can now say what a \LAF\ instance with eigenlabels is:

\begin{definition}[\LAF\ instance with eigenlabels]\label{def:LAFwE}

  A \LAF\ \Index{instance with eigenlabels} is given by the following tuple:
  \[(\Sorts,\vare,\Terms,\decs{},\Atms,\Moles,\atmEq{}{},\var[+],\var[-],\TContexts,\Data,\decf{},\Crename{\mathcal V}{\Delta},\Cst{\mathcal V}{\Delta})\]
  where 

  \begin{itemize}
  \item
    $\Sorts$ is as in Definition~\ref{def:quantstruct} (a set of
    elements called \Index[sort]{sorts}, denoted $s$, $s_1$, etc);
  \item
    $\vare$ is a set of elements called \Index[eigenlabel]{eigenlabels}, denoted
    $x$, $x'_1$, etc;
  \item
    $\Terms$ is a set of elements called \Index[term]{terms}, denoted $r$,
    $r_1$, etc,
    \begin{itemize}
    \item that extends the set $\vare$ of eigenlabels,
    \item and with a systematic way of lifting a function $\vare\rightarrow\vare$ to $\Terms\rightarrow\Terms$;
    \end{itemize}
    We can then apply a function $\pi:\vare\rightarrow\vare$ to lists of terms (using the standard map function on lists);
  \item $\decs{}$ is a \Index{sorting relation}, \ie a set of elements $(\Ders{\_}{\XcolY \_{\_}}):((\vare\pfunspace\Sorts)\times\Terms\times\Sorts)$,
    with
    \begin{itemize}
    \item
      $\Ders{\Sigma}{\XcolY x s}$ \iff\ $s=\Sigma(x)$
    \item for all $\pi:\vare\rightarrow\vare$,
      if $\Ders{\Sigma\circ\pi}{\XcolY r s}$ then $\Ders{\Sigma}{\XcolY {\pi(r)} s}$;
    \end{itemize}
  \item
    $\Atms$, $\Moles$, $\var[+]$ and $\var[-]$ are as in
    Definitions~\ref{def:atommolecules1o}
    and~\ref{def:typingcontexts};

    We can apply a function $\pi:\vare\rightarrow\vare$ to instantiated
    atoms and molecules using 
    \ctr{$\pi(a^l,{\bf r})\eqdef(a^l,\pi({\bf
      r}))$ and $\pi(M^l,{\bf r})\eqdef(M^l,\pi({\bf r}))$;}

    We then impose that equality on instantiated atoms be stable under
    any such function $\pi:\vare\rightarrow\vare$:\qquad If
    $\atmEq{(a,{\bf r})}{(a',{\bf r'})}$ then
    $\atmEq{\pi(a,{\bf r})}{\pi(a',{\bf r'})}$.

  \item $\TContexts$ is an $(\IAtms,\IMoles,\Sorts)$-context algebra,
    called the \Index{typing context algebra}, equipped with a
    \Index{map} operation that associates, to a context $\Gamma$ and
    two functions $f_1:\IAtms\rightarrow\IAtms$,
    $f_2:\IMoles\rightarrow\IMoles$, a context $\CMap{f_1,f_2}\Gamma$
    that is a map of $\Gamma$ according to $f_1$, $f_2$ and the
    identity on sorts;

  \item $\Data$ and $\decf{}$ are as in Definition~\ref{def:patterns1o}.

  \item We have two functions, respectively called the \Index{renaming policy} and the \Index{fresh naming policy}, of the form
    \[
    \left(\begin{array}{c@{}l@{}l}
      \powerset{\var}\times\Dstruct
      &\rightarrow
      &(\vare\rightarrow\vare)\\
      (\mathcal V,\Delta)&\mapsto&\Crename{\mathcal V}{\Delta}
    \end{array}\right),
    \left(\begin{array}{c@{}l@{}l}
      \powerset{\var}\times\Dstruct
      &\rightarrow
      &\DecompType[\unitt,\unitt,\vare]\\
      (\mathcal V,\Delta)&\mapsto&\Cst{\mathcal V}{\Delta}
    \end{array}
    \right)
    \]

    We abbreviate $\Cst{\domAll{\Gamma}}{\abs\Delta}$ as $\Cst{\Gamma}{\Delta}$ and 
    $\Crename{\domAll{\Gamma}}{\abs\Delta}$ as $\Crename{\Gamma}{\Delta}$.
  \end{itemize}
\end{definition}

Most of the above definition is rather straightforward when thinking
of sorting contexts as assigning sorts to eigenlabels.  What is
probably more cryptic is the naming policies $\Crename{\mathcal
  V}{\Delta}$ and $\Cst{\mathcal V}{\Delta}$ (as well as the map
operation of typing contexts): they compensate for the fact that the
extension operation of the typing context algebra $\TContexts$ is more
basic than in Definition~\ref{def:patterns1o}. In short,
$\Cst{\Gamma}{\Delta}$ and $\Crename{\Gamma}{\Delta}$ describe
which labels are used in an extended typing context
$\Gamma\Cextend\Delta$ (especially regarding the labels used in
$\Gamma$). Section~\ref{sec:LAFe2LAF} explains this in details, but we
first start with the example of \LAF[K1].

\subsection{\LAF[K1] is a \LAF\ instance with eigenlabels}

In this section we illustrate the above concept by giving an
alternative definition for system \LAF[K1] (from
Section~\ref{sec:LAFexbind}), this time as a \LAF\ instance with
eigenvariables.

Among the parameters 
\[(\Sorts,\vare,\Terms,\decs{},\Atms,\Moles,\atmEq{}{},\var[+],\var[-],\TContexts,\Data,\decf{})\]
of a \LAF\ instance with eigenvariables, the context algebra
$\TContexts$ is perhaps the least obvious to identify for \LAF[K1].
We do this now, by going via the definition of a generic family of
context algebras as we had done in the quantifier-free version of
\LAF[K1] (Sections~\ref{sec:LAFK1qf} and~\ref{sec:LAFlabelsK}).

\begin{definition}[A generic family of context algebras]\label{def:tcontextsLAFK1}\nopagesplit

  Given three sets $\mathcal A$, $\mathcal B$ and $\mathcal C$, we
  define $\Contexts[\mathcal A,\mathcal B,\mathcal C]$ as the set of
  elements of the form $(\Gamma^+,\Gamma^-,\Gamma^e)$ where $\Gamma^+$ (\resp $\Gamma^-$, $\Gamma^e$)
  is a list of elements of $\mathcal A$ (\resp $\mathcal B$,  $\mathcal C$).

  As in Definition~\ref{def:TcontextsK1wQ}, three disjoint copies
  $\var[+]$, $\var[-]$ and $\vare$ of the set of natural numbers are used for
  positive labels, negative labels and eigenlabels, respectively denoted $n^+$,
  $n^-$ and $n^e$, and we define\\ $\varRead[n^+]{(\Gamma^+,\Gamma^-,\Gamma^e)}$
  as the $(n+1)^{th}$ element of
  $\Gamma^+$\\ $\varRead[n^-]{(\Gamma^+,\Gamma^-,\Gamma^e)}$ as the
  $(n+1)^{th}$ element of $\Gamma^-$\\
  $\varRead[n^e]{(\Gamma^+,\Gamma^-,\Gamma^e)}$ as the
  $(n+1)^{th}$ element of $\Gamma^e$.

  We turn the resulting structure
  \[
  \left(\Contexts[\mathcal A,\mathcal B,\mathcal C],
       {\footnotesize
         \left(\begin{array}{c@{}l@{}l}
           \Contexts[\mathcal A,\mathcal B,\mathcal C]\times\var[+]&\pfunspace&\mathcal A\\
           (\Gam,n^+)&\mapsto&\varRead[n^+]\Gam
         \end{array}\right),
         \left(\begin{array}{c@{}l@{}l}
           \Contexts[\mathcal A,\mathcal B,\mathcal C]\times\var[-]&\pfunspace&\mathcal B\\
           (\Gam,n^-)&\mapsto&\varRead[n^-]\Gam
         \end{array}\right),
         \left(\begin{array}{c@{}l@{}l}
           \Contexts[\mathcal A,\mathcal B,\mathcal C]&\rightarrow&(\vare\pfunspace\mathcal C)\\
           \Gam&\mapsto&(n^e\mapsto\varRead[n^e]\Gam)
         \end{array}\right)
       }\right)
  \]
  into an $(\mathcal A,\mathcal B,\mathcal C)$-context algebra, by
  defining the notion of extension as follows:
  \[
  \left(\begin{array}{c@{}l@{}l}
    \Contexts[\mathcal A,\mathcal B,\mathcal C]\times\DecompType[\mathcal A,\mathcal B,\mathcal C]&\rightarrow&\Contexts[\mathcal A,\mathcal B,\mathcal C]\\
    (\Gam,\Delta)&\mapsto&\Gam\Cextend\Delta
  \end{array}\right)
  \]
  \[
  \begin{array}{llll}
    (\Gamma^+,\Gamma^-,\Gamma^e)\Cextend a&\eqdef(\cons a {\Gamma^+},\Gamma^-,\Gamma^e)&  
    (\Gamma^+,\Gamma^-,\Gamma^e)\Cextend \Drefute b &\eqdef(\Gamma^+,\cons b {\Gamma^-},\Gamma^e)\\
    (\Gamma^+,\Gamma^-,\Gamma^e)\Cextend \Dunit &\eqdef(\Gamma^+,\Gamma^-,\Gamma^e)
    &(\Gamma^+,\Gamma^-,\Gamma^e)\Cextend(\Delta_1\Dand\Delta_2)
    &\eqdef(\Gamma^+,\Gamma^-,\Gamma^e)\Cextend\Delta_1\Cextend\Delta_2\\
    (\Gamma^+,\Gamma^-,\Gamma^e)\Cextend \Dex[c]{}{\Delta} &\eqdef(\Gamma^+,\Gamma^-,\cons c {\Gamma^e})\Cextend\Delta
  \end{array}
  \]
\end{definition}

We can now give the parameters that present \LAF[K1]  as a \LAF\ instance with eigenlabels:

\begin{definition}[{\LAF[K1]} as a \LAF\ instance with eigenlabels]\label{def:LAFK1as}\nopagesplit

$\Sorts$ is the set of sorts as in Definition~\ref{def:K1struct}.

$\vare$ is a copy of the set of natural numbers and $\Terms$ is the
  set of terms, as in Definition~\ref{def:K1terms}. Notice that
  $\Terms$ does extend $\vare$, and substitution of eigenlabels for
  terms gives a systematic way to lift a function
  $\vare\rightarrow\vare$ to a function $\Terms\rightarrow\Terms$.

$\decs{}$ is the sorting relation as in
  Definition~\ref{def:K1terms}. Notice that $\Ders{\Sigma}{\XcolY x
    s}$ \iff\ $s=\Sigma(x)$, and that for all
  $\pi:\vare\rightarrow\vare$, if $\Ders{\Sigma\circ\pi}{\XcolY r s}$
  then $\Ders{\Sigma}{\XcolY {\pi(r)} s}$.

$\Atms$, $\Moles$ and $\atmEq{}{}$ are as in
  Definition~\ref{def:K1struct}.

$\var[+]$ and $\var[-]$ are as in Definition~\ref{def:TcontextsK1wQ},
  and the context algebra $\TContexts$ is the instance
  $\Contexts[\IAtms,\IMoles,\Terms]$ of the generic family of contexts
  from Definition~\ref{def:tcontextsLAFK1}.

Given two functions $f_1:\IAtms\rightarrow\IAtms$,
$f_2:\IMoles\rightarrow\IMoles$ and an
$(\IAtms,\IMoles,\Terms)$-context $(\Gamma^+,\Gamma^-,\Gamma^e)$, the
result of the map operation
$\CMap{f_1,f_2}{(\Gamma^+,\Gamma^-,\Gamma^e)}$ is defined as
$(f_1(\Gamma^+),f_2(\Gamma^-),\Gamma^e)$, where $f_1(\Gamma^+)$ and
$f_2(\Gamma^-)$ are defined with the standard map operation on lists.

We define $\Crename{\mathcal V}{\Delta}$ as the identity (note that we have
$\varRead[x]{\Gamma}=\varRead[x]{(\Gamma\Cextend\Delta)}$).

We define $\Cst{\mathcal V}{\Delta}$ as the element $\Cstaux{{\sf
    sup}(\mathcal V)}{\Delta}{(n,\Pi)\mapsto\Pi}$ of
$\DecompType[\unitt,\unitt,\vare]$, where $\Cstaux n {\Delta}{f}$ is
defined in continuation-passing style\footnote{\ie for a continuation
  $f:(\N\times\DecompType[\unitt,\unitt,\vare])\rightarrow
  \DecompType[\unitt,\unitt,\vare]$} as follows:
\[\begin{array}{lllllll}
\Cstaux{n}{\uniti}f&\eqdef f ({n},\uniti)\\
\Cstaux{n}{\Drefute \uniti}f&\eqdef f ({n}, {\Drefute\uniti})\\
\Cstaux{n}\Dunit f&\eqdef f ({n},\Dunit)\\
\Cstaux{n}{(\Dex[\uniti]{}{\Delta})}f&\eqdef\Cstaux{n+1}{\Delta}{(n',\Pi)\mapsto f(n',\Dex[(n+1)^e]{}{\Pi})}\\
\Cstaux{n}{({\Delta_1}\Dand {\Delta_2})} f
&\eqdef\Cstaux{n}{\Delta_1}{(n_1,\Pi_1)\mapsto\Cstaux{n_1}{\Delta_2}{(n_2,\Pi_2)\mapsto f(n_2,(\Pi_1\Dand\Pi_2))}}
\end{array}
\]


Finally, $\Data$ and $\decf{}$ are as in Definition~\ref{def:K1struct}.
\end{definition}

The only subtle things in the above definition are that:
\begin{itemize}
\item we defined $\Crename{\mathcal V}{\Delta}$ as the identity, since
  the use of De Bruijn's levels avoids the need to update labels with
  new names every time a context is extended;\footnote{With De
    Bruijn's indices we would there have the opportunity to specify
    how the eigenlabels in a context $\Gamma$ should be updated when
    $\Gamma$ is extended into $\Gamma\Cextend\Delta$; namely, the
    indices should be raised by the number of new eigenlabels that $\Delta$
    introduces.}
\item we defined $\Cst{\mathcal V}{\Delta}$ as a data-structure that
  does nothing but remember the fresh eigenlabels that will be used
  for each construct $\Dex[s]{}{\Delta'}$ within $\Delta$.
\end{itemize}

From this alternative definition of \LAF[K1] we now have to describe
how the original definition of \LAF[K1] can be recovered. More
precisely, the context algebra $\Contexts[\IAtms,\IMoles,\Terms]$ of
Definitions~\ref{def:tcontextsLAFK1} and~\ref{def:LAFK1as} yields the
typing context algebra of Definition~\ref{def:TcontextsK1wQ}.

This is a particular case of a more general construction that turns
every \LAF\ instance with eigenlabels into a \LAF\ instance, which we
now present.

\subsection{\LAF\ instances with eigenlabels are \LAF\ instances}
\label{sec:LAFe2LAF}

We now show how a \LAF\ instance with eigenlabels forms a \LAF\ 
instance.

As expected, \Index[sorting context]{sorting contexts} are now partial
functions from eigenlabels to sorts (\ie
$\SoContexts=\vare\pfunspace\Sorts$).

What remains to do is to turn the $(\IAtms,\IMoles,\Sorts)$-context
algebra $\TContexts$ into a proper typing context algebra in the sense
of Definition~\ref{def:typingcontexts}. What is missing is the notion
of extension:
\[
\left(\begin{array}{c@{}l@{}l}
  \TContexts\times\ITDecs&\rightarrow&\TContexts\\
  (\Gam,(\Delta^l,{\bf r}))&\mapsto&\Gam\Textend[{\bf r}]{\Delta^l}
\end{array}\right)
\]

We define such an extension from the notion of extension that is available in
the $(\IAtms,\IMoles,\Sorts)$-context algebra $\TContexts$
\[
\left(\begin{array}{c@{}l@{}l}
  \TContexts\times\DecompType[\IAtms,\IMoles,\Sorts]&\rightarrow&\TContexts\\
  (\Gam',\Delta')&\mapsto&\Gam'\Cextend{\Delta'}
\end{array}\right)
\]
and from the naming policies $(\mathcal
V,\Delta)\mapsto\Crename{\mathcal V}{\Delta}$ and $(\mathcal
V,\Delta)\mapsto\Cst{\mathcal V}{\Delta}$.

More precisely, $\Gam\Textend[{\bf r}]{\Delta^l}$ is defined as
$\Gamma'\Cextend\Delta'$, where $\Gamma'$ is an
$(\IAtms,\IMoles,\Sorts)$-context and $\Delta'$ is an
$(\IAtms,\IMoles,\Sorts)$-decomposition, obtained from $\Gamma$ and
$(\Delta^l,{\bf r})$ by using the two new functions. These functions
allow us to describe the intricacies of the operation
$\Gam\Textend[{\bf r}]{\Delta^l}$ that is completely unspecified in
the abstract \LAF\ system:
\begin{itemize}
\item for instance, imagining that the eigenlabel $x$ is mapped to $s$
  by $\Cfune\Gamma$, we will need to know what happens to this mapping when
  $\Gamma$ is extended into $\Gam\Textend[{\bf r}]{\Delta^l}$
\item also, when $\Delta^l$ contains a typing decomposition of the
  form ${\Dex[s]{}\Delta^{\cons s l}}$, we expect a new eigenlabel to
  be mapped to $s$ and we will need to know which one it is.
\end{itemize}
The two naming policies provide this information:
\begin{itemize}
\item in the first example, the renaming policy $\Crename{\Gamma}{{\Delta^l}}(x)$
  provides the eigenlabel corresponding to $x$ and mapped to $s$ in
  the extended environment $\Gam\Textend[{\bf r}]{\Delta^l}$;\footnote{In other
    words, the eigenlabel $x$ has been renamed
    $\Crename{\Gamma}{{\Delta^l}}(x)$ in the extended environment;
    depending on how labels are implemented, it might be the case that
    $x$ keeps its name and $\Crename{\Gamma}{{\Delta^l}}$ is
    simply the identity.}
\item in the second example, the fresh naming policy $\Cst{\Gamma}{{\Delta^{l}}}$
  provides the names of the newly introduced eigenlabels, placed in a
  $(\unitt,\unitt,\vare)$-decomposition with the same structure as
  ${\Delta^{l}}$; hence, it will contain a decomposition of the form
  $\Dex[x]{}{\Pi}$ to indicate that $x$ is the eigenlabel we are
  looking for (mapped to $s$ in the extended environment).
\end{itemize}

Building the $(\IAtms,\IMoles,\Sorts)$-decomposition $\Delta'$ from
$(\Delta^l,{\bf r})$ thus relies on the following instantiation
mechanism:

\begin{definition}[Instantiation of typing decompositions]

  The \Index{instantiation} $\Iproj{\bf r}{\Pi}{\Delta^l}$ of a typing
  decomposition $\Delta^l$ is defined for a list of terms ${\bf r}$ of length
  $\abs l$ and a $(\unitt,\unitt,\vare)$-decomposition $\Pi$ that has
  the same structure as $\Delta^l$, as follows:
  \[\begin{array}{llll}
  \Iproj{\bf r}\uniti {a^l}&\eqdef(a^l,{\bf r})& \Iproj{\bf r}\uniti {(\Drefute {M^l})}&\eqdef\Drefute{(M^l,{\bf r})}\\
  \Iproj{\bf r}\Dunit\Dunit&\eqdef\Dunit
  &\Iproj{\bf r}{\Pi_1\Dand\Pi_2}{({\Delta^l_1}\Dand {\Delta^l_2})}
  &\eqdef(\Iproj{\bf r}{\Pi_1}{\Delta^l_1})\Dand(\Iproj{\bf r}{\Pi_1}{\Delta^l_2})\\
  \Iproj{\bf r}{\Dex[x]{}\Pi}{(\Dex[s]{}{\Delta^{\cons s l}})}&\eqdef\Dex[s]{}{(\Iproj{\cons {x} {\bf r}}\Pi{\Delta^{\cons s l}})}
  \end{array}
  \]
\end{definition}

\begin{definition}[Typing contexts in the sense of \LAF\ instances]


  The $(\IAtms,\IMoles,\Sorts)$-context algebra $\TContexts$ of a
  \LAF\ instance with eigenlabels, is turned into a typing context in
  the sense of \LAF\ instances by defining the following extention
  operation:

  Given a typing context $\Gamma$, a typing decomposition $\Delta^l$
  of arity $l$ and a list of terms ${\bf r}$ of length $\abs l$, we
  define
  \[\Gamma\Textend[{{\bf r}}]{\Delta^l}\ \eqdef
  \ (\CMap{\Crename{\Gamma}{{\Delta^l}},\Crename{\Gamma}{{\Delta^l}}}\Gamma)
  \Cextend
  ({\Iproj{\Crename{\Gamma}{{\Delta^l}}({\bf r})}{\Cst{\Gamma}{{\Delta^l}}}{\Delta^l}})\]
\end{definition}

The way we perform this extension can be explained as follows:
\begin{itemize}
\item first, the extension will rename the eigenlabels that were
  declared in $\Gamma$; these eigenlabels are mentioned in the
  parameters of the instantiated atoms and molecules in $\Gamma$, so
  we use the renaming policy $\Crename{\Gamma}{{\Delta^l}}$ to update
  with the new names these instantiated atoms and molecules; the
  result is the context
  \[\Gamma'\eqdef\CMap{\Crename{\Gamma}{{\Delta^l}},\Crename{\Gamma}{{\Delta^l}}}\Gamma\]
\item second, we turn $\Delta^l$ into a
  $(\IAtms,\IMoles,\Sorts)$-decomposition as follows: the instantiated
  atoms and molecules at the leaves of this decomposition to produce
  will have their parameters based on ${\bf r}$; but the terms in
  ${\bf r}$ may mention the eigenlabels declared in $\Gamma$, which
  are now renamed, so we update ${\bf r}$ into
  $\Crename{\Gamma}{{\Delta^l}}({\bf r})$; then a parameterised atom $a$
  (\resp molecule $M$) at a leaf of $\Delta^l$ has an arity of the
  form $\cons{s_1}{\ldots\cons{s_n}{l}}$, and turns into the
  instantiated atom $(a,\cons{x_1}{\ldots\cons{x_n}{\Crename{\Gamma}{{\Delta^l}}({\bf r})}})$ (\resp
  molecule $(M,\cons{x_1}{\ldots\cons{x_n}{\Crename{\Gamma}{{\Delta^l}}({\bf r})}})$), where
  $x_1,\ldots,x_n$ are new eigenlabels whose names we get from the fresh naming policy
  $\Cst{\Gamma}{{\Delta^l}}$; this results in the
  $(\IAtms,\IMoles,\Sorts)$-decomposition 
  \[\Delta'\eqdef{\Iproj{\Crename{\Gamma}{{\Delta^l}}({\bf r})}{\Cst{\Gamma}{{\Delta^l}}}{\Delta^l}}\]
\item third, we extend $\Gamma'$ with $\Delta'$.
\end{itemize}

\begin{theorem}[The case of {\LAF[K1]}]
  The \LAF\ instance \LAF[K1], defined according to the above
  methodology from its definition as a \LAF\ instance with eigenlabels
  (Definition~\ref{def:LAFK1as}), coincides with the direct definition
  of \LAF[K1] as a \LAF\ instance (Sections~\ref{sec:LAFK1qf}
  and~\ref{sec:LAFlabelsK}).
\end{theorem}
\begin{proof}Clearly we have
  \[
  \Gamma\Textend[{{\bf r}}]{\Delta^l}\ = \ \Gamma \Cextend
  ({\Iproj{\bf r}{\Cst{\Gamma}{{\Delta^l}}}{\Delta^l}})\] with the
  left-hand side being defined in Definition~\ref{def:TcontextsK1wQ}
  and the right-hand side being defined in
  Definitions~\ref{def:tcontextsLAFK1} and~\ref{def:LAFK1as}.
\end{proof}

As we have seen, $\Crename{\mathcal V}{\Delta}(x)$ and $\Cst{\mathcal
  V}{\Delta}(x)$ form a naming policy for the eigenlabels used
after a (typing) context extension. More generally, the fact that an
$(\mathcal A,\mathcal B,\mathcal C)$-context algebra respects this
naming policy can be expressed as follows:

\begin{definition}[Respecting naming policies]
  An $(\mathcal A,\mathcal B,\mathcal C)$-context algebra $\Contexts$
  \Index[respecting naming policies]{respects the naming policies}
  $({\mathcal V},{\Delta})\mapsto\Crename{\mathcal V}{\Delta}$ and
  $({\mathcal V},{\Delta})\mapsto\Cst{\mathcal V}{\Delta}$ if for all $\rho$ and $v$ we have
  \begin{enumerate}
  \item
    $\varRead[x]{\rho}=\varRead[\Crename{\domAll{\rho}}{\abs v}(x)]{(\rho\Cextend v)}$ for all eigenlabel $x\in\domE\rho$;
  \item and $\Cst{\domAll{\rho}}{\abs v}$ relates to
    $ v$ according to $(\unitt\times\IAtms)$,
    $(\unitt\times\IMoles)$,\\\strut\hfill and $\{(x,\varRead[x]{(\rho\Cextend v)})\mid
    x\in\domE{\rho\Cextend v}\}$.
  \end{enumerate}
\end{definition}

\section{A more concrete class of realisability algebras}
\label{sec:RealConcreteA}

Now the whole point of introducing the subclass of \LAF\ instances
that we call ``with eigenlabels'', is to have a tighter Adequacy Lemma
that relies on a weaker (and more systematically derivable)
correlation property than typing correlation.

For this we identify a class of realisability algebras that naturally
form models for \LAF\ instances with eigenlabels.

In brief, a realisability algebra with eigenlabels is a realisability
algebra where valuations are functions mapping eigenlabels to term
denotations.

Assume we have a \LAF\ instance with eigenlabels
\[(\Sorts,\vare,\Terms,\decs{},\Atms,\Moles,\atmEq{}{},\var[+],\var[-],\TContexts,\Data,\decf{},\Crename{\mathcal V}{\Delta},\Cst{\mathcal
  V}{\Delta})\]

\begin{definition}[Realisability algebras with eigenlabels]

  A \Index{model structure with eigenlabels} is a model structure where $\SSoContexts = \vare\pfunspace\STerms$, satisfying
  \begin{itemize}
  \item for all $x\in\vare$, $\sigma:\vare\pfunspace\STerms$, we have $\sem[\sigma] x=\sigma(x)$;
  \item for all $r\in\Terms$, $\sigma:\vare\pfunspace\STerms$ and $\pi:\vare\pfunspace\vare$, we have $\sem[\sigma\circ\pi] r = \sem[\sigma]{\pi(r)}$;
  \end{itemize}
  and where the semantic context algebra respects the naming policies.

  A \Index{realisability algebra with eigenlabels} is a realisability
  algebra whose model structure is a model structure with eigenlabels
  and where, for all $\Sigma:\vare\pfunspace\Sorts$ and
  $\sigma:\vare\pfunspace\STerms$,
  \begin{centre}
    $\sigma\in\sem\Sigma$ \iff\ for all $x\in\vare$ we have
    $\sigma(x)\in\sem{\Sigma(x)}$ (and $\dom \sigma=\dom\Sigma$).
  \end{centre}
\end{definition}

\begin{definition}[Generic correlation]

  Given three relations $\mathcal R_1\subseteq\mathcal A\times\mathcal
  A'$, $\mathcal R_2\subseteq\mathcal B\times\mathcal B'$ and
  $\mathcal R_3\subseteq\mathcal C\times\mathcal C'$,\\ we say that an
  $(\mathcal A,\mathcal B, \mathcal C)$-context algebra $\Contexts$
  and an $(\mathcal A',\mathcal B', \mathcal C')$-context algebra
  $\Contexts'$ satisfy the \Index{correlation property} for $\mathcal
  R_1$, $\mathcal R_2$ and $\mathcal R_3$ if the following holds:
  \begin{centre}
For all $\Gamma\in\Contexts$, $\Gamma'\in\Contexts'$, $\Delta\in\DecompType[\mathcal A,\mathcal B, \mathcal C]$ and $\Delta'\in\DecompType[\mathcal A',\mathcal B', \mathcal C']$\\
    if $\Gamma$ relates to $\Gamma'$ according to $\mathcal R_1$,
    $\mathcal R_2$ and $\mathcal R_3$\\ and $\Delta$ relates to
    $\Delta'$ according to $\mathcal R_1$, $\mathcal R_2$ and
    $\mathcal R_3$\\ then $\Gamma\Cextend\Delta$ relates to
    $\Gamma'\Cextend\Delta'$ according to $\mathcal R_1$,
    $\mathcal R_2$ and $\mathcal R_3$;
  \end{centre}
\end{definition}

\begin{definition}[Correlation with eigenlabels]
  Given a realisability algebra with eigenlabels (for our \LAF\ instance
  with eigenlabels), we define three relations\footnote{In this definition we implicitly require $\SemTy[\sigma]{(a,{\bf r})}$ and $\SemTy[\sigma]{(M,{\bf r})}$ to be defined.}
  \[
  \begin{array}{lll}
    \mathcal R^\sigma_1
    &\eqdef\{(\frak l,(a,{\bf r}))\mid \frak l\in\SemTy[\sigma]{(a,{\bf r})}\}
    &\subseteq \SPrim\times\IAtms\\
    \mathcal R^\sigma_2
    &\eqdef\{(\frak n,(M,{\bf r}))\mid \frak n\in\SemTy[\sigma]{(M,{\bf r})}\}
    &\subseteq \SNeg\times\IMoles\\
    \mathcal R_3 &\eqdef\{(\frak r,s)\mid \frak r\in\sem{s}\}
    &\subseteq \STerms\times\Sorts
  \end{array}
  \]
  for any given $\sigma:\vare\pfunspace\STerms$.

  We say that $\SContexts$ and $\TContexts$ satisfy the
  \Index[correlation with eigenlabels]{correlation with eigenlabels}
  property if for all $\sigma:\vare\pfunspace\STerms$, they satisfy
  the correlation property for $\mathcal R^\sigma_1$, $\mathcal
  R^\sigma_2$ and $\mathcal R_3$.
\end{definition}

\begin{remark}
  $\rho\in\SemTy\Gamma$ \iff\ $\rho$ relates to $\Gamma$ according to $\mathcal R^{\Cfune\rho}_1$, $\mathcal R^{\Cfune\rho}_2$ and $\mathcal R_3$.
\end{remark}

\begin{lemma}[Correlation with eigenlabels implies typing correlation]\label{lem:corr2}

  If $\SContexts$ and $\TContexts$ satisfy the correlation with
  eigenlabels property, then they satisfy the typing correlation property:
  if $\rho\in\SemTy\Gamma$ and $\frak \Delta\in\SemTy[\Cfune\rho]{(\Delta^l,{\bf
      r})}$ then
  $(\rho\Cextend\frak\Delta)\in\SemTy{\Gamma\Textend{\Delta^l}}$.
\end{lemma}
\begin{proof}See the proof in \Coq~\cite{LengrandHDRCoq}. The main lines are as follows:

From $\rho\in\SemTy\Gamma$ we get that $\rho$ relates to $\Gamma$
according to $\mathcal R^{\Cfune\rho}_1$, $\mathcal R^{\Cfune\rho}_2$
and $\mathcal R_3$.

Then $\rho$ relates to
$\CMap{\Crename{\Gamma}{{\Delta^l}},\Crename{\Gamma}{{\Delta^l}}}\Gamma$
according to $\mathcal R^{\Cfune{(\rho\Cextend{\frak\Delta})}}_1$,
$\mathcal R^{\Cfune{(\rho\Cextend{\frak\Delta})}}_2$ and $\mathcal
R_3$.

From $\frak \Delta\in\SemTy[\Cfune\rho]{(\Delta^l,{\bf r})}$
we get that $\frak \Delta$ relates to $\Iproj{\Crename{\Gamma}{{\Delta^l}}({\bf
  r})}{\Cst{\Gamma}{{\Delta^l}}}{\Delta^l}$ according to $\mathcal
R^{\Cfune{(\rho\Cextend{\frak\Delta})}}_1$, $\mathcal
R^{\Cfune{(\rho\Cextend{\frak\Delta})}}_2$ and $\mathcal R_3$.

Then correlation with eigenlabels provides that
$\rho\Cextend\frak\Delta$ relates to $\Gam\Textend[{\bf r}]{\Delta^l}$
according to $\mathcal R^{\Cfune{(\rho\Cextend{\frak\Delta})}}_1$,
$\mathcal R^{\Cfune{(\rho\Cextend{\frak\Delta})}}_2$ and $\mathcal
R_3$, which means that
$(\rho\Cextend\frak\Delta)\in\SemTy{\Gamma\Textend{\Delta^l}}$.

At some point in the above proof we use the fact that the semantic
context algebra respects the naming policies.
\end{proof}

\begin{lemma}[Adequacy for \LAF\ with eigenlabels]\label{lem:adequacyWeigen}\nopagesplit

We assume the following hypotheses:
  \begin{enumerate}
  \item[Well-foundedness:]\\
    The \LAF\ instance with eigenlabels is well-founded.
  \item[Correlation with eigenlabels:]\\
    $\SContexts$ and $\TContexts$ satisfy the correlation with eigenlabels property.
  \item[Stability:]\\
        If $\frak d\in\SemTy[\sigma]{(\Delta^l,{\bf r})}$ for some $\Delta^l,\sigma,{\bf r}$ and $\SemTe{f(p)}{\rho\Cextend \frak d}\in\orth{}{}$, then $\orth{\sem[\rho] f}{\spat p(\frak d)}$.
  \end{enumerate}

  We conclude that, for all $\rho\in\SemTy\Gamma$,
  \begin{enumerate}
  \item if $\DerF\Gamma {\XcolY {t^+} {(M^l,{\bf r})}}{}$ then $\SemTe{t^+}\rho\in\SemTyP {(M^l,{\bf r})}$; 
  \item if $\Der\Gamma {\XcolY {d} {(\Delta^l,{\bf r})}}{}$ then $\SemTe{d}\rho\in\SemTy {(\Delta^l,{\bf r})}$; 
  \item if $\Der\Gamma {{t} {}}{}$ then $\SemTe{t}\rho\in\orth{}{}$ 
  \end{enumerate}
\end{lemma}
\begin{proof}Corollary of Lemmata~\ref{lem:adequacy} and~\ref{lem:corr2}.
\end{proof}

This Adequacy Lemma looks similar to Lemma~\ref{lem:adequacy}, but the
correlation assumption is much ``weaker'': all the job is done in the
extra structure with eigenlabels that we have required from terms,
sorting contexts, typing contexts and valuations (and the
finer-grained specifications we have imposed on them).

Indeed, correlation with eigenlabels often holds as a particular case
of the more general correlation property for all relations $\mathcal
R_1$, $\mathcal R_2$, $\mathcal R_3$, typically when $\TContexts$ and
$\SContexts$ are respectively defined as the two instances
$\Contexts[\IAtms,\IMoles,\Terms]$ and
$\Contexts[\SPrim,\SNeg,\STerms]$ of a generic family
$(\Contexts[\mathcal A,\mathcal B,\mathcal C])_{\mathcal A,\mathcal
  B,\mathcal C}$ of $(\mathcal A,\mathcal B,\mathcal C)$-context
algebras whose definition is ``sufficiently parametric''. In
particular for \LAF[K1]:

\begin{remark}
  Generic correlation always holds for the family
  $(\Contexts[\mathcal A,\mathcal B,\mathcal C])_{\mathcal A,\mathcal
    B,\mathcal C}$ of  $(\mathcal A,\mathcal B,\mathcal C)$-context
  algebras defined for \LAF[K1] (Definition~\ref{def:tcontextsLAFK1}).

  In particular, correlation with eigenlabels holds for that system
  and for any of its realisability algebras where
  $\SContexts=\Contexts[\SPrim,\SNeg,\STerms]$.  If stability also
  holds for that instance and that realisability algebra, then the
  conclusions of the Adequacy Lemma hold.
\end{remark}

The same remark would hold of any \LAF\ instance and any realisability
algebra where $\TContexts$ and $\SContexts$ are defined from a
similarly parametric family of context algebras.

\section{Example: boolean models to prove Consistency}
\label{sec:RealConsistency}

We now exhibit models to prove the consistency of \LAF\ systems.




Assume we have a \LAF\ instance with eigenlabels
\[(\Sorts,\vare,\Terms,\decs{},\Atms,\Moles,\atmEq{}{},\var[+],\var[-],\TContexts,\Data,\decf{})\]

\begin{definition}[Boolean realisability algebras]

  A \Index{boolean realisability algebra} is a realisability algebra where $\orth\ \ =\emptyset$.
\end{definition}

The terminology comes from the remark that in a boolean realisability
algebra, $\SemTyN[\sigma]{(M^l,{\bf r})}$ can only take one of two
values: $\emptyset$ or $\SNeg$, depending on whether
$\SemTyP[\sigma]{(M^l,{\bf r})}$ is empty or not.

\begin{remark}
  A boolean realisability algebra satisfies Stability.
\end{remark}

\begin{theorem}[Consistency of \LAF\ instances with eigenvariables]

Assume the \LAF\ instance with eigenlabel is well-founded and assume
that we have a boolean realisability algebra with eigenlabels where
\begin{itemize}
\item there is an empty semantic context $\rho_\emptyset$;
\item $\SContexts$ and $\TContexts$ satisfy the correlation with eigenlabels property.
\end{itemize}
Then there is no empty typing context $\Gamma_\emptyset$ and command $t$ such that
$\Der{\Gamma_\emptyset} {{t} {}}{}$.
\end{theorem}
\begin{proof}
The previous remark provides Stability. If there was such a $\Gamma_\emptyset$ and $t$, then we would have $\rho_\emptyset\in\SemTy{\Gamma_\emptyset}$, and the Adequacy Lemma (Lemma~\ref{lem:adequacyWeigen}) would conclude $\SemTe t{{\rho_\emptyset}}\in\emptyset$.
\end{proof}

We provide such a realisability model that works with all parametric \LAF\ instances with eigenlabels:

\begin{definition}[Trivial model for parametric \LAF\ instances with eigenlabels]

  Assume that $\TContexts$ is the instance
  $\Contexts[\IAtms,\IMoles,\Terms]$ of a family of context algebras
  $(\Contexts[\mathcal A,\mathcal B,\mathcal C])_{\mathcal A,\mathcal
    B,\mathcal C}$.

  The \Index{trivial boolean model} for it is:
  \[
  \begin{array}{ll}
    \STerms\eqdef\SPrim\eqdef\SPos\eqdef\SNeg\eqdef\unitt\\
    \orth \ \ \eqdef \emptyset\\
    \SContexts\eqdef \Contexts[\unitt,\unitt,\unitt]\\
    \mbox{and therefore}\\
    \begin{array}{lllll}
    &\forall\rho\in\SContexts,\forall x^+\in\domP\rho,& \varRead[x^+]\rho&\eqdef \uniti\\
    &\forall\rho\in\SContexts,\forall x^-\in\domN\rho,& \varRead[x^-]\rho&\eqdef \uniti\\
    &\forall\rho\in\SContexts,\forall x\in\domE\rho,& \varRead[x]\rho&\eqdef \uniti\\
  \end{array}\\
\begin{array}{llll}
    \forall \frak{\Delta}\in\SDecs,&\spat p(\frak{\Delta})&\eqdef \uniti\\
    \forall r\in\Terms, \forall \sigma\in\SSoContexts,&\sem[\sigma] r&\eqdef\uniti\\
    \forall f:\Data\pfunspace\PTerms, \forall \rho\in\SContexts,& \sem[\rho] f&\eqdef \uniti\\\\
    \forall s\in\Sorts,& \sem s&\eqdef\unitt\\
    \forall a^l\in\Atms[l],\forall {\frak{rl}}\in\STerms^l,
    &\sem {a^l}({\frak{rl}})
    &\eqdef\unitt
  \end{array}
  \end{array}
\]
\end{definition}

It is straightforward to check that the above definition does satisfy
the specification of a realisability algebra with eigenlabels.

Note that, not only can $\SemTyN[\sigma]{(M^l,{\bf r})}$ only take one
of the two values $\emptyset$ or $\unitt$, but
$\SemTyP[\sigma]{(M^l,{\bf r})}$ can also only take one of the two
values $\emptyset$ or $\unitt$.

We can now use such a structure to derive consistency for a large class of systems:

\begin{corollary}[Consistency for parametric \LAF\ instances with eigenlabels]

Assume that the \LAF\ instance with eigenlabels is well-founded and that
\begin{itemize}
\item $\TContexts$ is the instance $\Contexts[\IAtms,\IMoles,\Terms]$
  of a family of context algebras $(\Contexts[\mathcal A,\mathcal
    B,\mathcal C])_{\mathcal A,\mathcal B,\mathcal C}$,
\item Any two context algebras of the family $(\Contexts[\mathcal
  A,\mathcal B,\mathcal C])_{\mathcal A,\mathcal B,\mathcal C}$
  satisfy the correlation property for all $\mathcal R_1$, $\mathcal
  R_2$, $\mathcal R_3$.
\item There is an empty $(\unitt,\unitt,\unitt)$-context in $\Contexts[\unitt,\unitt,\unitt]$,
\end{itemize}
Then there is no empty typing context $\Gamma_\emptyset$ and command $t$ such that
$\Der{\Gamma_\emptyset} {{t} {}}{}$.

In particular, this is the case for \LAF[K1].
\end{corollary}

The system \LAF[J] does not fall in the above category since the
operation of context extension is not parametric enough: when
computing $\Gamma\Textend{a^l}$ (\resp
$\Gamma\Textend{\Drefute{M^l}}$), we have to make a case analysis on
whether $a^l$ is of the form $(l^+,\righths)$ or $(v,\lefths)$ (\resp whether $M^l$ is of
the form $(N,\lefths)$ or $(P,\righths)$).

But we can easily adapt the above trivial model into a
not-as-trivial-but-almost model:

\begin{definition}[{{Trivial model for {\LAF[J]}}}]
  The \Index{trivial boolean model for \LAF[J]} is:
  \[
  \begin{array}{ll}
    \STerms\eqdef\SPos\eqdef\unitt\\
    \SPrim\eqdef\SNeg\eqdef\{\lefths,\righths\}\\
    \orth \ \ \eqdef \emptyset\\
    \SContexts\mbox{ has semantics contexts of the form $(m^+,m^-,m^e,R)$,}\\
    \flush\mbox{where $m^+,m^-,m^e\in\N$ and $R\in\{0,1\}$}\\
    \mbox{and an extension operation defined as follows}\\
    {\small
    \begin{array}{ll@{\qquad}ll}
      (m^+,m^-,m^e,R)\Cextend \righths&\eqdef({m^+}+1,m^-,R)&  
      (m^+,m^-,m^e,R)\Cextend \lefths&\eqdef(m^+,m^-,m^e,0)\\
      (m^+,m^-,m^e,R)\Cextend \Drefute {\lefths} &\eqdef(m^+,{m^-}+1,R)&  
      (m^+,m^-,m^e,R)\Cextend \Drefute {\righths} &\eqdef(m^+,m^-,m^e,1)
    \end{array}}\\
    {\small
    \begin{array}{ll@{\qquad}ll}
      (m^+,m^-,m^e,R)\Cextend \Dunit &\eqdef(m^+,m^-,m^e,R)\\
      (m^+,m^-,m^e,R)\Cextend \Dex[\uniti]{}{\Delta}&\eqdef(m^+,m^-,m^e+1,R)\\
      (m^+,m^-,m^e,R)\Cextend(\Delta_1\Dand\Delta_2)
      &\eqdef(m^+,m^-,m^e,R)\Cextend\Delta_1\Cextend\Delta_2
    \end{array}}\\\\
    \mbox{and we define}\\
    {\small
    \begin{array}{ll@{\quad}llllll}
      \varRead[n^+]{(m^+,m^-,m^e,R)}&\eqdef\righths\mbox{ if }n^+< m^+
      &\varRead[n^-]{(m^+,m^-,m^e,R)}&\eqdef\lefths\mbox{ if }n^-< m^-\\
      \varRead[n^+]{(m^+,m^-,m^e,R)}&\mbox{undefined otherwise}
      &\varRead[n^-]{(m^+,m^-,m^e,R)}&\mbox{undefined otherwise}\\
      \varRead[\star^+]{(m^+,m^-,m^e,0)}&\eqdef \lefths
      &\varRead[\star^-]{(m^+,m^-,m^e,0)}&\mbox{undefined}\\
      \varRead[\star^+]{(m^+,m^-,m^e,1)}&\mbox{undefined}
      &\varRead[\star^-]{(m^+,m^-,m^e,1)}&\eqdef\righths\\
      \varRead[x^+_{(\falseN,\lefths)}]{(m^+,m^-,m^e,R)}&\eqdef\lefths\\\\
      \varRead[n^e]{(m^+,m^-,m^e,R)}&\eqdef\uniti\mbox{ if }n^e< m^e\\
      \varRead[n^e]{(m^+,m^-,m^e,R)}&\mbox{undefined otherwise}
    \end{array}
    }\\\\
\begin{array}{llll}
    \forall \frak{\Delta}\in\SDecs,&\spat p(\frak{\Delta})&\eqdef \uniti\\
    \forall r\in\Terms, \forall \sigma\in\SSoContexts,&\sem[\sigma] r&\eqdef\uniti\\
    \forall f:\Data\pfunspace\PTerms, \forall \rho\in\SContexts,& \sem[\rho] f&\eqdef \lefths
\mbox{ if every $p\in\dom f$ is of the form}\\
&&\hfill\Ppos_\righths \mid \Pneg_\righths \mid \Ptrue_\righths\mid\paire {p_1}{p_2}\mid\inj i p\\
&&\eqdef\righths\mbox{ if not}\\\\
    \forall s\in\Sorts,& \sem s&\eqdef\unitt\\
    \forall (l^+,{\sf r})\in\Atms[l],\forall {\frak{rl}}\in\STerms^l,
    &\sem {(l^+,{\sf r})}({\frak{rl}})&\eqdef\{{\sf r}\}\\
    \forall (v,{\sf l})\in\Atms[l],\forall {\frak{rl}}\in\STerms^l,
    &\sem {(v,{\sf l})}({\frak{rl}})&\eqdef\{{\sf l}\}
  \end{array}
  \end{array}
\]
\end{definition}

It is straightforward to check that the above definition does satisfy
the specification of a realisability algebra with eigenlabels.
Moreover, $\TContexts$ and $\SContexts$ satisfy the correlation
property with eigenlabels.

We can now use such a structure to derive consistency for \LAF[J]:

\begin{theorem}[Consistency of {{\LAF[J]}}]

  There is no command $t$ such that
  $\Der{(\Gamma^+,\el,\Cfune\Gamma,(v,\lefths,{\bf r}))} {{t} {}}{}$ in \LAF[J].
\end{theorem}
\begin{proof}
  Take the trivial boolean model for \LAF[J]; we have Stability.
  Take $\rho\eqdef(\abs{\Gamma^+},0,\abs{\Cfune\Gamma},\lefths)$; clearly $\rho\in\SemTy{(\Gamma^+,\el,\Cfune\Gamma,(v,\lefths,{\bf r}))}$, and the Adequacy Lemma (Lemma~\ref{lem:adequacyWeigen}) would conclude $\SemTe t{{\rho}}\in\emptyset$.
\end{proof}

\cchapter[Transforming proofs in \LAF]{Transforming proofs in the abstract focussed sequent calculus}
\label{ch:trans}

\minitoc 

In this chapter we investigate how \LAF\ proofs can be transformed.

First and foremost, we have in mind the key process of structural
proof theory: \emph{cut-elimination}.
The process is all the more interesting as it relates, through the
Curry-Howard correspondence~\cite{How:fortnc}, to the paradigm of computation in
functional programming; and in the case of \LAF\ systems,
cut-elimination strongly relates to the very concept of
pattern-matching, following~\cite{ZeilbergerPhD}.

Let us also remember that originally, admissibility of cuts was a
property used by\linebreak Gentzen~\cite{Gentzen35} to relate the sequent
calculus with cuts, which can easily be proved complete, to the cut-free
sequent calculus, which is easily proved consistent.
Even though we already have consistency results for \LAF\ systems
(with cuts) obtained by semantical methods (see
Section~\ref{sec:RealConsistency}), we are still interested in cut
admissibility to get completeness of cut-free \LAF\ systems.
Indeed, we identifiedde \LAF\ systems with the perspective of using them
as the basis of proof-search implementations, and knowing this
property will help organising the exploration of the search-space.

The prospect of implementing proof-search also motivates the study of
another kind of proof-transformation: As we shall seek to memoise the
proof-search process (tabling all the proofs and sub-proofs we
complete to re-use them as often as possible), we will often seek to
\emph{adapt} a previously obtained proof to a new sequent to be proved
(provided of course this new sequent contains all the necessary
ingredients for the proof to be replayed).

In Section~\ref{sec:headreduction}, we identify a notion of
\emph{abstract machine} to reduce the proof-terms of \LAF,
implementing in effect a notion of \emph{head reduction}. 
In Section~\ref{sec:headnormalistion} we prove that this reduction
terminates on typed terms, for which the realisability models of
Chapter~\ref{ch:real} will play a key role.
In Section~\ref{sec:pruning} we investigate the re-usability of
proofs, by identifying with the concept of \emph{free label} the atoms
and molecules of a proved sequent that are necessary for the proof
to be replayed on another sequent to prove.
In Section~\ref{sec:cut-elimination}, we will investigate how the
transformations explored in the previous sections can be used to prove
cut-elimination in \LAF.
In Section~\ref{sec:SN} we discuss the possibility of more general
notions of reduction and the issue of Strong Normalisation.

The proof transformations explored in this chapter will prove
particularly useful when using \LAF\ in automated reasoning (see the
third part of this dissertation).

\section{Head reduction}
\label{sec:headreduction}

It is natural to want to reduce $\cutc{\THO f}{pd}$ to $f(p)$
``substituted by $d$''. Indeed, this would be the evaluation rule of
pattern-matching: we can think of $p$ as a pattern and $d$ as a way to
fill its holes, while $f$ is a pattern-matching function; the rule
then selects the branch of $f$ corresponding to $p$ and depending on
the pattern's holes, and computation continues with the code in that
branch where the holes have been substituted according to $d$.

Such a notion of substitution, however, is not yet defined. And so far
$d$ is a decomposition term: we can easily imagine using it to extend
a context, but it is not a context itself.

Now following the view that ``there is no such thing as a free
variable'' (what is thought of as free in in fact bound somewhere else), we
can accept that reducing $\cutc{\THO f}{pd}$ is in fact done in a
context $\rho$ that assigns ``values'' to the ``free labels'' of
$\THO f$ and $d$. This view is actually quite natural when thinking of
evaluating programs by an abstract machine: evaluation is
performed within an ``environment'' that maps variables to
values such as \emph{closures}.

In the case of \LAF, this view helps understanding how the term
decomposition $d$ can be involved in reductions, as it can now be used
to extend the local context $\rho$ in which the command is evaluated:
\[\machine[\rho]{\cutc{\THO f}{pd}}\Rew{}\machine[\rho\Cextend d']{f(p)}\]
where $d'$ is ``$d$ in the context $\rho$''.
This we could think as simply the pairing $(d,\rho)$, were it not for
the fact that the extension $\rho\Cextend d'$ needs $d'$ to be a
decomposition, not a pair.
Hence, $d'$ will rather be the distribution of $\rho$ down to each
leaf of $d$.

This is formalised as follows:

\begin{definition}[Abstract machine for \LAF]\label{def:absmachine}\nopagesplit

  Assume we have four sets $\Values[+]$, $\Values[-]$, $\ValuesE$,
  $\ValuesSC$, and a $(\Values[+],\Values[-],\ValuesE,\ValuesSC)$-context
  algebra with support set $\Contexts$ such that the set
  $\Closures\eqdef(\Data\pfunspace\PTerms)\times\Contexts$ is a subset
  of $\Values[-]$.

  Elements of $\Closures$ are called \Index[closure]{closures} and
  denoted $\closure[\rho]f$ (where $f:\Data\pfunspace\PTerms$ and
  $\rho\in\Contexts$), while elements of $\Contexts$ are called
  \Index[evaluation context]{evaluation contexts}.

  An \Index{evaluation decomposition} is a
  $(\Values[+],\Values[-],\ValuesE)$-decomposition.

  An \Index{evaluation triple} is a triple denoted $\cutc {v} {pd}$
  (overloading the notation for commands) where $v\in\Values[-]$,
  $p\in\Data$ and $d$ is an evaluation decomposition.

  A \Index{contextualised command} is a pair denoted $\machine [\rho]
  t$ where $t$ is a command and $\rho$ is an evaluation context.

  We assume we have an \Index{instantiation function}
  \[
  \left(\begin{array}{c@{}l@{}l}
    \Terms\times\ValuesSC&\pfunspace&\ValuesE\\
    (r,\sigma)&\mapsto&\halfsubst[\sigma]r
  \end{array}\right)
  \]

  We define the \Index[distribution]{distribution of an evaluation
    context $\rho$ over a term decomposition $d$}, denoted
  $\halfsubst[\rho]{d}$, as the following evaluation decomposition:
  \[
  \begin{array}{ll}
    \halfsubst[\rho]{x^+}&\eqdef\varRead[x^+]\rho\\
    \halfsubst[\rho]{\THO f}&\eqdef\closure[\rho] f\\
    \halfsubst[\rho]{\Tunit}&\eqdef\Dunit\\
    \halfsubst[\rho]{d_1\Tand d_2}&\eqdef\halfsubst[\rho]{d_1}\Dand\halfsubst[\rho]{d_2}\\
    \halfsubst[\rho]{\Tex r d}&\eqdef\Dex[{\halfsubst[\Cfune\rho]r}]{}{\halfsubst[\rho]{d}}
  \end{array}
  \]

  The reduction relation is defined in two steps:
  the reduction of a contextualised command to an evaluation triple, and the reduction of an evaluation triple to a contextualised command:
  \[
  \begin{array}{lll}
    \rulenamed{\headred[1]}
    &\machine[\rho]{\cutc{x^-}{pd}}
    &\Rew{}\cutc{\varRead[x^-]\rho}{p\,\halfsubst[\rho] d}\\
    \rulenamed{\headred[2]}
    &\machine[\rho]{\cutc{\THO f}{pd}}
    &\Rew{}\cutc{\closure[\rho] f} {p\,\halfsubst[\rho] d}\\\\
    \rulenamed{\headred[3]}
    &\cutc{\closure[\rho]{f}}{p d}
    &\Rew{}\machine[\rho\Cextend {d}]{f(p)}\\
  \end{array}
  \]

  We will write $\Rewn{\headred[123]}$ for
  $\Rewn{\headred[1],\headred[2],\headred[3]}$, which will always be
  an alternation of $\Rew{\headred[1],\headred[2]}$ and
  $\Rew{\headred[3]}$.

  There are no contextualised commands in normal form and evaluation
  triples in normal form are those of the form $\cutc {x^-}{pd}$.

  If a contextualised command or an evaluation triple reduces by
  $\Rewn{\headred[123]}$ to such a normal form, we say that it
  \Index[head-normalise]{head-normalises}.
\end{definition}

\begin{example}[Syntactic abstract machine]\strut

  Standard examples of abstract machine are \Index{syntactic abstract
    machines}, where $\Values[+]\eqdef\var[+]$ and
  $\ValuesE\eqdef\Terms$, and $\Values[-]=\var[-]\cup\Closures$. In
  other words, computation can substitute positive labels for positive
  labels, and substitute either negative labels or closures for
  negative labels.

  Note however that this makes $\Values[-]$ and $\Closures$ mutually
  dependent,\footnote{Remember that $\Closures$ is
    $(\Data\pfunspace\PTerms)\times\Contexts$, where $\Contexts$ is
    (the support set of) a
    $(\Values[+],\Values[-],\ValuesE,\ValuesSC)$-context algebra.} so
  their exact definition can hardly be defined at this abstract level.

  But for instance with \LAF[K1], we can adapt
  Definition~\ref{def:tcontextsLAFK1} to define $\Closures$,
  $\Values[-]$ and the set $\Contexts$ of evaluation contexts by
  simultaneous induction:
  \begin{itemize}
  \item $\Closures\eqdef(\Data\pfunspace\PTerms)\times\Contexts$
  \item $\Values[-]\eqdef\var[-]\cup\Closures$
  \item $\Contexts$ is the set of elements of the form
    $(\Gamma^+,\Gamma^-,\Gamma^e)$ where $\Gamma^+$ (\resp $\Gamma^-$,
    $\Gamma^e$) is a list of elements of $\var[+]$ (\resp
    $\Values[-]$, $\Terms$).
  \end{itemize}
  Once the set $\Contexts$ of evaluation contexts is defined, the full
  evaluation context algebra is simply
  $\Contexts[{\var[+]},{\Values[-]},\Terms]$ (using the notation of
  Definition~\ref{def:tcontextsLAFK1}).

  Similarly, the set $\ValuesSC$ and the function
  \[
  \left(\begin{array}{c@{}l@{}l}
    \Terms\times\ValuesSC&\pfunspace&\Terms\\
    (r,\sigma)&\mapsto&\halfsubst[\sigma]r
  \end{array}\right)
  \]
  can hardly be defined at the abstract level. But for first-order
  logic it is natural to define $\ValuesSC$ as the set
  $\vare\rightarrow \Terms$ of substitutions, and
  $\halfsubst[\sigma]r$ is simply the application of substitution
  $\sigma$ to the first-order term $r$.
\end{example}

\section{Head normalisation}
\label{sec:headnormalistion}

In this section we show that the abstract machine from
Definition~\ref{def:absmachine} terminates, when starting from typed
proof-terms.

Mimicking the use of orthogonality models to prove strong
normalisation result as in Chapter~\ref{ch:ortho}, we prove
normalisation of the abstract machine by the use of a realisability
model, in the sense of Chapter~\ref{ch:real}.

\begin{definition}[A realisability model for head-normalisation]\label{def:model4headnorm}

  Assume we have an abstract machine defined by four sets
  $\Values[+]$, $\Values[-]$, $\ValuesE$, $\ValuesSC$, an evaluation
  context algebra $\Contexts$, and an instantiation function
  \((r,\sigma)\mapsto\halfsubst[\sigma]r\).

  The \Index{head-normalisation model} for this abstract machine is
  \[
  \begin{array}{l}
  \begin{array}{ll}
    \SSoContexts&\eqdef\ValuesSC\\ 
    \STerms&\eqdef\ValuesE\\ 
    \SPrim&\eqdef\Values[+]\\ 
    \SPos&\eqdef\Data\times\DecompType[{\Values[+]},{\Values[-]},\ValuesE]\\ 
    \SNeg&\eqdef\Values[-]\\ 
    \orth
    v {pd}&\mbox{if the evaluation triple }\cutc{v}{p d} \mbox{
      head-normalises\footnote{for the reduction relation defined in
      Definition~\ref{def:absmachine}}}\\
    \SContexts&\eqdef \Contexts
  \end{array}\\
    \begin{array}{llll}
    \forall \frak{\Delta}\in\SDecs,&\spat p(\frak{\Delta})&\eqdef p\frak{\Delta}\\
    \forall r\in\Terms, \forall \sigma\in\SSoContexts,&\sem[\sigma] r&\eqdef\halfsubst[\sigma]r\\
    \forall f:\Data\pfunspace\PTerms, \forall \rho\in\SContexts,& \sem[\rho] f&\eqdef \closure[\rho]{f}\\
    \forall s\in\Sorts,& \sem s&\eqdef\ValuesE\\
    \forall \Sigma\in\SoContexts,& \sem \Sigma&\eqdef\ValuesSC\\
    \forall a^l\in\Atms[l],\forall {\frak{rl}}\in\STerms^l,
    &\sem {a^l}({\frak{rl}})
    &\eqdef\Values[+]
  \end{array}
  \end{array}
\]
\end{definition}

\begin{remark}Notice that 
  $\machine[\rho] t\Rew{\headred[1],\headred[2]}\SemTe t \rho$.
\end{remark}

\begin{theorem}[Head-normalisation of an abstract machine]

  We assume the following hypotheses:
  \begin{enumerate}
  \item[Well-foundedness:]\\
    The \LAF\ instance is well-founded.
  \item[Typing correlation:]\\
    If $\rho\in\SemTy\Gamma$ and $\frak \Delta\in\SemTy{(\Delta^l,{\bf r})}$ then $(\rho\Cextend\frak\Delta)\in\SemTy{\Gamma\Textend{\Delta^l}}$.
  \end{enumerate}

  We conclude that, for all $\rho\in\SemTy\Gamma$, if $\Der\Gamma {{t}
    {}}{}$ then $\machine[\rho] t$ head-normalises.
\end{theorem}
\begin{proof}
Stability is obvious for the head-normalisation model:

Assume $\SemTe{f(p)}{\rho\Cextend d}\in\orth{}{}$. Following
the previous remark, this entails that
$\machine[{\rho\Cextend d}]{f(p)}$
head-normalises. Hence, $\cutc{\closure[\rho]{\THO
    f}}{pd}$ head-normalises, which is literally what
$\in\orth{\SemTe{f}\rho}{\spat p(d)}$ means.

So we can apply the Adequacy Lemma (Lemma~\ref{lem:adequacy}), and
obtain that $\SemTe t \rho$ head-normalises, from which get that
$\machine[\rho] t$ head-normalises.
\end{proof}

We now see how this applies to a syntactic abstract machine.  Assume
we have a well-founded \LAF\ instance, and a syntactic abstract
machine for it that features identity evaluation contexts, \ie a
family of contexts ${\sf id}$ satisfying $\varRead[x^+]{{\sf
    id}}=x^+$ and $\varRead[x^-]{{\sf id}}=x^-$.

\begin{corollary}[Head normalisation]
  Assume that the evaluation context algebra $\Contexts$ and $\TContexts$ satisfy the typing correlation.

  If $\Der\Gamma {{t} {}}{}$ then $\machine[{\sf id}] t$ head-normalises.\footnote{for the evaluation context ${\sf id}$ with $\domP{{\sf id}}=\domP\Gam$ and $\domP{{\sf id}}=\domP\Gam$}
\end{corollary}
\begin{proof}
  The valuation $\Cfune{{\sf id}}$ is in $\sem{\Cfune\Gamma}=\ValuesSC$.

  Every positive label $x^+$ is in $\SemTy[\sigma]{(a^l,{\bf
      r})}=\Values[+]=\var[+]$ (for every $\sigma$, $a^l$ and ${\bf r}$).

  Every negative label $x^-$ is in $\SemTyN[\sigma]{(M^l,{\bf r})}$
  (for every $\sigma$, $M^l$ and ${\bf r}$),\\
  since $x^-$ is in
  $\SNeg=\Values[-]=\var[-]\cup\Closures$ and $\orth{x^-}{pd}$ for
  all $(p,d)\in\SemTyP[\sigma]{(M^l,{\bf r})}$.\footnote{Indeed,
    $\cutc{x^-}{pd}$ is head-normalising since it cannot be reduced by the
    abstract machine.}
  
  Hence, the identity evaluation context ${\sf id}$ is in $\SemTy\Gamma$.
\end{proof}

In particular, \LAF[K1], \LAF[K2], \LAF[J] are all head normalising.

\section{Re-using proofs}
\label{sec:pruning}

Now, in order to have strong normalisation, and even just
cut-elimination itself, our notion of abstract machine above is too
weak, as it only (and deterministically) performs ``head reduction''.

A state of a syntactic machine such as $\cutc{v}{p\, d}$ could almost
be read back as a real command, if only we could \emph{compute}
closures such as $\closure [\rho]f$, which we never do: just as in
the weak reduction in $\lambda$-calculus, we never propagate the
evaluation context $\rho$ (which can be seen as a substitution) into
$f$ (in other words propagate it under the abstraction represented by
the meta-level function $f$).

We could compute a closure $\closure [\rho]f$ as a function
$\halfsubst[\rho]f\col\Data\pfunspace\PTerms$ such that
\[\halfsubst[\rho]f(p) = \halfsubst[\rho'\Cextend {\sf idd}]{f(p)}\]
with the recursively defined propagation of an evaluation context $\rho$
into a command $c$ denoted $\halfsubst[\rho]c$, and where
\begin{itemize}
\item ${\sf idd}$ is an ``identity decomposition term'', to create
  identity bindings for the labels in $f(p)$ introduced by the
  application of $f$ to $p$;
\item $\rho'$ is the update of $\rho$, providing the same bindings as
  $\rho$ but taking care that the labels might have changed after the
  context extension with ${\sf idd}$.
\end{itemize}
But so far a \LAF\ instance does not tell us how to infer $\sf idd$
and $\rho'$ from $\rho$.

This is exactly the same situation as with the eigenlabels for which a
\LAF\ instance with eigenlabels provided two functions
$\Cst{\Gamma}{\Delta}$ and $\Crename{\Gamma}{\Delta}$ to do exactly
that.

We therefore enrich the concept of a \LAF\ instance with eigenlabels
as follows:

\begin{definition}[\LAF\ instance with explicit label updates]

  A \LAF\ \Index{instance with explicit label updates} is given by the
  following tuple:
  \[(\Sorts,\vare,\Terms,\decs{},\Atms,\Moles,\atmEq{}{},\var[+],\var[-],\TContexts,\RCA,\Data,\decf{}, 
  \Crename{\mathcal V}{\Delta},\Cst{\mathcal
    V}{\Delta})\]
  whose components are exactly as in the definition of a
  \LAF\ instance with eigenlabels, except that
  \begin{itemize}
  \item The map operation of the typing context algebra $\TContexts$ satisfies the following property:

    For all $f_1:\IAtms\rightarrow\IAtms$ and
    $f_2:\IMoles\rightarrow\IMoles$, all
    $(\IAtms,\IMoles,\Sorts)$-decompositions $\Delta$ and $\Delta'$,
    and all typing contexts $\Gamma$,

    If $\Delta$ relates to $\Delta'$ according to $\{(a,f_1(a))\mid
    a\in\IAtms\}$ $\{(m,f_2(m))\mid m\in\IMoles\}$ and the identity
    relation on sorts,\\
    then 
    $\CMap{f_1,f_2}(\Gamma\Cextend\Delta)$ relates to 
    $(\CMap{f_1,f_2}\Gamma)\Cextend{\Delta'}$ according to the identity relations.

  \item There is a $(\var[+],\var[-],\vare)$-context algebra $\RCA$
    called the \Index{renaming context algebra}, and equipped with a
    \Index{renaming composition} that combines two renaming contexts
    $\pi$ and $\pi'$ into $\CwR{\pi}{\pi'}$ so that
    $\varRead[x]{\CwR{\pi}{\pi'}}=\varRead[{\varRead[x]{{\pi'}}}]{\pi}$
    (\resp
    $\varRead[x^+]{\CwR{\pi}{\pi'}}=\varRead[{\varRead[x^+]{{\pi'}}}]{\pi}$
    and
    $\varRead[x^-]{\CwR{\pi}{\pi'}}=\varRead[{\varRead[x^-]{{\pi'}}}]{\pi}$);
  \item we require the naming policies $\Crename{\mathcal V}{\Delta}$ and $\Cst{\mathcal
      V}{\Delta}$ to give information not only on eigenlabels, but
    also on positive and negative labels:
    \[
    \left(\begin{array}{c@{\quad}l@{\quad}l}
      \powerset{\var}\times\Dstruct
      &\rightarrow
      &\RCA\\
      (\mathcal V,\Delta)&\mapsto&\Crename{\mathcal V}{\Delta}
    \end{array}\right)
    \qquad
    \left(\begin{array}{c@{\quad}l@{\quad}l}
      \powerset{\var}\times\Dstruct
      &\rightarrow
      &\DecompType[{\var[+],\var[-],\vare}]\\
      (\mathcal V,\Delta)&\mapsto&\Cst{\mathcal V}{\Delta}
    \end{array}
    \right)
    \]

    Clearly, we can extract from those naming policies the policies in
    the sense of Definition~\ref{def:LAFwE} (with types
    \(
    \left(\powerset{\var}\times\Dstruct\rightarrow(\vare\rightarrow\vare)\right)
    \) and 
    \(\left(\powerset{\var}\times\Dstruct\rightarrow \DecompType[{\unit,\unit,\vare}]\right)
    \)).

    Finally, we require that $\RCA$ respect those naming policies.
  \end{itemize}
\end{definition}

\begin{remark}
  It is straightforward to define $\RCA$,
  $\Crename{\domAll{\Gamma}}{\abs\Delta}$, and
  $\Cst{\domAll{\Gamma}}{\abs\Delta}$ in \LAF[K1] and \LAF[K2]
  to make them \LAF\ instances with explicit label updates.
\end{remark}

With this information, we can now properly define the free labels of a
proof-term, something which we surprinsingly did not need so far, but
that will indicate which parts of a typing environment are actually
used in a proof.

\begin{definition}[Free labels]
  The free labels of a positive term (\resp decomposition term,
  command) that is typed in a typing context $\Gamma$, are defined by
  the rules of Fig.~\ref{fig:freelabels}.
\end{definition}
\begin{bfigure}[!h]
  \[
  \begin{array}{lll@{\recdef}l}
    \FL{pd}&\eqdef \FL{d}\\\\

    \FL{x^+}&\eqdef\{x^+\}\\
    \FL{\THO f}&\eqdef \bigcup_{p\in\dom f}\pi^{-1}(\FL{f(p)})\\
    \FL\Tunit&\eqdef\emptyset\\
    \FL{d_1\Tand d_2}&\eqdef \FL{d_1}\cup\FL{d_2}\\
    \FL{\Tex r  d}&\eqdef\FL{r}\cup\FL{d}\\\\
    \FL{\cutc{x^-}{t^+}}&\eqdef\{x^-\}\cup\FL{t^+}\\
    \FL{\cutc{f}{t^+}}&\eqdef\FL{\THO f}\cup\FL{t^+}\\\strut
  \end{array}
  \]
  where $\pi$ is the function in
  $(\var[+]\rightarrow\var[+])\cup(\var[-]\rightarrow\var[-])\cup(\vare\rightarrow\vare)$
  mapping every $x^+\in\var[+]$ to
  $\varRead[x^+]{\Crename{\domAll{\Gamma}}{\abs p}}$ (\resp
  $x^-\in\var[-]$ to $\varRead[x^-]{\Crename{\domAll{\Gamma}}{\abs p}}$,
  and $x\in\vare$ to $\varRead[x]{\Crename{\domAll{\Gamma}}{\abs p}}$).
  \caption{Free labels}
  \label{fig:freelabels}
\end{bfigure}

Knowing what free variables are, we are now able, given a proof of a
sequent, to replay the proof for any other sequent whose typing
context contains the atoms and molecules that type the free variables
of the original proof.

For this we define the renaming of a term:

\begin{definition}[Renaming]
  The renaming, denoted $\rename\pi{t^+}$ (\resp $\rename\pi{d}$,
  $\rename\pi{t}$), by a renaming context $\pi$, of a positive term
  (\resp decomposition term, command) that is typed in a typing
  context $\Gamma$, is defined by the rules of
  Fig.~\ref{fig:renaming}.
\end{definition}
\begin{bfigure}[!h]
  \[
  \begin{array}{lll@{\recdef}l}
    \rename\pi{pd}&\eqdef p(\rename\pi{d})\\\\

    \rename\pi{x^+}&\eqdef\varRead[x^+]\pi\\
    \rename\pi{\THO f}&\eqdef p\mapsto \rename{\left(\left(\CwR{\Crename{\domAll{\Gamma}}{\abs p}}\pi\right)\Cextend\Cst{\domAll{\Gamma}}{\abs p}\right)}{f(p)}\\
    \rename\pi\Tunit&\eqdef\Tunit\\
    \rename\pi{(d_1\Tand d_2)}&\eqdef (\rename\pi{d_1})\Tand(\rename\pi{d_2})\\
    \rename\pi{(\Tex r  d)}&\eqdef\Tex{\Cfune\pi(r)}{(\rename\pi{d})}\\\\
    \rename\pi{\cutc{x^-}{t^+}}&\eqdef\cutc{\varRead[x^-]\pi}{(\rename\pi{t^+})}\\
    \rename\pi{\cutc{f}{t^+}}&\eqdef\cutc{(\rename\pi{\THO f})}{(\rename\pi{t^+})}
  \end{array}
  \]
  \caption{Renaming}
  \label{fig:renaming}
\end{bfigure}

In the renaming of a function $f$, $\CwR{\Crename{\domAll{\Gamma}}{\abs
    p}}\pi$ updates the co-domain of $\pi$ as we went ``through a
binding'', and composing with $\Cst{\domAll{\Gamma}}{\abs p}$ adds the
``identity bindings'' for the labels introduced by the application of
$f$ to $p$.

Now as mentioned before, when we have a proof for a particular
sequent, we want to identify when it can be replayed for another
sequent. For this we define what it means for a typing context
$\Gamma'$ to at least contain the instantiated atoms and molecules of
a typing context $\Gamma$: this is done by identifying a renaming
$\pi$ that will map the labels in $\Gamma'$ to some labels in $\Gamma'$
that have the same type.

\begin{definition}[Context embedding]

  We say that $\Gamma$ embeds into $\Gamma'$ along a renaming context
  $\pi$, written $\Gamma\rembeds[\pi]\Gamma'$, if for all $x$ (\resp $x^+$, $x^-$) in $\domE\pi$ (\resp
  $\domP\pi$, $\domN\pi$) we have
  $\varRead[x]\Gamma=\varRead[{\varRead[x] \pi}]{\Gamma'}$ (\resp $\varRead[x^+]\Gamma=\varRead[{\varRead[x^+] \pi}]{\Gamma'}$, $\varRead[x^-]\Gamma=\varRead[{\varRead[x^-] \pi}]{\Gamma'}$).
\end{definition}

Notice that the domain of $\pi$ might be smaller than that of
$\Gamma$, so that $\pi$ does not necessarily map \emph{every} label
declared in $\Gamma$. This is a feature (rather than a bug) that will
allow us to ignore those instantiated atoms and molecules in $\Gamma$
that are not used in the proof that we want to replay (the renaming
$\pi$ may only be defined on those labels that are free in the
proof-term).

\begin{theorem}[Replaying a proof]\label{th:renaminglemma}
  Assume the following property:
  \begin{itemize}
  \item[Renaming correlation:]\\ For all $\Gamma$, $\Gamma'$, $\pi$,
    if $\Gamma\rembeds[\pi]\Gamma'$ then
    $\Gamma\Cextend\Delta\rembeds[\pi']\Gamma'\Cextend\Delta$, where
    $\pi' = \left(\CwR{\Crename{\domAll{\Gamma}}{\abs
        p}}\pi\right)\Cextend\Cst{\domAll{\Gamma}}{\abs p}$.
  \end{itemize}

  We conclude that, for all $\pi\in\RCA$ such that $(\CMap{\Cfune\rho,\Cfune\rho}\Gamma)\rembeds[\pi]\Gamma'$,
  \begin{enumerate}
  \item if $\FL{t^+}\subseteq\dom\pi$ and $\DerF\Gamma {\XcolY {t^+} {(M^l,{\bf r})}}{}$ then $\DerF{\Gamma'} {\XcolY {(\rename\pi{t^+})} {(M^l,{\bf r})}}{}$
  \item if $\FL{d}\subseteq\dom\pi$ and $\Der\Gamma {\XcolY {d} {(\Delta^l,{\bf r})}}{}$ then $\Der{\Gamma'} {\XcolY {(\rename\pi d)} {(\Delta^l,{\bf r})}}{}$
  \item if $\FL{t}\subseteq\dom\pi$ and $\Der\Gamma {{t} {}}{}$ then $\Der{\Gamma'} {{(\rename\pi t)} {}}{}$
  \end{enumerate}
\end{theorem}
\begin{proof}See the \Coq\ proof~\cite{LengrandHDRCoq}.\end{proof}

A \LAF\ instance with explicit label updates thus allows us to apply a
renaming to a proof to get a proof of a new sequent. This will be used
heavily in an implementation of proof-search that memoises proofs in
order to paste them as often as possible.

\ignore
{\subsection{Substituting proofs}

Now we can also perform the substitution, not of labels for labels,
but of terms for labels, so as to compute closures into real functions
as described in the introduction of this section:

\begin{definition}[Substitution]
  Consider an
  $(\var[+],\var[-]\cup(\Data\pfunspace\PTerms),\Terms)$-context
  algebra $\SCA$ called the \Index{substitution context algebra},
  equipped with a \Index{renaming operation} that associates, to a
  renaming context $\pi$ and a substitution context $\rho$, a
  substitution context $\CwR{\pi}\rho$ such that 
  \begin{itemize}
  \item $\varRead[x]{\CwR{\pi}\rho}=\Cfune\pi({\varRead[x]{\rho}})$ for all $x\in\domE\rho$
  \item $\varRead[x^+]{\CwR{\pi}\rho}=\varRead[{\varRead[x^+]{\rho}}]{\pi}$ for all $x^+\in\domP\rho$
  \item $\varRead[x^+]{\CwR{\pi}\rho}=\rename\pi{(\varRead[x^-]{\rho})}$ for all $x^-\in\domN\rho$
  \end{itemize}

  The substitution, denoted $\substitute\rho{t^+}$ (\resp $\substitute\rho{d}$,
  $\substitute\rho{t}$), by a renaming context $\pi$, of a positive term
  (\resp decomposition term, command) that is typed in a typing
  context $\Gamma$, is defined by the rules of
  Fig.~\ref{fig:substitution}.
\end{definition}
\begin{bfigure}[!h]
  \[
  \begin{array}{lll@{\recdef}l}
    \substitute\rho{pd}&\eqdef p(\substitute\rho{d})\\\\

    \substitute\rho{x^+}&\eqdef\varRead[x^+]\pi\\
    \substitute\rho{\THO f}&\eqdef p\mapsto \substitute{\left(\left(\CwR{\Crename{\dom{\Gamma}}{\abs p}}\rho\right)\Cextend\Cst{\dom{\Gamma}}{\abs p}\right)}{f(p)}\\
    \substitute\rho\Tunit&\eqdef\Tunit\\
    \substitute\rho{(d_1\Tand d_2)}&\eqdef (\substitute\rho{d_1})\Tand(\substitute\rho{d_2})\\
    \substitute\rho{(\Tex r  d)}&\eqdef\Tex{\Cfune\pi(r)}{(\substitute\rho{d})}\\\\
    \substitute\rho{\cutc{x^-}{t^+}}&\eqdef\cutc{\varRead[x^-]\pi}{(\substitute\rho{t^+})}\\
    \substitute\rho{\cutc{f}{t^+}}&\eqdef\cutc{(\substitute\rho{\THO f})}{(\substitute\rho{t^+})}
  \end{array}
  \]
  \caption{Substitution}
  \label{fig:substitution}
\end{bfigure}

Now we generalise the notion of context embedding:

\begin{definition}[Context embedding]

  We say that $\Gamma$ embeds into $\Gamma'$ along a substitution context
  $\rho$, written $\Gamma\sembeds[\rho]\Gamma'$, if
  \begin{itemize}
  \item for all $x$ in $\domE\rho$
    we have
    $\Ders{\Cfune{\Gam'}}{\XcolY {\varRead[x] \rho} {\varRead[x]\Gamma}}$
  \item for all $x^+$ in $\domP\rho$ we have
    $\varRead[{\varRead[x^+] \rho}]{\Gamma'}=\Cfune\rho(\varRead[x^+]\Gamma)$
  \item for all $x^-$ in $\domN\rho$ we have
    \begin{itemize}
    \item if $\varRead[x^-] \rho = y^-\in\var[-]$ then $\varRead[{\varRead[x^+] \rho}]{\Gamma'}=\Cfune\rho(\varRead[x^+]\Gamma)$
    \item if $\varRead[x^-] \rho = f\in\Data\pfunspace\Terms$ then
    $\Der{\Gam'} {\XcolY {\THO f} {(\Drefute M^l,\Cfune\rho({\bf r}))}}$, where $(M^l,{\bf r})={\varRead[x^-]\Gamma}$
  \end{itemize}
  \end{itemize}
\end{definition}

Again, notice that the domain of $\rho$ might be smaller than that of
$\Gamma$.

\begin{definition}[Substitution Lemma]
  Assume the following property:
  \begin{itemize}
  \item[Substitution correlation:]\\ For all $\Gamma$, $\Gamma'$,
    $\rho$, if $\Gamma\sembeds[\rho]\Gamma'$ then
    $\Gamma\Cextend\Delta\sembeds[\rho']\Gamma'\Cextend\Delta$, where
    $\rho' = \left(\CwR{\Crename{\dom{\Gamma}}{\abs
      p}}\rho\right)\Cextend\Cst{\dom{\Gamma}}{\abs p}$.
  \end{itemize}

  We conclude that, for all $\rho\in\RCA$ such that $\Gamma\sembeds[\rho]\Gamma'$,
  \begin{enumerate}
  \item if $\FL{t^+}\subseteq\dom\rho$ and $\DerF\Gamma {\XcolY {t^+} {(M^l,{\bf r})}}{}$ then $\DerF{\Gamma'} {\XcolY {(\substitute\rho{t^+})} {(M^l,{\bf r})}}{}$
  \item if $\FL{d}\subseteq\dom\rho$ and $\Der\Gamma {\XcolY {d} {(\Delta^l,{\bf r})}}{}$ then $\Der{\Gamma'} {\XcolY {(\substitute\rho d)} {(\Delta^l,{\bf r})}}{}$
  \item if $\FL{t}\subseteq\dom\rho$ and $\Der\Gamma {{t} {}}{}$ then $\Der{\Gamma'} {{(\substitute\rho t)} {}}{}$
  \end{enumerate}
\end{definition}
\begin{proof}See the \Coq\ proof for the quantifier-free case~\cite{LengrandHDRCoq}.\end{proof}

}

\section{Cut-elimination}
\label{sec:cut-elimination}

We now show how to use substitution and the substitution lemma to
define a normalisation procedure, in a \LAF\ instance with explicit label
updates, to produce cut-free terms.

\begin{definition}[Normalisation]
  We take a syntactic abstract machine, whose evaluation context
  algebra is equipped with a \Index{renaming operation} that
  associates, to a renaming context $\pi$ and an evaluation context
  $\rho$, an evaluation context $\CwR{\pi}\rho$ such that
  \begin{itemize}
  \item for all $x\in\domE\rho$, we have $\varRead[x]{\CwR{\pi}\rho}=\Cfune\pi({\varRead[x]{\rho}})$ 
  \item for all $x^+\in\domP\rho$, we have $\varRead[x^+]{\CwR{\pi}\rho}=\varRead[{\varRead[x^+]{\rho}}]{\pi}$ 
  \item for all $x^-\in\domN\rho$, we have
    \begin{itemize}
    \item if $\varRead[x^-]{\rho}=y^-$ then $\varRead[x^-]{\CwR{\pi}\rho}=\varRead[y^-]{\pi}$
    \item if $\varRead[x^-]{\rho}=\closure[\rho']f$ then $\varRead[x^-]{\CwR{\pi}\rho}= \closure[\CwR\pi{\rho'}]{f}$
    \end{itemize}
  \end{itemize}

  We define the \Index{big-step semantics} of the \LAF\ instance with
  explicit label updates as two relations
  \begin{itemize}
    \item one denoted $d\NormTo d'$ between an evaluation
      decomposition $d$ and a cut-free decomposition term $d'$,
    \item one denoted $\machine[\rho] t \NormTo t'$ between a
      contextualised command $\machine [\rho]t$ and a cut-free command
      $t'$,
  \end{itemize}
  defined by simultaneous induction by the rules of
  Fig.~\ref{fig:cut-elim}.

  We say that a contextualised command $\machine[\rho]t$ (\resp an
  evaluation decomposition $d$) \Index[normalise]{normalises} if there
  is some $t'$ such that $\machine[\rho] {t}\NormTo t'$ (\resp some
  $d'$ such that ${d}\NormTo d'$). Notice in that case
  that $t'$ (\resp $d'$) is cut-free.  We also say that an evaluation
  triple $\cutc{v}{pd}$ \Index[normalise]{normalises} if
  $\cutc{v}{pd}\Rew{\headred[123]}\cutc{x^-}{p'd'}$ and $d'$
  normalises.
\end{definition}
\begin{bfigure}[!h]
  \[\begin{array}c
  \infer{\closure[\rho] f\NormTo f'}{
    \forall p\in\dom f,\qquad\machine[{\left(\left(\CwR{\Crename{\domAll{\rho}}{\abs p}}\rho\right)\Cextend\Cst{\domAll{\rho}}{\abs p}\right)}] {f(p)}\NormTo  f'(p)
  }
  \qquad  
  \infer{x^-\NormTo x^-}{\strut}  
  \\\\
  \infer{x^+\NormTo x^+}{\strut}  
  \qquad
  \infer{\Tunit\NormTo \Tunit}{\strut}
  \qquad  
  \infer{{d_1}\Tand{d_2}\NormTo {d'_1}\Tand{d'_2}}{
    d_1\NormTo d'_1
    \qquad
    d_2\NormTo d'_2}
  \qquad
  \infer{\Tex r d\NormTo \Tex r {d'}}{d\NormTo d'}\\\\
  \hline\\
  \infer{\machine[\rho] {t}\NormTo  \cutc {x^-}{pd'}}{
    \machine[\rho] {t}\Rewn{\headred[123]} \cutc{x^-}{p d}
    \qquad
    d\NormTo d'
  }
  \end{array}
  \]
  \caption{Cut-elimination}
  \label{fig:cut-elim}
\end{bfigure}

In order to show that this forms a cut-elimination procedure, we need to
\begin{itemize}
\item give typing rules for contextualised commands, evaluation decomposition, and evaluation triples;
\item show that $\Rewn{\headred[123]}$ satisfies Subject Reduction with these rules;
\item show that every typed contextualised command normalises. 
\end{itemize}

\begin{definition}[Typing the elements of a syntactic abstract machine]

  The typing rules for the elements of a syntactic abstract machine
  are given in Fig.~\ref{def:LAFext}, where $\Der[\mbox{\scriptsize
      \LAF}]{}{}$ denotes the derivability of sequents in
  \LAF\ (Fig.~\ref{def:LAF}).
\end{definition}
\begin{bfigure}[!h]
  \[
  \begin{array}{c}
    \mbox{Evaluation decompositions}\hfill
    \\
    \infer{\Der\Gam{\XcolY{\Tunit}{(\Dunit,\lv r)}}}{\strut}
    \qquad
    \infer{\Der\Gam{\XcolY{ d_1\Tand d_2}{((\Del_1\Dand\Del_2),\lv r)}}}{
      \Der\Gam{\XcolY{ d_1}{(\Del_1,\lv r)}}
      \quad
      \Der\Gam{\XcolY{ d_2}{(\Del_2,\lv r)}}
    }
    \qquad
    \infer{\Der\Gam{\XcolY{\Tex {r'} d}{\Dex[s]{} {(\Del,\lv r)}}}}{
      \Ders{\Cfune\Gam}{\XcolY {r'} s}
      \quad
      \Der\Gam{\XcolY{ d}{(\Del,\cons {r'}{\lv r})}}
    }
    \\\\
    \infer{\Der{\Gam}{\XcolY{x^+}{(a,\lv r)}}}{
      \atmEq{\varRead[x^+]\Gam} {(a,\lv r)}
    }
    \qquad
    \infer{\Der\Gam {\XcolY {\THO {\closure[\rho]f}} {(\Drefute M,\lv r)}}}{
      \Der{\Gam}{\XcolY \rho {\Gam'}}
      \quad
      \Der[\mbox{\scriptsize \LAF}]{\Gam'} {\XcolY {\THO f} {(\Drefute M,\lv r)}}
    }
    \\
    \midline
    \mbox{Evaluation triples}\hfill
    \\
    \infers{\Der\Gam{\cutc{x^-}{pd}}}[{\varRead[x^-]\Gam}=(M,\lv r)]{
      \DerDec{}{\Delta}{\XcolY p M}
      \quad
      \Der{\Gamma}{\XcolY  d{(\Delta,\lv r)}}
    }
    \\\\
    \infer{\Der\Gam{\cutc{\closure[\rho]f}{pd}}}{
      \Der\Gam {\XcolY {\closure[\rho]f}{(\Drefute M,\lv r)}}
      \quad
      \DerDec{}{\Delta}{\XcolY p{M}}
      \quad
      \Der{\Gamma}{\XcolY  d{(\Delta,\lv r)}}
    }
    \\
    \midline
    \mbox{Contextualised commands}\hfill
    \\
    \infer{\Der\Gam {\machine[\rho]t}}{
      \Der{\Gam}{\XcolY \rho {\Gam'}}
      \quad
      \Der[\mbox{\scriptsize \LAF}]{\Gam'} {t}
    }
    \\
    \midline
    \mbox{Evaluation contexts}\hfill
    \\
    \infers{\Der{\Gam}{\XcolY \rho {\Gam'}}}{
      \begin{array}c
      (\forall x\in\domE{\Gam'},\quad  \Ders{\Cfune\Gam}{\XcolY{\varRead[x]\rho}{\varRead[x]{\Gam'}}})\\
      (\forall x^+\in\domP{\Gam'},\quad \atmEq{\varRead[{\varRead[x^+]\rho}]\Gam} {\varRead[x^+]{\Gam'}}\\
      \left(
      \begin{array}{lll}
        \forall x^-\in\domP{\Gam'},\\
        \qquad\mbox{either } \varRead[y^-]\Gamma=\varRead[x^-]{\Gam'}&\mbox{ if  }\varRead[x^-]\rho=y^-\\
        \qquad \mbox{ or } \Der\Gam{\XcolY{\closure[\rho']f}{(\Drefute M,{\bf r})}}&\mbox{ if  }\varRead[x^-]\rho=\closure[\rho']f\mbox{ and }\varRead[x^-]{\Gam'}=(M,{\bf r})
      \end{array}
      \right)
      \end{array}
    }
  \end{array}
  \]
  \caption{Typing a syntactic abstract machine}
  \label{def:LAFext}
\end{bfigure}

\begin{theorem}[Subject Reduction]\label{th:SRLAFwelu}
  \begin{enumerate}
    \item If $\Der\Gam {\machine[\rho]t}$ and $\machine[\rho]t\Rew{\headred[1],\headred[2]}\cutc{v}{pd}$ then $\Der\Gam {\cutc{v}{pd}}$.
    \item If $\Der\Gam {\cutc{v}{pd}}$ and $\cutc{v}{pd}\Rew{\headred[3]}\machine[\rho]t$ then $\Der\Gam {\machine[\rho]t}$.
    \item If $\Der\Gam {\XcolY d{(\Del,{\bf r})}}$ and $d\NormTo d'$ then $\Der[\mbox{\scriptsize \LAF}]\Gam {\XcolY {d'}{(\Del,{\bf r})}}$.
    \item If $\Der\Gam {\machine[\rho]t}$ and $\machine[\rho]t\NormTo t'$ then $\Der[\mbox{\scriptsize \LAF}]\Gam {t'}$.
  \end{enumerate}
\end{theorem}
\begin{proof}
  The first two points are given by a simple rearrangement of the sub-derivation trees.
  The last two points are proved by induction on the normalisation derivations.
\end{proof}

Finally, we adapt the realisability model for head normalisation
(Definition~\ref{def:model4headnorm}) to prove cut-elimination:

\begin{definition}[A realisability model for normalisation]\label{def:model4norm}

  The \Index{normalisation model} for this syntactic abstract machine is
  \[
  \begin{array}{l}
  \begin{array}{ll}
    \SSoContexts&\eqdef\ValuesSC\\ 
    \STerms&\eqdef\Terms\\ 
    \SPrim&\eqdef\var[+]\\ 
    \SPos&\eqdef\Data\times\DecompType[{\var[+]},{\var[-]\cup\Closures},\Terms]\\ 
    \SNeg&\eqdef\var[-]\cup\Closures\\ 
    \orth
    v {(p,d)}&\mbox{if the evaluation triple }\cutc{v}{p d} \mbox{ normalises}\\
    \SContexts&\eqdef \Contexts
  \end{array}\\
    \begin{array}{llll}
    \forall \frak{\Delta}\in\SDecs,&\spat p(\frak{\Delta})&\eqdef (p,\frak{\Delta})\\
    \forall r\in\Terms, \forall \sigma\in\SSoContexts,&\sem[\sigma] r&\eqdef\halfsubst[\sigma]r\\
    \forall f:\Data\pfunspace\PTerms, \forall \rho\in\SContexts,& \sem[\rho] f&\eqdef \closure[\rho]{f}\\
    \forall s\in\Sorts,& \sem s&\eqdef\Terms\\
    \forall \Sigma\in\SoContexts,& \sem \Sigma&\eqdef\ValuesSC\\
    \forall a^l\in\Atms[l],\forall {\frak{rl}}\in\STerms^l,
    &\sem {a^l}({\frak{rl}})
    &\eqdef\var[+]
  \end{array}
  \end{array}
\]
\end{definition}

\begin{remark}

  Notice this is the same definition as
  Definition~\ref{def:model4headnorm}, except for the orthogonality
  relation which we have strengthened by requiring normalisation
  instead of head-normalisation.

  Of course we still have $\machine[\rho] t\Rew{\headred[12]}\SemTe t
  \rho$.
\end{remark}

\begin{theorem}[Normalisation of a syntactic abstract machine]

  We assume the following hypotheses:
  \begin{enumerate}
  \item[Well-foundedness:]\\
    The \LAF\ instance is well-founded.
  \item[Correlation with eigenlabels:]\\
    $\SContexts$ and $\TContexts$ satisfy the correlation with eigenlabels property.
  \end{enumerate}

  We conclude that, for all $\rho\in\SemTy\Gamma$, if $\Der\Gamma {{t}
    {}}{}$ then $\machine[\rho] t$ normalises.
\end{theorem}
\begin{proof}
Stability is obvious for the normalisation model:

Assume $\SemTe{f(p)}{\rho\Cextend d}\in\orth{}{}$. Following the
previous remark, this entails that $\machine[{\rho\Cextend d}]{f(p)}$
normalises. Hence, $\cutc{\closure[\rho]{\THO f}}{pd}$ normalises,
which is literally what $\in\orth{\SemTe{f}\rho}{\spat p(d)}$ means.

So we can apply the Adequacy Lemma (Lemma~\ref{lem:adequacyWeigen}), and
obtain that $\SemTe t \rho$ normalises, from which get that
$\machine[\rho] t$ normalises.
\end{proof}

Again, assume we have a \LAF\ instance with explicit label updates, a
syntactic machine for it that features identity evaluation contexts,
\ie a family of contexts ${\sf id}$ satisfying $\varRead[x^+]{{\sf
    id}}=x^+$, $\varRead[x^-]{{\sf id}}=x^-$ and $\varRead[x]{{\sf
    id}}=x$.

\begin{lemma}[Normalisation and renaming]

  If $d$ (\resp $t$) normalises then $\rename \pi d$ (\resp $\rename \pi t$) normalises.
\end{lemma}
\begin{proof}
  By induction on the normalisation derivation.
\end{proof}



\begin{lemma}[Evaluation decompositions in the model are normalising]
  \begin{enumerate}
  \item For all typing decomposition $\Delta^l$, for all ${\bf r}$,
    $\sigma$ and all $d\in\DecompType[{\var[+],\var[-],\vare}]$ with
    the same structure as $\Delta^l$,
    $d\in\SemTy[\sigma]{(\Delta^l,{\bf r})}$.
  \item For all molecules $M^l$, for all ${\bf r}$, $\sigma$ and all
    $\closure[\rho]f$ in $\SemTyN[\sigma]{(M^l,{\bf r})}$, $\closure[\rho]f$ normalises.
  \item For all typing decomposition $\Delta^l$, for all ${\bf r}$,
    $\sigma$ and all $d$ in $\SemTy[\sigma]{(\Delta^l,{\bf r})}$,
    $d$ normalises.
  \item For all molecules $M^l$, for all ${\bf r}$, $\sigma$ and all
    negative labels $x^-$, $x^-\in \SemTyN[\sigma]{(M^l,{\bf r})}$.
  \end{enumerate}
\end{lemma}
\begin{proof}
  By simultaneous induction on $\Delta^l$ and $M^l$, using the
  well-founded property of the \LAF\ instance.

  For point 1: by induction on $\Delta^l$, the base case being point 4.

  For point 2: by unfolding the definition of $\SemTyN[\sigma]{(M^l,{\bf r})}$,\\ 
  $\closure[\CwR{\Crename{\domAll{\rho}}{\abs p}}\rho]f$ is orthogonal to $(p,\Cst{\domAll{\rho}}{\abs p})$ (using point 1).

  For point 3: by induction on $\Delta^l$, the base case being point 2.

  For point 4: for all
  $(p,d)\in\SemTyP[\sigma]{(M^l,{\bf r})}$, we have the evaluation
  decomposition $d$ in some $\SemTy[\sigma]{(\Delta^l,{\bf r})}$, and
  by point 3 $d$ normalises; hence $\cutc{x^-}{pd}$ 
  normalises (\ie $\orth{x^-}{pd}$), so $x^-$ is in
  $\SemTyN[\sigma]{(M^l,{\bf r})}$.
\end{proof}

\begin{corollary}[Cut-elimination]
  Assume that the evaluation context algebra $\Contexts$ and
  $\TContexts$ satisfy the correlation with eigenlabels property.

  If $\Der\Gamma {{t} {}}{}$ then $\machine[{\sf id}] t$
  normalises.\footnote{for an identity evaluation context ${\sf
      id}$ with the same domains as $\Gamma$}

  Therefore the \LAF\ instance with explicit label updates admits
  cuts.
\end{corollary}
\begin{proof}
  Every eigenlabel $x$ is in $\sem s=\ValuesE=\Terms$ (for every $s$).

  Every positive label $x^+$ is in $\SemTy[\sigma]{(a^l,{\bf
      r})}=\Values[+]=\var[+]$ (for every $\sigma$, $a^l$ and ${\bf r}$).

  Every negative label $x^-$ is in $\SemTyN[\sigma]{(M^l,{\bf r})}$ (previous lemma).
  
  Hence, the identity evaluation context ${\sf id}$ is in
  $\SemTy\Gamma$, and we can apply the previous theorem.

  Combined with Subject Reduction (Theorem~\ref{th:SRLAFwelu}), we can
  transform every proof with cuts into a cut-free proof.
\end{proof}

In particular, \LAF[K1] and \LAF[K2] admit cuts.

\section{Conclusion and further work: Strong normalisation}
\label{sec:SN}

Now, we have proved cut-elimination but not strong normalisation, as
we have used a big-step operational semantics to reduce proof-terms to
cut-free forms, but we still have not defined a non-deterministic
reduction relation for which strong normalisation might be
interesting. For this we would definitely need to compute closures
(which we still have avoided so far), by pushing down evaluation
contexts with the rules of Fig.~\ref{fig:substitution}.
\begin{bfigure}[!h]
  \[
  \begin{array}{lll@{\recdef}l}
    \substitute\rho{pd}&\eqdef p(\substitute\rho{d})\\\\

    \substitute\rho{x^+}&\eqdef\varRead[x^+]\pi\\
    \substitute\rho{\THO f}&\eqdef p\mapsto \substitute{\left(\left(\CwR{\Crename{\dom{\Gamma}}{\abs p}}\rho\right)\Cextend\Cst{\dom{\Gamma}}{\abs p}\right)}{f(p)}\\
    \substitute\rho\Tunit&\eqdef\Tunit\\
    \substitute\rho{(d_1\Tand d_2)}&\eqdef (\substitute\rho{d_1})\Tand(\substitute\rho{d_2})\\
    \substitute\rho{(\Tex r  d)}&\eqdef\Tex{\Cfune\pi(r)}{(\substitute\rho{d})}\\\\
    \substitute\rho{\cutc{x^-}{t^+}}&\eqdef\cutc{\varRead[x^-]\pi}{(\substitute\rho{t^+})}\\
    \substitute\rho{\cutc{f}{t^+}}&\eqdef\cutc{(\substitute\rho{\THO f})}{(\substitute\rho{t^+})}
  \end{array}
  \]
  \caption{Substitution}
  \label{fig:substitution}
\end{bfigure}

Using this to define a non-deterministic reduction relation, Subject
Reduction for the latter would rely on a typability result for the
substitution operation, similar to Theorem~\ref{th:renaminglemma} for
renamings. On the other hand, we would then avoid introducing all the
extra typing rules of Fig.~\ref{def:LAFext}, as the reduction relation
would not rely on the constructs of an abstract machine but would
directly operate on terms and commands.

We conjecture that the normalisation model of
Definition~\ref{def:model4norm} would work, exactly as it is, to show
that typed terms and commands are strongly normalising.\footnote{Well,
  not exactly ``as it is'', since instead of closures we would
  directly take functions from $\Data\pfunspace\PTerms$.}  However, we
would probably need to prove the equivalent, for \LAF, of the
substitution lemma in $\lambda$-calculus:
\[\subst{\subst M x N}y P=\subst{\subst M y P}x {\subst N y P}\]
which we would also need if we are to prove confluence of the
(non-deterministic) reduction relation. Such a lemma would broach the
topic of equality between \LAF\ proof-terms, a question that we
carefully managed to avoid so far as it involves considering which
equality we take on the meta-level functions
$f\col\Data\pfunspace\PTerms$ (extensional, intensional?).

We therefore leave all of these questions for future work.

\part{Theorem proving}
\label{partIII}

\chapter*{Introduction}

The Sequent Calculus, even in Gentzen's original
formulation~\cite{Gentzen35}, is not only a formalism to represent
complete proofs, but it also specifies a natural, non-deterministic
proof-search procedure: the gradual completion of incomplete
proof-trees, starting from the one-node tree carrying a sequent to be
proved, and extending the incomplete branches step-by-step until a
complete proof-tree is obtained. This is called \Index{bottom-up
  proof-search}, or \Index{root-first proof-search}. It is the basic
mechanism of \eg \Index[tableaux method]{tableaux methods} (see
\eg\cite{TableauxHandbook,BG01}), and it was also used to describe and
extend the logic programming paradigm~\cite{Miller91apal}.

As mentioned in Chapter~\ref{ch:polarfocus}, focussing was originally
introduced~\cite{AndreoliP89,andreoli92focusing} in the framework of
linear logic~\cite{girard-ll}, with motivations for logic programming.
In other words, focussing helped designing proof-search procedures.
As described in Chapter~\ref{ch:polarfocus} (and in
Part~\ref{partII}), focussing had a important impact on the theory of
complete proofs (and their semantics). We now come back to the view of
focussing as an algorithmic methodology for completing incomplete
proof-trees.

In~\cite{lengrand11lmcs} we used a focussed sequent calculus to
describe type inhabitation / proof-construction in Pure Type
Systems~\cite{Barendregt:hlcs1992} (and higher-order unification),
which provides (the basis for) the type theory behind several proof
assistants such as \Coq~\cite{Coq} or Twelf~\cite{Twelf}.

In this dissertation, we illustrate how the above methodology can
apply to theorem proving in classical logic. This was mostly the
object of our PSI project~\cite{PSI}, with contributions shared with
my student Mahfuza
Farooque~\cite{farooqueTR12,farooqueTR12b,farooqueTR13,farooque13,FarooquePhD}.

\Index[analytic tableau]{Analytic tableaux} probably form the
proof-search procedures that are closest to the sequent calculus.
Being much more procedure-oriented than the sequent calculus (whose
theory handles complete proofs), tableaux offer an important
difference in their explicit management of existential variables
during search (variables that may be instantiated to conclude
provability or refutability of the input), for instance via
\Index{first-order unification} in the case of pure first-order logic.

\Index[clause tableau]{Clause tableaux} provide variants of tableaux
procedures that exploit a clausal formulation of the formulae to be
refuted (which they share with resolution-based techniques or even
SAT-solving techniques). In Farooque's Ph.D.~\cite{FarooquePhD},
clause tableaux were shown to be simulated (in a strong sense) by
root-first proof-search in the focussed sequent calculus \LKF\ (see
\eg Chapter~\ref{ch:polarfocus} or~\cite{liang09tcs}), when the latter
is extended with the ability to change the polarity of atoms
\emph{on-the-fly} during proof-search. More interestingly, clause
tableaux that satisfy \Index{connection} properties (strong and weak
connections) were shown to correspond to the construction of
\LKF\ proofs that abide by specific polarisation policies:

Connections require that, when a branch of an incomplete proof-tree
(or tableau) is extended by expanding on a clause $l_1\vee\cdots\vee
l_n$, thus creating $n$ new sub-branches, then at least one of them is
closed immediately by connecting its corresponding literal $l_i$ with
a literal that was obtained earlier on the branch.

Similarly, when root-first proof-search in \LKF\ focusses on (the
negation of)\footnote{In sequent calculus we try to prove the negation
  of the formulae that tableaux methods seek to refute.} the clause
$l_1\vee\cdots\vee l_n$, a policy that forces the polarity of $\vee$
to be negative and forces the polarity of one literal among $l_1\ldots
l_n$ to also be negative, may be used so that the synchronous phase of
\LKF\ forces the immediate closing of the branch corresponding to that
literal, by ``connecting it'' to a previously obtained literal.

This was formalised in~\cite{FarooquePhD}, which also broached the
topic of reasoning \emph{modulo a theory}. Building on the idea of
(clause) \Index{tableaux-modulo-theories} suggested by
Tinelli~\cite{Tinelli07} in connection with SAT-modulo-theories
solving (SMT-solving), we developed in the PSI project~\cite{PSI} an
extension of \LKF\ with a decision procedure, and showed its
application to SMT-solving.

This is what is presented in Chapter~\ref{ch:dpll}. The motivation
behind it is to propose a focussed sequent calculus framework where
different techniques for automated (or interactive!) theorem proving
can be simulated: tuning the polarities or the polarisation policies
determines (or contributes to determining) the proof-search strategies
that capture the said techniques, switching for instance from a
tableau procedure to an SMT-procedure (such as \DPLLTh) by a simple
change of polarity policy.

This aim gave rise to the implementation of the \Psyche\ prototype,
which is still in the early development phase, which is the object of
the system description~\cite{GLPsyche13} and which is presented in
Chapter~\ref{ch:Psyche}. The system is designed as a platform for
implementing the proof-search strategies that capture different
theorem proving techniques. Doing so raises the question of trust, and
of the correctness of an output produced by any of these implemented
strategies. What the platform offers is an architecture that lets
various strategies and techniques be experimented, and implemented as
\Index[plugin]{plugins} via an API with \Psyche's \Index{kernel},
while guaranteeing the correctness of the output. This is obtained by
a somewhat transformed LCF-architecture~\cite{GMW79}. A potential
application of such a platform is to offer it as a backend prover for
the proof obligations produced by verification tools such as
Why3~\cite{Why3,FilliatreP13}; the strategy programming and
experimenting facilities of \Psyche\ could then be used to tune the
behaviour of the proof-search to the specific kind of proof
obligations that need to be proved, without worrying about
correctness.

We then conclude this dissertation in Chapter~\ref{ch:quant} with the
perspectives of \Psyche's development as impacted by the material
developed in this dissertation, and an opening to the numerous
connections with the automated reasoning literature that remain to be
investigated.

\cchapter[\DPLLTh\ as proof-search]{\DPLLTh\ as proof-search in a focussed sequent calculus}
\label{ch:dpll}

\minitoc

This chapter focusses on automated techniques for solving the
\emph{Satisfiability Modulo Theories (SMT)} family of problems,
illustrating how these can be available in a system based on
goal-directed proof-search.  Such problems generalise propositional
SAT-problems: instead of considering the satisfiability of conjunctive
normal forms (CNF) over propositional variables, SMT problems concern
the satisfiability of CNF over atomic propositions from a theory
$\mathcal T$ such as linear arithmetic or bit vectors. Given a
procedure deciding the consistency -with respect to $\mathcal T$- of a
conjunction of atoms or negated atoms, SMT-solving organises a
cooperation between this procedure and SAT-solving techniques, thus
providing a decision procedure for SMT-problems. This smart extension
of the successful SAT-solving techniques opened a prolific area of
research and led to the implementation of ever-improving tools, namely
SMT-solvers, now crucial to a number of applications in software
verification.  The architecture of SMT-solvers is based on the
extension of the Davis, Putnam, Logemann and Loveland (\DPLL)
procedure~\cite{DavisP60,DavisLL62} for solving SAT-problems to a
procedure called \DPLLTh~\cite{Nieuwenhuis06} addressing SMT-problems.

This chapter does not try to improve the \DPLLTh\ technique itself, or
current SMT-solvers based on it, but makes a step towards the
integration of the technique into a sequent calculus framework.
More precisely, 
we investigate how we can perform each of the \emph{steps} of
\DPLLTh\ as bottom-up proof-search in sequent calculus. This allows
the \DPLLTh\ algorithm to be applied \emph{up-to-a-point}, where a
switch to another technique can be made (depending on the newly
generated goals).\footnote{In contrast, other approaches to
  integrating SAT- or SMT-solving to a wider theorem proving framework
  usually rely on the automated technique to perform a \emph{full}
  run; this is the case
  of~\cite{weber11smt,ArmandFGKTW11,BCP:CPP2011,BlanchetteBP11} and
  Lescuyer's solver~\cite{Lescuyer09} that runs within the \Coq\ proof
  assistant thanks to its \Index{reflection} ability.}
This simulation can be seen as a first step toward a better
proof-theoretical understanding of how different proof-search
strategies (\eg tableaux, resolution, \DPLLTh,...), geared toward
different logical fragments, could efficiently cooperate inside a
common platform for theorem proving.


The polarities and the focussing properties of the sequent calculus we
use allow us to derive a stronger result than the mere simulation of
\DPLLTh: the proofs that are the images of \DPLLTh\ runs finishing on
\unsat can be characterised by a simple criterion only involving the
way polarities are assigned to literals and the way formulae are
placed into the focus of sequents. From this criterion we directly get
a simple proof-search strategy that is bisimilar to \DPLLTh\ runs:
that which performs the depth-first completion of incomplete
proof-trees (starting with the leftmost open leaf), using any
inference steps satisfying the given criterion on polarities and
focusing. The bisimulation ensures that bottom-up proof-search in
sequent calculus can be as efficient as the \DPLLTh\ procedure.

Section~\ref{sec:LKThp} presents the variant of System \LKF\ (from
Section~\ref{sec:LKF}) that we use to describe \DPLLTh\ in terms of
proof-search. Section~\ref{sec:DPLLThLKThp} describes the details of
how \DPLLTh\ is captured: we first identify an \emph{elementary}
version of \DPLLTh\ that is the direct extension of the
\Index[classical \DPLL]{Classical \DPLL\ procedure} to a background
theory $\mathcal T$, as well as being a restriction of the full
\Index[abstract \DPLL\ modulo theories]{Abstract \DPLL\ DPLL Modulo
  Theories} system,\footnote{that allows more advanced features such
  as \Index{backjumping} and \Index{clause learning}} both of which
can be found in~\cite{Nieuwenhuis06}; then we prove the bisimulation
result and discuss the \DPLLTh\ mechanisms that are not in our
elementary version. Section~\ref{sec:future} concludes by connecting
the above to the abstract \LAF\ system(s) developed in
Part~\ref{partII} of this dissertation.

\section{A version of \LKF\ to work modulo a theory: \LKThp}
\label{sec:LKThp}

\subsection{Background}

Clearly in root-first proof-search, asynchronous rules can be applied
eagerly (\ie can be chained, without creating backtrack points and
losing completeness), since they are invertible. Focussing says that
applying synchronous rules (although possibly creating backtrack
points) can also be chained without losing completeness. This forced
chaining of synchronous rules can be seen for instance in the
\LKF\ system of Section~\ref{sec:LKF}, where sequent may feature a
formula in its \Index{focus}.

A sequent with a positive atom in focus must be proved immediately by
an axiom on that atom; hence, the polarity of atoms greatly affects
the shape of proofs. As illustrated in \eg\cite{liang09tcs}, the
following sequent expresses the \emph{Fibonacci} logic program (in
some language where addition is primitive) and a goal
$\textsf{fib}(n,p)$ (where $n$ and $p$ are closed terms):
\[ \begin{array}{l}
  \textsf{fib}(0,0),\\
  \textsf{fib}(1,1),\\
  \forall
  ip_1p_2(\textsf{fib}(i,p_1)\imp
  \textsf{fib}(i+1,p_2)\imp\textsf{fib}(i+2,p_1+p_2))\\
  \vdash\textsf{fib}(n,p) 
\end{array} \]

The goal will be proved with backward-reasoning if the
\textsf{fib} atoms are negative (yielding a proof of exponential size
in $n$), and forward-reasoning if they are positive (yielding many
proofs, one of which being linear).

In classical logic, polarities of connectives and atoms do not affect
the provability of formulae, but still greatly affect the shape of
proofs, and hence the basic proof-construction steps. This chapter
shows how the \DPLLTh\ steps correspond to proof-construction steps
for an appropriate management of polarities. For this we use a variant
of the \LKF\ system~\cite{liang09tcs} presented in
Section~\ref{sec:LKF}: the sequent calculus \LKThp~\cite{farooque13,FarooquePhD}.

In order to make logical sense of \eg the primitive addition in the
Fibonacci example above, we only enrich \LKF\ with the ability to call
a decision procedure to decide the consistency of conjunctions of
literals \wrt a theory (\ie the same as for \DPLLTh): for a theory
that equates $1+1$ and $2$, a call to the procedure proves $p(2),\non
p(1+1)\vdash $ in one step (unlike \LKF's syntactic checks).

System \LKF\ also assumes that all atoms come with a pre-determined
polarity, whereas \LKThp\ allows \emph{on-the-fly} polarisation of
atoms: the root of a proof-tree might have none of its atoms
polarised, but atoms may become positive or negative as progress is
made in the proof-search.

\subsection{Definitions}

In this section we present the quantifer-free fragment of the focussed
sequent calculus\linebreak \LKThp~\cite{farooque13,FarooquePhD}. This fragment
concerns propositional classical logic~\emph{modulo a theory} and will
be sufficent for the simulation of \DPLLTh.

This sequent calculus (and this logic) involves a notion of
\emph{literal} and a notion of \emph{theory}. The reader can safely
see behind this terminology the standard notions from proof theory and
automated reasoning. However at this point,
very little is required from or assumed about those two notions.

\begin{definition}[Literals]

  Let $\mathcal L$ be a set of elements called \Index[literal]{literals},
  equipped with an involutive function called \Index{negation} from
  $\mathcal L$ to $\mathcal L$. In the rest of this chapter, a possibly
  primed or indexed lowercase $l$ always denotes a literal, and $\non
  l$ its negation.
\end{definition}

Another ingredient of \LKThp\ is a \emph{theory} $\mathcal T$, given
in the form of an \emph{inconsistency predicate}, a notion that we now
introduce:

\begin{definition}[Inconsistency predicates]
  \label{def:inconsistency}

  An \Index{inconsistency predicate} is a predicate over sets of literals 
  \begin{itemize}
  \item satisfied by the set $\{l,\non l\}$ for every literal $l$;
  \item that is upward closed (if a subset of a set satisfies the predicate, so does the set);
  \item such that if the sets $\mathcal P,l$ and $\mathcal P,\non l$ satisfy it then so does $\mathcal P$.
  \end{itemize}

  The smallest inconsistency predicate is called the \Index{syntactical
    inconsistency} predicate\footnote{It is the predicate that is true of a set $\mathcal P$ of literals iff $\mathcal P$ contains both $l$ and $\non l$ for some $l\in\mathcal L$.}.   If a set $\mathcal P$ of literals
  satisfies the syntactically inconsistency predicate, we say that
  $\mathcal{P}$ is \Index{syntactically inconsistent}, denoted $\mathcal
  P\models$. Otherwise $\mathcal P$ is \Index{syntactically consistent}.

  The theory $\mathcal T$ in the notation \LKThp\ is described by means of
  an(other) inconsistency predicate, called the \Index{semantical
    inconsistency predicate}, which will be a formal parameter of the
  inference system defining \LKThp.

  If a set $\mathcal P$ of literals satisfies the semantical
  inconsistency predicate, we say that $\mathcal P$ is
  \Index{semantically inconsistent} or \Index{inconsistent modulo theory},
  denoted by $\mathcal P\models_{\mathcal T}$. Otherwise $\mathcal P$ is
  \Index{semantically consistent} or  \Index{consistent modulo theory}.
\end{definition}

\begin{definition}[Formulae, negation]
  \label{def:seminconsistency}

   Let $\mathcal L$ be a set of literals. The formulae of propositional polarised classical logic are given by the following grammar:
  \[
  \begin{array}{lrl}
    \mbox{Formulae }&A,B,\ldots\ \strut\recdef &l\hfill\mbox{where $l$ ranges over $\mathcal L$}\\
    &\sep &A\andP B\sep A\orP B\sep \top^+\sep \bot^+\qquad\strut\\
    &\sep &A\andN B\sep A\orN B \sep \top^-  \sep \bot^-
  \end{array}
  \]

  The size of a formula $A$, denoted $\size A$, is its size as a tree (number of nodes).

  Let $\mathcal P\subseteq\mathcal L$ be syntactically consistent.
  Intuitively, it represents the set of literals declared to be
  positive.

  We define $\mathcal P$-positive formulae and $\mathcal P$-negative formulae as the formulae generated by the following grammars:
  \[
  \begin{array}{l@{}l@{}l}
    \mathcal P\mbox{-positive formulae }&P,\ldots&\recdef p\sep A\andP B \sep A\orP B\sep \top^+ \sep \bot^+\\
    \mathcal P\mbox{-negative formulae }&N,\ldots&\recdef \non p\sep A\andN B\sep A\orN B\sep \top^- \sep \bot^-\\
  \end{array}
  \]
where $p$ ranges over $\mathcal P$.

Let $\UP$ be the set of all \Index[$\mathcal P$-unpolarised literal]{$\mathcal P$-unpolarised literals}, \ie literals that are neither $\mathcal P$-positive nor $\mathcal P$-negative.

Negation is recursively extended into an involutive map on formulae as follows:
  \[
  \begin{array}{|ll|ll|}
    \hline
    \non{(A\andP B)}&\eqdef \non A \orN\non B&\non{(A\andN B)}&\eqdef \non A \orP\non B    \\
    \non{(A\orP B)}&\eqdef \non A \andN\non B&\non{(A\orN B)}&\eqdef \non A \andP\non B    \\
    \non{(\top^+)} &\eqdef \bot^- & \non{(\top^-)} &\eqdef \bot^+    \\
    \non{(\bot^+)} &\eqdef \top^- & \non{(\bot^-)} &\eqdef \top^+    \\
    \hline
  \end{array}
  \]
\end{definition}
\begin{remark}
  Note that, given a syntactically consistent set $\mathcal P$ of
  literals, negations of $\mathcal P$-positive formulae are $\mathcal
  P$-negative and vice versa.
\end{remark}

\begin{notation}
A possibly primed or indexed $\Gam$ always denotes a set of formulae.
By $\atmCtxt\Gam$ we denote the subset of elements of $\Gam$ that are
literals, and we write $l\sqin\Gam$ if $l$ or $\non l$ appears in
$\Gamma$.  By $\atmCtxtP{\mathcal P}\Gam$ we denote the sub-multiset
of $\Gam$ consisting of its $\mathcal P$-positive literals (\ie
$\mathcal P\cap\Gamma$ as a set).
\end{notation}


\begin{definition}[System \LKThp]

  The system \LKThp\ is the sequent calculus defined by the rules of
  Fig.~\ref{fig:LKThp}, which fall into three categories:
  \emph{synchronous}, \emph{asynchronous}, and
  \emph{structural} rules, and manipulate two kinds of sequents:\\
    \begin{tabular}{ l r }
      $\DerPosTh \Gamma A {}{\mathcal P}$   &  where the formula $A$ is in the \emph{focus} of the sequent\\
      $\DerNegTh \Gamma {\Gam'}{}{\mathcal P}$ & 
    \end{tabular}\\
where $\mathcal P$ is a syntactically consistent set of literals declared to be positive.\\
A sequent of the second kind where $\Gam'$ is empty is called \Index[developed sequent]{developed}.
\end{definition}

\begin{bfigure}[!ht]
  \[
  \begin{array}{c}
    \textsf{Synchronous rules}
    \hfill\strut\\[3pt]
    \infer[{[(\andP)]}]{\DerPos{\Gamma}{A\andP B}{}{\mathcal{P}}}
    {\DerPos{\Gamma}{A}
      {}{\mathcal P} \qquad \DerPos{\Gamma}{B}{}{\mathcal{P}}}
    \qquad
    \infer[{[(\orP)]}]{\DerPos{\Gamma}{A_1\orP A_2}{}{\mathcal{P}}}
    {\DerPos{\Gamma}{A_i}{}{\mathcal{P}}}
    \\[15pt]
      \infer[{[{(\top^+)}]}]{\DerPos{\Gamma}{\top^+} {} {\mathcal{P}}} {\strut}
       \qquad 
\infer[{[({\Init[1]})]l\in\mathcal P}]
	{\DerPos{\Gamma  }{l}{} {\mathcal{P}}} {\atmCtxtP{\mathcal P}\Gam,\non l\models_{\mathcal T}}    \qquad
    \infer[{[(\Release)]N \mbox{ is not $\mathcal P$-positive}}]{\DerPos {\Gam} {N} {} {\mathcal{P}}}
    {\DerNeg {\Gam} {N} {} {\mathcal{P}}}\\
    \midline[0pt]
    \textsf{Asynchronous rules}
    \hfill\strut\\[3pt]
    \infer[{[(\andN)]}]{\DerNeg{\Gamma}{A\andN B,\Delta} {} {\mathcal{P}}}
    {\DerNeg{\Gamma}{A,\Delta} {} {\mathcal{P}} 
      \qquad \DerNeg{\Gamma}{B,\Delta} {} {\mathcal{P}}}
    \qquad
    \infer[{[(\orN)]}]{\DerNeg {\Gamma} {A_1\orN A_2,\Delta} {} {\mathcal{P}}}
    {\DerNeg {\Gamma} {A_1,A_2,\Delta} {} {\mathcal{P}}}
    \\\\
    \infer[{[{(\bot^-)}]}]{\DerNeg{\Gamma} {\Del,\bot^-} {} {\mathcal{P}}}
    {\DerNeg{\Gamma} {\Del} {} {\mathcal{P}}}
     \qquad
    \infer[{[{(\top^-)}]}]{\DerNeg{\Gamma} {\Del,\top^-} {} {\mathcal{P}}}{\strut}
     \qquad
    \infer[{[({\Store})]%
      \begin{array}l%
        A\mbox{ is a literal}\\
        \mbox{or is $\mathcal P$-positive}\\
      \end{array}}]{\DerNeg \Gam {A,\Del} {} {\mathcal{P}}} 
    {\DerNeg {\Gam,\non A} {\Del} {} {\polar{\non A}}}  
    \\\midline[0pt]
      \textsf{Structural rules}
    \hfill\strut\\[3pt]
    \infer[{[(\Select)]\mbox{$P$ is not $\mathcal P$-negative}}]
    {\DerNeg {\Gam,\non P} {}{} {\mathcal{P}}} 
    {\DerPos {\Gam,\non P} {P} {} {\mathcal{P}}}
    \qquad
    \infer[{[({\Init[2]})]}]{\DerNeg {\Gam} {}{} {\mathcal{P}}}{\atmCtxtP{\mathcal P}\Gam\models_{\mathcal T}}
  \end{array}
  \]
  \medskip
  \begin{tabular}{lll}%
    where& $\polar A \eqdef \mathcal P,A$& if $A\in\UP$\\
    &$\polar A \eqdef \mathcal P$& if not
  \end{tabular}%
  \caption{System \LKThp}
  \label{fig:LKThp}
\end{bfigure}

The gradual proof-tree construction defined by the bottom-up
application of the inference rules of \LKThp, is a goal-directed
mechanism whose intuition can be given as follows:

Asynchronous rules are invertible: $(\andN)$ and $(\orN)$ are applied
eagerly when trying to construct the proof-tree of a given sequent;
$(\Store)$ is applied when hitting a positive formula or a negative
literal on the right-hand side of a sequent, storing its negation on the
left.

When the right-hand side of a sequent becomes empty (\ie the sequent
is \emph{developed}), a sanity check can be made with $({\Init[2]})$
to check the semantical consistency of the stored literals (\wrt the
theory), otherwise a choice must be made to place a positive formula
in focus, using rule $(\Select)$, before applying synchronous rules like $(\andP)$ and
$(\orP)$. Each such rule decomposes the formula in focus, keeping the
revealed sub-formulae in the focus of the corresponding premises,
until a positive literal or a non-positive formula is obtained: the
former case must be closed immediately with $({\Init[1]})$ calling the
decision procedure, and the latter case uses the $(\Release)$ rule to
drop the focus and start applying asynchronous rules again. The
synchronous and the structural rules are in general not
invertible,\footnote{(but they may be so, \eg $(\andP)$)} so each
application of those yields in general a backtrack point in the
proof-search.

Notice that an invariant of such a proof-tree construction process is
that the left-hand side of a sequent only contains negative formulae
and positive literals.


\begin{notation}\label{entailment}
  When $F$ is a formula of unpolarised propositional logic and $\Psi$
  is a set of such formulae, $\Psi\models F$ means that
  $\Psi$ entails $F$ in propositional classical logic.
  Given a theory $\mathcal T$ (given by a semantical inconsistency predicate), we define the set of all \emph{theory lemmas} as
  \(\Psi_{\mathcal{T}} \eqdef \{l_1\vee \cdots \vee l_n \mid \non l_1 , \cdots , \non l_n \models_{\mathcal T} \}\)
  and generalise the notation $\models_{\mathcal T}$ to write
  $\Psi\models_{\mathcal T} F$ when 
  $\Psi_{\mathcal T},\Psi\models F$. In that case we say that $F$
  is a \emph{semantical consequence} of $\Psi$.
  For any polarised formula $A$, let $\underline{A}$ be the unpolarised formula obtained by removing all polarities on connectives.
\end{notation}

\begin{theorem}[Cut-elimination and Completeness of \LKThp]
\label{clkdpllth}
\begin{itemize}
\item The following rules are admissible in \LKThp:
\[
\infers[(\Pol)]{
  \DerNeg{\Gamma}{}{}{\mathcal P}
}[
  l\sqin\Gamma \mbox{ and }\atmCtxtP{\mathcal {P}}{\Gamma},\non l\models_{\mathcal T}
]{
  \DerNeg{\Gamma}{}{}{\mathcal {P},l}
}
\qquad
\infers[(\cut)]{
  \DerNeg {\Gamma}{}{}{\mathcal P}
}[
  l\sqin\Gamma
]{
  \DerNeg{\Gamma}{l}{}{\mathcal P} 
  \quad 
  \DerNeg{\Gamma}{\non l}{}{\mathcal P}
}
\]
provided the bottom sequent satisfies some property called
\Index{safety}~\cite{farooqueTR13,FarooquePhD}.
\item If $\models_{\mathcal T} {F}$, then for all $A$ such that
  $\underline{A}=F$, we can prove $\DerNegLKpp {} {A}{\emptyset}$ in
  \LKThp.
\end{itemize}
\end{theorem}
The meta-theory of \LKThp, in particular the proofs of the above, can be found in~\cite{farooqueTR13,FarooquePhD}.

\renewcommand\neg[1]{\non{#1}}

\section{Bisimulation with the \DPLLTh\ procedure}
\label{sec:DPLLThLKThp}

\subsection{The elementary \DPLLTh\ procedure}
\label{DPLLTh}

Intuitively, \DPLLTh\ aims at proving the inconsistency of a set of
\emph{clauses} with respect to a theory. We therefore retain from the
previous section the notion of literal and inconsistencies, and
introduce clauses:
\begin{definition}[Clause]

  A \Index{clause} is a finite set of literals, which can be seen as their disjunction.

  In the rest of the chapter, a possibly indexed upper cased $C$ always
  denotes a clause. The empty clause is denoted by $\bot$. The number
  of literals in a clause $C$ is denoted $\size C$. The possibly
  indexed symbol $\phi$ always denotes finite sets of clauses
  $\{C_1,\ldots,C_n\}$, which can also be seen as a Conjunctive Normal
  Form (CNF).  We use $\size \phi$ to denote the sum of the sizes of
  the clauses in $\phi$. Finally $\atm \phi$ denotes the set of
  literals that appear in $\phi$ or whose negations appear in $\phi$.

  Viewing clauses as disjunctions of literals and sets of clauses as
  CNF, we will generalise Notation~\ref{entailment}, writing for
  instance
  $\phi \models \neg C$ or $\phi \models C$, as well as $\phi
  \models_{\mathcal T} \neg C$  or $\phi \models_{\mathcal T}
  C$.
\end{definition}

\begin{definition}[Decision literals and sequences]

  We consider a (single) copy of the set $\mathcal L$ of
  literals, denoted $\mathcal L^d$, whose elements are called
  \Index[decision literal]{decision literals}, which are just
  tagged clones of the literals in $\mathcal L$. Decision literals
  are denoted\footnote{This exponent tag is a standard notation,
    standing for ``decision''.} by $l^d$.

  We use the possibly indexed symbol $\Delta$ to denote a finite
  sequence of possibly tagged literals, with $\emptyset$ denoting the
  empty sequence.  We also use $\Delta_1, \Delta_2$ and $\Delta_1, l,
  \Delta_2$ to denote the suggested concatenation of sequences.

  For such a sequence $\Delta$, we write $\forget{\Delta}$ for the
  subset of $\mathcal L$ containing all the literals in $\Delta$ with
  their potential tags removed. The sequences that \DPLLTh\ will
  construct will always be duplicate-free, so the difference between
  $\Delta$ and $\forget \Delta$ is just a matter of tags and ordering.
  When the context is unambiguous, we will sometimes use $\Delta$ when
  we mean $\forget \Delta$.


  We define $\mSat \Del \eqdef \{ l \mid \Del,\non l \models_{\mathcal
    T} \}$, the closure of a sequence $\Delta$ by semantical
  entailment. For any set of clauses $\phi$, the set of literals
  occuring in $\phi$ that are semantically entailed by $\Delta$ is
  denoted by $\nSat \phi \Del \eqdef \mSat \Del \cap \atm \phi $.
\end{definition}

\begin{remark} Semantical consequences are the analogues of the
  consequences of a partial boolean assignment in the context of 
  a \DPLL\ procedure for propositional logic without theory.\\
  Obviously, if $l \in \Del$, then $l \in \mSat \Del$.
  If $\phi_1 \subseteq \phi_2$, then for any $\Delta$, 
  $\nSat {\phi_1} \Del \subseteq \nSat {\phi_2} {\Del}$.
\end{remark}

We can now describe the elementary \DPLLTh\ procedure as a
transition system between  states. 
\begin{definition}[Elementary \DPLLTh]\label{def:dpllth}\strut

  A \emph{state} of the 
  \DPLLTh\ procedure is either the state $\unsat$, or a pair denoted
  $\Del \| \phi$, where $\phi$ is a set of clauses and $\Del$ is a
  sequence of possibly tagged literals. The \emph{transition rules} of the 
  elementary \DPLLTh\ procedure are given in Fig.~\ref{fig:dpllth}.
\end{definition}
\begin{bfigure}[!h]
  \centering
  \small
  \begin{tabular}{l@{\quad}l@{\ $\Rightarrow$\ }l@{\quad}l@{\ }l}


    $\Decide$
    &
    $\Delta  \| \phi$&$\Delta, l^d   \| \phi$
    &
    where $l\in\atm \phi$& and $l \not \in \Delta$ and $\non l \not \in \Delta$.
    \\
      {$\Propagate$}
      &
      $\Delta \| \phi, C \vee l $&$
      \Delta, l \| \phi, C \vee l$
      & where $\Delta \models \neg C$& and $l \not \in \Delta$ and $\non l \not \in \Delta$.
      \\
        {$\Propagate[\mathcal T]$}
      &
      $\Delta \| \phi $&$ \Delta, l \|
      \phi$ & where $l\in\nSat\phi{\Delta}$& and $l \not \in \Delta$ and $\non l \not \in \Delta$.
      \\{$\Fail$}
      &
      $\Del \| \phi,C $& \unsat, & where
      $\Del \models \neg C$&  and there is no decision literal in
      $\Delta$.
      \\
        {$\Fail[\mathcal T]$}
        &
        $\Del \| \phi $& \unsat, & where
        $\Del \models_{\mathcal T}$& and there is no decision literal in
        $\Delta$.
        \\
          {$\Backtrack$}
          &
          $\Delta_1, l^d, \Delta_2 \| \phi, C $&$
          \Delta_1, \non l \| \phi, C$ 
          & where $\Delta_1, l, \Delta_2
          \models \neg C$& and there is no decision literal in $\Delta_2$.
          \\{$\Backtrack[\mathcal T]$}
          &
          $\Delta_1, l^d, \Delta_2 \| \phi $&$
          \Delta_1, \non l \| \phi$
          & where $\Delta_1, l, \Delta_2
          \models_{\mathcal T}$&  and there is no decision literal in $\Delta_2$.
  \end{tabular}
  \caption{Elementary \DPLLTh}
  \label{fig:dpllth}
\end{bfigure}

This transition system is an extension of the \emph{Classical
  \DPLL\ procedure}, as presented in~\cite{Nieuwenhuis06}, to the
background theory $\mathcal T$.\footnote{We removed the $\textsf{Pure
    Literal}$ rule, in general unsound in presence of a theory
  $\mathcal T$.}  The first four rules are explicitly taken from the
\textsf{Abstract DPLL Modulo Theories system}
of~\cite{Nieuwenhuis06}.\footnote{Unit Propagate and Theory Propagate
  are renamed as ${\Propagate}$ and ${\Propagate[\mathcal T]}$ for
  consistency with the other rule names.} The other rules of that
system (namely $\Backjump$, $\Learn$, $\Forget$, etc), are not
considered here in their full generality, but specific cases and
combinations are covered by the rest of our elementary
\DPLLTh\ system, so that it is logically
complete.\footnote{$\Backtrack$ is a restricted version of $\Backjump$
  (this holds on the basis that the full system satisfies some basic
  invariant -Lemma 3.6 of~\cite{Nieuwenhuis06}), $\Fail[\mathcal T]$
  (\resp $\Backtrack[\mathcal T]$) is a combination of $\Learn$,
  $\Fail$ (\resp $\Backtrack$), and $\Forget$ steps.}  Note that this
transition system is not deterministic: for instance the $\Decide$
rule can be applied from any state and it furthermore does not enforce
a strategy for picking the literal to be tagged among the eligible
elements of $\atm \phi$. At the level of implementation, this (non
deterministic) transition system is turned into a deterministic
algorithm, whose efficientcy crucially relies on the strategies
adopted to perform the choices left unspecified by \DPLLTh.

We illustrate those rules, in the theory $\mathcal T$ of \emph{Linear
  Rational Arithmetic}, with the two basic examples of elementary
\DPLLTh\ runs presented in Fig.~\ref{ex:FwithoutTh} (where $\Delta$
and $\phi$ always refer to the current state $\Del \| \phi $).
\begin{bfigure}[!h]
  \small
  \[    
  \begin{array}{r || cl}
    \emptyset  
    &  {x>0,(x+y>0)^\bot,(y>0 \vee x=-1)} & \Propagate\\
    
    x>0        
    &  x>0,(x+y>0)^\bot,(y>0 \vee x=-1) & \Propagate\\
    
    x>0, (x+y>0)^\bot         
    &  x>0,(x+y>0)^\bot,(y>0 \vee x=-1) & \Propagate[\mathcal T]\\
    &&((y>0)^\bot\in\nSat\phi{\Delta})\\

    x>0,(x+y>0)^\bot,  (y>0)^\bot          
    &  x>0,(x+y>0)^\bot,(y>0 \vee x=-1) & \Propagate[\mathcal T]\\
    &&((x=-1)^\bot\in\nSat\phi{\Delta})\\
    
    x>0,(x+y>0)^\bot,  (y>0)^\bot, (x=-1)^\bot          
    &  x>0,(x+y>0)^\bot,(y>0 \vee x=-1) & \Fail \mbox{ on clause}\\
    &&(y>0 \vee x=-1)\\
    \unsat         &   
  \end{array}
  \]
  \strut
  \[    \begin{array}{r || cl}
        \emptyset         
        &  x>0,(x+y>0)^\bot,(y>0 \vee x=-1)  & \Propagate\\
        
  	x>0         
        &  x>0,(x+y>0)^\bot,(y>0 \vee x=-1)  & \Propagate\\
        
        x>0, (x+y>0)^\bot         
        &  x>0,(x+y>0)^\bot,(y>0 \vee x=-1)  & \Propagate[\mathcal T]\\
        &&\quad\hfill((y>0)^\bot\in\nSat\phi{\Delta})\\

        x>0,(x+y>0)^\bot,  (y>0)^\bot          
        &  x>0,(x+y>0)^\bot,(y>0 \vee x=-1)  & \Propagate \mbox{ ($x=-1$)}\\
        &&\mbox{from $(y>0 \vee x=-1)$}\\

        x>0,(x+y>0)^\bot,  (y>0)^\bot, (x=-1)
        &  x>0,(x+y>0)^\bot,(y>0\vee x=-1)   & {\Fail[\mathcal T]}\\
        &&\mbox{$x>0, x=-1$}\\ 
        &&\mbox{inconsistent with $\mathcal T$}\\ 
        \unsat         &    
      \end{array}
\]
\caption{Examples of elementary \DPLLTh\ runs}
\label{ex:FwithoutTh}
\label{ex:FwithTh}
\end{bfigure}

A reason to introduce rule $\Fail[\mathcal T]$ is to allow the second run to finish with the same output as the first:
Indeed, the last $\Propagate$ step has created a $\mathcal T$-inconsistency from which we could not derive \unsat\ without a $\Fail[\mathcal T]$ step.\footnote{(or, alternatively, a $\Learn$ step in~\cite{Nieuwenhuis06})}



\subsection{Simulation of the elementary \DPLLTh\ procedure in \LKThp}
\label{DPLLTh-LKpTh}

The aim of this section is to describe how the elementary
\DPLLTh\ procedure can be transposed into a proof-search process for
sequents of the \LKThp\ calculus. A complete and successful run of the
\DPLLTh\ procedure is a sequence of transitions $\emptyset \| \phi
\Rightarrow^* \unsat$, which ensures that the set of clauses $\phi$ is
inconsistent modulo the theory. Hence, we are devising a proof-search
process aiming at building an \LKThp\ proof-tree for sequents of the
form $\phi' \vdash$, where $\phi'$ represents the set of clauses
$\phi$ as a sequent calculus structure, in the following sense:
\begin{definition}[Representation of clauses as formulae]\label{def:repr}\label{def-dpl1}\strut

  An \LKThp\ formula $C'$ \Index{represents} a \DPLLTh\ clause $\{l_{j}\}_{j=1\ldots p}$ if $C'=l_1 \orN \ldots \orN {l_{p}}\orN\bot^-$.

  A set of formulae $\phi'$ \Index{represents} a set of clauses $\phi$ if there is a bijection $f$ from $\phi$ to $\phi'$ such that for all clauses $C$ in $\phi$, $f(\phi)$ represents $C$.
\end{definition}

\begin{remark}
  If $C'$ represents $C$, then $\size {C'}\leq 2\size C$ (there are
  fewer symbols $\orN$ than there are literals in $C$).
\end{remark}

Note here that we carefully use the negative disjunction connective to
translate \DPLLTh\ clauses. This is crucial not only to mimic \DPLLTh\ without duplicating formulae but more generally to control the search space.

Now, in order to construct a proof of $\phi' \vdash$ from a run $\emptyset \| \phi \Rightarrow^* \unsat$, we proceed gradually by considering the intermediate steps of the \DPLLTh\ run:
\[\emptyset \| \phi \Rightarrow^* \Del\|\phi \Rightarrow^* \unsat\]\vskip3pt

In the intermediate \DPLLTh\ state $\Del\|\phi$, the sequence $\Delta$
is a log of both the search space explored so far (in $\emptyset \|
\phi \Rightarrow^* \Del\|\phi$) and the search space that remains to
be explored (in $\Del\|\phi \Rightarrow^* \unsat$).  In this log, a
tagged decision literal $l^d$ indicates a point where the procedure
has made an exploratory choice (the case where $l$ is true has been/is
being explored, the case where $\non l$ is true remains to be
explored), while untagged literals in $\Delta$ are predictable
consequences of the decisions made so far and of the set of clauses
$\phi$ to be falsified.

If we are to express the \DPLLTh\ procedure as the gradual
construction of a \LKThp\ proof-tree, we should get from $\emptyset \|
\phi \Rightarrow^* \Del\|\phi$ a proof-tree that is not yet complete
and get from $\Del\|\phi \Rightarrow^* \unsat$ some (complete)
proof-tree(s) that can be ``plugged into the holes'' of the incomplete
tree.  We should read in $\Delta$ the ``interface'' between the
incomplete tree that has been constructed and the complete sub-trees
to be constructed.

We use the plural here since there can be more than one sub-tree left
to construct: $\Del\|\phi \Rightarrow^* \unsat$ contains the
information to build not only a proof of $\forget\Delta,\phi' \vdash$,
but also proofs of the sequents corresponding to the other parts of
the search space to be explored, characterised by the tagged literals
in $\Delta$.  For instance, a run from $l_1,l_2^d,l_3,l_4^d\|\phi
\Rightarrow^* \unsat$ contains the information to build a proof of
$\DerNeg{l_1,l_2,l_3,l_4,\phi'}{}{}{}$ but also the proofs of
$\DerNeg{l_1,l_2,l_3,\non l_4,\phi'}{}{}{}$ and $\DerNeg{l_1,\non
  l_2,\phi'}{}{}{}$.  Those extra sequents are obtained by collecting
from a sequence $\Delta$ its ``backtrack points'' as follows:

\begin{definition}[Backtrack points]

  The \Index[backtrack point]{backtrack points} $\backstrict{\Delta}$
  of a sequence $\Delta$ of possibly tagged literals is the list of
  sets of untagged literals recursively defined by the following
  rules, where $\emptylist$ and $\cons{}{}$ are the standard list
  constructors.
    \[
    \begin{array}{|c|}
      \upline
      \begin{array}{ll}
        \backstrict{ \emptyenv } &\eqdef  \emptylist\\
        \backstrict{ \Delta,l } &\eqdef \backstrict \Delta\\
        \backstrict{ \Delta,l^d} &\eqdef \cons{\forget{\Delta,\non l}}{\backstrict {\Delta}}
      \end{array}
      \downline
    \end{array}
    \]
\end{definition}

\begin{remark} The length of $\backstrict{\Delta}$ is the number of decision literals in $\Delta$.
\end{remark}

Now, coming back to the \DPLLTh\ transition sequence $\emptyset \| \phi \Rightarrow^* \Del\|\phi$ and its intuitive counterpart in sequent calculus, we have to formalise the notion of \emph{incomplete} proof-tree together with the notion of ``filling its holes'':

\begin{definition}[Incomplete proof-tree, extension]

  An \Index{incomplete proof-tree} in \LKThp\ is a tree labelled with sequents,
  \begin{itemize}
  \item whose leaves are tagged as either \emph{open} or \emph{closed};
  \item whose open leaves are labelled with developed sequents;
  \item and such that every node that is not an open leaf, together
    with its children, forms an instance of the \LKThp\ rules.
  \end{itemize}

  The \Index{size} of an incomplete proof-tree is its number of nodes.

  An incomplete proof-tree $\pi'$ is an \Index{extension} of $\pi$, if
  there is a tree (edge and nodes preserving) homomorphism from $\pi$ to $\pi'$.
  It is an \Index{$n$-extension} of $\pi$, if moreover the difference of size between $\pi'$ and $\pi$ is less than or equal to $n$.
\end{definition}

\begin{remark}
  An incomplete proof-tree that has no open leaf is (isomorphic to) a
  well-formed complete \LKThp\ proof of the sequent labelling its
  root. In that case, we say the proof-tree is \Index{complete}.
\end{remark}

The intuition that an intermediate \DPLLTh\ state describes an
``interface'' between an incomplete proof-tree and the complete
proof-trees that should be plugged into its holes, is formalised as
follows:
\begin{definition}[Correspondence]
\label{def:cor}

  An incomplete proof-tree $\pi$ \Index{corresponds to} a \DPLLTh\ state $\Del \| \phi$ if:
  \begin{itemize}
  \item the length of $\cons{\forget \Del}{\backstrict \Del}$ is the
    number of open leaves of $\pi$;
  \item if $\Del_i$ is the $i^{\mbox{\scriptsize th}}$ element of
    $\cons{\forget \Del}{\backstrict \Del}$, then the
    $i^{\mbox{\scriptsize th}}$ open leaf of $\pi$ (taken
    left-to-right) is labelled by a developed sequent of the form
    $\DerNegTh {\Delta_i',\phi_i'} {}{} {\Del_i}$, where:
    \begin{itemize}
    \item $\phi_i'$ represents $\phi$ (in the sense of Definition \ref{def:repr});
    \item $\nSat {\phi}{\Del_i} = \nSat {\phi} {\Del_i'}$.
    \end{itemize}
  \end{itemize}

  An incomplete proof-tree $\pi$ \Index{corresponds to} the state \unsat\
  if it has no open leaf.
\end{definition}

\begin{remark}
  In the general case, different incomplete proof-trees might
  correspond to the same \DPLLTh\ state (just like different
  \DPLLTh\ runs may reach that state from the initial one).

  Note that we do not require anything from the conclusion of an
  incomplete proof-tree corresponding to $\Del \| \phi$: just as our
  correspondence says nothing about the \DPLLTh\ transitions taking
  place after $\Del \| \phi$ (nor about the trees to be plugged into
  the open leaves), it says nothing about the transitions taking place
  before $\Del \| \phi$ (nor about the incomplete proof-tree, except
  for its open leaves).

  If an incomplete proof-tree $\pi$ corresponds to a \DPLLTh\ state
  $\Del \| \phi$ where there are no decision literals in $\Del$, then
  there is exactly one open leaf in $\pi$, and it is labelled by a
  sequent of the form $\DerNegTh {\Delta',\phi'} {}{} {\forget\Del}$,
  where $\phi'$ represents $\phi$ and $\nSat {\phi}{\Del} = \nSat
  {\phi} {\Del'}$.

  To the initial state $\emptyset \|{\phi}$ of a run of the
  \DPLLTh\ procedure corresponds the incomplete proof-tree consisting
  of one node (both root and open leaf) labelled with the sequent
  $\DerNegTh {\phi'} {}{} {}$, where $\phi'$ represents $\phi$.
\end{remark}

The simulation theorem below provides a systematic way of interpreting
any \DPLLTh~transition as a completion of incomplete proof-trees that
preserves the correspondence given in Definition~\ref{def:cor} and
controls the growth of the proof trees.

\begin{theorem}[Simulation of \DPLLTh\ in \LKThp] \label{TTFail}

  If $ \Del \| \phi \Rightarrow \mathcal S_2$ is a valid
  \DPLLTh\ transition, and $\pi_1$ is an incomplete proof tree in
  \LKThp\ corresponding to $\Del \| \phi$, then there exists a
  $(2\size \phi+3)$-extension $\pi_2$ of $\pi_1$ that corresponds to
  $\mathcal S_2$.
\end{theorem}
\begin{proof}
  See~\cite{farooque13,FarooquePhD}. By case analysis on the nature
  of the transition, completing the leftmost open leaf of $\pi_1$. Basically:
  \begin{itemize}
  \item{$\Fail$ using clause $C$\flush corresponds to\flush $\Select$ on $\non C$}
  \item{$\Fail[\mathcal T]$\flush corresponds to\flush $\Init[2]$ rule}
  \item{$\Backtrack$ using clause $C$\flush corresponds to\flush $\Select$ on $\non C$}
  \item{$\Backtrack[\mathcal T]$\flush corresponds to\flush $\Init[2]$ rule}
  \item{$\Propagate$ using clause $C$\flush corresponds to\flush $\Select$ on $\non C$}
  \item{$\Fail[\mathcal T]$\flush corresponds to\flush $\Pol$ rule}
  \item{$\Decide$\flush corresponds to\flush $\cut$ rule}
  \end{itemize}
\end{proof}

\begin{corollary}\label{cor:DPLL-LK}

  If $\emptyset \| \phi \Rightarrow^n \unsat$ and $\phi'$ represents
  $\phi$ then there is a complete proof in \LKThp\ of $\DerNegTh
  {\phi'} {}{} {\empty}$, of size smaller than $(2\size \phi+3)n$.
\end{corollary}

\subsection{Completing the bisimulation}
\label{sec:LK-DPLL}

Now the point of having mentioned quantitative information in
Theorem~\ref{TTFail}, via the notion of $n$-extension, is to motivate
the idea that performing proof-search directly in \LKThp\ is in
essence not less efficient than running \DPLLTh: we have a linear
bound in the length of the \DPLLTh\ run (and the proportionality ratio
is itself an affine function of the size of the original problem).

We also need to make sure that this final proof-tree is indeed found
as efficiently as running \DPLLTh, which can be done by identifying,
in \LKThp, a (complete) search space that is isomorphic to (and hence
no wider than) that of \DPLLTh. We analyse for this a proof-search
strategy, in \LKThp, that exactly captures the proof-extensions that
we have used in the simulation of \DPLLTh, \ie the proof of
Theorem~\ref{TTFail}:

\begin{definition}[\DPLLTh-extensions]\label{def:dpllext}\nopagesplit

  An incomplete proof tree $\pi_2$ is a \Index{\DPLLTh-extension} of
  an incomplete proof tree $\pi_1$ if
  \begin{enumerate}
  \item it extends $\pi_1$ by replacing its leftmost open leaf with an incomplete proof-tree of one of the forms:
    \[
	\begin{array}{c}
    \infer[(a)]{\DerNegTh {\Gamma,\non A} {}{} {\mathcal P}} 
    {
      \infer[(b)]{\DerPosTh {\Gamma,\non A} {A}{} {\mathcal P}} {\ldots}
    }
    \qquad
    \infer[l\sqin\Gamma]{\DerNegTh {\Gamma} {}{} {\mathcal P}}
    { 
      {\DerNegTh {\Gamma} {l}{} {\mathcal P}} 
      \quad
      {\DerNegTh {\Gamma} {\non {l}}{} {\mathcal P}}
    }
   \\\\ 
    \infer[(c)]{\DerNegTh {\Gamma} {}{} {\mathcal P}}
    { 
      {\DerNegTh {\Gamma} {}{} {\mathcal P,l}} 
    }
    \qquad
    \infer[\atmCtxt\Gam\models_{\mathcal T}]{\DerNegTh {\Gamma} {}{} {\mathcal P}}{\strut}
    \end{array}
    \]
    where
    \begin{enumerate}
    \item $A$ is a (positive) conjunction of literals that are all in $\mathcal P$ except maybe one that is $\mathcal P$-unpolarised
    \item the only instances of $(\Pol)$ in the above proof are of the form
      \(
      \infer{\DerNegTh {\Gamma} {l}{} {\mathcal P}}
      { 
        {\DerNegTh {\Gamma} {l}{} {\mathcal P,\non l}} 
      }
      \)
    \item $l\sqin\Gamma$ with $\atmCtxt\Gam,\non l\models_{\mathcal T}$
    \end{enumerate}
  \item any incomplete proof-tree satisfying point 1.\ and extended by $\pi_2$ is $\pi_2$ itself.
  \end{enumerate}
\end{definition}

Given a \DPLLTh-extension, we can now identify a \DPLLTh\ transition
that the extension simulates, in the sense of Theorem~\ref{TTFail}:

\begin{theorem}[Simulation of the strategy back into \DPLLTh]

  If $\pi_2$ is a \DPLLTh-extension of $\pi_1$, and $\pi_1$
  corresponds to $\Del \| \phi$, then there is a (unique)
  \DPLLTh\ transition $ \Del \| \phi \Rightarrow \mathcal S_2$ such
  that $\pi_2$ corresponds to $\mathcal S_2$.
\end{theorem}
\begin{proof}See~\cite{farooque13,FarooquePhD}.
\end{proof}

If a complete proof-tree of \LKThp{}, whose conclusion is an
SMT-problem,\footnote{\ie it corresponds to an initial state of
  \DPLLTh} systematically uses the rules in the way described by the
above shapes, then it is the image of a \DPLLTh\ run.

While it could be envisaged to simulate \DPLLTh\ in a Gentzen-style
sequent calculus (with a variant of Theorem~\ref{TTFail}), the above
definition and theorem reveal the advantage of using a focused sequent
calculus for polarised logic: Definition~\ref{def:dpllext}
presents\footnote{mostly by specifying the management of polarities}
different ways of \emph{starting} the extension of an open branch (whose
leaf sequent is developed), each one of them corresponding to a
specific \DPLLTh\ transition; then \emph{focussing} takes care of the
following steps of the extension so that, when hitting developed
sequents again, the exact simulation of the \DPLLTh\ transition has
been performed. 

In order for proof-search mechanisms to exactly match
\DPLLTh\ transitions, focussing therefore provides the right level of
granularity and (together with an appropriate management of
polarities) the right level of determinism.

\begin{corollary}[Bisimulation]\nopagesplit
  The correspondance relation (see Definition~\ref{def:cor}) between 
  incomplete proof trees and \DPLLTh\ states is a bisimulation for the
  transition system defined on incomplete proof-trees of
  \LKThp\ by the strategy of \DPLLTh-extensions and on states by \DPLLTh. 
\end{corollary}

\subsection{More advanced features}

Finally, obtaining this tight result is the reason why we identified
the \emph{elementary \DPLLTh} system, a restriction of the
\textsf{Abstract DPLL Modulo Theories system} of~\cite{Nieuwenhuis06}:

Modern SMT-solvers feature some mechanisms that are not part of our
(logically complete) \emph{elementary \DPLLTh} system but increase
efficiency, such as \emph{backjumping} and \emph{lemma learning} (\cf
rules $\Backjump$, $\Learn$ in~\cite{Nieuwenhuis06}).

It is possible to simulate those rules in \LKThp\ by using general
cuts, by extending with identical steps several open branches of
incomplete proof-trees, and possibly by using explicit weakenings
(depending on whether we adapt the correspondence between
\DPLLTh\ states and incomplete proof-trees). Again, the details of
this can be found in~\cite{farooque13,FarooquePhD}.

However, with such ``parallel extensions'' of incomplete proof-trees,
it is not clear how to count the \emph{sizes} of proofs and extensions
in a meaningful way, so the quantitative aspects of
Theorem~\ref{TTFail} and Corollary~\ref{cor:DPLL-LK} are compromised;
neither is it clear which criterion on proof-trees (and on how to
extend them) identifies the proof-construction strategy that is the
exact image of a \DPLLTh\ procedure featuring those advanced
mechanisms. In other words, it is not clear how to obtain such a tight
correspondence.

Nonetheless, understaning backjumping and lemma learning in terms of
``parallel extensions'' of incomplete proof-trees, gives some concrete
leads on how to integrate these features to a root-first proof-search
procedure as described in this chapter. One of them is to use
\Index{memoisation} for the proof-search function. This is used to
close, in one single step, any branch that would otherwise be closed
by repeating the same steps as in a subproof that has already been
found. In particular, doing this avoids repeating, several times, the
proof-construction steps of a ``parallel extension'' corresponding to a
single backjump.

Memoisation is also a way of performing clause-learning: a learnt
clause $C$ is a clause for which we know that $\phi \models_{\mathcal
  T} C$, and that is made available for ${\Fail}$, ${\Backtrack}$ or
${\Propagate}$. Such a clause corresponds to a key $\DerNegTh
{\phi',\non {C}}{} {}{}$ of the memoisation table, with its proof as
value. A state where $C$ can be used for ${\Fail}$ or ${\Backtrack}$
is necessarily a sequent weakening $\DerNegTh {\phi',\non {C}}{} {}{}$
with extra formulae or literals, so the proof recorded in the
memoisation table can be plugged there to close the current
branch. When $C$ can be used for ${\Propagate}$, it suffices to make a
cut on the missing literal: one branch will be closed by plugging-in
the proof recorded in the memoisation table, while the other branch
will continue the simulation.

In the next chapter, we expand on the implementation of the above
results in the form of a specific \Index{plugin} for the
\Psyche\ system.

\section{Future work: Relation to abstract focussing}
\label{sec:future}

In this section, we give some hints as to how we could develop, in the
abstract \LAF\ system, the methodology of simulating \DPLLTh\ as
bottom-up proof-search in a focussed sequent calculus.  We already
know that we can capture \LKF\ as the \LAF\ instance \LAF[K1]. Since
we have slightly modified \LKF\ for the purpose of capturing \DPLLTh,
it is natural to ask whether that fits the \LAF\ framework as well.

\subsection{On-the-fly polarisation}

The first difference between \LKF\ and \LKThp, is that we have
on-the-fly polarisation of atoms. This is a feature that can easily be
integrated as another \LAF\ instance:

\begin{definition}[{{The \LAF\ instance for on-the-fly polarisation}}]

  The definition of the instance \LAF[K1p] is the same as that of
  \LAF[K1], except that
  \begin{itemize}
  \item molecules are now pairs $(A,\mathcal P)$ made of a formula $A$
    and a polarisation set $\mathcal P$ such that $A$ is $\mathcal
    P$-positive;
  \item atoms are literals.
  \end{itemize}
  The decomposition relation is defined in Fig.~\ref{fig:decompK1p} (which adapts Fig.~\ref{fig:decompK1}).

  Typing contexts are defined similarly to Definition~\ref{def:DBlabels}, but with the extra information about polarities:\\
  A typing context is given by $(\Gamma^+,\Gamma^-,\mathcal P)$, where
  $\Gamma^+$ is a list of literals and $\Gamma^-$ is a list of
  formulae;\\ $\varRead[n^+]{(\Gamma^+,\Gamma^-,\mathcal P)}$ is the
  $(n+1)^{th}$ element of
  $\Gamma^+$\\ $\varRead[n^-]{(\Gamma^+,\Gamma^-,\mathcal P)}$ is
  $(A,\mathcal P)$, where $A$ is the $(n+1)^{th}$ element of
  $\Gamma^-$.\\
  Context extension updates $\mathcal P$ just as in rule $\Store$ of \LKThp, so that (for instance)
  \[(\Gamma^+,\Gamma^-,\mathcal P)\Cextend a = ((\cons a{\Gamma^+}),\Gamma^-,(\polar a))\]
\end{definition}
\begin{bfigure}
    \[
    \begin{array}c
      \infer{\DerDec{}{\Dunit}{\XcolY{\Ptrue}{(\trueP,\mathcal P)}}}{}
      \qquad
      \infers{\DerDec{}{\Drefute {(\non N,\mathcal P)}}{\XcolY{\Pneg}{(N,\mathcal P)}}}[N\mbox{ is $\mathcal P$-negative}]{}
      \qquad
      \infers{\DerDec{}{a}{\XcolY{\Ppos}{(a,\mathcal P)}}}[a\in \mathcal P]{}\\\\
      \infer{\DerDec{}{\Delta_1,\Delta_2}{\XcolY{\paire{p_1}{p_2}}{(A_1\andP A_2,\mathcal P)}}}{
        \DerDec{}{\Delta_1}{\XcolY{p_1}{(A_1,\mathcal P)}}
        \quad
        \DerDec{}{\Delta_2}{\XcolY{p_2}{(A_2,\mathcal P)}}
      }
      \qquad
      \infer{\DerDec{}{\Delta}{\XcolY{\inj i p}{(A_1\orP A_2,\mathcal P)}}}{\DerDec{}{\Delta}{\XcolY{p}{(A_i,\mathcal P)}}}  
    \end{array}
    \]
  \caption{Decomposition relation for \LAF[K1p]}
  \label{fig:decompK1p}
\end{bfigure}

That instance being defined, it is however not completely clear how to
integrate to \LAF[K1p] the two admissible rules of \LKThp\ (at least
in their current form):
\[
\infers[(\Pol)]{
  \DerNeg{\Gamma}{}{}{\mathcal P}
}[
  l\sqin\Gamma \mbox{ and }\atmCtxtP{\mathcal {P}}{\Gamma},\non l\models_{\mathcal T}
]{
  \DerNeg{\Gamma}{}{}{\mathcal {P},l}
}
\qquad
\infers[(\cut)]{
  \DerNeg {\Gamma}{}{}{\mathcal P}
}[
  l\sqin\Gamma
]{
  \DerNeg{\Gamma}{l}{}{\mathcal P} 
  \quad 
  \DerNeg{\Gamma}{\non l}{}{\mathcal P}
}
\]
and how to use them in a bottom-up proof-search procedure based on
\LAF[K1p]. This is future work.

\subsection{Extending \LAF\ to \LAFTh{\mathcal T}}

The second difference between \LKF\ and \LKThp, is of course the
theory $\mathcal T$ and its decision procedure, used in rules
$\Init[1]$ and $\Init[2]$.

Notice that \LAF\ can accommodate a weak form of ``modulo theory'', at
least according to the definition of Chapter~\ref{ch:LAFwQ}: The
equality on atoms is a parameter that we can use to identify for
instance $a(3+4)$ with $a(7)$, in particular in the rule typing positive labels
\[
    \infer[\textsf{init}]{\Der{\Gam}{\XcolY{x^+}{(a,\lv r)}}}{
      \atmEq{\varRead[x^+]\Gam} {(a,\lv r)}
    }
\]

However in \LAF, we cannot close a branch in one step by involving
several atoms of $\Gam$, as we do for instance in \LKThp\ when we call
\eg a simplex algorithm to check the consistency of (the positive
literals of) $\Gamma$ (which in \LAF\ would be $\Gamma^+$).

For this we would need to extend \LAF\ with a decision procedure. We
could think of doing it in the following way:
\begin{itemize}
\item replace the notion of positive label by a notion of
  \Index{focussed justification}, and abstract away the part
  $\Gamma^+$ of a typing context $\Gamma$, which is no longer a
  function from positive labels to instantiated atoms but an abstract
  data structure called a \Index{positive typing context};
\item replace the notion of equality between instantiated atoms by a
  typing relation of the form $\DerTF{\Gam^+}{\XcolY{s^+}{(a,{\bf
        r})}}{}$, where $\Gam^+$ is a positive typing context, and
  $s^+$ is a positive justification.
\item add a notion of \Index{justification} and a typing relation of
  the form $\DerT{\Gam^+}{s}{}$, where $\Gam^+$ is a positive typing
  context, and $s$ is a justification.
\end{itemize}

We would then get:

\begin{definition}[Proof-Terms]
  Let \(\Just\) be a set of elements called \Index[justification]{justifications},
  and \(\Just^+\) be a set of elements called \Index[focussed justification]{focussed justifications}.

  Proof-terms are defined by the following syntax:
  \[
  \begin{array}{lll@{\recdef}l}
    \mbox{Positive terms }&\PTerms^+&t^+&pd\\
    \mbox{Decomposition terms }&\Decomp&d&s^+ \sep \THO f\sep\Tunit\sep d_1\Tand d_2 \sep \Tex r  d \\
    \mbox{Commands}&\PTerms&c& \just s\sep \cutc{x^-}{t^+} \sep \cutc{f}{t^+}
  \end{array}
  \]
  where $p$ ranges over patterns, $s$ ranges over justifications, $s^+$ ranges over focused justifications, $x^-$ ranges over $\var[-]$, $f$ ranges over functions from patterns to commands.
\end{definition}

And now we can give the \LAF\ system parameterised by a ``theory''
given by the pair of typing relations $\DerTF{\_} {\XcolY{\_}{\_}}{}$
and $\DerT{\_} {\_}{}$, which we may call $\mathcal T$, and which plays the
same role as the semantical inconsistency predicate in \LKThp.

\begin{definition}[\LAFTh{\mathcal T}]
  Let $\TContexts^+$ be the family of positive typing contexts.

  Assume we are given a pair $\mathcal T$ of two relations\\
  $(\DerTF{\_}
  {\XcolY{\_}{\_}}{}):(\TContexts^+\times\Just^+\times\IAtms)$ and
  $(\DerT{\_} {\_}{}):(\TContexts^+\times\Just)$.

  We define in Fig.~\ref{def:LAFTh} the derivability of three typing judgements
  \begin{itemize}
  \item $(\DerF\_ {\XcolY {\_} \_}{})\quad:\quad(\TContexts\times \PTerms^+\times\IMoles)$
  \item $(\Der\_ {\XcolY \_ \_})\quad:\quad(\TContexts\times \Decomp\times\ITDecs)$
  \item $(\Der\_ \_)\quad:\quad(\TContexts\times \PTerms)$
  \end{itemize}
\end{definition}

\begin{bfigure}[!h]
  \[
  \begin{array}{c}
    \infer[\textsf{sync}]{\DerF\Gam {\XcolY {p d} {(M,\lv r)}}{}}{
      \DerDec{}{\Delta}{\XcolY p{M}}
      \quad
      \Der{\Gamma}{\XcolY  d{(\Delta,\lv r)}}
    }\\
    \midline
    \\
    \infer{\Der\Gam{\XcolY{\Tunit}{(\Dunit,\lv r)}}}{\strut}
    \qquad
    \infer{\Der\Gam{\XcolY{ d_1\Tand d_2}{((\Del_1\Dand\Del_2),\lv r)}}}{
      \Der\Gam{\XcolY{ d_1}{(\Del_1,\lv r)}}
      \quad
      \Der\Gam{\XcolY{ d_2}{(\Del_2,\lv r)}}
    }
    \qquad
    \infer{\Der\Gam{\XcolY{\Tex {r'} d}{\Dex[s]{} {(\Del,\lv r)}}}}{
      \Ders{\Cfune\Gam}{\XcolY {r'} s}
      \quad
      \Der\Gam{\XcolY{ d}{(\Del,\cons {r'}{\lv r})}}
    }
    \\\\
    \infers
    {\Der{\Gam}{\XcolY{s^+}{(a,{\bf r})}}}[{{\Init[1]}}]{\DerTF{\Gam^+} {\XcolY{s^+}{(a,{\bf r})}}{}}
    \qquad
    \infer[\textsf{async}]{\Der\Gam {\XcolY {\THO f} {(\Drefute M,\lv r)}}}{
      \forall p,\forall\Delta,\quad
      \DerDec{}{\Delta}{\XcolY p{M}}\quad\imp\quad\Der{\Gamma\Textend[{\bf r}]{\Delta}}{f(p)}
    }
    \\
    \midline
    \\
    \infers{\Der\Gam{\just s}}[{{\Init[2]}}]{\DerT{\Gam^+}{s}}
    \qquad
    \infer[\Select]{\Der\Gam{\cutc{x^-}{t^+}}}{
      \DerF\Gam {\XcolY{t^+}{\varRead[x^-]\Gam}} {}}
    \qquad
    \infer[\cut]{\Der\Gam{\cutc{f}{t^+}}}{
      \Der\Gam {\XcolY {f}{(\Drefute M,\lv r)}}
      \qquad
      \DerF\Gam {\XcolY{t^+}{(M,\lv r)}} {}
    }
  \end{array}
  \]
\caption{\LAFTh{\mathcal T}}
\label{def:LAFTh}
\end{bfigure}

The empty theory could be recovered by having positive labels, having
positive typing contexts as maps from positive labels to instantiated
atoms, having focussed justifications be exactly positive labels, and
by setting setting $\DerTF{\Gam^+} {\XcolY{s^+}{(a,{\bf r})}}{}$
\iff\ $\atmEq{\Gam^+(s^+)} {(a,\lv r)}$ and never having
$\DerT{\Gam^+} {s}{}$.

Linear arithmetic could be defined by a relation $\DerT{\Gam^+} {s}{}$
that would check the consistency of the instantiated atoms in
$\Gam^+$, and a relation $\DerTF{\Gam^+} {\XcolY{s^+}{(a,{\bf r})}}{}$
that would check the consistency of the instantiated atoms in $\Gam^+$
together with $(\non a,{\bf r})$.

For congruence closure we could have the same approach, or we could
give a special role to $(a,{\bf r})$ in defining when $\DerTF{\Gam^+}
{\XcolY{s^+}{(a,{\bf r})}}{}$ holds.

The abstract focussing system could be seen as a \emph{functor} (in
the programming language sense) $\mathcal T\mapsto
\mbox{\LAFTh{\mathcal T}}$ that takes a pair of typing relations
(focussed, unfocussed) and returns a new pair of typing relations
(focussed, unfocussed). In that view, the functor could be composed
with others, and iterated. We conjecture that second-order logic or
higher-order logic could be captured by the fixpoint of this functor,
together with one that can convert an atom into a molecule, etc.

In every theory, the justifications could be dummy objects, if we do
not have proof objects to produce when running the decision procedure.
Or they could be as informative as one would like; in particular, it
would be useful if $s$ (\resp $s^+$) could at least indicate which
part of $\Gamma^+$ is actually used to derive $\DerTF{\Gam^+}
{\XcolY{s^+}{(a,{\bf r})}}{}$ (\resp $\DerT{\Gam^+} {s}{}$). This
could be done via a notion of free labels, so that we can apply the
same methodology as that of Section~\ref{sec:pruning} to re-use proofs
in different contexts.

The formal study of such a \LAF\ system, together with the adaptation
of its realisability models, is left for future work.

\chapter{The \Psyche\ system}
\label{ch:Psyche}

\minitoc

In this chapter, we describe \Psyche~\cite{Psyche}, a system
programmed in OCaml that implements, among other things, the ideas
developed in the previous chapter(s). In particular, it uses
polarities and focussing as a way to organise proof-search.

\Psyche\ is a highly modular proof-search engine designed as a
platform for either interactive or automated theorem proving, and the
acronym stands for the \emph{Proof-Search factorY for Collaborative
  Heuristics}. By platform, we mean that its architecture is organised
around a \Index{kernel} that interacts with \Index[plugin]{plugins} to
be programmed via a specific API. The goal of this architecture is to
allow the implementation of various theorem proving techniques while
guaranteeing correctness of the output: whether an input formula is
provable or not provable. As a platform, it can also be used to
implement the collaboration of various techniques which, once
programmed as plugins, share the same notion of \Index{proof-search
  state}.

The aim is therefore to provide a high level of confidence about the
output of the theorem proving process, no matter how programmers have
implemented their plugins, which is done by adopting and somewhat
transforming the LCF architecture~\cite{GMW79}.

Finally, \Psyche\ features the ability to call \Index[decision
  procedure]{decision procedures} such as those used in
Sat-Modulo-Theories provers. We therefore illustrate \Psyche\ by using
it for SMT-solving.

In brief:
\begin{itemize}
\item The kernel is based on a proof-search engine \emph{à la} Prolog,
  offering an API to perform incremental and goal-directed
  constructions of proof-trees in a focussed Sequent Calculus, which
  can be seen as a \Index{tableaux method}~\cite{TableauxHandbook}.
\item \Psyche\ can produce proof objects.
\item Plugins can be programmed to drive the kernel, using its API,
  through the search space towards an answer \emph{provable} or
  \emph{not provable}; correctness of the answer only relies on the
  kernel via the use of a private type for answers (similar to LCF's
  \emph{theorem} type).
\item Plugins can be interactive.
\item \Psyche\ offers a \Index{memoisation} feature to help programming
  efficient plugins.
\item The kernel is parameterised by a procedure deciding the
  consistency of collections of literals with respect to a background
  theory, just as in SAT-modulo-theories (SMT) solvers.
\end{itemize}

\medskip

The current version 2.0 of \Psyche\  is distributed
\begin{itemize}
\item with a kernel designed for first-order logic modulo a theory $\mathcal T$;
\item with a plugin whose behaviour on quantifier-free problems is
  \DPLLTh, using \Index{watched literals} to propagate literals or
  close branches, and \Psyche's memoisation feature to learn lemmas;
\item with decision procedures for: pure propositional logic (for
  SAT-solving), pure first-order logic, quantifier-free Linear
  Rational Arithmetic (LRA), and Congruence Closure;
\item with a \textsf{DIMACS} parser and an \textsf{SMTLib2}
  parser\footnote{The latter is taken without modification from the
    Alt-Ergo SMT prover.};
\item as a program of about 6700 lines of OCaml 4.00 (the kernel
  itself is only 800 lines), using hash-consing and Patricia tries for
  efficiency reasons.
\end{itemize}

\Psyche\ does not claim to be a better SAT- or SMT-solver or
first-order theorem prover than any existing one: for instance the
heuristics for applying \DPLLTh\ rules in the aforementioned plugin
are still basic, and so is the decision procedure for LRA (it is not
incremental).  What we offer here is a platform and its modularity:
anyone with better (or different) heuristics and decision procedures
can simply write them as OCaml modules of our predefined module types,
and \Psyche\ will seamlessly run with them, keeping the same LCF-style
guarantees.




In Section~\ref{sec:Psychemotivation}, we give more motivation for the
development of \Psyche. In Section~\ref{sec:psyche}, we describe the
general architecture of the system, in particular we explain how the
guarantee of correctness is enforced, using the kernel API and a
private type for answers. In Section~\ref{sec:kernel}, we briefly
review how the kernel works, connecting to the theory decribed in the
previous chapters.  Section~\ref{sec:plugins} then describes what the
specifications required of a plugin, and the way our distributed
plugin simulates \DPLLTh\ according to the results from
Chapter~\ref{ch:dpll}. Section~\ref{sec:decprocs} describes the
specifications of decision procedures and parsers, while
Section~\ref{sec:Psycheconclusion} concludes with some tests and
perspectives.

\section{Motivation}
\label{sec:Psychemotivation}

\Psyche's architecture is designed for the ambition of allowing various
theorem proving techniques (generic or problem-specific) to
collaborate on a common platform, whilst giving high confidence in the
answers produced.

Interfacing the numerous techniques and tools available for theorem
proving is legitimately receiving a lot of attention: Automated
Theorem Provers, SAT-solvers, SMT-solvers, Proof assistants, etc.
While trust is already an issue even for a single tool running on its
own, it becomes even more of an issue when different tools interact.
\emph{Proof-checking} is one way of addressing this, \ie being very
permissive in the algorithms used for theorem proving, as long as they
output some proof objects that can be checked. Another way is the
LCF-style~\cite{GMW79}, where only a small kernel of primitives needs to be
trusted, and anything smarter (\eg the interaction between
sophisticated techniques) boils down to calls to the kernel's primitives.

In the context of proof-checking (such as in Coq~\cite{Coq}), a
natural way to interact with different (already implemented)
techniques, is the black box approach, where an external tool is
called and its output is converted back into a proof that can be
checked by the system~\cite{ArmandFGKTW11,BCP:CPP2011}. It is somewhat
more surprising that, despite the highly programmable possibilities of
the LCF architecture (from which the ML languages come), the most
successful integration of automated reasoning techniques in an
LCF-based proof assistant such as
Isabelle~\cite{Nipkow-Paulson-Wenzel:2002,Isabelle} seems to also use
variants of the black box approach (as very impressively demonstrated
by Sledgehammer)~\cite{weber11smt,sledgehammer10,BlanchetteBP11}.


\Psyche\  aims at producing answers that are correct by construction, not
having to rely on proof-checking; it therefore adopts the LCF
philosophy (although it can produce proof objects), also because
having a simple trusted kernel is a convenient starting point for
different techniques to collaborate.
But the goal here is to open the black boxes and program their algorithms
directly with calls to the kernel's API, as plugins for \Psyche.

Such a deeper level of integration opens up the perspective of
interleaving the use of different techniques: An external tool
requires an input problem that it can entirely treat; but implementing
the \emph{steps} of its algorithm as small progressions in the
search-space covered by the main system, allows more possibilities,
such as running the technique \emph{up-to-a-point}, where a switch to
another technique may be appropriate (\eg depending on newly generated
goals).

The challenge is for the kernel to offer an appropriate API of
proof-search or proof-construction primitives, to allow the efficient
implementation of theorem proving techniques as plugins.  Most
LCF-style systems offer primitives corresponding to the inference
rules of Natural deduction, or a Hilbert-style system.  This is a very
fine-grained level, that leaves most (if not all) of the work to the
plugin; requiring it to use the kernel's primitives is less of an aid
and more of a constraint: it does ensure that, in case the output is
provable, a proof has been constructed (at least theoretically), but
it is a computational overhead for the plugin's work.

\Psyche\ makes the choice of a bigger grain, and leaves to the kernel
some real proof-search computation, but where no decision needs to be
made. For this we use the focussed sequent
calculus~\LKThp~\cite{farooqueTR13,farooque13}, whose quantifier-free
version has been presented in Chapter~\ref{ch:dpll}.  Not only can
polarities and focussing be used to describe effective proof-search
strategies in Sequent Calculus (narrowing the search-space offered by
Gentzen's original rules), but in our case, they also specify a
sensible division of labour between \Psyche's kernel and \Psyche's
plugins, re-designing the standard LCF-style API.

This new design makes \Psyche\ guarantee the correcteness of both
types of answers: \emph{provable} or \emph{not provable}, while the
traditional LCF style only guarantees the correctness of answers of
the form \emph{provable}.

\section{Overview and general architecture}
\label{sec:psyche}

The kernel contains the mechanisms for exploring the proof-search
space, taking into account branching and backtracking. It has no
\emph{a priori} regarding the order in which branches are explored,
and this lack of intelligence makes its code rather short. If it
reaches a proof, then that proof is correct by construction, and if
the entire search space is explored and no proof is found, then the
kernel correctly outputs that no proof exists.

The plugins then drive the kernel by specifying in which order the
branches of the search space should be explored and to which depth,
something that is expected to depend on the kind of problem that is
being treated. The quality of the plugin is how fast it drives the
kernel towards a answer \emph{provable} or \emph{not provable}.

This already departs from the traditional LCF-style in that some
actual proof-search computation is performed in the kernel, not
just atomic steps of proof-construction: 

In traditional LCF, each inference rule of the logic
\[\infer[\textsf{name}]{\textsf{conc}}{\textsf{prem}_1\quad\ldots\quad\textsf{prem}_n}\]
gives rise to a primitive of the kernel's API, whose type declares $n$ arguments:
\ctr{\texttt{name: thm -> $\cdots$ -> thm -> thm}}

In \Psyche's kernel, such an inference rule is ``wrapped'' in the
kernel's unique API primitive: 
\ctr{\texttt{machine: statement -> output}}
such that \texttt{search(conc)} will trigger the
recursive calls \texttt{search(prem\_1)},\ldots,
\texttt{search(prem\_n)}, as bottom-up proof-search should do.

\Psyche's general architecture is illustrated by its main top-level call (slightly reworded for clarity):
\ctr{\texttt{Plugin.solve(Kernel.machine(Parser.parse input))}}

\Psyche\  has a collection of parsers (currently one for \textsf{DIMACS} and one for \textsf{SMTLib2}) and calls the appropriate one on \Psyche's input.
The resulting abstract syntax tree is fed to the kernel's \texttt{machine} function that will initiate the search.\footnote{In fact, the kernel module is created with the choice of a background theory that is either guessed from the input or specified by the user on the command line.}
This produces a value of type \texttt{output} that is given to the plugin to work with, and the plugin must solve the problem by outputting an answer \emph{provable} or \emph{not provable}.

This could give the impression that the plugin performs computation after the kernel has finished his, but this is not quite true, as illustrated by the nature of type \texttt{output}: \ctr{\texttt{type output = Jackpot of answer | InsertCoin of coin -> output}}
which describes the kernel as a \emph{slot machine}:
when it is run, it outputs 
\begin{itemize}
\item either a definitive answer \emph{provable} or \emph{not provable}
\item or an intermediate output that represents unfinished computation: in order for computation to continue, the plugin needs to ``insert another coin in the slot machine'';
  depending on the kind of coin inserted, proof-search will resume in a certain way.
\end{itemize}
To summarise, the kernel performs proof-search as long as there is no decision to be made (on which backtrack may later be needed), and when it hits such a point, it stops and asks for another coin to indicate how to proceed next. The plugin drives the kernel in the exploration of the proof-search space by inserting carefully chosen coins, hoping that one day the machine will stop with the ``jackpot'': a value of the form \texttt{Jackpot(...)}.

Now while this architecture somewhat departs from LCF, it does share with it the distrust of anything outside the kernel: when concerned with the soundness of the answer (whichever it be), the plugin is here considered as an adversary, so \Psyche\  defines the type \texttt{answer} as a private type that only the kernel can inhabit (just like the \texttt{thm} type of LCF). \Psyche's type
\ctr{\texttt{answer = private Provable of statement*proof | NotProvable of statement}}
can be read by the plugin and the top-level if need be, but cannot be inhabited by them.
That way, a plugin cannot cheat about \Psyche's answer: the worst it can do is of course to crash \Psyche's runs or diverge.
In \Psyche\  as in traditional LCF, inhabitation of the abstract type (in case of \Psyche, with a value of the form \texttt{Provable(...)}) explicitly or implicitly constructs a proof of the statement.
But contrary to LCF, \Psyche\  also gives guarantees when the output is \emph{not provable}: it can only occur when the kernel has entirely explored the search-space unsuccessfully.

Such a use of typing prompted for an ML-language to implement \Psyche, and we chose OCaml (4.0).

\section{\Psyche's Kernel}
\label{sec:kernel}

As described above, the kernel's API has the slot machine as its only
primitive, controlled by the \emph{coins} that are inserted in it.  In
order for efficient plugins to be conveniently programmed, the
kernel's primitive needs to accept a rather expressive range of coins
that can specify a smart exploration of the search-space.  This
depends on the inference system that is used in the kernel for the
incremental and bottom-up construction of proof-trees, and on
identifying the inference rules that the kernel will perform
automatically from those that will pause computation and prompt the
plugin for new directions.

This is where focussed sequent calculi for polarised logic(s) come in.
Focussing is what we use for the division of labour between \Psyche's
kernel and \Psyche's plugins:

The kernel applies the asynchronous steps automatically without any
instruction from the plugin, and then stops and asks for another coin
describing the next synchronous phase, where smart choices may have to
be made (starting with the choice of the positive formula to work on).

An important consequence of this division of labour is that
\textbf{every kernel call terminates}, because the length of each
phase is bounded by the size of the formula(e) being decomposed.
Therefore, infinite proof-search has to go through an infinite
interaction between the kernel and the plugin (unless the plugin itself
loops before inserting the next coin).

The choice of polarities on connectives and literals affects the
kernel-plugin interaction. For instance the polarity of $\vee$ will
determine whether it will be decomposed automatically by the kernel
(second rule, asynchronous) or with a smart choice by the plugin
(first rule, synchronous):
\[
\infer{\Der{\Gamma}{A_1\orP A_2}}
{\Der{\Gamma}{A_i}}
\qquad
\infer{\Der{\Gamma} {A_1\orN A_2,\Gam'}}
{\Der {\Gamma} {A_1,A_2,\Gam'}}
\]
The polarity of literal being also crucial, \Psyche\ offers the
plugin the possibility to polarise literals \emph{on-the-fly}, during
the search (which is very useful for the plugins we implemented).

In \Psyche\ 2.0, the kernel implements the sequent calculus
\LKThp~\cite{farooque13,FarooquePhD}, whose quantifier-free version
was presented in Chapter~\ref{ch:dpll}. But \Psyche\ does implement
the full system with quantifiers, with specific mechanisms dealing
with eigenvariables (introduced when proving a universal formula) and
meta-variables (introduced when proving an existential formula).

The different coins that the plugin can insert thus correspond to the
smart application of the non-asynchronous inference rules of \LKThp: a
formula to select, a side to choose when decomposing $\orP$, a literal
to polarise in a certain way, a cut to be made (\LKThp\ admits cuts),
or a consistency check of the current sequent with the given
background theory (a global parameter of the kernel).

Finally, the plugin can also instruct the kernel to move in the
search-space: when it gets tired of investigating the current branch,
it can abandon it temporarily and explore the next success/failure branch to
the left/right.

The code of the kernel is rather small (around 800 lines) and purely
functional. Continuation-Passing-Style (CPS) is used to minimise the
use of the stack and provide a natural way to represent the
progression of the kernel within the search space: the API
function
\ctr{\texttt{machine: statement->output}}
actually wraps a real (tail-)recursive function
\ctr{\texttt{search: statement->(output->'a)->'a}}
with the identity
continuation. Continuations are heavily used for branching and
backtracking (\eg when \texttt{search} applies a rule with several
premises, it makes a recursive call on one of the branches and stacks
up the others in the continuation that is passed; when the plugin
chooses to explore one branch, the kernel records in a similar way the
other branches that are not being explored yet -forcing in the end the
entire exploration of the search-space), and naturally provide the
values implementing a slot machine waiting for its coin.

\section{Plugins}
\label{sec:plugins}

\subsection{Specifications and implemented instances}

A plugin is any OCaml module implementing the following identified
module type (bearing in mind that \texttt{answer} is for the plugin a
private type that it cannot inhabit by itself):
\begin{verbatim}module type PluginType = sig
  ...
  solve: output->answer
end
\end{verbatim}

However, it is likely that the sophisticated strategies/heuristics that
the plugin is meant to implement rely on some clever choice of
data-structures for formulae, sets of formulae, sets of literals. So
the plugin and the kernel have to agree on those three data-structures
that are communicated both ways during the interaction.  In
\Psyche\ 2.0, the kernel provides the data-structure to represent
formulae, but the plugin can embark in the data-structure the
information that it needs to treat them efficiently. The
data-structures implementing sets of formulae and sets of literals, on
the other hand, are parameters of the kernel, and the plugin provides
them.\footnote{This is admittedly a security problem, since a bug in
  the plugin's data-structure for sets could affect the way a sequent
  is transformed by the kernel when it applies an inference rule
  bottom-up. The next version of \Psyche\ will adopt a double
  representation of sets (one for the kernel, one for the plugin) to
  completely avoid the kernel relying on plugin code.}

\medskip

We first tested \Psyche's architecture with a basic plugin \texttt{Naive}, which
\begin{itemize}
\item implements sets (of formulae, literals) with OCaml's lists;
\item inserts the first available coin in the slot machine, whenever asked.
\end{itemize}
This works fine for small tautologies, printable on a screen.

\medskip

More recently, Jean-Marc Notin provided a module for interactive
theorem proving, via a command-line interface: it still implements
sets using OCaml's lists, but every time a coin needs to be inserted
in the machine, the interface prompts the user for the coin to insert.

\medskip

A more ambitious aim for automated reasoning was to capture in
\Psyche\ some propositional SAT and SAT-Modulo-Theories solving
techniques, making \DPLLTh\ technology available in a generic
tableau-like / Prolog-like / goal-directed proof-search framework like
\Psyche.

For this we implemented the simulation of \DPLLTh, expressed rather
canonically as a transition system~\cite{Nieuwenhuis06}, as a simple
bottom-up proof-construction mechanism in \LKThp, as described in
Chapter~\ref{ch:dpll}. More practically, every rule of \DPLLTh\ can
be seen as the insertion of a particular coin in \Psyche's slot
machine.

We implemented this as two different plugins for \Psyche:
\texttt{DPLL\_Pat} and \texttt{DPLL\_WL}.  These remain toy plugins,
because, although it is now clear, from Chapter~\ref{ch:dpll}, how to
perform each rule of \DPLLTh\ in \Psyche, we still have to decide \emph{which} rule to apply and \emph{when}.  So the two plugins
\begin{itemize}
\item embark, in the kernel's representation of formulae, a flat representation of them as sets of literals when the formulae happen to be clauses;
\item implement sets (of formulae/clauses, literals) using Patricia tries;
\item implement a basic strategy to apply \DPLLTh\ rules; in the case
  of propositional logic: apply \textsf{Fail} or \textsf{Backtrack} if
  possible, if not try \textsf{Unit Propagate}, if not do
  \textsf{Decide} on some random literal.
\end{itemize}
The two plugins differ in the way they look up for the applicability
of \textsf{Fail} / \textsf{Backtrack} / \textsf{Unit Propagate}:
\texttt{DPLL\_Pat} looks it up using the Patricia tries implementing
sets of clauses, while \texttt{DPLL\_WL} looks it up using the
technique of \Index{watched literals}~\cite{MoskewiczMZZM01} (keeping
a small watching table in the plugin). This technique was originally
implemented in \Psyche\ by student Matthieu Vegreville, and the plugin
seems on average 1.5 faster than that using Patricia tries.

\subsection{Memoisation and lemma learning}

Such plugins would not be efficient at all if no backjumping and
clause learning was done while performing
\DPLLTh. In~\cite{farooque13} we also show how to do this using
general cuts, and either accept to extend several open branches of an
open proof-tree with identical steps or depart from the bottom-up
proof construction paradigm that we have used so far. We opted for a
generic mechanism to avoid re-doing, for some open branch, the same
steps as those used in a previously completed branch:
\Index{memoisation}. In Chapter~\ref{ch:dpll} we explained how the use
of memoisation emulates the use of a learnt clause for \textsf{Fail},
\textsf{Unit Propagate}, etc.

Indeed, nothing prevents a plugin from recording the sub-trees
completed by the kernel, and proposing them later for another branch
where the same proof-tree is relevant.  \Psyche\ 2.0 therefore offers
a memoisation module, to be used by plugins to record values of (the
abstract) type \texttt{answer}. And the kernel's slot machine accepts
from the plugin, as a special coin carrying such a value, ``here is an
already found answer that also applies to the current goal''. The
kernel accepts the value as closing the current branch (one way or
another) \textbf{without any proof-checking} (since the abstract type
ensures the value came as an earlier output of the kernel); it only
checks that the value applies to the current goal.

The memoisation table is filled-in by clause-learning: our plugin adds
an entry whenever it builds a complete proof of some sequent
$\DerNegTh {\Delta}{} {}{}$ and no previous entry $\DerNegTh
{\Delta'}{} {}{}$ exists with $\Delta'\subseteq\Delta$, or whenever it
concludes that some sequent $\DerNegTh {\Delta}{}{} {}$ is not
provable and no previous entry $\DerNegTh {\Delta'}{}{}{}$ exists with
$\Delta\subseteq\Delta'$.

Now for a memoised answer \emph{Provable} to be reusable as often as
possible, a pre-processing step is applied to a proof-tree before it
enters the table: it is pruned from every formula that is not used in
the proof. This is easy to do for the complete proofs
of \LKThp\ (eager weakening are applied a posteriori by inspection of
the inductive structure). \Psyche's kernel actually performs the
pruning on-the-fly whenever an inference is added to complete proofs,
so that, whenever it outputs \texttt{Jackpot(sequent,proof)}, the
sequent is already pruned.

Since proof-completion can be seen as finding a \Index{conflict} (a
situation where the current partial model contradicts the set of
clauses), pruning by eager weakening is a \Index{conflict analysis}
process naturally provided by structural proof theory:

Conflict anaylsis is a process used in SAT- and SMT-solving aims at
identifying, in a situation of \textsf{Fail}, \textsf{Backtrack}, etc,
which literals of the current model are sufficient to contradict, when
taken together, the set of clauses; the disjunction of their negations
forms a new clause that can be learnt and re-used later. Techniques to
compute this can be based on \emph{graph analysis}; the kind of
conflict analysis performed by the pruning mechanism of \Psyche\ turns
out to be a particular form the graph analysis mechanism.

Of course, just as in SMT-solving, the efficiency of conflict analysis
relies on the efficiency of the decision procedure in providing a
small inconsistent subset whenever it decides that a set of literals
is inconsistent.

\medskip

Another feature sometimes used in SMT-solving, in conjunction with
clause learning, is the use of \Index[restart]{restarts}: at some
point of the \DPLLTh\ run, computation resumes with the empty model:
\[\Delta  \| \phi \Rightarrow \emptyset \| \phi \]
This is only useful if the current set of clauses $\phi$ is different
from the original one, \ie some clauses have been learnt: in that
case, restarting from the empty model but with all the learnt clauses,
might be faster than closing all the branches that have been opened
(corresponding to the decision literals in $\Delta$). 

In \Psyche, restarts can be done the same way: since the plugin is in
charge of computation, it can record the first output that the kernel
produced, and later come back to it: the side effect that makes it
different from the first run is that the memoisation table has been
filled with valuable information; this information may allow the
search to find a proof more quickly than by closing all the open
branches of the current incomplete proof-tree.  This has been
implemented in \Psyche\ by students Zelda Mariet and Clément
Pit-Claudel, with convincing experimental results.

\section{Decision procedures}
\label{sec:decprocs}

Decision procedures and parsers integrate \Psyche's code the same way
as plugins: we offer a module type for decision procedures and one for
parsers. Someone with a decision procedure or a parser can implement a
module of the corresponding type and run \Psyche\ with it.

In the case of decision procedures for quantifier-free problems (\ie
with ground literals), the output of a decision procedure for the
background theory $\mathcal T$ should be able to decide whether a
conjunction of literals is consistent with $\mathcal T$ or not.
\begin{verbatim}
module type GroundDecProc = sig
  ...
  type literals
  ...
  consistency: literals set -> (literals set) option
end
\end{verbatim}
The decision procedure provides the type of literals, so as to run
efficiently, while the kernel accepts any type for literals since it
will not inspect its values.

The output of the \verb=consistency= function is not a boolean: it
should be \texttt{None} if the input is a set of literals consistent
with the theory $\mathcal T$, and \texttt{Some(s)} otherwise, with
\texttt{s} being a subset of the input that is already
inconsistent. Indeed, conflict analysis requires such subsets to be
produced when an inconsistency is found, and the smaller the subsets,
the more efficient clause learning and memoisation will be.

In the case of problems with quantifiers, the decision procedure
should answer whether there is a way to instantiate meta-variables so
as to make the conjunction of literals inconsistent with the theory:
only in this case will the current branch be immediately closed,
propagating the instantiation of meta-variables to the remaining open
branches. In case such an instantiation fails another branch, we
should backtrack to the current branch and propose another way of
closing it. Therefore, the decision should not only be able to decide
whether there is an instantiation of meta-variables that makes the
literals inconsistent, but it should be able to \emph{enumerate all}
possible instantiations that make the literals inconsistent.
Instead of a boolean answer, we thus expect a \Index{stream} of
solutions, each of which is a set of literals together with a working
instantiation:
\begin{verbatim}
module type DecProc = sig
  ...
  type literals
  type constraints
  ...
  val consistency : literals set -> (literals set,constraints) stream
end
\end{verbatim}

The idea of using streams of solutions is natural, and proposed in the
form of \Index[instance stream]{instance streams} in~\cite{Giese00} in
a proof-search methodology that deliberately avoids
backtracking. Although proof-search in \Psyche\ does backtrack, we
should investigate the connection between \Psyche's methodology and
that of~\cite{Giese00}.

As evoked by the module type above, instantiations in \Psyche\ are
actually called \Index[contraint]{constraints}: in pure first-order
logic, a constraint would simply be a (most general) unifier $\sigma$
that makes two literals $l_1$ and $l_2$ of the input set such that
$l_1\sigma=\non l_2\sigma$. It would be easy to enumerate all such
constraints, by enumerating all pairs $l_1$ and $l_2$ of the input set
that can be unified in the above sense: they are in finite numbers.

But for other theories we could imagine different kinds of constraints
on meta-variables: for instance in Linear Rational Arithmetic we could
imagine a constraint imposing that meta-variable \texttt{?X3} be in
the interval $\left[0;\frac 3 2\right]$. Therefore, the decision
procedure provides the notion of constraint that is appropriate for
the background theory $\mathcal T$, while the type for constraints is
abstract for the kernel, which will only propagate constraints from
branch to branch, but not inspect them.

The exact specifications that should be met by the constraint
structure so that proof-search using them is sound and complete with
respect to the formalisation of the theory $\mathcal T$ without
meta-variables, is the object of a paper being currently written with student
Damien Rouhling.

\bigskip

Finally, \Psyche\ is modular in its parsers: a \Index{parser} is any
module of a pre-defined module type, and should in particular
implement a function 
\ctr{\texttt{parse: string->((statement option)*(boolean option))}} that turns a string input into a
statement to be proved (or \texttt{None} if no statement was parsed in
the input), and possibly an expected result
\emph{Provable}/\emph{NotProvable} that the input string may indicate.

\sectionno{Conclusion: Testing and perspectives}
\label{sec:Psycheconclusion}

\Psyche\ 2.0 is run from the command-line, taking as input one or more file(s) or directory(ies), or, if none indicated, the standard input:

{\small
\begin{verbatim}
psyche [OPTION]... [FILE/DIR]...

Available options are:
  -theory selects theory (among empty, lra, cc, first-order; default is empty)
  -gplugin selects generic plugin (among naive, dpll_pat, dpll_wl; default is dpll_wl)
  -latex allows latex output of proof-trees
  -alphasort treats input files in alphabetical order (default is from smaller to bigger)
  -examples treats theory examples instead of standard input
  -nocuts disallows cuts
  -fair ensures fairness between formulae for focus
  -noweakenings disables conflict analysis
  -nomemo disables memoisation
  -restarts selects a restart strategy
  -help  displays this list of options
\end{verbatim}
} As illustrated by the options, \Psyche\ can produce proofs (of
\LKThp) and print them as inference trees in \LaTeX\ format (but
proofs can quickly get too big for \LaTeX).

We ran \Psyche\ on instances of SAT in \textsf{DIMACS} format, and
\texttt{QF\_LRA} instances in \textsf{SMTLib2} format and the results
are available on Psyche's website~\cite{Psyche}. Psyche works well on
small instances and its performance starts declining between 20Kb and
100Kb of input problem size (of course this is no appropriate measure
of difficulty). This is of course very far from current SAT
benchmarks, perhaps a bit less from SMTLib2 ones (our instances were
download from the up-to-date library).  But as we said, the current
plugins and decision procedures are illustrative toys. \Psyche\ is a
platform where people knowing good and efficient techniques should be
able to program them.

In the short-to-medium terms, we plan to 
\begin{itemize}
\item improve the decision procedure for LRA (making it incremental, and returning smaller sets);
\item implement other theories and combine them (congruence closure,
  Linear Integer Arithmetic, bit vectors, etc);
\item improve \DPLLTh\ plugins to better handle non-clausal formulae;
\item implement other theorem proving techniques as plugins:
  \Index[analytic tableau]{analytic tableaux} are the closest to our
  sequent calculus, but theoretical developments have already shown
  that \Index[connection tableau]{clausal tableaux} (including
  \Index[connection tableau]{connection tableaux}) can also be
  done~\cite{FarooquePhD}, as well as \emph{resolution} proofs.
\end{itemize}

Finally, we can imagine using a proof assistant to prove \Psyche's
correctness, since the kernel seems small enough (800 lines) and the
plugins need not be certified.

We will develop our long-term plans for \Psyche\ in the conclusion of
this dissertation.

\chapter{Conclusion and further work}
\label{ch:quant}

\minitoc

\section{Summary of the topics covered by this dissertation}

In the first part of this dissertation, we reviewed the computational
interpretation of proofs in terms of Call-by-Name and Call-by-Value
evaluation of
programs~\cite{CurienHerbelinDuality99,SelingerControlCat99}; we
approached realisability semantics by a systematic construction of
orthogonality
models~\cite{Parigot97,DanosKrivine00,Krivine01,MiquelLMCS11}, used
for instance to prove strong normalisation results or classical
witness extraction. From this the concepts of polarities and focussing
naturally emerged~\cite{MunchCSL09}, and a computational
interpretation of focussed proofs was given in terms of
\linebreak pattern-matching~\cite{ZeilbergerPOPL08,Zeilberger08}.

In the second part of this dissertation, we developed this approach
into an abstract focussed sequent calculus \LAF\ with proof-terms, of
which several focussed calculi of the literature are instances, such
as \LKF\ and \LJF~\cite{liang09tcs}. We used this framework to
formally relate classical realisability with the computational
interpretation of focussing as pure pattern-matching, again via the
construction of orthogonality models.

In the third part of the thesis we explored a specific approach to
theorem proving benefitting from the use of polarities and focussing.
We described how these concepts can contribute to the description of
the \DPLLTh\ procedure for SMT-solving~\cite{Nieuwenhuis06} as a
specific strategy for the bottom-up proof-search process specified by
the sequent calculus. For this we extended the focussed sequent calculus
\LKF\ into \LKThp, equipped with the ability to polarise atoms
\emph{on-the-fly} and call a decision procedure specific to the
background theory $\mathcal T$.

We then described the implementation of a proof-search engine called
\Psyche~\cite{Psyche} whose architecture is based on a kernel that
interacts with plugins to be programmed via a specific API. This
allows the implementation and experimentation of various reasoning
techniques (among which \DPLLTh) and heuristics, without worrying
about breaking the correctness of \Psyche's output: this is guaranteed
to be correct by the architecture, which develops a new variant of the
LCF style~\cite{GMW79}.

\section[Further work]{Impact of this dissertation of the development of \Psyche, and further work}

In conclusion of this dissertation we proffer two main directions in
which the material of this dissertation will be developed and
integrated to the next releases of \Psyche.

\bigskip

The first one is a rather major change in the kernel of \Psyche:
instead of implementing bottom-up proof-search in the particular
focussed sequent calculus called \LKThp, which is specific to
classical logic, \Psyche\ 3.0 will implement proof-search in the
abstract focussed sequent calculus \LAF\ developed in
Part~\ref{partII} of this dissertation. This will allow the kernel to
be decomposed into smaller components: the main module will have much
fewer rules to implement than in \LKThp, taking advantage of the
``big-step presentation'' of focussing; the decomposition of formulae
into smaller formulae, given by the specific relation $\decf{}$ of
\LAF, will be moved to a specific module where the inductive syntax of
formulae is implemented; another module will implement typing contexts
with their notion of context extension that crucially determines which
logic is being implemented, etc.

The advantages of modularising the kernel in this way are numerous: 
\begin{itemize}
\item It will allow \Psyche\ to run on different instances of \LAF,
  thus handling different systems and logics;
\item \Psyche\ will then be equipped with proof-terms, which may be
  used as compact representations of proofs in memory (\eg in the
  memoisation table); this will also allow the extraction of programs
  from proofs (in different logics);
\item the code will also be simpler to understand and formally prove
  correct, as most of the specifications describing the roles of each
  component have already been identified in Part~\ref{partII} of this
  dissertation, with most of the theorems already formalised in
  \Coq~\cite{LengrandHDRCoq};
\end{itemize}

In retrospect, this next move in the development of \Psyche\ is also
what motivated the detailed study of \LAF, at the cost of presenting it
in a rather technical way. However, theoretical work still needs to be
done before this new basis for \Psyche's implementation replaces the
current one. Indeed, as described in details in
Section~\ref{sec:future}, it is not clear how \Psyche\ can take
advantage of rules that are admissible for specific instances of
\LAF\ but not generically (\eg specific forms of cuts, on-the-fly
polarisation rules, etc); more importantly, in order to supersede the
current implementation, \LAF\ needs to be generalised into a system
\LAFTh{\mathcal T}\ that can call a decision procedure for a theory
$\mathcal T$. What conditions are required of such procedures for
cut-elimination to work, etc, remains to be identified.

\bigskip

The second direction for further development is exploiting the
machinery for quantifiers that has newly been introduced with the
release of \Psyche\ 2.0 on 20th September 2014. 

As briefly described in Section~\ref{sec:decprocs}, this machinery
involves meta-variables which are introduced when breaking an
existential quantifier sitting on top of a formula to be proved, and
eigenvariables which are introduced when breaking a universal
quantifier. Dependencies between them are recorded in the
proof-search, so as to avoid the production of incorrect instances for
meta-variables, from which no actual proof could be re-constituted.

On the note of dependencies, Skolemisation is often described as the
transformation of a formula $\forall x_1\ldots \forall x_n\exists y A$
to be \emph{refuted} (by tableaux methods, resolution, etc\ldots) into
the formula $\forall x_1\ldots \forall x_n\subst A y{{\sf
    sk}_y(x_1,\ldots,x_n)}$, where ${\sf sk}_y$ is a (new)
\Index{Skolem symbol} (specific to $y$). The correctness of this
transformation is often justified by semantical models involving the
axiom of choice, and Skolemisation is often applied as a
pre-processing step before refutational methods are applied.  In
bottom-up proof-search, Skolemisation occurs \emph{on-the-fly} when
the universal quantifier of $\exists x_1\ldots \exists x_n\forall y
{\non A}$ is broken, and the skolem symbol ${\sf sk}_y$ is merely the
eigenvariable $Y$ introduced for $y$; the fact that the Skolem symbol
is \emph{applied} to $x_1$,\ldots,$x_n$ (or in proof-search, to their
corresponding meta-variables $?X_1$,\ldots,$?X_n$) is a mere
implementation trick to record the dependencies between eigenvariables
and meta-variables: writing $Y(?X_1,\ldots,?X_n)$ simply records that
$?X_1$,\ldots,$?X_n$ were introduced before $Y$ and any correct
instances for them cannot mention $Y$. This is a smart implementation
of the dependencies in the case of pure first-order logic, inasmuch
first-order unification will rule out incorrect instances ``for free''
thanks to the \verb=occurs_check=.

As \Psyche\ aims at working modulo theories, it is no longer clear
that this specific implementation of dependencies will be as
appropriate when another algorithm than first-order unification is run
to close branches. A dual implementation of dependencies would
consist for instance in recording, whenever a meta-variable $?X$ is
introduced, the eigenvariables that existed at that point, among which
any correct instance for $?X$ would need to find its free
variables. This is for instance the choice in \Coq~\cite{Coq}: instead
of recording the dependencies that are disallowed (as in
Skolemisation), one records the dependencies that are allowed.  To
avoid making any commitment on that choice of implementation,
\Psyche\ 2.0 is modular in the data-structure that implements
dependencies.

Similarly, \Psyche's kernel is agnostic in regard of the
\Index[constraint]{constraints} imposed on the instantiation of
meta-variables by closing branches: proving a sequent $\DerNegTh
\Gamma {\Gam'}{}{}$ mentioning meta-variables should output (if
successful) on constraint $\sigma$ on these meta-variables, which in
pure first-order logic could simply be a \Index{first-order unifier},
but in other theories could be of a different nature (one could think
of convex polytops for arithmetic, for instance). A branching rule
such as
\[
\infer[{[(\andN)]}]{\DerNeg{\Gamma}{A\andN B,\Delta} {} {}}
    {\DerNeg{\Gamma}{A,\Delta} {} {}
      \qquad \DerNeg{\Gamma}{B,\Delta} {} {}}
\]
should eventually produce the \Index{meet} $\sigma\et\sigma'$ of the
two constraints $\sigma$ and $\sigma'$ returned by the recursive calls
of the proof-search function on the two premisses, something we can write as
\[
\infer[{[(\andN)]}]{\DerNeg{\Gamma}{A\andN B,\Delta} {} {}\;\Rightarrow\sigma\et\sigma'}
    {\DerNeg{\Gamma}{A,\Delta} {} {}\;\Rightarrow\sigma
      \qquad \DerNeg{\Gamma}{B,\Delta} {} {}\;\Rightarrow\sigma'}
\]
The meet should of course exists (otherwise we should try to find
other ways to prove the premisses), in other words $\sigma\et\sigma'$
should not be the unsatisfiable constraint $\bot$. There we see an
algebraic structure like a semi-lattice appear as the natural concept
to spell out the abstract specifications of how constraints work.  To
avoid the independent exploration of branches before realising they
produce incompatible constraints, \Psyche\ 2.0 implements the
propagation of constraints from one branch to the next, which we could
write as
\[
\infer[{[(\andN)]}]{\sigma_0\Rightarrow\;\DerNeg{\Gamma}{A\andN B,\Delta} {} {}\;\Rightarrow\sigma'}
    {\sigma_0\Rightarrow\;\DerNeg{\Gamma}{A,\Delta} {} {}\;\Rightarrow\sigma
      \qquad \sigma\Rightarrow\;\DerNeg{\Gamma}{B,\Delta} {} {}\;\Rightarrow\sigma'}
\]
where $\sigma_0\geq\sigma\geq\sigma'$ for the semi-lattice ordering,
breaking the symmetry of the introduction rule for conjunction
depending on which of the two branches is actually explored first.

A paper with Damien Rouhling is currently being written on such
systems of constraint propagation, which still allow proof-search to
backtrack, with persistent data-structures for constraints.  Besides
justifying the implementation of \Psyche\ 2.0 and identifying the
algebraic specifications that an implemented constraint module should
satisfy, further development should investigate the connections with
\Index[constraint tableau]{constraint tableaux}~\cite{GieseHaehnle03}
and \Index[constraint logic programming]{constraint logic
  programming}~\cite{Jaffar1994503}.

\bigskip

Now, what to do with the quantifier features of Psyche\ 2.0?

First, test them on standard benchmarks and possibly add a TPTP parser
to \Psyche. 

Secondly, it was shown in~\cite{FarooquePhD} how to simulate, as
proof-search in \LKThp[\emptyset]\ with quantifiers, the techniques of
\Index[clause tableau]{clause tableaux} and strong or weak
\Index[connection tableau]{connection tableaux} (see
\eg\cite{RobinsonV01}) for pure first-order logic (hence the empty
theory $\emptyset$). How to simulate resolution is also known. So
turning these simulations into the implementation of plugins for
\Psyche\ is the obvious next step, which would allow us to run tests
and compare the plugins with other implementations.

Thirdly, we have an approach for mixing first-order reasoning with
theories (or, said differently, perform instantiations in presence
of a theory); with this we can:
\begin{itemize}
\item Investigate to what extent the standard
  \Index[trigger]{triggers-based} mechanisms of SMT-solvers for
  instantiations, can be described in our setting; in particular,
  Dross~\cite{DrossCKP12} provides theoretical foundations for the use
  of triggers:

  Intuitively, an existential formula $\exists x[k]A$ with a
  \Index{trigger} $[k]$ allows, as only instantiations of $x$, those
  which turn $k$ into a term that is already \emph{known}, as if the
  notation represented a formula $\exists x({\sf known}(k)\et A)$
  together with a proof-search policy that forces to prove the left
  branch ${\sf known}(k)$ before the branch on formula $A$ is
  explored. This strongly suggests a focussing approach in which the
  predicate ${\sf known}(\_)$ is positive, so that in the above case
  ${\sf known}(k)$ has to immediately be proved by an axiom. This
  should be formalised in our focussed calculi.

\item More generally look at the various extensions of \DPLLTh, in
  particular the systems with full first-order logic and/or equality
  as developed by
  \eg\cite{Baumgartner00,BaumgartnerT08,BaumgartnerT11}, and compare
  them to \LKThp\ with quantifiers and meta-variables. Note that the
  abstract setting of \LAF\ might be appropriate for equality itself:
  since we have abstracted away from connectives, it may be the case
  that sequent calculi with equality just fit as \LAF\ instances.
\end{itemize}





\def\homedir{\textasciitilde}

\ifdefined\url
\else
\def\url[#1]#2{\texttt{#2}}
\fi

\let\oldurl\url

\makeatletter
\ifdefined\href
\def\myurl{\@ifnextchar[\@myurlwith\@myurlwithout}
\long\def\@myurlwith[#1]#2{\mbox{\href{#2}{#1}}}
\long\def\@myurlwithout#1{\mbox{\href{#1}{#1}}}
\else
\def\myurl{\@ifnextchar[\@myurlwith\@myurlwithout}
\long\def\@myurlwith[#1]#2{\mbox{\oldurl[#1]{#2}}}
\long\def\@myurlwithout#1{\mbox{\oldurl[]{#1}}}
\fi
\makeatother

\def\url{\myurl}



\newcommand\emphasizeformat[1]{\emph{#1}}
\newcommand\authorformat[1]{{#1}}

\newcommand\monthdisplay[1]{\unskip}

\bibliographystyle{Common/good}
\addcontentsline{toc}{chapter}{Bibliography}
\bibliography{Common/abbrev-short,Common/Main,Common/crossrefs}

\newcommand{\etalchar}[1]{$^{#1}$}
\begin{thebibliography}{MMZ{\etalchar{+}}01}

\bibitem[ADH{\etalchar{+}}12]{AriolaDownenHerbelinNakataSaurin12}
\authorformat{Z.~M. Ariola, P.~Downen, H.~Herbelin, K.~Nakata, and A.~Saurin}.
\newblock Classical call-by-need sequent calculi: The unity of semantic
  artifacts.
\newblock In T.~Schrijvers and P.~Thiemann, editors, \emphasizeformat{Proc.{}
  of the 11th Int.{} Symp.{} Functional and Logic Programming (FLOPS'12)},
  volume 7294 of \emphasizeformat{LNCS}, pages 32--46. Springer-Verlag,
  \monthdisplay{May} 2012.%
\parfillskip=0pt\penalty7000\null\hfill

\bibitem[AF97]{AriolaF97}
\authorformat{Z.~M. Ariola and M.~Felleisen}.
\newblock The call-by-need lambda calculus.
\newblock \emphasizeformat{J. Funct. Programming}, 7(3):265--301, 1997.%
\parfillskip=0pt\penalty7000\null\hfill

\bibitem[AFG{\etalchar{+}}11]{ArmandFGKTW11}
\authorformat{M.~Armand, G.~Faure, B.~Gr{\'e}goire, C.~Keller, L.~Th{\'e}ry,
  and B.~Werner}.
\newblock A modular integration of {SAT/SMT} solvers to {Coq} through proof
  witnesses.
\newblock In \emphasizeformat{Proc.{} of the 1st Int.{} Conf.{} on Certified
  Programs and Proofs (CPP'11)}, volume 7086 of \emphasizeformat{LNCS}, pages
  135--150. Springer, \monthdisplay{December} 2011.%
\parfillskip=0pt\penalty7000\null\hfill

\bibitem[AH03]{HerbelinAriolaH03}
\authorformat{Z.~M. Ariola and H.~Herbelin}.
\newblock Minimal classical logic and control operators.
\newblock In J.~C.~M. Baeten, J.~K. Lenstra, J.~Parrow, and G.~J. Woeginger,
  editors, \emphasizeformat{Proc.{} of the 30th Intern.\ Col.\ on Automata,
  Languages and Programming (ICALP)}, volume 2719 of \emphasizeformat{LNCS},
  pages 871--885. Springer-Verlag, \monthdisplay{July} 2003.%
\parfillskip=0pt\penalty7000\null\hfill

\bibitem[AH08]{AriolaHerbelin07}
\authorformat{Z.~M. Ariola and H.~Herbelin}.
\newblock Control reduction theories: the benefit of structural substitution.
\newblock \emphasizeformat{J. Funct. Programming}, 18(3):373--419, 2008.%
\parfillskip=0pt\penalty7000\null\hfill

\bibitem[AHS09]{AriolaHerbelinSabry07}
\authorformat{Z.~M. Ariola, H.~Herbelin, and A.~Sabry}.
\newblock A type-theoretic foundation of delimited continuations.
\newblock \emphasizeformat{Higher-Order and Symbolic Computation},
  22(3):233--273, 2009.
\newblock online from 2007.%
\parfillskip=0pt\penalty7000\null\hfill

\bibitem[AHS11]{AriolaHS11}
\authorformat{Z.~M. Ariola, H.~Herbelin, and A.~Saurin}.
\newblock Classical call-by-need and duality.
\newblock In C.~L. Ong, editor, \emphasizeformat{Proc.{} of the 10th Int.{}
  Conf.{} on Typed Lambda Calculus and Applications (TLCA'11)}, volume 6690 of
  \emphasizeformat{LNCS}, pages 27--44. Springer-Verlag, \monthdisplay{June}
  2011.%
\parfillskip=0pt\penalty7000\null\hfill

\bibitem[And92]{andreoli92focusing}
\authorformat{J.~M. Andreoli}.
\newblock Logic programming with focusing proofs in linear logic.
\newblock \emphasizeformat{J. Logic Comput.}, 2(3):297--347, 1992.%
\parfillskip=0pt\penalty7000\null\hfill

\bibitem[AP89]{AndreoliP89}
\authorformat{J.-M. Andreoli and R.~Pareschi}.
\newblock Logic programming with sequent systems, a linear logic approach.
\newblock In P.~Schroeder-Heister, editor, \emphasizeformat{Proc.{} of the
  Int.{} Work.{} on Extensions of Logic Programming}, LNCS, pages 1--30.
  Springer-Verlag, \monthdisplay{December} 1989.%
\parfillskip=0pt\penalty7000\null\hfill

\bibitem[Bar84]{Bar84}
\authorformat{H.~P. Barendregt}.
\newblock \emphasizeformat{The {L}ambda-{C}alculus, its syntax and semantics}.
\newblock Studies in Logic and the Foundation of Mathematics. Elsevier, 1984.
\newblock Second edition.%
\parfillskip=0pt\penalty7000\null\hfill

\bibitem[Bar91]{Bar:intgts}
\authorformat{H.~P. Barendregt}.
\newblock Introduction to generalized type systems.
\newblock \emphasizeformat{J. Funct. Programming}, 1(2):125--154, 1991.%
\parfillskip=0pt\penalty7000\null\hfill

\bibitem[Bar92]{Barendregt:hlcs1992}
\authorformat{H.~P. Barendregt}.
\newblock Lambda calculi with types.
\newblock In S.~Abramsky, D.~M. Gabby, and T.~S.~E. Maibaum, editors,
  \emphasizeformat{Hand. Log. Comput. Sci.}, volume~2, chapter~2, pages
  117--309. Oxford University Press, 1992.%
\parfillskip=0pt\penalty7000\null\hfill

\bibitem[Bau00]{Baumgartner00}
\authorformat{P.~Baumgartner}.
\newblock {FDPLL} - a first order {Davis}-{Putnam}-{Longeman}-{Loveland}
  procedure.
\newblock In \emphasizeformat{Proc.{} of the 17th Int.{} Conf.{} on Automated
  Deduction (CADE'00)}, volume 1831 of \emphasizeformat{LNCS}, pages 200--219.
  Springer-Verlag, \monthdisplay{June} 2000.%
\parfillskip=0pt\penalty7000\null\hfill

\bibitem[BB96]{BBsymm}
\authorformat{F.~Barbanera and S.~Berardi}.
\newblock A symmetric lambda-calculus for classical program extraction.
\newblock \emphasizeformat{Inform. and Comput.}, 125(2):103--117, 1996.%
\parfillskip=0pt\penalty7000\null\hfill

\bibitem[BBP11]{BlanchetteBP11}
\authorformat{J.~C. Blanchette, S.~B{\"o}hme, and L.~C. Paulson}.
\newblock Extending {Sledgehammer} with {SMT} solvers.
\newblock In \emphasizeformat{Automated Deduction}, volume 6803 of
  \emphasizeformat{LNCS}, pages 116--130. Springer-Verlag, 2011.%
\parfillskip=0pt\penalty7000\null\hfill

\bibitem[BCP11]{BCP:CPP2011}
\authorformat{F.~Besson, P.-E. Cornilleau, and D.~Pichardie}.
\newblock Modular {SMT} proofs for fast reflexive checking inside {Coq}.
\newblock In J.-P. Jouannaud and Z.~Shao, editors, \emphasizeformat{Certified
  Programs and Proofs}, volume 7086 of \emphasizeformat{LNCS}, pages 151--166.
  Springer-Verlag, 2011.%
\parfillskip=0pt\penalty7000\null\hfill

\bibitem[BFM{\etalchar{+}}13]{Why3}
\authorformat{F.~Bobot, J.-C. Filli{\^a}tre, C.~March{\'e}, G.~Melquiond, and
  A.~Paskevich}.
\newblock {The Why3 platform 0.81}.
\newblock \monthdisplay{March} 2013.
\newblock Tutorial and Reference Manual.
\newblock Available at \url{http://hal.inria.fr/hal-00822856}%
\parfillskip=0pt\penalty7000\null\hfill

\bibitem[BG01]{BG01}
\authorformat{H.~Barendregt and H.~Geuvers}.
\newblock Proof-assistants using dependent type systems.
\newblock In Robinson and Voronkov~\cite{RobinsonV01}, pages 1149--1238.%
\parfillskip=0pt\penalty7000\null\hfill

\bibitem[BGL12]{bernadetlengrand12b}
\authorformat{A.~Bernadet and S.~Graham-Lengrand}.
\newblock A simple presentation of the effective topos.
\newblock Technical report, Laboratoire d'informatique de l'{\'E}cole
  {Polytechnique} - {CNRS}, France, \monthdisplay{April} 2012.
\newblock Available at \url{http://hal.archives-ouvertes.fr/hal-00844250}%
\parfillskip=0pt\penalty7000\null\hfill

\bibitem[BGL13]{bernadetlengrand13}
\authorformat{A.~Bernadet and S.~Graham-Lengrand}.
\newblock Non-idempotent intersection types and strong normalisation.
\newblock \emphasizeformat{Logic. Methods Comput. Science}, 9(4), 2013.%
\parfillskip=0pt\penalty7000\null\hfill

\bibitem[BL11a]{bernadetleng11}
\authorformat{A.~Bernadet and S.~Lengrand}.
\newblock Complexity of strongly normalising $\lambda$-terms via non-idempotent
  intersection types.
\newblock In M.~Hofmann, editor, \emphasizeformat{Proc.{} of the 14th Int.{}
  Conf.{} on Foundations of Software Science and Computation Structures
  (FOSSACS'11)}, volume 6604 of \emphasizeformat{LNCS}. Springer-Verlag,
  \monthdisplay{March} 2011.%
\parfillskip=0pt\penalty7000\null\hfill

\bibitem[BL11b]{bernadetleng11b}
\authorformat{A.~Bernadet and S.~Lengrand}.
\newblock Filter models: non-idempotent intersection types, orthogonality and
  polymorphism.
\newblock In M.~Bezem, editor, \emphasizeformat{Proc.{} of the 20th Annual
  Conf.{} of the European Association for Computer Science Logic (CSL'11)},
  LIPIcs. Schloss Dagstuhl LCI, \monthdisplay{September} 2011.%
\parfillskip=0pt\penalty7000\null\hfill

\bibitem[BL11c]{bernadetleng11blong}
\authorformat{A.~Bernadet and S.~Lengrand}.
\newblock Filter models: non-idempotent intersection types, orthogonality and
  polymorphism - long version.
\newblock Technical report, LIX, CNRS-INRIA-Ecole Polytechnique,
  \monthdisplay{June} 2011.
\newblock Available at \url{http://hal.archives-ouvertes.fr/hal-00600070/en/}%
\parfillskip=0pt\penalty7000\null\hfill

\bibitem[BMS10]{BaeldeMS10}
\authorformat{D.~Baelde, D.~Miller, and Z.~Snow}.
\newblock Focused inductive theorem proving.
\newblock In J.~Giesl and R.~H{\"{a}}hnle, editors, \emphasizeformat{Proc.{} of
  the 5th Int.{} Joint Conf.{} on Automated Reasoning (IJCAR'10)}, volume 6173
  of \emphasizeformat{LNCS}, pages 278--292. Springer-Verlag,
  \monthdisplay{July} 2010.%
\parfillskip=0pt\penalty7000\null\hfill

\bibitem[BT08]{BaumgartnerT08}
\authorformat{P.~Baumgartner and C.~Tinelli}.
\newblock The model evolution calculus as a first-order {DPLL} method.
\newblock \emphasizeformat{Artificial Intelligence}, 172(4-5):591--632, 2008.%
\parfillskip=0pt\penalty7000\null\hfill

\bibitem[BT11]{BaumgartnerT11}
\authorformat{P.~Baumgartner and C.~Tinelli}.
\newblock Model evolution with equality modulo built-in theories.
\newblock In N.~Bj{\o}rner and V.~Sofronie-Stokkermans, editors,
  \emphasizeformat{Proc.{} of the 23rd Int.{} Conf.{} on Automated Deduction
  (CADE'11)}, volume 6803 of \emphasizeformat{LNCS}, pages 85--100.
  Springer-Verlag, \monthdisplay{July} 2011.%
\parfillskip=0pt\penalty7000\null\hfill

\bibitem[CD78]{CD78:newtal}
\authorformat{M.~Coppo and M.~Dezani{-}Ciancaglini}.
\newblock A new type assignment for lambda-terms.
\newblock \emphasizeformat{Arch. Math. Log.}, 19:139--156, 1978.%
\parfillskip=0pt\penalty7000\null\hfill

\bibitem[CF58]{CurryFeys:1958}
\authorformat{H.~B. Curry and R.~Feys}.
\newblock \emphasizeformat{Combinatory Logic}, volume~I.
\newblock {N}orth-{H}olland, 1958.%
\parfillskip=0pt\penalty7000\null\hfill

\bibitem[CH00]{CurienHerbelinDuality99}
\authorformat{P.-L. Curien and H.~Herbelin}.
\newblock The duality of computation.
\newblock In \emphasizeformat{Proc.{} of the $5^{th}$ ACM SIGPLAN Int.{}
  Conf.{} on Functional Programming (ICFP'00)}, pages 233--243. ACM Press,
  2000.%
\parfillskip=0pt\penalty7000\null\hfill

\bibitem[Chu41]{Church}
\authorformat{A.~Church}.
\newblock \emphasizeformat{The Calculi of Lambda Conversion}.
\newblock Princeton University Press, 1941.%
\parfillskip=0pt\penalty7000\null\hfill

\bibitem[CMM10]{Munch11}
\authorformat{P.-L. Curien and G.~Munch-Maccagnoni}.
\newblock The duality of computation under focus.
\newblock In C.~S. Calude and V.~Sassone, editors, \emphasizeformat{Theoretical
  Computer Science}, volume 323 of \emphasizeformat{IFIP Advances in
  Information and Communication Technology}, pages 165--181. Springer-Verlag,
  2010.%
\parfillskip=0pt\penalty7000\null\hfill

\bibitem[Coq]{Coq}
{The Coq Proof Assistant}.
\newblock Available at \url{http://coq.inria.fr/}%
\parfillskip=0pt\penalty7000\null\hfill

\bibitem[Cro04]{Crolard04}
\authorformat{T.~Crolard}.
\newblock A formulae-as-types interpretation of subtractive logic.
\newblock \emphasizeformat{J. Logic Comput.}, 14(4):529--570, 2004.%
\parfillskip=0pt\penalty7000\null\hfill

\bibitem[CS09]{CockettS09}
\authorformat{J.~R.~B. Cockett and L.~Santocanale}.
\newblock On the word problem for {$\Sigma\Pi$}-categories, and the properties
  of two-way communication.
\newblock In Gr{\"a}del and Kahle~\cite{csl09}, pages 194--208.%
\parfillskip=0pt\penalty7000\null\hfill

\bibitem[Cur09]{tlca09}
P.~Curien, editor.
\newblock \emphasizeformat{Proc.{} of the 9th Int.{} Conf.{} on Typed Lambda
  Calculus and Applications (TLCA'09)}, volume 5608 of \emphasizeformat{LNCS}.
  Springer-Verlag, \monthdisplay{July} 2009.%
\parfillskip=0pt\penalty7000\null\hfill

\bibitem[dC05]{Carvalho05}
\authorformat{D.~de~Carvalho}.
\newblock Intersection types for light affine lambda calculus.
\newblock \emphasizeformat{ENTCS}, 136:133--152, 2005.%
\parfillskip=0pt\penalty7000\null\hfill

\bibitem[dC09]{Carvalho09corr}
\authorformat{D.~de~Carvalho}.
\newblock Execution time of lambda-terms via denotational semantics and
  intersection types.
\newblock \emphasizeformat{CoRR}, abs/0905.4251, 2009.%
\parfillskip=0pt\penalty7000\null\hfill

\bibitem[DCKP12]{DrossCKP12}
\authorformat{C.~Dross, S.~Conchon, J.~Kanig, and A.~Paskevich}.
\newblock Reasoning with triggers.
\newblock In P.~Fontaine and A.~Goel, editors, \emphasizeformat{10th Int.{}
  Work.{} on Satisfiability Modulo Theories, {SMT} 2012}, volume~20 of
  \emphasizeformat{EPiC Series}, pages 22--31. EasyChair, \monthdisplay{June}
  2012.%
\parfillskip=0pt\penalty7000\null\hfill

\bibitem[DF89]{danvy1989functional}
\authorformat{O.~Danvy and A.~Filinski}.
\newblock A functional abstraction of typed contexts.
\newblock Technical Report 89/12, DIKU, University of Cophenhagen, 1989.%
\parfillskip=0pt\penalty7000\null\hfill

\bibitem[DF90]{danvy1990abstracting}
\authorformat{O.~Danvy and A.~Filinski}.
\newblock Abstracting control.
\newblock In \emphasizeformat{Proc.{} of the 1990 {ACM} Conf.{} on LISP and
  functional programming}, pages 151--160. ACM Press, 1990.%
\parfillskip=0pt\penalty7000\null\hfill

\bibitem[DGHP99]{TableauxHandbook}
\authorformat{M.~D'Agostino, D.~M. Gabbay, R.~{H\"ahnle}, and J.~Posegga}.
\newblock \emphasizeformat{Handbook of Tableau Methods}.
\newblock Kluwer Academic Publishers, 1999.%
\parfillskip=0pt\penalty7000\null\hfill

\bibitem[DJS95]{DJS95}
\authorformat{V.~Danos, J.-B. Joinet, and H.~Schellinx}.
\newblock {\uppercase{\sf LKQ}} and {\uppercase{\sf lkt}}: sequent calculi for
  second order logic based upon dual linear decompositions of classical
  implication.
\newblock In J.-Y. Girard, Y.~Lafont, and L.~Regnier, editors,
  \emphasizeformat{Proc.{} of the Work.{} on Advances in Linear Logic}, volume
  222 of \emphasizeformat{London Math. Soc. Lecture Note Ser.}, pages 211--224.
  Cambridge University Press, 1995.%
\parfillskip=0pt\penalty7000\null\hfill

\bibitem[DJS97]{DJS97}
\authorformat{V.~Danos, J.-B. Joinet, and H.~Schellinx}.
\newblock A new deconstructive logic: Linear logic.
\newblock \emphasizeformat{J. of Symbolic Logic}, 62(3):755--807, 1997.%
\parfillskip=0pt\penalty7000\null\hfill

\bibitem[DK00]{DanosKrivine00}
\authorformat{V.~Danos and J.-L. Krivine}.
\newblock Disjunctive tautologies as synchronisation schemes.
\newblock In P.~Clote and H.~Schwichtenberg, editors, \emphasizeformat{Proc.{}
  of the 9th Annual Conf.{} of the European Association for Computer Science
  Logic (CSL'00)}, volume 1862 of \emphasizeformat{LNCS}, pages 292--301.
  Springer-Verlag, \monthdisplay{August} 2000.%
\parfillskip=0pt\penalty7000\null\hfill

\bibitem[DL07]{DL:JLC07}
\authorformat{R.~Dyckhoff and S.~Lengrand}.
\newblock Call-by-value $\lambda$-calculus and {LJQ}.
\newblock \emphasizeformat{J. Logic Comput.}, 17:1109--1134, 2007.%
\parfillskip=0pt\penalty7000\null\hfill

\bibitem[DLL62]{DavisLL62}
\authorformat{M.~Davis, G.~Logemann, and D.~W. Loveland}.
\newblock A machine program for theorem-proving.
\newblock \emphasizeformat{Communications of the ACM}, 5(7):394--397, 1962.%
\parfillskip=0pt\penalty7000\null\hfill

\bibitem[DP60]{DavisP60}
\authorformat{M.~Davis and H.~Putnam}.
\newblock A computing procedure for quantification theory.
\newblock \emphasizeformat{J. of the ACM Press}, 7(3):201--215, 1960.%
\parfillskip=0pt\penalty7000\null\hfill

\bibitem[DP04]{DosenPetric04book}
\authorformat{K.~Do\v{s}en and Z.~Petri\'c}.
\newblock \emphasizeformat{{Proof-theoretical Coherence}}.
\newblock {King's College Publications}, 2004.%
\parfillskip=0pt\penalty7000\null\hfill

\bibitem[Far13]{FarooquePhD}
\authorformat{M.~Farooque}.
\newblock \emphasizeformat{Automated reasoning techniques as proof-search in
  sequent calculus}.
\newblock PhD thesis, Ecole Polytechnique, 2013.%
\parfillskip=0pt\penalty7000\null\hfill

\bibitem[Fel87]{Felleisen:phd1987}
\authorformat{M.~Felleisen}.
\newblock \emphasizeformat{The Calculi of $\lambda$-v-CS Conversion: A
  Syntactic Theory of Control and State in Imperative Higher-Order Programming
  Languages}.
\newblock PhD thesis, Department of Computer Science, Indiana University,
  Bloomington, Indiana, 1987.%
\parfillskip=0pt\penalty7000\null\hfill

\bibitem[FGL13]{farooqueTR13}
\authorformat{M.~Farooque and S.~Graham-Lengrand}.
\newblock Sequent calculi with procedure calls.
\newblock Technical report, Laboratoire d'informatique de l'{\'E}cole
  {Polytechnique} - {CNRS}, {Parsifal} - {INRIA} Saclay, France,
  \monthdisplay{January} 2013.
\newblock Available at \url{http://hal.archives-ouvertes.fr/hal-00779199}%
\parfillskip=0pt\penalty7000\null\hfill

\bibitem[FGLM13]{farooque13}
\authorformat{M.~Farooque, S.~Graham-Lengrand, and A.~Mahboubi}.
\newblock A bisimulation between {DPLL(T)} and a proof-search strategy for the
  focused sequent calculus.
\newblock In A.~Momigliano, B.~Pientka, and R.~Pollack, editors,
  \emphasizeformat{Proc.{} of the 2013 Int.{} Work.{} on Logical Frameworks and
  Meta-Languages: Theory and Practice (LFMTP 2013)}. ACM Press,
  \monthdisplay{September} 2013.%
\parfillskip=0pt\penalty7000\null\hfill

\bibitem[Fil89]{Filinski:1989}
\authorformat{A.~Filinski}.
\newblock Declarative continuations and categorical duality.
\newblock Master's thesis, DIKU, Computer Science Department, University of
  Copenhagen, 1989.
\newblock DIKU Rapport 89/11.%
\parfillskip=0pt\penalty7000\null\hfill

\bibitem[Fis72]{Fischer:pacp1972}
\authorformat{M.~J. Fischer}.
\newblock Lambda calculus schemata.
\newblock In \emphasizeformat{Proc.{} of the {ACM} Conf.{} on {\it Proving
  Assertions about Programs}}, pages 104--109. SIGPLAN Notices, Vol.~7, No~1
  and SIGACT News, No~14, \monthdisplay{January} 1972.%
\parfillskip=0pt\penalty7000\null\hfill

\bibitem[FL11]{farooqueTR11}
\authorformat{M.~Farooque and S.~Lengrand}.
\newblock {A sequent calculus with procedure calls}.
\newblock Technical report, Laboratoire d'informatique de l'{\'E}cole
  {Polytechnique} - {CNRS}, {Parsifal} - {INRIA} Saclay, France,
  \monthdisplay{December} 2011.
\newblock Available at \url{http://hal.archives-ouvertes.fr/hal-00690577}%
\parfillskip=0pt\penalty7000\null\hfill

\bibitem[FLM12a]{farooqueTR12b}
\authorformat{M.~Farooque, S.~Lengrand, and A.~Mahboubi}.
\newblock {Simulating the DPLL(T) procedure in a sequent calculus with
  focusing}.
\newblock Technical report, Laboratoire d'informatique de l'{\'E}cole
  {Polytechnique} - {CNRS}, Microsoft Research - {INRIA Joint Centre},
  {Parsifal} \& {TypiCal} - {INRIA} Saclay, France, \monthdisplay{March} 2012.
\newblock Available at \url{http://hal.inria.fr/hal-00690392}%
\parfillskip=0pt\penalty7000\null\hfill

\bibitem[FLM12b]{farooqueTR12}
\authorformat{M.~Farooque, S.~Lengrand, and A.~Mahboubi}.
\newblock {Two simulations about DPLL(T)}.
\newblock Technical report, Laboratoire d'informatique de l'{\'E}cole
  {Polytechnique} - {CNRS}, Microsoft Research - {INRIA Joint Centre},
  {Parsifal} \& {TypiCal} - {INRIA} Saclay, France, \monthdisplay{March} 2012.
\newblock Available at \url{http://hal.archives-ouvertes.fr/hal-00690044}%
\parfillskip=0pt\penalty7000\null\hfill

\bibitem[FP06]{PymFu04}
\authorformat{C.~F{\"u}rmann and D.~Pym}.
\newblock Order-enriched categorical models of the classical sequent calculus.
\newblock \emphasizeformat{J. Pure Appl. Algebra}, 204(1):21--78, 2006.%
\parfillskip=0pt\penalty7000\null\hfill

\bibitem[FP13]{FilliatreP13}
\authorformat{J.-C. Filliâtre and A.~Paskevich}.
\newblock Why3 - where programs meet provers.
\newblock In M.~Felleisen and P.~Gardner, editors, \emphasizeformat{ESOP},
  volume 7792 of \emphasizeformat{LNCS}, pages 125--128. Springer-Verlag,
  2013.%
\parfillskip=0pt\penalty7000\null\hfill

\bibitem[FR94]{FleuryR94}
\authorformat{A.~Fleury and C.~Retor{\'{e}}}.
\newblock The mix rule.
\newblock \emphasizeformat{Math. Structures in Comput. Sci.}, 4(2):273--285,
  1994.%
\parfillskip=0pt\penalty7000\null\hfill

\bibitem[Fre79]{Frege:1879}
\authorformat{G.~Frege}.
\newblock \emphasizeformat{{Begriffsschrift, eine der arithmetischen
  nachgebildete Formelsprache des reinen Denkens}}.
\newblock Verlag von Louis Nebert, 1879.%
\parfillskip=0pt\penalty7000\null\hfill

\bibitem[Gen35]{Gentzen35}
\authorformat{G.~Gentzen}.
\newblock Investigations into logical deduction.
\newblock In \emphasizeformat{Gentzen collected works}, pages 68--131. Ed M. E.
  Szabo, North Holland, (1969), 1935.%
\parfillskip=0pt\penalty7000\null\hfill

\bibitem[GGL14]{DyckhoffJLC}
\authorformat{D.~Galmiche and S.~Graham-Lengrand}.
\newblock Special issue on computational logic in honour of {Roy} {Dyckhoff}.
\newblock \emphasizeformat{Journal of Logic and Computation}, 2014.%
\parfillskip=0pt\penalty7000\null\hfill

\bibitem[GH03]{GieseHaehnle03}
\authorformat{M.~Giese and R.~H{\"a}hnle}.
\newblock Tableaux + constraints.
\newblock In M.~C. Mayer and F.~Pirri, editors, \emphasizeformat{Proc.{} of the
  16th Int.{} Conf.{} on Automated Reasoning with Analytic Tableaux and Related
  Methods (Tableaux'03)}, volume 2796 of \emphasizeformat{LNCS}.
  Springer-Verlag, \monthdisplay{September} 2003.%
\parfillskip=0pt\penalty7000\null\hfill

\bibitem[Gie00]{Giese00}
\authorformat{M.~Giese}.
\newblock Proof search without backtracking using instance streams, position
  paper.
\newblock In P.~Baumgartner and H.~Zhang, editors, \emphasizeformat{3rd Int.\
  Workshop on First-Order Theorem Proving (FTP), St.~Andrews, Scotland, TR
  5/2000 Univ.\ of Koblenz}, pages 227--228, 2000.%
\parfillskip=0pt\penalty7000\null\hfill

\bibitem[Gir72]{Girard72}
\authorformat{J.-Y. Girard}.
\newblock \emphasizeformat{Interpr{\'e}tation fonctionelle et {\'e}limination
  des coupures de l'arithm{\'e}tique d'ordre sup{\'e}rieur}.
\newblock Th{\`e}se d'{\'e}tat, Universit\'e Paris 7, 1972.%
\parfillskip=0pt\penalty7000\null\hfill

\bibitem[Gir87]{girard-ll}
\authorformat{J.-Y. Girard}.
\newblock Linear logic.
\newblock \emphasizeformat{Theoret. Comput. Sci.}, 50(1):1--101, 1987.%
\parfillskip=0pt\penalty7000\null\hfill

\bibitem[Gir91]{Gir:newclc}
\authorformat{J.-Y. Girard}.
\newblock A new constructive logic: Classical logic.
\newblock \emphasizeformat{Math. Structures in Comput. Sci.}, 1(3):255--296,
  1991.%
\parfillskip=0pt\penalty7000\null\hfill

\bibitem[GK09]{csl09}
E.~Gr{\"a}del and R.~Kahle, editors.
\newblock \emphasizeformat{Proc.{} of the 18th Annual Conf.{} of the European
  Association for Computer Science Logic (CSL'09)}, volume 5771 of
  \emphasizeformat{LNCS}. Springer-Verlag, \monthdisplay{September} 2009.%
\parfillskip=0pt\penalty7000\null\hfill

\bibitem[GL08]{GabbayLengrand08}
\authorformat{M.~Gabbay and S.~Lengrand}.
\newblock The lambda-context calculus.
\newblock volume 196 of \emphasizeformat{ENTCS}, pages 19--35, 2008.
\newblock Revision from the Second Int.{} Work.{} on Logical Frameworks and
  Meta-Languages: Theory and Practice (LFMTP 2007) (was {\em A(nother) NEW
  Calculus of Contexts}).%
\parfillskip=0pt\penalty7000\null\hfill

\bibitem[GL09]{gablen08}
\authorformat{M.~Gabbay and S.~Lengrand}.
\newblock The $\lambda$-context calculus.
\newblock \emphasizeformat{Inform. and Comput.}, 207(12):1369--1400, 2009.%
\parfillskip=0pt\penalty7000\null\hfill

\bibitem[GL13]{GLPsyche13}
\authorformat{S.~Graham-Lengrand}.
\newblock {Psyche}: a proof-search engine based on sequent calculus with an
  {LCF}-style architecture.
\newblock In D.~Galmiche and D.~Larchey-Wendling, editors,
  \emphasizeformat{Proc.{} of the 22nd Int.{} Conf.{} on Automated Reasoning
  with Analytic Tableaux and Related Methods (Tableaux'13)}, volume 8123 of
  \emphasizeformat{LNCS}, pages 149--156. Springer-Verlag,
  \monthdisplay{September} 2013.%
\parfillskip=0pt\penalty7000\null\hfill

\bibitem[GL14]{LengrandHDRCoq}
\authorformat{S.~Graham-Lengrand}.
\newblock Polarities \& focussing: a journey from realisability to automated
  reasoning -- {Coq} proofs of {Part II}.
\newblock 2014.
\newblock Available at \url{http://www.lix.polytechnique.fr/\homedir
  lengrand/Work/HDR/}%
\parfillskip=0pt\penalty7000\null\hfill

\bibitem[GMW79]{GMW79}
\authorformat{M.~Gordon, R.~Milner, and C.~Wadsworth}.
\newblock \emphasizeformat{Edinburgh {LCF}: a mechanized logic of computation},
  volume~78 of \emphasizeformat{LNCS}.
\newblock Springer-Verlag, 1979.%
\parfillskip=0pt\penalty7000\null\hfill

\bibitem[Gri90]{Griffin:popl90}
\authorformat{T.~G. Griffin}.
\newblock A formulae-as-type notion of control.
\newblock In P.~Hudak, editor, \emphasizeformat{17th Annual ACM Symp.{} on
  Principles of Programming Languages (POPL'90)}, pages 47--58. ACM Press,
  \monthdisplay{January} 1990.%
\parfillskip=0pt\penalty7000\null\hfill

\bibitem[GTL89]{GTL:prot}
\authorformat{J.-Y. Girard, P.~Taylor, and Y.~Lafont}.
\newblock \emphasizeformat{Proofs and Types}, volume~7 of
  \emphasizeformat{Cambridge Tracts in Theoret. Comput. Sci.}
\newblock Cambridge University Press, 1989.%
\parfillskip=0pt\penalty7000\null\hfill

\bibitem[Her05]{HerbHDR}
\authorformat{H.~Herbelin}.
\newblock \emphasizeformat{C'est maintenant qu'on calcule: au coeur de la
  dualité}.
\newblock Th\`ese d'habilitation à diriger des recherches, Universit{\'e}
  Paris 11, 2005.%
\parfillskip=0pt\penalty7000\null\hfill

\bibitem[HG08]{HerbelinGhilezan08}
\authorformat{H.~Herbelin and S.~Ghilezan}.
\newblock An approach to call-by-name delimited continuations.
\newblock In Necula and Wadler~\cite{popl08}, pages 383--394.%
\parfillskip=0pt\penalty7000\null\hfill

\bibitem[Hil28]{hilbert28}
\authorformat{D.~Hilbert}.
\newblock {Die Grundlagen der Mathematik}.
\newblock \emphasizeformat{Abhandlungen aus dem Mathematischen Seminar der
  Universität Hamburg}, 6(1):65--85, 1928.%
\parfillskip=0pt\penalty7000\null\hfill

\bibitem[Hil31]{hilbert31}
\authorformat{D.~Hilbert}.
\newblock {Die Grundlegung der elementaren Zahlenlehre}.
\newblock \emphasizeformat{Mathematische Annalen}, 104:485--494, 1931.%
\parfillskip=0pt\penalty7000\null\hfill

\bibitem[How80]{How:fortnc}
\authorformat{W.~A. Howard}.
\newblock The formulae-as-types notion of construction.
\newblock In J.~P. Seldin and J.~R. Hindley, editors, \emphasizeformat{To
  {H}.~{B}.~{C}urry: Essays on Combinatory Logic, Lambda Calculus, and
  Formalism}, pages 479--490. Academic Press, 1980.
\newblock Reprint of a manuscript written 1969.%
\parfillskip=0pt\penalty7000\null\hfill

\bibitem[HS97]{HofmanStreicher97}
\authorformat{M.~Hofmann and T.~Streicher}.
\newblock Continuation models are universal for $\lambda\mu$-calculus.
\newblock In \emphasizeformat{Proc.{} of the 12th Annual {IEEE} Symp.{} on
  Logic in Computer Science}, pages 387--397. IEEE Computer Society Press,
  \monthdisplay{July} 1997.%
\parfillskip=0pt\penalty7000\null\hfill

\bibitem[Hyl82]{HylandJ:efft}
\authorformat{J.~Hyland}.
\newblock The effective topos.
\newblock In A.~Troelstra and D.~V. Dalen, editors, \emphasizeformat{The
  {L.E.J. Brouwer} Centenary Symposium}, pages 165--216. North Holland
  Publishing Company, 1982.%
\parfillskip=0pt\penalty7000\null\hfill

\bibitem[HZ09]{HerbelinZimmermann09}
\authorformat{H.~Herbelin and S.~Zimmermann}.
\newblock An operational account of call-by-value minimal and classical
  $\lambda$-calculus in ``natural deduction'' form.
\newblock In Curien~\cite{tlca09}, pages 142--156.%
\parfillskip=0pt\penalty7000\null\hfill

\bibitem[Isa]{Isabelle}
The {Isabelle} theorem prover.
\newblock Available at \url{http://isabelle.in.tum.de/}%
\parfillskip=0pt\penalty7000\null\hfill

\bibitem[JM94]{Jaffar1994503}
\authorformat{J.~Jaffar and M.~J. Maher}.
\newblock Constraint logic programming: a survey.
\newblock \emphasizeformat{J. Logic Programming}, 19–20, Supplement 1(0):503
  -- 581, 1994.
\newblock Special Issue: Ten Years of Logic Programming.%
\parfillskip=0pt\penalty7000\null\hfill

\bibitem[Joh36]{johansson36minimal}
\authorformat{I.~Johansson}.
\newblock {Der Minimalkalkül, ein reduzierter intuitionistischer Formalismus}.
\newblock \emphasizeformat{Compositio Math.}, 4:119--136, 1936.%
\parfillskip=0pt\penalty7000\null\hfill

\bibitem[KL08]{kikleng08}
\authorformat{K.~Kikuchi and S.~Lengrand}.
\newblock Strong normalisation of cut-elimination that simulates
  {\(\beta\)}-reduction.
\newblock In R.~Amadio, editor, \emphasizeformat{Proc.{} of the 11th Int.{}
  Conf.{} on Foundations of Software Science and Computation Structures
  (FOSSACS'08)}, volume 4962 of \emphasizeformat{LNCS}, pages 380--394.
  Springer-Verlag, \monthdisplay{March} 2008.%
\parfillskip=0pt\penalty7000\null\hfill

\bibitem[Kle45]{KleeneSC:intint}
\authorformat{S.~Kleene}.
\newblock On the interpretation of intuitionistic number theory.
\newblock \emphasizeformat{J. of Symbolic Logic}, 10:109--124, 1945.%
\parfillskip=0pt\penalty7000\null\hfill

\bibitem[Kri71]{KrivineSetTheory71}
\authorformat{J.-L. Krivine}.
\newblock \emphasizeformat{Introduction to axiomatic set theory}.
\newblock Dordrecht, Reidel, 1971.%
\parfillskip=0pt\penalty7000\null\hfill

\bibitem[Kri01]{Krivine01}
\authorformat{J.-L. Krivine}.
\newblock Typed lambda-calculus in classical {Zermelo}-{Fr{\ae}nkel} set
  theory.
\newblock \emphasizeformat{Arch. Math. Log.}, 40(3):189--205, 2001.%
\parfillskip=0pt\penalty7000\null\hfill

\bibitem[Lam07]{lamarche07}
\authorformat{F.~Lamarche}.
\newblock {Exploring the Gap between Linear and Classical Logic}.
\newblock \emphasizeformat{Theory and Applications of Categories},
  18(17):473--535, 2007.%
\parfillskip=0pt\penalty7000\null\hfill

\bibitem[Lau02]{phdlaurent}
\authorformat{O.~Laurent}.
\newblock \emphasizeformat{Etude de la polarisation en logique}.
\newblock Th\`ese de doctorat, {U}niversit\'e {A}ix-{M}arseille~{II}, 2002.%
\parfillskip=0pt\penalty7000\null\hfill

\bibitem[LC09]{Lescuyer09}
\authorformat{S.~Lescuyer and S.~Conchon}.
\newblock Improving {Coq} propositional reasoning using a lazy {CNF} conversion
  scheme.
\newblock In \emphasizeformat{Proc.{} of the 7th Int.{} Conf.{} on Frontiers of
  combining systems (FroCoS'09)}, pages 287--303. Springer-Verlag, 2009.%
\parfillskip=0pt\penalty7000\null\hfill

\bibitem[LDM11]{lengrand11lmcs}
\authorformat{S.~Lengrand, R.~Dyckhoff, and J.~McKinna}.
\newblock A focused sequent calculus framework for proof search in {Pure}
  {Type} {Systems}.
\newblock \emphasizeformat{Logic. Methods Comput. Science}, 7(1), 2011.%
\parfillskip=0pt\penalty7000\null\hfill

\bibitem[Len03]{lengrand03call-by-value}
\authorformat{S.~Lengrand}.
\newblock Call-by-value, call-by-name, and strong normalization for the
  classical sequent calculus.
\newblock volume~86(4) of \emphasizeformat{ENTCS}. Elsevier, 2003.
\newblock Revision from the 3rd Int.{} Work.{} on Reduction Strategies in
  Rewriting and Programming (WRS'03).%
\parfillskip=0pt\penalty7000\null\hfill

\bibitem[Len06]{LengrandPhD}
\authorformat{S.~Lengrand}.
\newblock \emphasizeformat{Normalisation \& Equivalence in Proof Theory \& Type
  Theory}.
\newblock PhD thesis, Universit{\'e} {Paris} 7 \& University of {St} {Andrews},
  2006.%
\parfillskip=0pt\penalty7000\null\hfill

\bibitem[Len08]{lengrand07tr}
\authorformat{S.~Lengrand}.
\newblock Termination of lambda-calculus with the extra call-by-value rule
  known as assoc.
\newblock Technical report, LIX, CNRS-INRIA-Ecole Polytechnique,
  \monthdisplay{January} 2008.
\newblock Available at \url{http://hal.inria.fr/inria-00292029}%
\parfillskip=0pt\penalty7000\null\hfill

\bibitem[LM08]{LM:APAL07}
\authorformat{S.~Lengrand and A.~Miquel}.
\newblock Classical {$F_\omega$}, orthogonality and symmetric candidates.
\newblock \emphasizeformat{Ann. Pure Appl. Logic}, 153:3--20, 2008.%
\parfillskip=0pt\penalty7000\null\hfill

\bibitem[LM09]{liang09tcs}
\authorformat{C.~Liang and D.~Miller}.
\newblock Focusing and polarization in linear, intuitionistic, and classical
  logics.
\newblock \emphasizeformat{Theoret. Comput. Sci.}, 410(46):4747--4768, 2009.%
\parfillskip=0pt\penalty7000\null\hfill

\bibitem[LM11]{liang11apal}
\authorformat{C.~Liang and D.~Miller}.
\newblock A focused approach to combining logics.
\newblock \emphasizeformat{Ann. Pure Appl. Logic}, 162(9):679--697, 2011.%
\parfillskip=0pt\penalty7000\null\hfill

\bibitem[LQdF05]{LaurentQF05}
\authorformat{O.~Laurent, M.~Quatrini, and L.~T. de~Falco}.
\newblock Polarized and focalized linear and classical proofs.
\newblock \emphasizeformat{Ann. Pure Appl. Logic}, 134(2-3):217--264, 2005.%
\parfillskip=0pt\penalty7000\null\hfill

\bibitem[LS86]{LambekJ:inthoc}
\authorformat{J.~Lambek and P.~J. Scott}.
\newblock \emphasizeformat{Introduction to Higher Order Categorical Logic}.
\newblock Cambridge University Press, 1986.%
\parfillskip=0pt\penalty7000\null\hfill

\bibitem[LS05]{LamStr05LICS}
\authorformat{F.~Lamarche and L.~Stra{\ss}burger}.
\newblock Constructing free boolean categories.
\newblock In Panangaden~\cite{lics05}, pages 209--218.%
\parfillskip=0pt\penalty7000\null\hfill

\bibitem[Miq09]{MiquelTLCA09}
\authorformat{A.~Miquel}.
\newblock Relating classical realizability and negative translation for
  existential witness extraction.
\newblock In Curien~\cite{tlca09}, pages 188--202.%
\parfillskip=0pt\penalty7000\null\hfill

\bibitem[Miq11]{MiquelLMCS11}
\authorformat{A.~Miquel}.
\newblock Existential witness extraction in classical realizability and via a
  negative translation.
\newblock \emphasizeformat{Logic. Methods Comput. Science}, 7(2), 2011.%
\parfillskip=0pt\penalty7000\null\hfill

\bibitem[ML82]{ML82}
\authorformat{P.~Martin-L{\"{o}}f}.
\newblock Constructive mathematics and computer programming.
\newblock In \emphasizeformat{Proc.{} of the Sixth Int.{} Congress for Logic,
  Methodology, and Philosophy of Science}, pages 153--175. North-Holland,
  1982.%
\parfillskip=0pt\penalty7000\null\hfill

\bibitem[ML84]{ML84}
\authorformat{P.~Martin-L{\"o}f}.
\newblock \emphasizeformat{Intuitionistic Type Theory}.
\newblock Number~1 in Studies in Proof Theory, Lecture Notes. Bibliopolis,
  1984.%
\parfillskip=0pt\penalty7000\null\hfill

\bibitem[MM09]{MunchCSL09}
\authorformat{G.~Munch-Maccagnoni}.
\newblock Focalisation and classical realisability.
\newblock In Gr{\"a}del and Kahle~\cite{csl09}, pages 409--423.%
\parfillskip=0pt\penalty7000\null\hfill

\bibitem[MM13]{MunchPhD}
\authorformat{G.~Munch-Maccagnoni}.
\newblock \emphasizeformat{Syntax and Models of a Non-Associative Composition
  of Programs and Proofs}.
\newblock PhD thesis, Université Paris Diderot – Paris 7, 2013.%
\parfillskip=0pt\penalty7000\null\hfill

\bibitem[MMZ{\etalchar{+}}01]{MoskewiczMZZM01}
\authorformat{M.~W. Moskewicz, C.~F. Madigan, Y.~Zhao, L.~Zhang, and S.~Malik}.
\newblock Chaff: Engineering an efficient {SAT} solver.
\newblock In \emphasizeformat{DAC}, pages 530--535. ACM Press, 2001.%
\parfillskip=0pt\penalty7000\null\hfill

\bibitem[MNPS91]{Miller91apal}
\authorformat{D.~Miller, G.~Nadathur, F.~Pfenning, and A.~Scedrov}.
\newblock Uniform proofs as a foundation for logic programming.
\newblock \emphasizeformat{Ann. Pure Appl. Logic}, 51:125--157, 1991.%
\parfillskip=0pt\penalty7000\null\hfill

\bibitem[Mog89]{Moggi:lics1989}
\authorformat{E.~Moggi}.
\newblock Computational lambda-calculus and monads.
\newblock In \emphasizeformat{Proc.{} of the 4th Annual {IEEE} Symp.{} on Logic
  in Computer Science}, pages 14--23. IEEE Computer Society Press,
  \monthdisplay{June} 1989.%
\parfillskip=0pt\penalty7000\null\hfill

\bibitem[MP08]{McLaughlinP08}
\authorformat{S.~McLaughlin and F.~Pfenning}.
\newblock Imogen: Focusing the polarized inverse method for intuitionistic
  propositional logic.
\newblock In I.~Cervesato, H.~Veith, and A.~Voronkov, editors,
  \emphasizeformat{Proc.{} of the the 15th Int.{} Conf.{} on Logic for
  Programming Artificial Intelligence and Reasoning (LPAR'08)}, volume 5330 of
  \emphasizeformat{LNCS}, pages 174--181. Springer-Verlag,
  \monthdisplay{November} 2008.%
\parfillskip=0pt\penalty7000\null\hfill

\bibitem[MV05]{mellies05recursive}
\authorformat{P.-A. {Melli\`es} and J.~Vouillon}.
\newblock Recursive polymorphic types and parametricity in an operational
  framework.
\newblock In Panangaden~\cite{lics05}, pages 82--91.%
\parfillskip=0pt\penalty7000\null\hfill

\bibitem[Nig09]{nigam09phd}
\authorformat{V.~Nigam}.
\newblock \emphasizeformat{Exploiting non-canonicity in the sequent calculus}.
\newblock PhD thesis, Ecole Polytechnique, 2009.%
\parfillskip=0pt\penalty7000\null\hfill

\bibitem[NM10]{nigam10jar}
\authorformat{V.~Nigam and D.~Miller}.
\newblock A framework for proof systems.
\newblock \emphasizeformat{J. of Automated Reasoning}, 45(2):157--188, 2010.%
\parfillskip=0pt\penalty7000\null\hfill

\bibitem[NOT06]{Nieuwenhuis06}
\authorformat{R.~Nieuwenhuis, A.~Oliveras, and C.~Tinelli}.
\newblock Solving {SAT} and {SAT Modulo Theories}: From an abstract
  {Davis}--{Putnam}--{Logemann}--{Loveland} procedure to {DPLL({\it T})}.
\newblock \emphasizeformat{J. of the ACM Press}, 53(6):937--977, 2006.%
\parfillskip=0pt\penalty7000\null\hfill

\bibitem[NPW02]{Nipkow-Paulson-Wenzel:2002}
\authorformat{T.~Nipkow, L.~C. Paulson, and M.~Wenzel}.
\newblock \emphasizeformat{Isabelle/HOL --- A Proof Assistant for Higher-Order
  Logic}, volume 2283 of \emphasizeformat{LNCS}.
\newblock Springer-Verlag, 2002.%
\parfillskip=0pt\penalty7000\null\hfill

\bibitem[NW08]{popl08}
G.~C. Necula and P.~Wadler, editors.
\newblock \emphasizeformat{Proc.{} of the 35th Annual ACM Symp.{} on Principles
  of Programming Languages (POPL'08)}. ACM Press, \monthdisplay{January} 2008.%
\parfillskip=0pt\penalty7000\null\hfill

\bibitem[Pan05]{lics05}
P.~Panangaden, editor.
\newblock \emphasizeformat{Proc.{} of the 20th Annual {IEEE} Symp.{} on Logic
  in Computer Science}. IEEE Computer Society Press, \monthdisplay{June} 2005.%
\parfillskip=0pt\penalty7000\null\hfill

\bibitem[Par92]{Parigot92}
\authorformat{M.~Parigot}.
\newblock $\lambda\mu$-calculus: An algorithmic interpretation of classical
  natural deduction.
\newblock In A.~Voronkov, editor, \emphasizeformat{Proc.{} of the Int.{}
  Conf.{} on Logic Programming and Automated Reasoning (LPAR'92)}, volume 624
  of \emphasizeformat{LNCS}, pages 190--201. Springer-Verlag,
  \monthdisplay{July} 1992.%
\parfillskip=0pt\penalty7000\null\hfill

\bibitem[Par97]{Parigot97}
\authorformat{M.~Parigot}.
\newblock Proofs of strong normalisation for second order classical natural
  deduction.
\newblock \emphasizeformat{J. of Symbolic Logic}, 62(4):1461--1479, 1997.%
\parfillskip=0pt\penalty7000\null\hfill

\bibitem[PB12]{sledgehammer10}
\authorformat{L.~C. Paulson and J.~C. Blanchette}.
\newblock Three years of experience with {Sledgehammer}, a practical link
  between automatic and interactive theorem provers.
\newblock In G.~Sutcliffe, S.~Schulz, and E.~Ternovska, editors,
  \emphasizeformat{IWIL 2010}, volume~2 of \emphasizeformat{EPiC Series}, pages
  1--11. EasyChair, 2012.%
\parfillskip=0pt\penalty7000\null\hfill

\bibitem[Pit03]{PittsAM:nomlfo-jv}
\authorformat{A.~M. Pitts}.
\newblock Nominal logic, a first order theory of names and binding.
\newblock \emphasizeformat{Inform. and Control}, 186:165--193, 2003.%
\parfillskip=0pt\penalty7000\null\hfill

\bibitem[Plo75]{Plotkin75}
\authorformat{G.~D. Plotkin}.
\newblock Call-by-name, call-by-value and the lambda-calculus.
\newblock \emphasizeformat{Theoret. Comput. Sci.}, 1:125--159, 1975.%
\parfillskip=0pt\penalty7000\null\hfill

\bibitem[Pol04]{Polo04}
\authorformat{E.~Polonovski}.
\newblock Strong normalization of {$\lambda\mu\tilde{\mu}$}-calculus with
  explicit substitutions.
\newblock In I.~Walukiewicz, editor, \emphasizeformat{Proc.{} of the 7th Int.{}
  Conf.{} on Foundations of Software Science and Computation Structures
  (FOSSACS'04)}, volume 2987 of \emphasizeformat{LNCS}, pages 423--437.
  Springer-Verlag, \monthdisplay{March} 2004.%
\parfillskip=0pt\penalty7000\null\hfill

\bibitem[PSI]{PSI}
The {PSI} project.
\newblock 2009-2013. \url{http://www.lix.polytechnique.fr/\homedir
  lengrand/PSI}.%
\parfillskip=0pt\penalty7000\null\hfill

\bibitem[Psy]{Psyche}
Psyche: the {Proof-Search factorY for Collaborative HEuristics}.
\newblock Available at \url{http://www.lix.polytechnique.fr/\homedir
  lengrand/Psyche}%
\parfillskip=0pt\penalty7000\null\hfill

\bibitem[Rea00]{Read00}
\authorformat{S.~Read}.
\newblock Harmony and autonomy in classical logic.
\newblock \emphasizeformat{J. of Philosophical Logic}, 29(2):123--154, 2000.%
\parfillskip=0pt\penalty7000\null\hfill

\bibitem[Rea10]{Read10}
\authorformat{S.~Read}.
\newblock General-elimination harmony and the meaning of the logical constants.
\newblock \emphasizeformat{J. of Philosophical Logic}, 39(5):557--576, 2010.%
\parfillskip=0pt\penalty7000\null\hfill

\bibitem[Rey72]{Reynolds72}
\authorformat{J.~C. Reynolds}.
\newblock Definitional interpreters for higher-order programming languages.
\newblock In \emphasizeformat{Proc.{} of the {ACM annual} Conf.{} ~}, pages
  717--740, 1972.%
\parfillskip=0pt\penalty7000\null\hfill

\bibitem[Roc05]{Rocheteau05}
\authorformat{J.~Rocheteau}.
\newblock lambda-{\(\mathrm{\mu}\)}-calculus and duality: Call-by-name and
  call-by-value.
\newblock In J.~Giesl, editor, \emphasizeformat{Proc.{} of the 16th Int.{}
  Conf.{} on Rewriting Techniques and Applications (RTA'05)}, volume 3467 of
  \emphasizeformat{LNCS}, pages 204--218. Springer-Verlag, \monthdisplay{April}
  2005.%
\parfillskip=0pt\penalty7000\null\hfill

\bibitem[RV01]{RobinsonV01}
J.~A. Robinson and A.~Voronkov, editors.
\newblock \emphasizeformat{Handbook of Automated Reasoning (in 2 volumes)}.
\newblock Elsevier and The MIT Press, 2001.%
\parfillskip=0pt\penalty7000\null\hfill

\bibitem[Sau05]{Saurin05}
\authorformat{A.~Saurin}.
\newblock Separation with streams in the lambda{\(\mathrm{\mu}\)}-calculus.
\newblock In Panangaden~\cite{lics05}, pages 356--365.%
\parfillskip=0pt\penalty7000\null\hfill

\bibitem[Sau08]{Saurin08}
\authorformat{A.~Saurin}.
\newblock On the relations between the syntactic theories of lambda-mu-calculi.
\newblock In \emphasizeformat{Proc.{} of the 17th Annual Conf.{} of the
  European Association for Computer Science Logic (CSL'08)}, volume 5213 of
  \emphasizeformat{LNCS}, pages 154--168. Springer-Verlag, 2008.%
\parfillskip=0pt\penalty7000\null\hfill

\bibitem[Sau10a]{Saurin10c}
\authorformat{A.~Saurin}.
\newblock A hierarchy for delimited continuations in call-by-name.
\newblock In C.~L. Ong, editor, \emphasizeformat{Proc.{} of the 13th Int.{}
  Conf.{} on Foundations of Software Science and Computation Structures
  (FOSSACS'10)}, volume 6014 of \emphasizeformat{LNCS}, pages 374--388.
  Springer-Verlag, \monthdisplay{March} 2010.%
\parfillskip=0pt\penalty7000\null\hfill

\bibitem[Sau10b]{Saurin10b}
\authorformat{A.~Saurin}.
\newblock Standardization and b{\"{o}}hm trees for
  lambda\emph{{\(\mathrm{\mu}\)}}-calculus.
\newblock In M.~Blume, N.~Kobayashi, and G.~Vidal, editors,
  \emphasizeformat{Proc.{} of the 10th Int.{} Symp.{} Functional and Logic
  Programming (FLOPS'10)}, volume 6009 of \emphasizeformat{LNCS}, pages
  134--149. Springer-Verlag, \monthdisplay{April} 2010.%
\parfillskip=0pt\penalty7000\null\hfill

\bibitem[Sau10c]{Saurin10}
\authorformat{A.~Saurin}.
\newblock Typing streams in the lambda{\(\mathrm{\mu}\)}-calculus.
\newblock \emphasizeformat{ACM Transactions on Computational Logic}, 11(4),
  2010.%
\parfillskip=0pt\penalty7000\null\hfill

\bibitem[Sau12]{Saurin12}
\authorformat{A.~Saurin}.
\newblock B{\"{o}}hm theorem and b{\"{o}}hm trees for the
  {\(\lambda\)}{\(\mu\)}-calculus.
\newblock \emphasizeformat{Theoret. Comput. Sci.}, 435:106--138, 2012.%
\parfillskip=0pt\penalty7000\null\hfill

\bibitem[Sch50]{schuette50}
\authorformat{K.~Sch\"utte}.
\newblock {Beweistheoretische Erfassung der unendlichen Induktion in der
  Zahlentheorie}.
\newblock \emphasizeformat{Mathematische Annalen}, 122:369--389, 1950.%
\parfillskip=0pt\penalty7000\null\hfill

\bibitem[Sel01]{SelingerControlCat99}
\authorformat{P.~Selinger}.
\newblock Control categories and duality: on the categorical semantics of the
  $\lambda\mu$-calculus.
\newblock \emphasizeformat{Math. Structures in Comput. Sci.}, 11(2):207--260,
  2001.%
\parfillskip=0pt\penalty7000\null\hfill

\bibitem[SR98]{StreicherReus98}
\authorformat{T.~Streicher and B.~Reus}.
\newblock Classical logic, continuation semantics and abstract machines.
\newblock \emphasizeformat{J. Funct. Programming}, 8(6):543--572, 1998.%
\parfillskip=0pt\penalty7000\null\hfill

\bibitem[Str11]{StrassburgerHDR}
\authorformat{L.~Stra{\ss}burger}.
\newblock \emphasizeformat{Towards a Theory of Proofs of Classical Logic}.
\newblock Habilitation thesis, Universit\'e {Denis} {Diderot} -- {Paris 7},
  2011.%
\parfillskip=0pt\penalty7000\null\hfill

\bibitem[SU06]{sor:CH}
\authorformat{M.~H.~B. S{\o}rensen and P.~Urzyczyn}.
\newblock \emphasizeformat{Lectures on the {C}urry-{H}oward Isomorphism}.
\newblock Studies in Logic and the Foundations of Mathematics. Elsevier, 2006.%
\parfillskip=0pt\penalty7000\null\hfill

\bibitem[SW00]{Strachey:2000}
\authorformat{C.~Strachey and C.~P. Wadsworth}.
\newblock Continuations: A mathematical semantics for handling fulljumps.
\newblock \emphasizeformat{Higher-Order and Symbolic Computation}, 13:135--152,
  2000.%
\parfillskip=0pt\penalty7000\null\hfill

\bibitem[Tai67]{tait67}
\authorformat{W.~W. Tait}.
\newblock Intensional interpretations of functionals of finite type {I}.
\newblock \emphasizeformat{J. of Symbolic Logic}, 32:198--212, 1967.%
\parfillskip=0pt\penalty7000\null\hfill

\bibitem[Tai75]{tait75}
\authorformat{W.~W. Tait}.
\newblock A realizability interpretation of the theory of species.
\newblock In \emphasizeformat{Logic Colloquium}, volume 453 of
  \emphasizeformat{LNM}, pages 240--251. Springer-Verlag, 1975.%
\parfillskip=0pt\penalty7000\null\hfill

\bibitem[Ten78]{TennantNatLogic}
\authorformat{N.~Tennant}.
\newblock \emphasizeformat{Natural logic}.
\newblock Edinburgh University Press, 1978.%
\parfillskip=0pt\penalty7000\null\hfill

\bibitem[Ter03]{Terese03}
\authorformat{Terese}.
\newblock \emphasizeformat{Term Rewriting Systems}, volume~55 of
  \emphasizeformat{Cambridge Tracts in Theoret. Comput. Sci.}
\newblock Cambridge University Press, 2003.%
\parfillskip=0pt\penalty7000\null\hfill

\bibitem[Tin07]{Tinelli07}
\authorformat{C.~Tinelli}.
\newblock An abstract framework for satisfiability modulo theories.
\newblock In N.~Olivetti, editor, \emphasizeformat{Proc.{} of the 16th Int.{}
  Conf.{} on Automated Reasoning with Analytic Tableaux and Related Methods
  (Tableaux'07)}, volume 4548 of \emphasizeformat{LNCS}. Springer-Verlag,
  \monthdisplay{July} 2007.
\newblock Invited talk, available at
  \url{http://ftp.cs.uiowa.edu/pub/tinelli/talks/TABLEAUX-07.pdf}.%
\parfillskip=0pt\penalty7000\null\hfill

\bibitem[TS00]{TS}
\authorformat{A.~S. Troelstra and H.~Schwichtenberg}.
\newblock \emphasizeformat{Basic Proof Theory}.
\newblock Cambridge University Press, 2000.%
\parfillskip=0pt\penalty7000\null\hfill

\bibitem[Twe]{Twelf}
{The Twelf Project}.
\newblock Available at \url{http://twelf.org}%
\parfillskip=0pt\penalty7000\null\hfill

\bibitem[Urb00]{UrbThes}
\authorformat{C.~Urban}.
\newblock \emphasizeformat{Classical Logic and Computation}.
\newblock PhD thesis, University of Cambridge, 2000.%
\parfillskip=0pt\penalty7000\null\hfill

\bibitem[VO02]{vanOosten02Realiz}
\authorformat{J.~VAN~OOSTEN}.
\newblock Realizability: a historical essay.
\newblock \emphasizeformat{Math. Structures in Comput. Sci.}, 12:239--263,
  2002.%
\parfillskip=0pt\penalty7000\null\hfill

\bibitem[Wad03]{Wadlerdual}
\authorformat{P.~Wadler}.
\newblock Call-by-value is dual to call-by-name.
\newblock In \emphasizeformat{Proc.{} of the 8th ACM SIGPLAN Int.{} Conf.{} on
  Functional programming (ICFP'03)}, volume~38(9), pages 189--201. ACM Press,
  \monthdisplay{September} 2003.%
\parfillskip=0pt\penalty7000\null\hfill

\bibitem[Web11]{weber11smt}
\authorformat{T.~Weber}.
\newblock {SMT} solvers: New oracles for the {HOL} theorem prover.
\newblock \emphasizeformat{International Journal on Software Tools for
  Technology Transfer (STTT)}, 13(5):419--429, 2011.%
\parfillskip=0pt\penalty7000\null\hfill

\bibitem[Zei08a]{ZeilbergerPOPL08}
\authorformat{N.~Zeilberger}.
\newblock Focusing and higher-order abstract syntax.
\newblock In Necula and Wadler~\cite{popl08}, pages 359--369.%
\parfillskip=0pt\penalty7000\null\hfill

\bibitem[Zei08b]{Zeilberger08}
\authorformat{N.~Zeilberger}.
\newblock On the unity of duality.
\newblock \emphasizeformat{Ann. Pure Appl. Logic}, 153(1-3):66--96, 2008.%
\parfillskip=0pt\penalty7000\null\hfill

\bibitem[Zei09]{ZeilbergerPhD}
\authorformat{N.~Zeilberger}.
\newblock \emphasizeformat{The Logical Basis of Evaluation Order and
  Pattern-Matching}.
\newblock PhD thesis, Carnegie Mellon University, 2009.%
\parfillskip=0pt\penalty7000\null\hfill

\bibitem[Zei10]{Zeilberger10}
\authorformat{N.~Zeilberger}.
\newblock Polarity and the logic of delimited continuations.
\newblock In J.-P. Jouannaud, editor, \emphasizeformat{Proc.{} of the 25th
  Annual {IEEE} Symp.{} on Logic in Computer Science}, pages 219--227. IEEE
  Computer Society Press, \monthdisplay{July} 2010.%
\parfillskip=0pt\penalty7000\null\hfill

\end{thebibliography}

\newpage
\addcontentsline{toc}{chapter}{Index}
\printindex
\listoffigures 

\appendix
\gettoappendix {partI}
\gettoappendix {partIproof}

\end{document}